\documentclass[useAMS,usenatbib,usegraphicx]{mn2e}

\title[The
$L_{CO(3-2)}$--$L_{FIR}$ correlation in the SINGS sample]{The 
JCMT Nearby Galaxies Legacy Survey VIII. 
CO data and the $L_{CO(3-2)}$--$L_{FIR}$ correlation in the SINGS sample}
\author[C. D. Wilson et al.]
    {C. D. Wilson$^1$, B. E. Warren$^{1,2}$, 
F. P. Israel$^3$, 
S. Serjeant$^4$, 
D. Attewell$^1$, 
G. J. Bendo$^{5}$, 
\newauthor   
H. M. Butner$^{6}$, 
P. Chanial$^{7}$, 
D. L. Clements$^8$, 
J. Golding$^1$,
V. Heesen$^{9}$ 
J. Irwin$^{10}$, 
J. Leech$^{11}$, 
\newauthor   
H. E. Matthews$^{12}$, 
    S. M\"uhle$^{13}$, 
A. M. J. Mortier$^{14}$,
G. Petitpas$^{15}$, 
J. R. S\'anchez-Gallego$^{16,17}$, 
\newauthor   
E. Sinukoff$^{1}$,
K. Shorten$^{1}$,
      B. K. Tan$^{11}$, 
R. P. J. Tilanus$^{18,19}$,
    A. Usero$^{20}$, 
M. Vaccari$^{21,43}$,  
\newauthor 
T. Wiegert$^{22}$,  
M. Zhu$^{23}$, 
D. M. Alexander$^{24}$,
P. Alexander$^{25,26}$,
M. Azimlu$^{27,15}$,
P. Barmby$^{27}$,
\newauthor
C. Borys$^{28}$,
R. Brar$^{10}$,
C. Bridge$^{28}$,
E. Brinks$^{9}$, 
S. Brooks$^1$,
K. Coppin$^{29}$,
S. C\^ot\'e$^{30}$, 
P. C\^ot\'e$^{30}$, 
\newauthor   
S. Courteau$^{10}$,
J. Davies$^{31}$,
 S. Eales$^{31}$,
M. Fich$^{32}$,
 M. Hudson$^{32}$, 
D. H. Hughes$^{33}$,
\newauthor   
R. J. Ivison$^{34,35}$,
J. H. Knapen$^{16,17}$,
M. Page$^8$,
T. J. Parkin$^1$,
A. Pope$^{36}$,
 D. Rigopoulou$^{11,37}$,
\newauthor   
E. Rosolowsky$^{38}$,
E. R. Seaquist$^{39}$,  
K. Spekkens$^{40}$,
N. Tanvir$^{9}$, 
J. M. van der Hulst$^{41}$, 
\newauthor   
P. van der Werf$^{3}$, 
C. Vlahakis$^{42}$,
T. M. Webb$^{29}$,
B. Weferling$^{18}$,
G. J. White$^{4,37}$
    \\
    $^1$Department of Physics \& Astronomy, McMaster University, Hamilton, 
        Ontario L8S 4M1, Canada\\
$^2$ International Centre for Radio Astronomy Research, M468, University of
Western Australia, 35 Stirling Hwy, Crawley, WA 6009, Australia \\
    $^3$Sterrewacht Leiden, Leiden University, PO Box 9513, 2300 RA Leiden,
        The Netherlands\\
    $^4$Department of Physics \& Astronomy, The Open University, 
        Milton Keynes MK7 6AA, United Kingdom\\
$^5$ UK ALMA Regional Centre Node, Jodrell Bank Centre for Astrophysics, School
of Physics and Astronomy, University of Manchester, Oxford Road, \\
Manchester M13 9PL, United Kingdom \\
    $^{6}$Department of Physics and Astronomy, James Madison
University, MSC 4502 - 901 Carrier Drive, Harrisonburg, VA 22807, U.S.A. \\ 
$^7$ Laboratoire AIM-Paris-Saclay, CEA/DSM/Irfu - CNRS - Universit\'e
Paris Diderot, CE-Saclay, pt courrier 131, F-91191 Gif-sur-Yvette,
France \\
    $^{8}$Astrophysics Group, Imperial College, Blackett Laboratory,
        Prince Consort Road, London SW7 2AZ, United Kingdom\\
    $^{9}$Centre for Astrophysics Research, University of Hertfordshire, 
        College Lane, Hatfield AL10 9AB, United Kingdom\\
    $^{10}$Department of Physics, Engineering Physics and Astronomy,
        Queen's University, Kingston, Ontario K7L 3N6, Canada\\
    $^{11}$Astrophysics, Oxford University, Keble Road, Oxford OX1
        3RH, United Kingdom\\
    $^{12}$National Research Council Canada, Herzberg Institute of Astrophysics,
        DRAO, P.O. Box 248, White Lake Road, Penticton, B.C.
        V2A 69J, Canada\\
    $^{13}$Joint Institute for VLBI in Europe, Postbus 2, 7990 AA Dwingeloo, 
        The Netherlands\\
$^{14}$Scottish Universities Physics
    Alliance,  Institute for
Astronomy, University of Edinburgh, Royal Observatory, Blackford Hill,\\
Edinburgh, EH9 3HJ, UK\\
    $^{15}$Harvard-Smithsonian Center for Astrophysics, 60 Garden St.,
        Cambridge, MA 02138, USA\\
    $^{16}$Instituto de Astrof\'isica de Canarias,  E-38205 La Laguna,
    Tenerife, Spain\\
    $^{17}$Departamento de Astrof\'\i sica, Universidad de La Laguna, E-38200 La
Laguna, Tenerife, Spain\\
    $^{18}$Joint Astronomy Centre, 660 N. A'ohoku Pl., University Park, Hilo, HI 
        96720, USA\\
    $^{19}$Netherlands Organisation for Scientific Research, Laan van Nieuw 
        Oost-Indie 300, NL-2509 AC The Hague, The Netherlands\\
    $^{20}$Observatorio de Madrid, OAN, Alfonso XII, 3, E-28014 Madrid, Spain\\
    $^{21}$Dipartimento di Astronomia, Universit\'a di Padova, Vicolo 
        dell'Osservatorio 5, 35122 Padua, Italy\\
    $^{22}$Department of Physics and Astronomy, University of Manitoba, Winnipeg, 
        Manitoba R3T 2N2, Canada\\
    $^{23}$National Astronomical Observatories, Chinese Academy of Sciences
20A Datun Road, Chaoyang District, Beijing, 100012, China \\
$^{24}$ Department of Physics \& Astronomy, Frederick Reines Hall,
University of California, Irvine, CA 92697-4575, USA \\
$^{25}$Astrophysics Group, Cavendish Laboratory, 19 J. J. Thomson
Avenue, Cambridge CB3 0HE, UK \\ 
$^{26}$Kavli Institute for Cosmology Cambridge, Madingley Road,
Cambridge CB3 0HA, UK\\ 
$^{27}$ Department of Physics \& Astronomy, University of Western Ontario
1151 Richmond St., London, ON N6A 3K7, Canada \\
$^{28}$ California Institute of Technology, 1200 E. California Blvd.,
Pasadena, CA 91125, USA Caltech (some flavour) \\
$^{29}$  Department of Physics, McGill University, 3600 rue
University, Montreal, QC H2A 2T8, Canada \\
$^{30}$ Herzberg Institute of Astrophysics, National Research Council of Canada, Victoria, BC V9E 2E7, Canada \\
$^{31}$ School of Physics and Astronomy, Cardiff University, Queens
Buildings, The Parade, Cardiff CF24 3AA \\
$^{32}$ Department of Physics and Astronomy
University of Waterloo
Waterloo, Ontario, Canada  N2L 3G1  \\
$^{33}$ Instituto Nacional de Astrofisica, Optica y Electronica
(INAOE), Aptdo. Postal 51 y 216, 72000 Puebla, Pue., Mexico  \\
$^{34}$ UK Astronomy Technology Centre, Royal Observatory, Blackford Hill,
Edinburgh EH9 3HJ \\
$^{35}$ Institute for Astronomy, University of Edinburgh, Royal Observatory,
Blackford Hill, Edinburgh EH9 3HJ \\
$^{36}$ Department of Astronomy, University of Massachusets, Amherst, MA01003, USA \\
$^{37}$ RALSpace, The Rutherford Appleton Laboratory, Chilton, Didcot OX11 0NL, UK  \\ 
$^{38}$ University of British Columbia, Okanagan Campus, 3333 University
Way, Kelowna, BC V1V 1V7 Canada \\
$^{39}$ Department of Astronomy and Astrophysics, University of
Toronto, 50 St. George Street, Toronto, ON M5S 3H4, Canada \\
$^{40}$ Department of Physics, Royal Military College of Canada, PO Box
17000, Station Forces, Kingston, ON K7K 7B4 \\
$^{41}$ Kapteyn Astronomical Institute, University of Groningen, P.O. Box 800, 9700 AV Groningen, The Netherlands \\
$^{42}$ Joint ALMA Office, Alonso de Cordova 3107, Vitacura, Santiago, Chile\\
$^{43}$ Astrophysics Group, Physics Department, University of the Western
Cape, Private Bag X17, 7535, Bellville, Cape Town, South Africa
}
\begin{document}

\date{2012 February 3}


\maketitle

\label{firstpage}

\begin{abstract}
The James Clerk Maxwell Telescope Nearby Galaxies Legacy Survey (NGLS)
comprises an H\,{\sc i}-selected sample 
of 155 galaxies spanning all morphological types with distances less
than 25 Mpc. We describe the scientific goals of the survey, the
sample selection, and the observing strategy. We also present an atlas
and analysis of the CO J=3-2 maps for the
47 galaxies in the NGLS which are also part of the Spitzer Infrared
Nearby Galaxies Survey.
We find a wide range of molecular gas mass fractions in the galaxies
in this sample and explore the correlation of the far-infrared
luminosity, which traces star formation, with the CO luminosity, which
traces the molecular gas mass. By comparing the NGLS data with merging
galaxies at low and high redshift which have also been observed in the
CO J=3-2 line, we show that the correlation of far-infrared and CO
luminosity shows a significant trend with luminosity. This trend is
consistent with a molecular gas depletion time which is more than an
order of magnitude faster in the merger galaxies than in nearby normal
galaxies. We also find
a strong correlation of the $L_{FIR}/L_{CO(3-2)}$ ratio with the atomic to
molecular gas mass ratio. This correlation suggests
that some of the far-infrared emission originates from dust associated
with atomic gas and that its
contribution is particularly important in galaxies where most of the
gas is in the atomic phase.
\end{abstract}


\section{Introduction}\label{sec-intro}

Star formation is one of the most important processes driving the
evolution of galaxies. The presence or absence
of significant star formation is one of the key characteristics which
distinguishes spiral and elliptical galaxies. The intense bursts
of star formation triggered by galaxy interactions and mergers produce
some of the most luminous galaxies in the local universe
\citep{sm96}. At high
redshift, many galaxies are seen to be forming stars at rates which
far exceed those of all but the most extreme local mergers
\citep{t08,tacc10}.
Since
stars form from gas, specifically from molecular gas, understanding the
properties of the interstellar medium (ISM) is critical to
understanding the rate and regulation of star formation in galaxies
\citep{l08,b08}.

At the most basic level, the amount of gas in a galaxy is an important
constraint on the amount of star formation the galaxy can sustain
\citep{k89,kenn07,l08,b08}.
Recent studies have shown that it is the
amount of molecular gas that is most important, rather than the total
gas content \citep{b08,l08,b11}.
This picture is consistent with
Galactic studies which show that stars form exclusively in molecular
clouds, and most commonly in the densest regions of those clouds
\citep{l91a,l91b,a11}.
This suggests that the properties of the molecular gas, in
particular its average density and perhaps the fraction of gas above a
critical density on sub-parsec scales, are likely to affect the
resulting star formation. Star formation in turn can affect the
properties of the dense gas, by increasing its temperature 
\citep{w97,meier01,t07}
and perhaps by triggering a subsequent generation
of stars \citep{z10}.
Unusual environments are also likely
to affect both the gas properties and the star formation process. High
shear in galactic bars, harassment in a dense
galaxy cluster, and galaxy mergers and interactions have the potential
to either dampen or enhance the star formation process. The wide range
of environmental processes at work, both on galactic and extragalactic
scales, implies that large samples of galaxies are required to tease
out the most important effects, while high resolution is required to
isolate individual star forming regions, separate arm from interarm
regions, and resolve galactic bars.

There have been a number of large surveys of the atomic gas content of
nearby galaxies, of which some of the most recent include 
The H\,{\sc i} Nearby Galaxy Survey, THINGS
\citep{w08}, the Arecibo Legacy Fast ALFA survey, ALFALFA \citep{giov05},
and the VLA Imaging of Virgo Spirals in Atomic Gas survey, VIVA \citep{c09}. 
However, only the 34 galaxies in the THINGS and the 53 galaxies in the
VIVA sample have 
sufficient spatial resolution to probe scales of one kiloparsec and
below.
Compared to the H\,{\sc i} 21 cm line,
the CO lines used to trace molecular gas are relatively more difficult
to observe due to their shorter wavelengths and the smaller field of
view of millimeter-wave radio telescopes equipped with single pixel
detectors. As a result,  
most CO extragalactic surveys have sampled a relatively small region
at the center of each galaxy
\citep{y95,h03,b93,d01}. Two recent surveys \citep{k07,l09}
have used array receivers to observe large unbiased
regions, although still in relatively small (18 and 40) samples of
galaxies. Finally, dust continuum observations in the submillimeter to
far-infrared present an alternative method of tracing the 
molecular gas that requires a mostly independent set of physical parameters,
such as dust emissivity, gas-to-dust mass ratio, and the
atomic gas mass 
\citep{t88,i97,e10,l11}. Two large surveys have
been made at 850 $\mu$m, one of an Infrared Astronomical Satellite
(IRAS) selected sample \citep{d00} 
and one of an optically selected sample \citep{v05}.
The galaxies selected for these samples were
relatively distant ($> 25$ Mpc)
and thus primarily global measurements of the dust luminosity were obtained.
The Herschel Reference Survey \citep{boselli10} is observing 323
galaxies with distances between 15 and 25 Mpc at 250, 350, and 500
$\mu$m and has significant overlap with the sample presented here.
Other Herschel surveys which overlap with the NGLS sample include the
Key Insights on Nearby Galaxies: A Far-Infrared Survey with Herschel
\citep{k11} and the Herschel Virgo Cluster Survey \citep{d10}.
Using the dust continuum to measure the star forming gas is a
promising avenue to explore, especially since this method is one that
can in principle be used for galaxies at higher redshifts.

Taking advantage of new instrumentation for both spectral line and
continuum data on the James Clerk Maxwell Telescope (JCMT), we are carrying
out the Nearby Galaxies Legacy Survey (NGLS)\footnote{http://www.jach.hawaii.edu/JCMT/surveys/}, a
large survey of 155 nearby galaxies using the CO J=3-2 line and
continuum observations at 850 and 450 $\mu$m. The survey is designed
to address four broad scientific goals.

\begin{enumerate}

\item {\it Physical properties of dust in galaxies.} The continuum
  data from this survey will probe a range of the dust spectral energy
  distribution (SED) that is critical to determining the total mass of dust
  as well as the relative proportion and physical properties of the
  different dust components (polycyclic aromatic hydrocarbons, very
  small grains, large 
  grains). Most importantly, these long wavelength data will trace any
  excess submillimetre emission that causes an upturn in the dust SED.
The
submillimetre excess could originate from: dust with $<$10 K
temperatures \citep{g03,g05,g11};
very small
grains with shallow dust emissivities that are not prominent in the
dust SED between 60 and 500 $\mu$m \citep{lis02,z09};
large dust grains with enhanced dust
emissivities at $>$850 $\mu$m \citep{b06,o10}; or spinning dust grains
\citep{bot10,ade11}.

\item {\it Molecular gas properties and the gas to dust ratio.} Our
  high-resolution data will allow us to compare the radial profiles of
  the dust, H\,{\sc i}, and CO emission within our well-selected sample. The
  CO J=3-2 line will effectively trace the warmer, denser molecular gas
  that is more directly involved in star formation \citep{i09}.
 In comparison, the CO J=1-0 line also traces more diffuse
  and low density gas \citep{ww94,r07}.
With observations of all three components of the ISM (molecular gas,
atomic gas, and dust), we
  will be able to determine accurate gas-to-dust mass ratios and
  provide constraints on the variation of the CO-to-H$_2$ conversion
  factor $X_{CO}$ \citep{s88} by fitting the data 
\citep{t88,i97,b97}.

\item {\it The effect of galaxy morphology.} The larger-scale galaxy
  environment can
play a significant role in the properties and structure of the
dense ISM. For example, an increase in the density in the ISM in the
centers of spiral galaxies has been attributed to an increased pressure
\citep{h93}. 
Elliptical galaxies have relatively low 
column densities of gas and dust \citep{knapp89,b98}
combined with an intense radiation field dominated by older,
low-mass stars and, in some cases, substantial X-ray halos. 
Early-type galaxies also show more compact and symmetric distributions
of dust at $< 70$ $\mu$m compared to late-type galaxies
\citep{bendo07,mm09}.
In spiral
galaxies, dust and gas properties may differ between arm and inter-arm
regions \citep{a02,f10}, while the role of spiral arms in the
star formation process is the subject of considerable debate
\citep{e86,v88}. With observations across the full range of galaxy
morphologies and with sub-kiloparsec resolution, we will be
able to study the effect of morphology on the molecular gas properties.

\item {\it The impact of unusual environments.} Metallicity has been
  shown to affect the structural properties of the ISM \citep{l11}.
  Lower self-shielding is expected to produce smaller regions
  of cold, dense gas that can be traced by CO emission with relatively
  larger and warmer photon-dominated regions \citep{m97,m06}.
Galaxies residing in rich clusters can be affected by ram
  pressure stripping \citep{l03} and gravitational harassment
\citep{m98}
 which reduces their ISM content relative to
  field galaxies. The halos of spiral galaxies can be
  populated with gas via superwinds or tidal interactions \citep{r10}
or even by relatively normal rates of star formation
\citep{l97}. With its galaxies spanning the full range of environment,
from isolated galaxies to small groups to the dense environment of the
Virgo cluster, the NGLS will be able to quantify the effect of
environment on the molecular ISM in galaxies.

\end{enumerate}

Most surveys of molecular gas in our own or other nearby
galaxies have used the ground state CO J=1-0 line as a tracer.
The choice of the CO J=3-2 line for the NGLS was driven by the
available instrumentation at the JCMT. Compared to the CO J=1-0 line,
(5.5 K above ground with a critical density of $1.1\times
10^3$ cm$^{-3}$), the CO J=3-2 line (33 K and $2.1\times 10^4$
cm$^{-3}$) traces relatively warmer and denser gas. (Since both lines
are usually optically thick, the effective critical density is likely
reduced by a factor of 10 or more.) There is growing evidence that the
CO J=3-2 emission correlates more tightly with the star formation rate
or star formation efficiency than does the CO J=1-0 line
\citep{w09,muraoka07}. \citet{komugi07} 
showed that the CO J=3-2 emission correlates linearly with star
formation rate derived from extinction-corrected H$\alpha$ emission
and that the correlation was tighter with the J=3-2 line than with the
J=1-0 line. Similarly, \citet{i09} showed that the CO J=3-2 emission
correlates nearly linearly with the far-infrared luminosity for a
sample of local luminous infrared galaxies and high redshift
submillimeter galaxies. Thus, it is appears that the CO J=3-2
emission is preferentially tracing the molecular gas associated
directly with star formation, such as high-density gas that is forming
stars or warm gas heated by star formation, rather than the total
molecular gas content of a galaxy.

In this paper, we describe the NGLS sample selection, the CO J=3-2
observations and data reduction, and the planned observing
strategy for continuum 450 and 850 $\mu$m observations (\S\ref{sec-obs}).
In \S~\ref{sec-disc}, we present the CO J=3-2 integrated intensity
images, as well as maps of 
the velocity field and velocity dispersion for those galaxies with
sufficiently strong signal. 
In \S~\ref{sec-comp}, we examine the molecular and atomic gas masses
for the Spitzer Infrared Nearby Galaxies Survey (SINGS) sample
\citep{k03} and compare
the CO J=3-2 luminosity with the
far-infrared luminosity for both the galaxies in the SINGS sample,
local luminous and ultraluminous
infrared galaxies, and high redshift quasars and submillimeter
galaxies which have also been observed in the CO J=3-2 transition
\citep{i09}.
We give our conclusions 
in \S\ref{sec-concl}.
Previous papers in this series have examined individual or small
samples of galaxies \citep{w09,warren10,b10,w11,i11,s-g11}. Future
papers will exploit the spatially resolved nature of these data by
examining the CO line ratios and excitation (Rosolowsky et al. in
prep.) and the gas depletion times (Sinukoff et al., in prep.).

\section{Observations and Data Processing}\label{sec-obs}

\subsection{Sample selection}\label{subsec-sample}

The galaxies in the NGLS are an H\,{\sc i}-flux selected sample. This
selection method was chosen to avoid biasing the survey towards
galaxies with higher star formation rates, as might be the case for
far-infrared or blue magnitude selection criteria, while still
targeting galaxies with a significant interstellar medium, as might
not be the case for a purely mass selected sample, e.g. 
\citet{boselli10}.
Our sample should also not be unduly biased by
dust content being either particularly hot or cold, and should be
reasonably representative of the local submillimetre universe.
H\,{\sc i} emission typically extends to much a larger
  radius than CO or stellar emission, so that our sample
 may include galaxies that are H\,{\sc i}-rich but H$_2$-poor.

To obtain good spatial resolution while still being able to obtain
sensitive maps in a reasonable period of time, the galaxies were
selected to have distances between 2 and 25 Mpc. This distance limit
excludes galaxies in the Local Group but does include galaxies in the
Virgo cluster, which allows us to test the effect of cluster
environment on the dense ISM.
We used the HyperLeda\footnote{http://leda.univ-lyon1.fr}
 data base in February 2005 
to extract all non-Virgo galaxies (see below) with H\,{\sc i}
fluxes $> 3.3$ Jy km s$^{-1}$ and Virgo-corrected galactocentric velocities
$v < 1875$ km s$^{-1}$. 
Out to these limits, the
HyperLeda database is essentially complete. 
We removed galaxies with declinations below -25$^o$; we further
removed galaxies with Galactic latitudes between -25$^o$ and
25$^o$ to minimize the impact of Galactic cirrus in complementary data
from Spitzer and Herschel.
Virgo cluster member galaxies were
selected to be galaxies
within an $8^o \times 16^o$ ellipse 
centered on M87 (RA=12.4h, 
DEC=12.4$^o$) and with velocities between 500 and 2500 km s$^{-1}$,
e.g. \citet{d04}; all Virgo galaxies with an H\,{\sc i} flux entry
were included in this initial selection.
These selection criteria gave us a sample of 1002 field galaxies
and 148 Virgo cluster galaxies.

Since we estimated that at least 1500 hours of telescope time would be required
to observe this complete sample of 1150 galaxies, we needed to apply
further selection criteria. We first selected all the SINGS galaxies
\citep{k03} that met our declination and Galactic latitude
criteria; the wealth of complementary data available on these galaxies
from the SINGS and other surveys \citep{w08} make them natural
targets. Forty-seven SINGS galaxies met our criteria
(Tables~\ref{tbl-large}
and~\ref{tbl-small}). Two SINGS galaxies, NGC 5194 (M51) and NGC 4789A
(DDO 154), were observed 
in separate programs and were not re-observed as part of our survey;
we present maps made from those data in this paper.

We then estimated
the minimum size of the field and
Virgo cluster samples that would allow
us to achieve the scientific goals of this study.
For statistical studies of galaxy properties, we want
to divide the galaxies into 4 morphological bins (E/S0, early-type spirals,
late-type spirals, and irregulars).
In addition, we want to compare the properties of galaxies in
the Virgo cluster to those in the field. We estimated that 18 galaxies
per bin were required to
obtain good statistics on
the average properties of each bin,
or 144 galaxies total. However, since our sample only contains 148 Virgo
galaxies,
we compromised on 9 galaxies per bin in Virgo and 18 per bin in
the field, giving a sample size of 108 galaxies.  In addition, we
limited the field and Virgo samples to galaxies with $D_{25} <
4^\prime$  so that we could use the jiggle-map mode with HARP-B 
\citep{bu09} and
still map the inner quarter of each galaxy.

To select the final sample, 
the field and Virgo lists were each divided into the four morphological
bins described above 
using the numerical Hubble stage, $T$ (E/S0: $T\le 0.5$; early-type
spirals: $0.5 < T \le 4.5$;
late-type spirals: $4.5 < T \le 9.5$; irregulars: $T > 9.5$). For the
field samples, we first applied 
an H\,{\sc i} flux cutoff of 6.3 Jy km s$^{-1}$ 
and then randomly selected 18 galaxies in each
of the four morphological bins (for the elliptical galaxies, there
were precisely 18 galaxies above this cutoff). 
The smaller Virgo sample meant that we
could only use this same process for the late-type spiral bin. The
brightest 9 in each category of 
E/S0 ($\ge 2.1$ Jy km s$^{-1}$), Irr ($\ge 4.1$ Jy km s$^{-1}$), and
early-type spiral galaxies ($\ge 6.2$ Jy km s$^{-1}$) from Virgo were included,
in addition to a random selection of 9 late-type spirals with H\,{\sc i}
fluxes
brighter than 6.3 Jy km s$^{-1}$.
The final field and Virgo source lists are given in
Tables~\ref{tbl-field}-\ref{tbl-virgo} in Appendix~\ref{members}. 

\renewcommand{\thefootnote}{\alph{footnote}}
\begin{table*}
 \centering
 \begin{minipage}{140mm}
  \caption{Large ($D_{25} > 4^\prime$) galaxies from the SINGS sample\label{tbl-large}}
  \begin{tabular}{lcccccc}
  \hline
Name & $\alpha$(J2000.0)\footnotemark[1] &
$\delta$(J2000.0)\footnotemark[1] &$D_{25,maj}/2$\footnotemark[2] &
$D_{25,min}/2$\footnotemark[2] & 
PA & $V_{hel}$\footnotemark[3] \\ 
& (h:m:s) & ($^\circ$:$'$:$''$) & ($^\prime$) & ($^\prime$) & ($^o$) & (km s$^{-1}$) \\
\\
 \hline
NGC 0024 & 00:9:56.6 & $-$24:57:47 & 3.1 & 0.8 & 46 & 563 \\
NGC 0628\footnotemark[4] & 01:36:41.8 & +15:47:00 & 5.2 & 4.8 & 25 & 648 \\
NGC 0925\footnotemark[4] & 02:27:17.0 & +33:34:44 & 5.6 & 3.2 & 102 & 540 \\
NGC 2403\footnotemark[4] & 07:36:50.7 & +65:36:10 & 11.7 & 6.2 & 127 & 121 \\
UGC 04305 & 08:19:03.9 & +70:43:09 & 4.2 & 3.1 & 15 & 158 \\
NGC 2841\footnotemark[4] & 09:22:02.5 & +50:58:36 & 3.8 & 1.8 & 147 & 641 \\
NGC 2976\footnotemark[4] & 09:47:15.5 & +67:55:03 & 2.9 & 1.5 & 143 &  22 \\
NGC 3031\footnotemark[4] & 09:55:33.1 & +69:03:56 & 11.2 & 5.7 & 157 & $-$48 \\
NGC 3034\footnotemark[4] & 09:55:52.4 & +69:40:47 & 5.2 & 2.2 & 65 & 254 \\
NGC 3184\footnotemark[4] & 10:18:17.0 & +41:25:28 & 4.0 & 3.8 & 135 & 574 \\
NGC 3198\footnotemark[4] & 10:19:55.0 & +45:32:59 & 3.9 & 1.3 & 35 & 666 \\
IC 2574\footnotemark[4]  & 10:28:23.5 & +68:24:44 & 6.7 & 3.2 & 50 &  28 \\
NGC 3351\footnotemark[4] & 10:43:57.8 & +11:42:13 & 3.8 & 2.2 & 13 & 768 \\
NGC 3521\footnotemark[4] & 11:05:48.9 & $-$00:02:06 & 5.4 & 2.7 & 163 & 778 \\
NGC 3627\footnotemark[4] & 11:20:14.9 & +12:59:30 & 4.4 & 2.0 & 173 & 697 \\
NGC 3938\footnotemark[4] & 11:52:49.3 & +44:07:17 & 2.5 & 2.4 &  0 & 829 \\
NGC 4236 & 12:16:42.0 & +69:27:46 & 11.4 & 3.6 & 162 &  10 \\
NGC 4254\footnotemark[4] & 12:18:49.6 & +14:25:00 & 2.7 & 2.3 &  0 & 2412 \\
NGC 4321\footnotemark[4] & 12:22:54.9 & +15:49:21 & 3.8 & 3.1 & 30 & 1599 \\
NGC 4450 & 12:28:29.7 & +17:05:06 & 2.6 & 1.9 & 175 & 1905 \\
NGC 4559 & 12:35:57.7 & +27:57:36 & 6.0 & 2.1 & 150 & 827 \\
NGC 4569\footnotemark[4] & 12:36:50.0 & +13:09:46 & 5.2 & 2.3 & 23 & $-$179 \\
NGC 4579\footnotemark[4] & 12:37:44.0 & +11:49:07 & 2.8 & 2.2 & 95 & 1540 \\
NGC 4594 & 12:39:59.3 & $-$11:37:22 & 4.2 & 2.0 & 90 & 1088 \\
NGC 4631\footnotemark[4] & 12:42:07.7 & +32:32:34 & 7.4 & 1.3 & 86 & 640 \\
NGC 4725 & 12:50:26.7 & +25:30:03 & 5.5 & 3.7 & 35 & 1188 \\
NGC 4736 & 12:50:53.1 & +41:07:12 & 6.3 & 5.5 & 105 & 295 \\
NGC 4826 & 12:56:43.9 & +21:41:00 & 5.4 & 2.6 & 115 & 390 \\
NGC 5033\footnotemark[4] & 13:13:27.6 & +36:35:38 & 5.1 & 1.9 & 170 & 873 \\
NGC 5055\footnotemark[4] & 13:15:49.4 & +42:01:46 & 6.6 & 4.0 & 105 & 495 \\
NGC 5194\footnotemark[4] & 13:29:52.4 & +47:11:41  & 4.9 & 3.4 & 163 & 452 \\
 \hline
\end{tabular}
\begin{tabular}{l}
\footnotemark[1] From RC3 \citep{deV91} \\
\footnotemark[2] From \citet{buta07} \\
\footnotemark[3] Systemic velocity from the H\,{\sc i} line (heliocentric)\\
\footnotemark[4] Member of high-priority SINGS subset for SCUBA-2 observations\\
\end{tabular}
\end{minipage}
\end{table*}

\renewcommand{\thefootnote}{\alph{footnote}}
\begin{table*}
 \centering
 \begin{minipage}{140mm}
  \caption{Small ($D_{25} < 4^\prime$)galaxies from the SINGS sample\label{tbl-small}}
  \begin{tabular}{lcccc}
  \hline
Name & $\alpha$(J2000.0)\footnotemark[1] & $\delta$(J2000.0)\footnotemark[1] & $V_{hel}$\footnotemark[2] & Alternative Name \\
& (h:m:s) & ($^\circ$:$'$:$''$) & (km s$^{-1}$) & \\
\\
 \hline
NGC 0337 & 00:59:50.1 & $-$07:34:41 & 1664 & ... \\
NGC 0584 & 01:31:20.8 & $-$06:52:05 & 1861 & ... \\
NGC 0855 & 02:14:03.6 & +27:52:38 & 583 & ... \\
PGC 023521 & 08:23:55.0 & +71:01:57 & 113 & M81DwA \\
UGC 05139 & 09:40:32.3 & +71:10:56 & 137 & HoI \\
NGC 3049 & 09:54:49.7 & +09:16:18 & 1458 & ... \\
UGC 05336 & 09:57:32.1 & +69:02:46 & 48 & HoIX \\
UGC 05423 & 10:05:30.4 & +70:21:54 & 344 & M81DwB \\
NGC 3190 & 10:18:05.3 & +21:49:58 & 1306 & ... \\
NGC 3265 & 10:31:06.8 & +28:47:47 & 1410 & ... \\
UGC 05720 & 10:32:32.0 & +54:24:04 & 1456 & Mrk33 \\
NGC 3773 & 11:38:13.0 & +12:06:43 & 1002 & ... \\
NGC 4625 & 12:41:52.7 & +41:16:26 & 590 & ... \\
NGC 4789A & 12:54:05.5 & +27:08:55 & 374 & DDO154 \\
UGC 08201 & 13:06:24.9 & +67:42:25 & 34 & DDO165 \\
NGC 5474 & 14:05:01.6 & +53:39:45 & 240 & ... \\
 \hline
\end{tabular}
\begin{tabular}{l}
\footnotemark[1] From RC3 \citep{deV91} \\
\footnotemark[2] Systemic velocity from the H\,{\sc i} line (heliocentric)\\
\end{tabular}
\end{minipage}
\end{table*}

\subsection{CO $J$=3-2 observations and data processing}\label{subsec-CO}

CO $J$=3-2 observations for all galaxies were obtained on the 
JCMT between 2007 November and 2009 November.
The angular resolution of the JCMT at this frequency is
14.5$^{\prime\prime}$, which corresponds to a linear resolution
ranging from 0.2 to 1.2 kpc for the galaxies in our sample.
The 30 large ($D_{25} > 4^\prime$) SINGS galaxies in the NGLS were observed
in raster map mode, while the smaller galaxies were observed in
jiggle-map mode \citep{bu09}. All galaxies were mapped over
a rectangular area
corresponding to $D_{25}/2$ on a side (with raster maps oriented with
the position angle 
of the galaxy's semi-major axis), with a 1 sigma sensitivity of better than
19 mK ($\rm T_A^*$) (32 mK $\rm T_{MB}$ for $\eta_{MB}=0.6$) at a
spectral resolution of 20 km 
s$^{-1}$.  We used the
16 pixel array receiver 
HARP-B \citep{bu09} with the Auto-Correlation Spectral Imaging System
(ACSIS) correlator configured to have a 
bandwidth of 1 GHz and a resolution of 0.488 MHz (0.43 km s$^{-1}$
at the frequency of the CO $J$=3-2 transition). 
The total time allocated to the CO $J$=3-2 portion of the NGLS,
including pointing and calibration observations, was 256 hours.

Calibration was checked each night by observing one or more of a set
of six calibrator sources in the $^{12}$CO J=3-2 transition. The
calibration sources used were W75N, CRL 618, IRC+10216, NGC 2071IR, CRL
2688, and IRAS 17293-2422. The peak intensities of the calibrator
spectra agreed to within $\pm8$\%, while the integrated intensities 
agreed to within $\pm$13\% from 2007 November to 2008 July and to
$\pm 9$\% from 2008 August onwards. We adopt 10\% as
the internal calibration uncertainty on the NGLS. 
Pointing was checked before starting a new source, and
every 1-2 hours (more frequently near sunrise and sunset). Any
scan where the pointing offset that was measured after the
scan differed by more than 4$^{\prime\prime}$ from the pointing
measured at the start of the scan was rejected in the final
data analysis. Less than 5\% of the data were rejected because of
large pointing changes. The  rms of the pointing measurements on a given
night was typically better than 2$^{\prime\prime}$.

Two sources in the
NGLS, NGC 5055 and NGC 337, were observed during Science Verification
runs in 2007 May and 2007 July. For these observing runs, the strength
of the spectral line calibrators was only 85\% of the expected values
and so the data for these two galaxies have been rescaled by a factor
of 1.18 to bring them onto the same calibration scale as the rest of
the survey. NGC 4789A (DDO 154) was observed as part of a separate
program (M07AC14, PI B. Warren), as was 
NGC 5194 (M51) (M06AN05, PI R. Tilanus). For both these galaxies, the
integration time per point in the map is longer than that of a typical
NGLS observation.

Details of the reduction of the raster map data are given in
the appendix in \citet{warren10} and so we discuss only the
jiggle map processing in detail here.
The individual raw data files were flagged to remove data from
any of the 16 individual receptors  with bad
baselines and then the scans were combined into a data cube using 
the ``nearest'' weighting function
to determine the contribution of individual receptors to each pixel in
the final map. This weighting function causes each measurement
from a raw data file to contribute to only a single pixel in the final cube.
The pixel size in the maps is 7.5$^{\prime\prime}$.
We fit and subtracted a first-order baseline from the data cube while
excluding the central 400 km s$^{-1}$ region from the baseline fit.

We used
the clumpfind
algorithm \citep{w94} implemented as part of the
Clump Identification and Analysis Package
(CUPID\footnote{CUPID 
is part of the Starlink \citep{c08}
  software package, which is available for download from
http://starlink.jach.hawaii.edu}) \citep{b07} task 
findclumps to identify regions with emission above a specified
signal-to-noise ratio. The algorithm measures the mean noise in the
data cube as part of the processing step. However,
the sensitivity of each of the sixteen individual receptors in the
HARP-B array varies quite significantly. In addition, a portion of the
data were obtained when four of the receptors were inoperative and a
rotation of the K-mirror between two observations was used to obtain a
complete map of the desired area. As a result, the noise in a final
data cube can vary quite significantly from region to region in a
jiggle map. Thus, the mean noise that findclumps measures will under
or over-estimate the noise in different regions of the cube.

To mitigate the effect of this noise variation, we calculated an image
of the noise in the map by averaging the noise in the line free
regions at the end of the spectrum. We then divided the cube by this
noise image to produce a cube which was now in units of
signal-to-noise, rather than Kelvins. This data cube was then boxcar smoothed
by 3 pixels and 25 velocity channels. 
We then applied the findclumps
algorithm to this signal-to-noise cube 
to identify regions with emission with signal-to-noise
greater than 2$\sigma$, 2.5$\sigma$, or 3$\sigma$.
In all cases, the spacing between contour levels (the $\Delta T$
paerameter) was set to 2$\sigma$.
The mask of regions of real 
emission produced by
findclumps was then applied to the original, unsmoothed data cube and
this masked cube was then collapsed in velocity to produce moment maps. 
For the raster maps, we produced masks with signal-to-noise cutoffs of
2$\sigma$, 2.5$\sigma$, and 3$\sigma$.

\begin{figure}
\includegraphics[width=60mm,angle=-90]{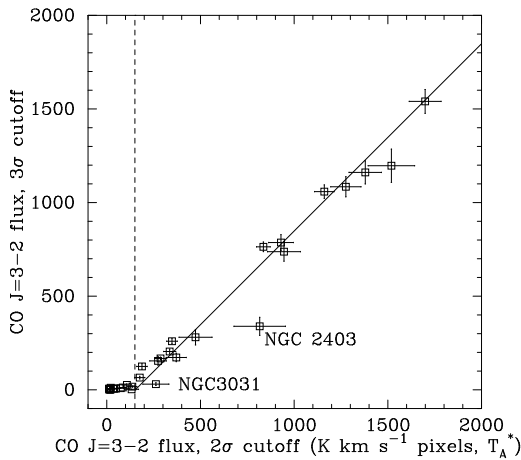}
\caption{CO J=3-2 flux for galaxies in the NGLS compared for a
  2$\sigma$ cutoff versus a 3$\sigma$ cutoff in making the moment
  maps. The solid line represents $2\sigma$ flux = 152 + $3\sigma$
  flux (see text).
NGC 3034 (M82) is much brighter than all other galaxies and is
  not included in this plot; its flux offset is consistent with the rest of the
  sample. Galaxies to the left of the vertical dashed line were not
  included in calculating the average offset (see text).
\label{fig-totint}}
\end{figure}

The zeroth moment map measures the integrated intensity in the data
cube, $\int T dv = \sum T_i \Delta v$ where $\Delta v$ is the channel
width, 
$T_i$ is the temperature in an individual channel. To obtain the final
integrated intensity map, the zeroth moment map needed to be
multiplied by the noise map (to recover units of K km s$^{-1}$) and
also to be divided by $\eta_{MB}=0.6$ to convert to the $T_{MB}$
temperature scale. For the zeroth moment maps, we use a noise cutoff
of 2$\sigma$; we chose to use this cutoff after comparing the CO
fluxes obtained with different noise cutoffs (Figure~\ref{fig-totint}). 
For bright galaxies with a flux greater than 150 K (T$_A^*$)
  km s$^{-1}$ pixels, the data show that fluxes measured using a
  2$\sigma$ cutoff are systematically offset from fluxes measured with
  a 3$\sigma$ cutoff, in the sense that the maps made withe the
  3$\sigma$ cutoff underestimate the total flux. Two galaxies,
  NGC 2403 \citep{b10} and NGC 3031 \citep{s-g11}, have much larger
  offsets and were not 
  included in deriving the average offset. These galaxies both
have  such low CO surface brightnesses such that an
increase in the noise 
cutoff causes large areas of the galaxy to be masked out and thus
has a large impact on the total flux that is
measured.

The first moment map measures the mean velocity of
the emission $\overline{v} = \sum T_i v_i / \sum T_i $
where $v_i$ is the velocity of a given velocity channel.
The mean velocity calculated in this way can be rather sensitive to
noise spikes, and so we used a higher signal-to-noise threshold of 2.5$\sigma$
in calculating the velocity field maps.

The second moment maps measure the velocity dispersion, $\sigma_v$, for each
pixel in the image using
$$\sigma_v = \sum T_i (v_i - \overline{v})^2 / \sum T_i \eqno(1) $$
This method of calculating the velocity
dispersion differs from those used in previous extragalactic CO
studies such as \citet{cb97}, which typically fit gaussian profiles to
the CO lines. However, in the limit of
gaussian lines with a high signal-to-noise ratio, the values from the
moment 2 maps should agree with the results from fitting a gaussian
directly to the line profiles. Note that for a gaussian line profile,
$\sigma_v$ is equal to the full-width at half maximum of the line
divided by a factor of 2.355. 
The second moment maps were calculated using a signal-to-noise
threshold of 2.5$\sigma$.
A more detailed discussion of the
possible systematics in second moment maps is given in \citet{w11}.

CO $J$=3-2 images for the SINGS sample are shown in
Appendix~\ref{images}. Images 
for the Virgo and field samples will be presented in a future
paper (Golding et al., in prep.). 
The reduced images, noise maps, and spectral cubes are available via the survey
website\footnote{http://www.physics.mcmaster.ca/$\sim$wilson/www\_xfer/NGLS/}
and will eventually be made available via the 
Canadian Astronomical Data Centre (CADC).
We note that some of the images show residual noise effects,
particularly around the outer edges of the map (see NGC 4736 and NGC
5033 for particularly obvious examples). These artifacts are generally
due to the increasing noise level in the data cubes towards the edges
of the map, which due to the scanning technique do not have as long an
integration time and are not sampled by as many of the individual HARP
detectors. Anyone interested in the reality of a particular faint
feature is encouraged to consult the images and noise maps available
on the survey web site.


\subsection{Future SCUBA-2 observations}

The
Submillimetre Common-User Bolometer Array-2 (SCUBA-2, Holland et
al. 2006) is a large-format bolometer camera
for the JCMT that is designed to produce simultaneous continuum images
at 450 and 850 $\mu$m. The camera consists of four $32\times 24$
transition-edge sensor arrays for each of the two wavelengths and has
a total field of view of $\sim 7^\prime\times 7^\prime$.
The SCUBA-2 portion of the NGLS began in December 2011 with
a period of
science verification observations for the legacy surveys.
In this section, we
briefly describe our planned observing strategy with SCUBA-2.

The ultimate goal of the survey is to map all the galaxies in the NGLS
with SCUBA-2 
out to $D_{25}$ to a 1$\sigma$ limit of 1.7 mJy at 850 $\mu$m; this
sensitivity limit includes both instrumental noise and confusion noise
from the high-redshift extragalactic background. Good 450 $\mu$m data
provide improved spatial resolution and are critical to tracing
the upturn of the dust spectral energy distribution that indicates
the presence of very 
cold dust or a change in dust properties (see \S~\ref{sec-intro}).
With similar (simultaneous) integration times at both 450 and 850
$\mu$m, we estimate that the 450 $\mu$m sensitivity will be about a
factor of two worse than at 850 $\mu$m for a given  surface density of
dust and smoothed to the same angular resolution. 

To put these sensitivities in physical terms, we can convert to an
equivalent mass surface density (gas plus dust) by using the formula
given in Johnstone et al. (2001). Assuming  $T_{dust} = 20$ K,
$\kappa_{850} = 0.0043$ cm$^2$ g$^{-1}$, and a circular beam with diameter
15$^{\prime\prime}$, a 4$\sigma$ detection at 850 $\mu$m corresponds
to an average visual extinction of $A_v = 3$ mag, which
corresponds to surface densities of $6\times 10^{21}$ H/cm$^2$ or 45
M$_\odot$/pc$^2$. For a galaxy at a distance of 10 Mpc, the equivalent
mass sensitivity is $2 \times 10^7$ M$_\odot$, equal to a few
massive giant molecular clouds within a 730 pc diameter region.

\section{CO Properties}\label{sec-disc}

\subsection{Measuring CO luminosities}

\renewcommand{\thefootnote}{\alph{footnote}}
\begin{table*}
 \centering
 \begin{minipage}{140mm}
  \caption{CO and far-infrared luminosities for the SINGS sample\label{tbl-Ls}}
  \begin{tabular}{lccccccc}
  \hline
Name & $D$\footnotemark[1] & $\Delta T$\footnotemark[2]  &
$L_{CO(3-2)}$\footnotemark[3] & $\log L_{FIR}$\footnotemark[4] &
$L_{FIR}/L_{CO(3-2)}$ 
& $M_{H_2}/M_{HI}$\footnotemark[5] \\ 
 & (Mpc) & (mK) & ($10^7$ K km s$^{-1}$ pc$^2$) & ($L_\odot$) &
 $L_\odot$/(K km s$^{-1}$ pc$^2$) & \\
\\
 \hline
NGC 0024    &      7.6 &  27 &  $<         0.4$                 &
8.19 &    $>$     40 & $<$ 0.10 \\  
NGC 0628    &      7.3 &  24 &  5.2        $\pm$      1.0        &
9.55 &         67 & 0.25 \\  
NGC 0925    &      9.1 &  25 &  0.9        $\pm$      0.2        &
9.31 &        221 & 0.04\\  
NGC 2403    &      3.2 &  23 &  1.7        $\pm$      0.3        &
9.09 &         72 & 0.12 \\  
UGC 04305   &      3.4 &  31 &  $<         0.09$                &
7.51 &        $>$  38 & $<$ 0.02 \\  
NGC 2841    &     14.1 &  24 &  $<         1.1$                 &
9.45 &      $>$   249 & $<$ 0.02 \\  
NGC 2976    &      3.6 &  26 &  0.50       $\pm$      0.08       &
8.72 &        103 & 0.65 \\  
NGC 3031    &      3.6 &  26 &  0.91       $\pm$      0.35       &
9.41 &        284 & 0.05 \\  
NGC 3034    &      3.6 &  31 &  39.2       $\pm$      0.4        &
10.60 &        102 & 8.6 \\  
NGC 3184    &     11.1 &  28 &  9.5        $\pm$      1.4        &
9.61 &         43 & 0.55 \\  
NGC 3198    &     13.7 &  31 &  6.9        $\pm$      1.1        &
9.61 &         60 & 0.12 \\  
IC 2574  & 3.8 & 23 & $< 0.08$ & ... & ... & $<$ 0.01 \\
NGC 3351    &      9.3 &  23 &  5.1        $\pm$      0.6        &
9.66 &         88 & 0.89\\  
NGC 3521    &      7.9 &  27 &  17.9       $\pm$      1.2        &
9.97 &         52 & 0.73 \\  
NGC 3627    &      9.4 &  23 &  31.1       $\pm$      1.7        &
10.18 &         49 & 6.5 \\  
NGC 3938    &     14.7 &  26 &  12.3       $\pm$      2.0        &
9.85 &         57 & 0.51 \\  
NGC 4236    &      4.4 &  21 &  0.53       $\pm$      0.15       &
8.19 &         29 & 0.05 \\  
NGC 4254    &     16.7 &  27 &  73.8       $\pm$      4.8        &
10.50 &         43 & 2.7 \\  
NGC 4321    &     16.7 &  28 &  54.7       $\pm$      5.3        &
10.36 &         42 & 3.1 \\  
NGC 4450    &     16.7 &  23 &  1.1        $\pm$      0.3        &
9.06 &        106 & 0.62\\  
NGC 4559    &      9.3 &  23 &  1.9        $\pm$      0.4        &
9.42 &        138 & 0.07 \\  
NGC 4569    &     16.7 &  26 &  20.1       $\pm$      1.7        &
9.95 &         44 & 5.3 \\  
NGC 4579    &     16.7 &  20 &  7.9        $\pm$      1.4        &
9.84 &         87 & 2.3 \\  
NGC 4594    &      9.8 &  28 &  $<         0.7$                 &
8.92 &       $>$  122 & $<$ 0.43 \\  
NGC 4631    &      7.7 &  20 &  14.6       $\pm$      0.7        &
10.10 &         85 & 0.31 \\  
NGC 4725 & 11.9 & 25 & $< 0.9$ & ... & ... & $<$ 0.05 \\
NGC 4736    &      5.2 &  30 &  5.2        $\pm$      0.4        &
9.65 &         86 & 1.9 \\  
NGC 4826    &      7.5 &  30 &  9.7        $\pm$      0.5        &
9.75 &         58 & 3.3 \\  
NGC 5033    &     16.2 &  24 &  23.1       $\pm$      3.0        &
10.19 &         67 & 1.6 \\  
NGC 5055    &      7.9 &  30 &  19.7       $\pm$      1.7        &
10.00 &         51 & 0.63 \\  
NGC 5194    &      7.7 &  15 &  53.2       $\pm$      2.0        &
10.21 &         31 & 4.0 \\  
 \hline
NGC 0337    &     23.1 &  37 &  $<         5.6$                 &
10.13 &       $>$  240 & $<$ 0.15 \\  
NGC 0584 & 20.1  & 19 & $< 2.5$ & .. & ... & $<$ 2.6 \\
NGC 0855    &      9.7 &  15 &  $<         0.5$                 &
8.29 &        $>$  41 & $<$ 0.54 \\  
PGC 023521 & 3.6  & 29 & $< 0.1$ & .. & ... & $<$ 1.4 \\
UGC 05139 & 3.6  & 22 & $< 0.09$ & .. & ... & $<$ 0.12 \\
NGC 3049    &     22.7 &  20 &  1.4        $\pm$      0.3        &
9.39 &        176 & 0.15 \\  
UGC 05336 & 3.6  & 21 & $< 0.09$ & .. & ... & $<$ 0.0.05 \\
UGC 05423 & 3.6  & 24 & $< 0.1$ & .. & ... & $<$ 1.5 \\
NGC 3190    &     21.3 &  19 &  $<         2.7$                 &
9.52 &       $>$  121 & $<$ 0.86 \\  
NGC 3265 & 22.7  & 20 & $< 3.3$ & .. & ... & $<$ 2.7 \\
UGC 05720 & 25.0 & 21 & 2.8 $\pm$ 0.6 & ... & ... & 0.70 \\
NGC 3773    &     10.5 &  16 &  0.20       $\pm$      0.05       &
8.38 &        122 & 0.54 \\  
NGC 4625    &      9.4 &  19 &  0.12       $\pm$      0.03       &
8.37 &        196 & 0.03 \\  
NGC 4789A & 3.8  & 7.3 & $< 0.03$ & .. & ... & $<$ 0.02 \\
UGC 08201 & 4.6  & 18 & $< 0.1$ & .. & ... & $<$ 0.19 \\
NGC 5474    &      6.5 &  26 &  $<         0.3$                 &
8.14 &        $>$  40 & $<$ 0.06 \\  
 \hline
\end{tabular}
\begin{tabular}{l}
\footnotemark[1] NGC 925, NGC 2403, NGC 3031,
NGC 3198, NGC 3351, NGC 3627, NGC 4725: \citet{f01}; 
NGC 2976, NGC 4236, \\
UGC04305, UGC08201: \citet{k02}
NGC 3034, PGC 023521, UGC05139, UGC05336,  UGC 05423: same distance
\\ as NGC 3031 \citep{f01};
NGC 584, NGC 855, NGC 4594, NGC 4736, NGC
  4826, NGC 5194: \citet{t01}; \\
Virgo Cluster: \citet{m07};  NGC 628: \citet{k04}; 
IC 2574: \citet{d09}
NGC 2841: \\\citet{m01}; 
NGC 3184: \citet{l02};  NGC 4631: \citet{s05}.
Remaining galaxies: Hubble flow distance with \\ 
velocity corrected  for Virgo infall \citep{m00} and $H_o = 70.5$ km s$^{-1}$
Mpc$^{-1}$. \\
\footnotemark[2] rms noise in individual spectra in the data cube at
20 km s$^{–1}$ resolution on $T_{MB}$ scale. \\
\footnotemark[3] Upper limits are 2$\sigma$ limits calculated over an
area of 1$'$ and a line width of 100 km s$^{-1}$.\\
\footnotemark[4] $L_{FIR}$ from \citet{s03} and \citet{l07} (see text)
adjusted for distances given here. \\
\footnotemark[5] Total H\,{\sc i} masses (including emission
  from areas outside the CO disk) from fluxes in
\citet{w08}, \\  \citet{c09}, 
or the HyperLeda database \citep{p03}; see text. 
$M_{H_2}$ calculated assuming a CO J=3-2/J=1-0 \\
line
ratio of 0.18, which may overestimate the total mass in some galaxies,
as well as 
a standard value for $X_{CO}$;
see text.\\
\end{tabular}
\end{minipage}
\end{table*}

For galaxies which were clearly detected,
the CO J=3-2 luminosity was measured directly from the moment 0 images
made with a 2$\sigma$ cutoff mask. We used apertures chosen by eye to
capture all the emission from the galaxy while excluding the
occasional noisy or doubtful emission towards the edges of the map.
Thus, the CO luminosity is given by
$$L_{CO(3-2)} = \sum I(CO)_i \times (4.848 D_{Mpc} \times L_{pix})^2 \eqno(2)$$
where $I(CO)_i$ is the CO intensity in an individual pixel of the map,
$D_{Mpc}$ is the distance to the galaxy in Mpc, and $L_{pix}$ is the
size of an individual pixel in the map (7.5$^{\prime\prime}$ for jiggle maps and
7.2761$^{\prime\prime}$ for raster maps).

We calculated a map of the uncertainty in $I(CO)$ for each pixel,
$\sigma_i$, as
$$\sigma_i = \sigma_{chan} \sqrt{\Delta v_{chan} \Delta v_{line}}
\sqrt{1 + \Delta v_{line} / \Delta v_{base} } \eqno(3)$$
where
$\Delta v_{chan}$ is the velocity width of a single channel,
$\sigma_{chan}$ is the standard deviation in K of the line-free channels,
$\Delta v_{line}$ is the velocity range used to measure the line, and
$\Delta v_{base}$ is the velocity range used to fit the baseline.
Now, both the raster and the jiggle observing modes use a shared off
position observed for a longer integration time than each of the
individual on positions. Because the off position is shared across
many pixels in the map, a simple combination of the noise for each
pixel in the map will underestimate the true uncertainty. Assuming an
on-source integration time per pixel of $t_{on}$ and a corresponding
integration time on the off position of $t_{off}$, the uncertainty in
the CO luminosity is given by
$$ \sigma_{L(CO)} =\sqrt{(1-a)\sum\sigma_i^2 + a(\sum\sigma_i)^2}
\times (4.848 D_{Mpc} \times L_{pix})^2 \eqno (4)$$
where $a = 1/(t_{off}/t_{on}+1$) (see Appendix~\ref{shared-offs}). 
For the jiggle map mode, $t_{off} = 4
t_{on}$ and so $a=0.2$. For the raster maps, the ratio of
$t_{off}/t_{on}$ varied with the size and dimensions of the map; we
have estimated an average value $a=1.07$ for the rasters and have used
this in the noise calculations for all the galaxies. 

For galaxies without detections, we calculated an uncertainty map
using equation (3) by adopting a line width of 100 km s$^{-1}$ and
then calculated $\sigma_{L(CO)}$ using equation (4). For galaxies
observed in jiggle map mode, the velocity width for
the baseline determination was always 400 km s$^{-1}$. For the galaxies
observed in raster mode, the actual region used to determine  the
baseline varied slightly from pixel to pixel as a result of weak
emission appearing from place to place in the spectrum. For these
galaxies, we used the actual region used for the baseline in
calculating $\sigma_i$ for each pixel.

Of the 31 large galaxies in the SINGS sample, 25 galaxies were
detected for a detection rate of 77\%.
Of the 16 smaller galaxies, only four galaxies (NGC3049, NGC 3773,
NGC 4625, and UGC 05720) are detected at $>4\sigma$ in the integrated
intensity maps. A fifth galaxy, NGC 3190, is marginally detected at the
$3\sigma$ level. Thus, the detection rate for the small SINGS galaxies
is 25\%, significantly lower than for the larger SINGS galaxies.
These smaller SINGS galaxies tend to be physically smaller galaxies
(often dwarf galaxies) and also have on average lower H\,{\sc i} fluxes than
their larger counterparts, both of which may impact the detection rate.

\subsection{Comparison with other surveys and the global CO line ratio}

\begin{figure*}
\includegraphics[width=120mm,angle=-90]{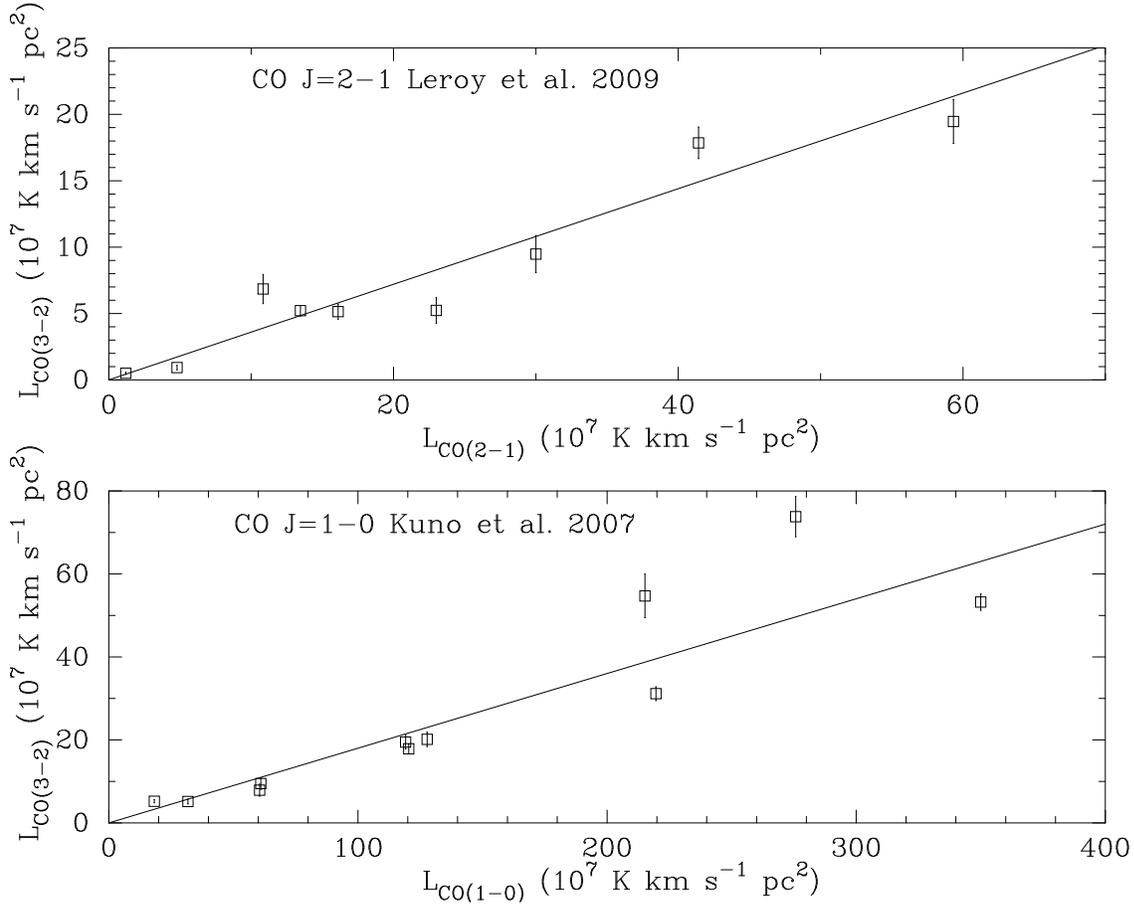}
\caption{(top) L$_{CO(3-2)}$ versus 
L$_{CO(2-1)}$ for all galaxies detected in both the NGLS and the
HERACLES survey \citep{l09}. The HERACLES luminosities have been
adjusted to use the distances adopted in this paper. Calibration
uncertainties in the HERACLES data are estimated at 20\% \citep{l09}.
The line shows
the mean CO J=3-2/2-1 ratio of 0.36 determined from the data points
(see text).
(bottom) L$_{CO(3-2)}$ versus 
L$_{CO(1-0)}$ for all galaxies detected in both the NGLS and the
survey of \citet{k07}. The CO J=1-0 luminosities have been
adjusted to use the distances adopted in this paper. The line shows
the mean CO J=3-2/1-0 ratio of 0.18 determined from the data points
(see text).
\label{fig-heracles}}
\end{figure*}

Because the different CO lines have different excitation temperatures
  and critical densities (\S\ref{sec-intro}), comparing the emission
  from different CO lines can provide 
  information on the physical conditions in the molecular gas. These
  different excitation conditions may also produce different radial
  distributions in the different CO lines.
The HERA CO-Line Extragalactic Survey, HERACLES, \citep{l09} mapped 18 nearby galaxies in the CO
J=2-1 line. The large area covered per galaxy and excellent
sensitivity make the data from this survey an excellent comparison to
the NGLS. Figure~\ref{fig-heracles} compares the CO luminosities for
the 9 galaxies that were detected in both surveys. (Three additional
galaxies in common had only upper limits in both surveys, and NGC 2841
was detected by HERACLES but not by us.) The correlation between the
two sets of global measurements is good, with a mean CO J=3-2/2-1 line
ratio of $0.36\pm0.04$ (where the quoted uncertainty is the standard
deviation of the mean; the standard deviation is 0.13). 
\citet{k07} mapped 40 nearby galaxies in the CO J=1-0 line, also
covering a large area with good sensitivity, although the fact
  that the maps are somewhat undersampled may introduce some scatter. 
Figure~\ref{fig-heracles} compares the CO luminosities for
the 11 galaxies that were detected in both surveys. The correlation
between the two sets of measurements is again good, with a mean
CO J=3-2/1-0 line ratio of $0.18 \pm 0.02$ (standard deviation 0.06).

This average CO J=3-2/1-0 line ratio is somewhat smaller than the
line ratios (0.4-0.8) seen in 
individual giant molecular clouds in M33 \citep{w97} and is at the low
end of the range (0.2-1.9) measured in the central regions of galaxies
\citep{m99,mao10}. However, the sample of \citet{mao10} includes roughly
50\% Seyfert, LINER, merging, or OH megamaser galaxies which tend to
show a higher than average line ratio.
Not surprisingly, our average line ratio is also 
at the small end of the line ratios (0.1-1.9) measured for
low-redshift luminous infrared galaxies \citep{l10,p11}.
Note also that we have not attempted in either comparison to correct for
differences in the fraction of the galaxy mapped in each survey.
Given the
differences in the processing techniques adopted between the three
surveys, the small scatter in this relationship is a good indicator
that both surveys are successfully measuring global CO luminosities in
the low signal-to-noise regime. A more detailed
analysis of spatially resolved images of the CO line ratios by
combining these various surveys with new $^{13}$CO data will be presented in
Rosolowsky et al. (in prep.).

\section{Comparison with other global properties}\label{sec-comp}

\subsection{The molecular mass fraction in the interstellar medium}

\begin{figure*}
\includegraphics[width=100mm,angle=-90]{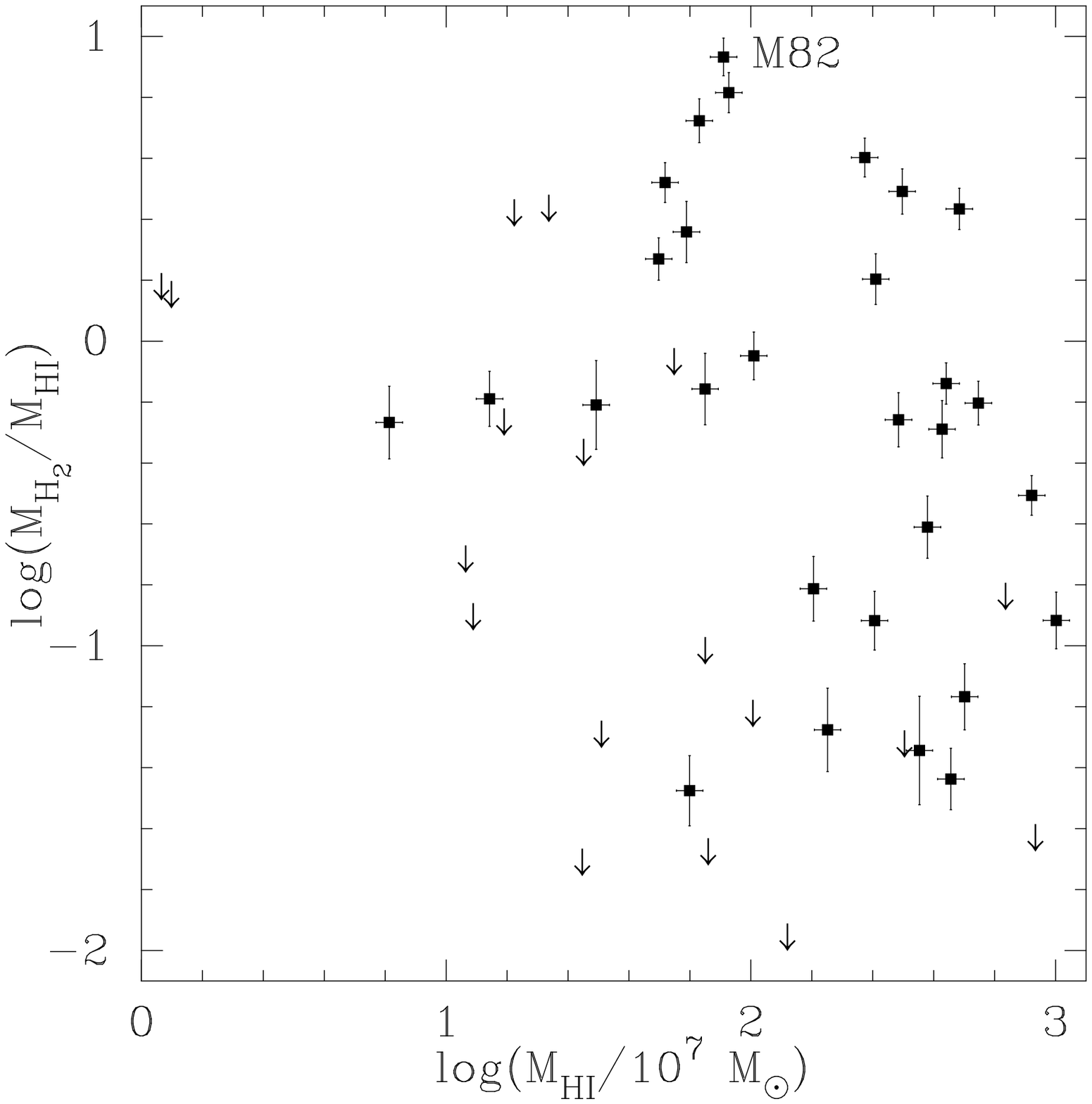}
\caption{The ratio of H$_2$ to H\,{\sc i} mass plotted as a function of H\,{\sc i}
  mass. Galaxies with only upper limits to the CO luminosity are shown
  as arrows. H$_2$ masses have been calculated assuming a CO
  J=3-2/J=1-0 line ratio of 0.18; this line ratio may not be
  appropriate for all galaxies, particularly the more luminous ones
  such as NGC 3034 (M82). 
\label{fig-H2fraction}}
\end{figure*}

To examine the molecular gas mass fraction in the interstellar medium
of the SINGS galaxies in our sample, we have converted the CO J=3-2
luminosities to 
molecular hydrogen mass adopting a CO-to-H$_2$ conversion factor of 
$X_{CO} = 2\times 10^{20}$ cm$^{-2}$ (K km s$^{-1}$)$^{-1}$
\citep{s88}. 
We have adopted a CO J=3-2/J=1-0 line ratio of 0.18 which is the mean
value derived from our comparison with the \citet{k07} sample. With
these assumptions, the molecular hydrogen mass is given by
$$ M_{H_2} = 17.8 (R_{31}/0.18)^{-1} L_{CO(3-2)}
\eqno (5)$$ where $R_{31}$ is the CO J=3-2/J=1-0 line ratio, 
$M_{H_2}$ is in M$_\odot$, and  
$L_{CO(3-2)}$ is in units of K km s$^{-1}$ pc$^2$. We note that the mean
value for the line ratio may not be appropriate for all galaxies; in
particular, it may result in an overestimate of the H$_2$ mass for
more luminous galaxies and an underestimate in less luminous galaxies.
We also note that adopting a single value for the CO-to-H$_2$
conversion factor, $X_{CO}$, will likely result in an underestimate of
the molecular gas mass in galaxies or regions of galaxies where the
metallicity is more than about a factor of two below solar
\citep{w95,a96,bol08}. 
Finally, our adopted value of $X_{CO}$ will likely result in an overestimate of
the gas mass in the starburst galaxy M82, as lower values of $X_{CO}$
are more appropriate in starburst and luminous infrared galaxies \citep{ds98}.

We have compiled H\,{\sc i} fluxes from the literature for all of our
sample. The prefered references were \citet{w08} 
and \citet{c09}, 
with most of the remaining galaxies retrieved from the HyperLeda database
\citep{p03}. The H\,{\sc i} flux for UGC 08201 is from \citet{c11}, while the
H\,{\sc i} flux for NGC 3190 is taken from \citet{martin98}. All fluxes have been
converted to H\,{\sc i} masses using the distances in Table~\ref{tbl-Ls} and
  the equations given in \citet{w08}.
Figure~\ref{fig-H2fraction} shows the H$_2$/H\,{\sc i} mass fraction as a
function of H\,{\sc i} mass for all the SINGS galaxies. Of the galaxies with
CO detections, 10 galaxies have
H$_2$/H\,{\sc i} mass fractions greater than 1, 9 galaxies of
$ 0.5 < M_{H_2}/M_{HI} < 1$ and 10 galaxies have 
H$_2$/H\,{\sc i} mass fractions less than 0.5. The 18 galaxies for which we
have only CO upper limits span a similar range of atomic gas masses,
although 78\% of these galaxies have H\,{\sc i}
masses less than $1.5\times 10^9$ M$_\odot$.
On average, these galaxies have significantly lower molecular gas
fractions than the detected galaxies.

The H$_2$/H\,{\sc i} mass fractions shown here are on average somewhat larger
than similar measurements for 14 galaxies given in \citet{l09}. For
the 9 galaxies in common with our survey, the  mass fraction given in
\citet{l09} is a factor of $\sim 1.5$ times smaller than the values
for the same galaxies given in Table~\ref{tbl-Ls}.  The primary reason
for this difference appears to be the CO line ratio. \citet{l09} adopt
a CO J=2-1/J=1-0 line ratio of 0.8 measured from comparing the peak line
temperatures with several surveys, including that of
\citet{k07}. On the other hand, comparing the global CO luminosities
for five galaxies in common between these two surveys gives an average
CO J=2-1/J=1-0 ratio of $0.52 \pm 0.06$. If we rescale  the mass
fractions given in \citet{l09} to this lower line ratio, the average
agreement between their mass fractions and our values is good. 

This comparison illustrates the importance of using the appropriate
value of the CO line ratio when observing galaxies in the higher CO
transitions. It is likely too simplistic to calculate gas masses using
a single value of the line ratio when the galaxy properties vary widely.
Indeed, the molecular gas mass calculated for NGC 3034 (M82) is
certainly an overestimate because the CO line ratios in this starburst
galaxy are much larger \citep{ward03}. Similarly, at least two of the
bright spirals in Virgo (NGC 4321 and NGC 4254) also show elevated
line ratios \citep{w09}. In contrast, for low
luminosity galaxies the molecular gas mass may be underestimated if
the CO excitation is very low. NGC 2841 has been detected in the CO
J=2-1 line by \citet{l09}; adopting a CO J=2-1/1-0 line ratio of 0.52
leads to a molecular mass fraction of 0.20, a factor of 10 above the
upper limit deduced from the CO J=3-2 upper limit. The fact that we
have measured upper limits to the CO luminosity over a 1$^\prime$
diameter aperture, which is roughly a factor of 7 smaller than the
area of the galaxy within $D_{25}/2$, may go some way to explaining
this discrepancy.

\subsection{The $L_{CO(3-2)}$--$L_{FIR}$ correlation}\label{sec-corr} 
\begin{figure*}
\includegraphics[width=100mm,angle=-90]{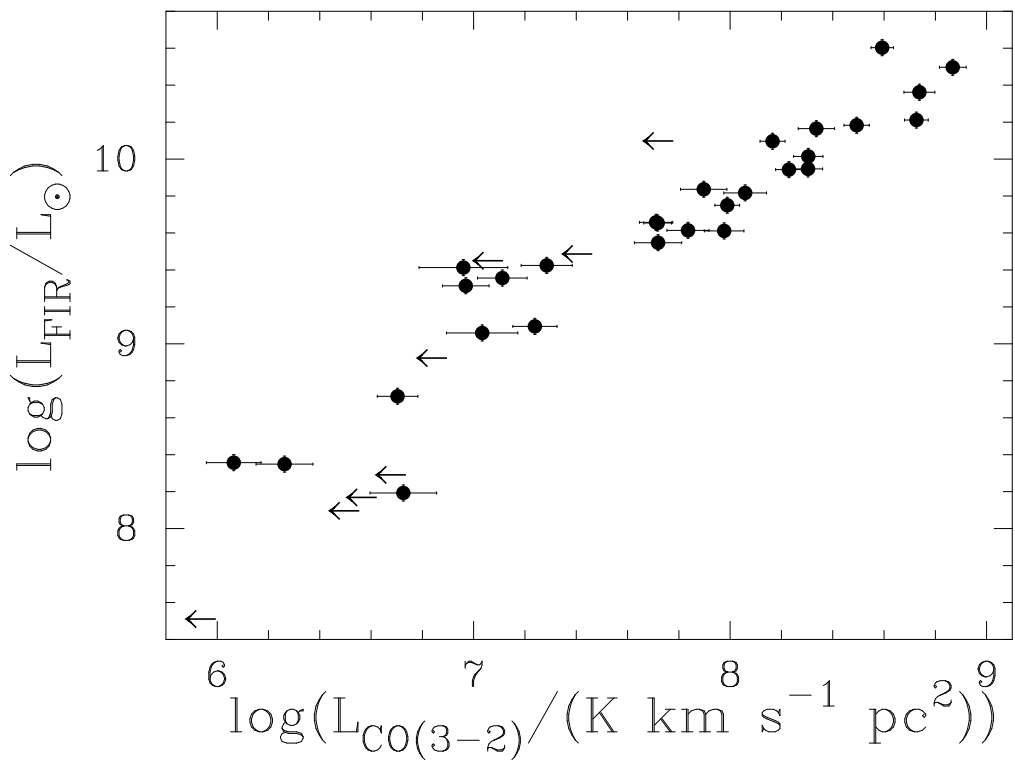}
\caption{$L_{FIR}$ plotted versus L$_{CO(3-2)}$ for all SINGS galaxies in the
  NGLS with far-infrared data. Galaxies which were not detected in CO
  are indicated by arrows.
\label{fig-LvL}}
\end{figure*}

\begin{figure*}
\includegraphics[width=100mm,angle=-90]{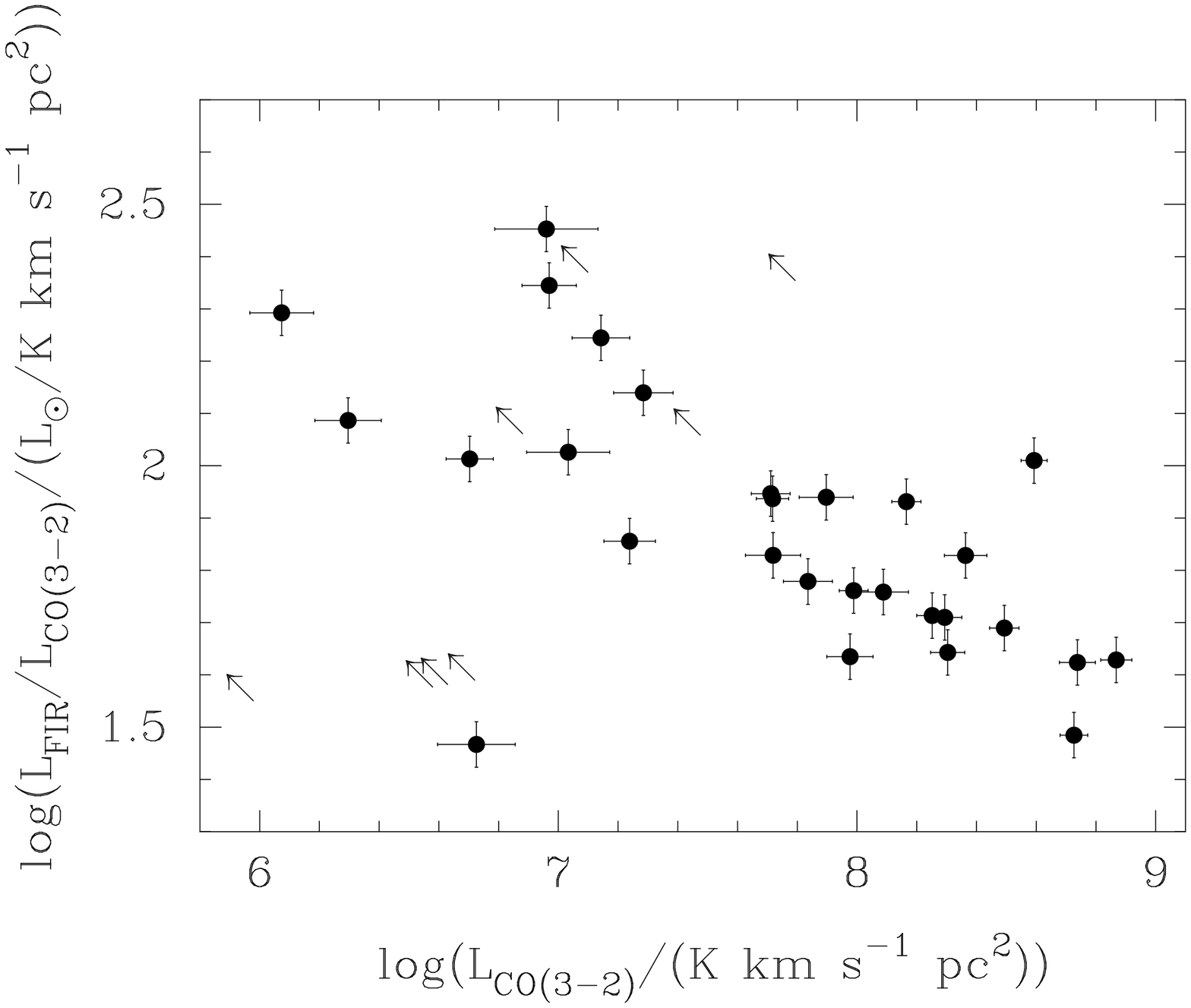}
\caption{$L_{FIR}/L_{CO(3-2)}$ plotted versus 
L$_{CO(3-2)}$ 
for all SINGS galaxies in the 
  NGLS with far-infrared data. Galaxies which were not detected in CO
  are indicated by arrows.
\label{fig-Lratios}}
\end{figure*}

We have used the NGLS CO measurements to examine the correlation between
the far-infrared luminosity and the CO J=3-2 luminosity. \citet{i09}
presented a nearly linear correlation between
these two luminosities, and suggested that the CO J=3-2 line could be
a good tracer of the dense gas involved in star formation, similar to
the HCN J=1-0 line \citep{g04}. We have collected
far-infrared luminosities derived from IRAS data for all the SINGS
galaxies for which such measurements are available. The preferred
source is the Revised Bright Galaxy Survey \citep{s03}
followed by \citet{l07}. For 8 additional galaxies we
retrieved the 60 and 100 $\mu$m fluxes from the IRAS Faint Source
Catalog and calculated the far-infrared luminosity using the formula
in \citet{l07}.  When necessary, published far-infrared
luminosities were adjusted for our adopted distances.

Figure~\ref{fig-LvL} plots the far-infrared luminosity as a function
of the CO J=3-2 luminosity, while Figure~\ref{fig-Lratios} plots the
ratio $L_{FIR}/L_{CO(3-2)}$ as a function of the CO
luminosity. These figures reveal a remarkably tight correlation for
the more luminous galaxies and increased scatter among the fainter galaxies.
Of 19 galaxies with $ 9.5 < \log L_{FIR} < 10.7$, only NGC 337
is not detected in the CO J=3-2 line.
In comparison to NGC 2841, which
has a similar lower limit to $L_{FIR}/L_{CO(3-2)}$ and which has been
detected in the CO 2-1 line \citep{l09}, NGC 337 is five times more
luminous in the infrared yet with a somewhat smaller atomic gas mass
and linear extent (as measured by the $D_{25}$ diameter). Despite its
inclusion in the SINGS sample \citep{k03}, this galaxy
has not been very well studied. One explanation of its unusually high
infrared luminosity would be the presence of an AGN. However, the GALEX
images \citep{g07} show no hint of a bright nuclear source while
the mid-infrared emission is extended and asymmetric \citep{bendo07}.
Its optical through infrared spectral energy distribution
is consistent with a normal spiral galaxy \citep{d07},
optical spectroscopy is dominated by star formation indicators,
and the metallicity is near solar \citep{m10}. This galaxy remains a
bit of a mystery and would be worthy of further study.

Excluding NGC 337, the mean $L_{FIR}/L_{CO(3-2)}$ ratio for the luminous galaxies
galaxies is $62 \pm 5$ with a standard deviation of 20.
The situation is more complicated for the fainter galaxies.
Of 12 galaxies with $ 8.3 < \log L_{FIR} < 9.5$, only 7 (58\%)  have
global CO luminosities that are 
detected at better than the $4\sigma$ level. 
For these 7 galaxies, the mean $L_{FIR}/L_{CO(3-2)}$ 
is $147 \pm 20$ with a standard deviation of 53. Some of the increased
scatter is likely due to the larger CO measurement uncertainties on
these fainter galaxies, but this seems unlikely to explain all the
scatter and the increased mean ratio. 
One possibility is that these lower luminosity and hence likely
lower mass galaxies may tend to have lower metallicities, in which
case the CO luminosity may systematically underestimate the molecular
hydrogen gas mass and cause them to appear underluminous in CO
\citep{i97,l11}.
However, the characteristic oxygen
abundance for these lower luminosity galaxies is not significantly
different from that of the higher luminosity systems
\citep{m10}. Another possibility is that
these fainter galaxies have lower average CO surface brightnesses, in
which case we could be 
systematically underestimating the CO luminosity due to the low
signal-to-noise ratio in the data. An example of this type of effect
is seen for NGC
2403, for which the CO luminosity increases by more than a factor of 2 when the
moment maps are made with a 2$\sigma$ cutoff compared to a 3$\sigma$
cutoff (see Section~\ref{subsec-CO}). This increase occurs because the
CO emission 
in NGC 2403 is very close to the noise limit over a large area of the galaxy.
Finally, if these lower luminosity galaxies 
tend to have systematically lower star formation rate surface
densities, the excitation of the CO 3-2 line relative to the ground
state line may be lower than in the more luminous galaxies, which
would again cause them to appear underluminous in CO J=3-2 (but
perhaps not in CO J=1-0).

We can use the average $L_{FIR}/L_{CO(3-2)}$ ratio to make a crude
estimate of the global molecular gas depletion time in the spiral
galaxies in our sample. We can convert $L_{CO(3-2)}$ to molecular gas
mass $M_{mol} = 1.36 M_{H_2}$ using equation (5) and $L_{FIR}$ to star
formation rate using equation (4) from \citet{k98}. We note that this
star formation rate equation is only strictly appropriate for
starburst  galaxies, in particular because the far-infrared luminosity
may include a contribution from dust heated by older stars; see
\citet{k98} for further details. We further convert from the $L_{IR}$
used in \citet{k98} to $L_{FIR}$ by adopting $L_{IR}=1.3L_{FIR}$
\citep{gc08} and convert to the double-power law initial mass function
used by \citet{c07} to give $SFR = 1.3\times 10^{-10} L_{FIR}$ with
$SFR$ in M$_\odot$ yr$^{-1}$ and $L_{FIR}$ in $L_\odot$.
With these two equations, the
average $L_{FIR}/L_{CO(3-2)}$ ratio of $62\pm 5$ corresponds to a
molecular gas
depletion time of $3.0\pm 0.3$ Gyr. Despite the uncertainties inherent
in this calculation, this mean value is in line with
with the recent estimate of the molecular gas depletion time of 2.35 Gyr
by \citet{b11}. The agreement is even more striking when we
consider that the two analyses use different CO data, different data
to trace the star formation rate, and one is a global measurement
while the other is based on resolved measurements.

\begin{figure*}
\includegraphics[width=100mm,angle=-90]{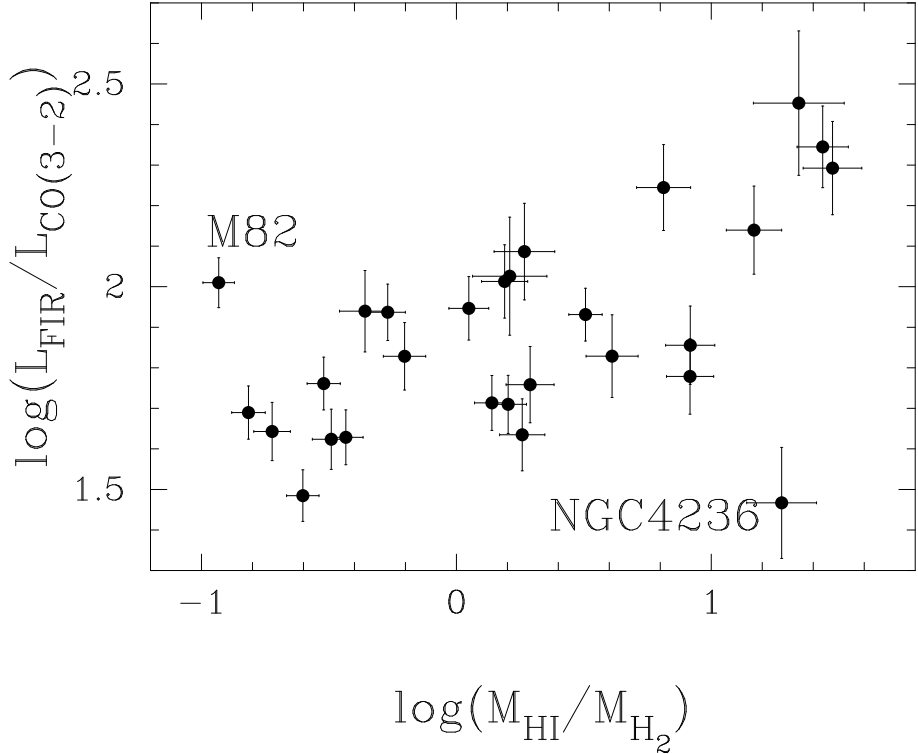}
\caption{$L_{FIR}/L_{CO(3-2)}$ plotted versus 
$M_{HI}/M_{H_2}$
for all SINGS galaxies in the 
  NGLS with far-infrared data. Galaxies which were not detected in CO
  are omitted from this figure
\label{fig-L_v_massratios}}
\end{figure*}

In contrast to this star formation rate analysis, \citet{bendo10,bendo11} and
others have presented evidence that much of the dust emission is at
least partially heated by a more quiescent stellar population. Even at
wavelengths as short as 60 and 100 $\mu$m,
there can be a significant contribution to dust heating from the
general interstellar radiation field; one of the most spectacular
examples of this phenomenon is M31, in which 90\% 
of the dust emission
is not directly associate with star formation \citep{w87}. Since the
far-infrared luminosity is also related to the total mass of dust, a
plot of $L_{FIR}$ versus $L_{CO}$ may be primarily probing the
molecular gas-to-dust mass ratio in galaxies. Figure~\ref{fig-L_v_massratios}
shows the $L_{FIR}/L_{CO(3-2)}$ ratio plotted as a function of
relative mass fractions of atomic and molecular gas,
$M_{HI}/M_{H_2}$. The two ratios are correlated at better than the
99\% level, even when outliers such as M82 and NGC 4236 are
included. If the far-infrared emission originated purely in dust
associated with the molecular phase of the ISM traced by the CO J=3-2
line, we would expect no  
correlation in this plot. At the opposite extreme, if the far-infrared
emission originated purely in dust associated with the atomic phase of the ISM,
we would expect a slope of unity. The observed slope of $\sim 0.2-0.3$
suggests that some of the far-infrared emission is originating in dust
associate with atomic gas or perhaps with more diffuse molecular gas
that does not emit strongly in the CO J=3-2 line, and that this is a
particularly important 
effect in galaxies where the ISM is predominantly atomic.  Galaxies in
the upper right portion of the figure tend to be either very early or
very late-type spiral galaxies. We note that if the H$_2$ mass is
underestimated in some of these galaxies due to low
metallicity, this effect would tend to steepen the slope and not
remove the observed trend. One additional complication in interpreting
this figure is that the spatial extent of the atomic gas tends to be
much larger than the CO or the far-infrared emission. The atomic gas
is distributed in a roughly constant surface density disk, while the
dust and molecular gas trace more of an exponential disk.

\subsection{$L_{CO(3-2)}$ and $L_{FIR}$ at low and high redshift\label{sec-highz} }

\citet{i09} presented a detailed comparison of the CO J=3-2 luminosity
for a sample of local ($D < 200$ Mpc) luminous and ultraluminous
infrared galaxies with a high-redshift sample of submillimeter
galaxies, quasars, and Lyman Break Galaxies. Including a small sample
of local star forming galaxies, they found a strong and nearly linear
correlation of $L_{CO(3-2)}$ and $L_{FIR}$ over nearly five orders of
magnitude, which suggested the star formation efficiency was constant
to within a factor of two across many different types of galaxies and
epochs. One limitation of the analysis was that the sample  of local
galaxies used was relatively small (14) and the data were limited to  a single
central pointing \citep{komugi07,m99}. In addition, the far-infrared
luminosity for the 22$^{\prime\prime}$ diameter aperture was calculated from an
H$_\alpha$-derived star formation rate. With the larger sample
available here, we are in a good position to re-examine this relation,
and to do so using global galaxy luminosities across all redshifts and
luminosities. 

\begin{figure*}
\includegraphics[width=100mm,angle=-90]{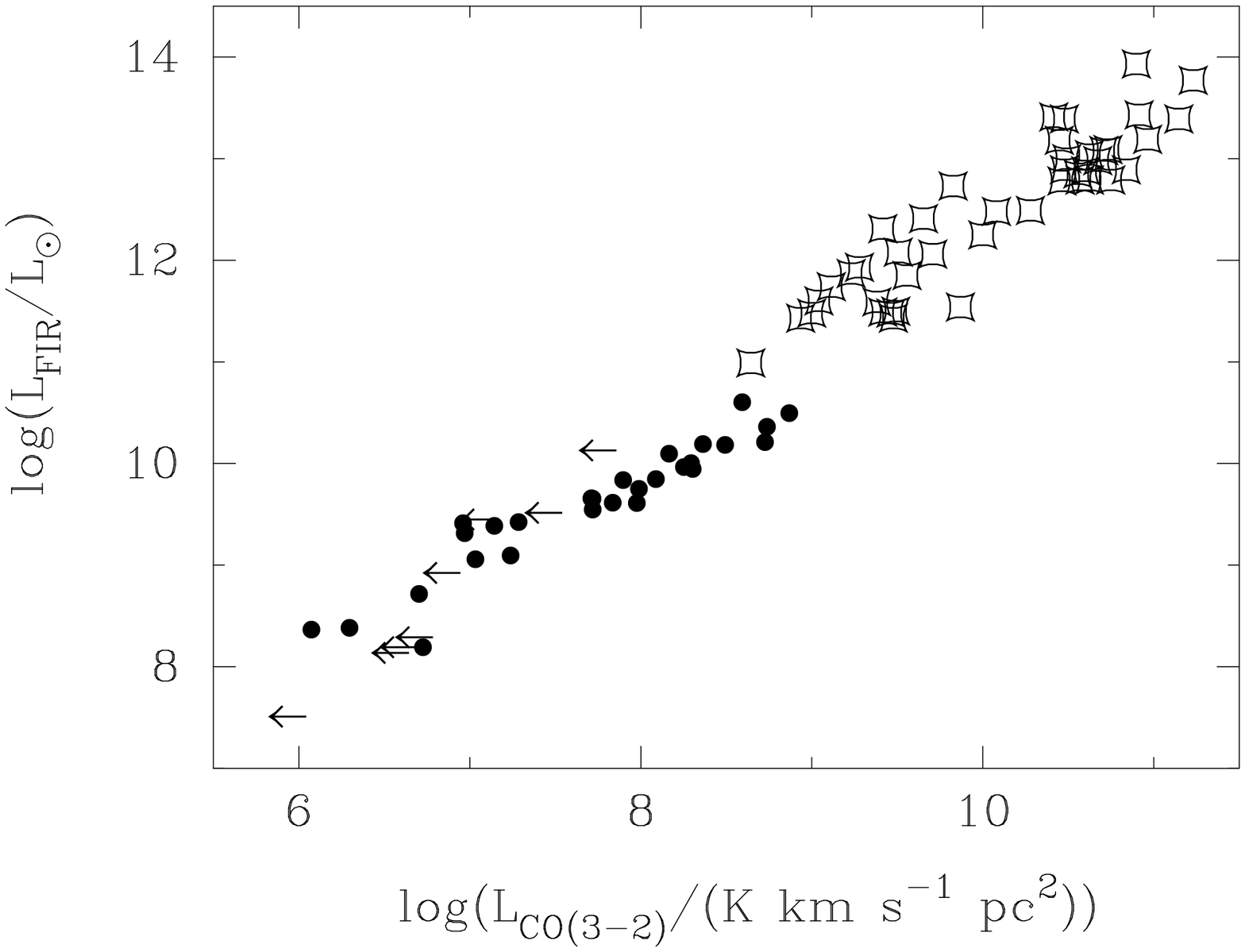}
\includegraphics[width=100mm,angle=-90]{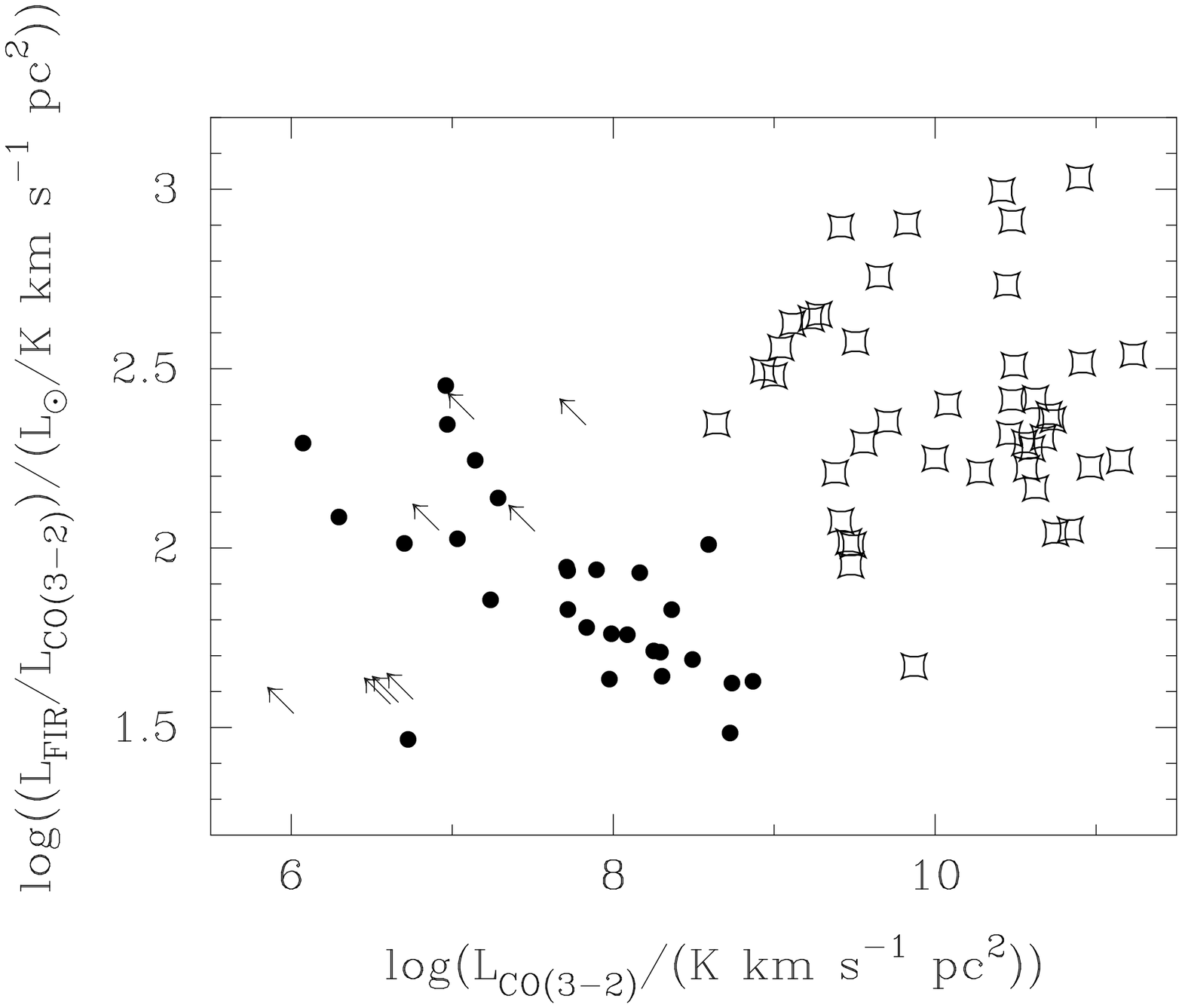}
\caption{(a) $\log L_{FIR}$ plotted versus 
$\log L_{CO(3-2)}$ 
for all SINGS galaxies in the 
  NGLS with far-infrared data (filled symbols and arrows; see 
Figure~\ref{fig-Lratios}). The local U/LIRG and high redshift samples from
\citep{i09} are plotted as the open symbols. 
(b) $\log L_{FIR}/L_{CO(3-2)}$ plotted versus 
$\log L_{CO(3-2)}$ for the same set of galaxies. 
\label{fig-iono}}
\end{figure*}

Figure~\ref{fig-iono} plots $L_{FIR}$ versus $L_{CO(3-2)}$ and
  also the ratio $L_{FIR}/L_{CO(3-2)}$ for both the
NGLS with the U/LIRG and high redshift galaxy sample from
\citet{i09}. This plot shows 
a significant trend of the
$L_{FIR}/L_{CO(3-2)}$ ratio as a function of CO
luminosity. The mean
$L_{FIR}/L_{CO(3-2)}$ for galaxies with $\log L_{FIR} > 11$ from the
local U/LIRG and high-redshift samples 
is $320 \pm 40$ with a standard deviation of 240. This is a factor of
five larger than the mean ratio of 60
measured for NGLS galaxies with
$9.5 < \log L_{FIR} < 10.7$. We can again convert the
$L_{FIR}/L_{CO(3-2)}$ to a molecular gas depletion time, but this time
using the CO-to-H$_2$ conversion factor (0.8) and CO J=3-2/J=1-0 line ratio
(0.5) appropriate for ULIRGs \citep{wils08}. Using equation (3) from
\citet{wils08} and the 
same star formation rate equation from \citet{k98} modified as
described above, the
$L_{FIR}/L_{CO(3-2)}$ ratio of 320 corresponds to a molecular gas
depletion time of only 50 Myr, more than a factor of 50 smaller than
the ratio for normal disk galaxies. A fit to the data for
galaxies with $L_{CO(3-2)} > 7.5$ (or $\log L_{FIR} > 9.5$) gives a slope
of $\sim 0.2$, which is consistent with 
a $\sim 3$ times higher line ratio for $10^{12}$ L$_\odot$ and
a $\sim 5$ times higher line ratio for $10^{13}$ L$_\odot$ than for a
typical local disk galaxy with $L_{FIR} = 10^{10}$ L$_\odot$.

Recently, \citet{g10} have analyzed the star formation properties of a
sample of low and high redshift galaxies, with one part of their
analysis examining the $L_{FIR}/L_{CO(1-0)}$ ratio. Their study combined CO
J=1-0 data for redshift zero galaxies with CO J=1-0 values for high
redshift galaxies that have been inferred from observations of the
CO J=2-1, J=3-2, and J=4-3 transitions assuming appropriate line
ratios for each type of galaxy. They also used a variety of star
formation rate tracers, which were then converted to $L_{FIR}$ using
the same equations as we have adopted above. A similar analysis with a
somewhat different high redshift sample has also been published by
\citet{daddi10}. Given the various
assumptions and conversions in these analyses of the
$L_{FIR}/L_{CO(1-0)}$ ratio, it is useful to compare their
results with our results which are based on direct measurements of
both the CO and the far-infrared luminosities without any need for
additional corrections.

The difference in the $L_{FIR}/L_{CO(3-2)}$ ratio of a factor of 5
between normal 
low-redshift galaxies and mergers at both low and high redshift is
consistent with the mean difference of 5.4 between the average
$L_{FIR}/L_{CO(1-0)}$ ratio of normal galaxies and merging galaxies
seen by \citet{g10}. The normal galaxy average value includes galaxies
at both low and high redshifts. When these data are converted to a gas
depletion time, the average for the low redshift galaxies is 1.5 Gyr
while the average for the high-redshift galaxies is 0.5 Gyr.
Using the average $L_{FIR}/L_{CO(1-0)}$ ratio given in Figure 2 of
\citet{g10}, we calculate 
an overall average depletion time of 1.2 Gyr for the normal
galaxies across all redshifts, which suggests that the 
$L_{FIR}/L_{CO(1-0)}$ ratio of the low and high redshift mergers is
$~\sim 7$ times higher than that of normal galaxies at z=0 alone.
Interestingly, the average $L_{FIR}/L_{CO(1-0)}$ ratio given in Figure
2 of \citet{g10} for the mergers implies a gas depletion time of only 70
Myr\footnote{Note that while \citep{g10} give a gas depletion time of
  200 Myr for these data, it appears that their calculation of this
  particular number neglected
  to take into account the smaller CO-to-H$_2$ conversion factor
  appropriate for
  merger galaxies.}, quite
consistent with our estimate of 50 Myr from the
$L_{FIR}/L_{CO(3-2)}$ ratio. Overall, the analysis presented here is
complementary to that of \citet{g10} and provides some evidence
that the conversions between CO transitions and different
star formation rate tracers used in \citet{g10} have not biased the
high-redshift analysis. 

\section{Conclusions}\label{sec-concl}

The James Clerk Maxwell Telescope Nearby Galaxies Legacy Survey (NGLS)
comprises an H\,{\sc i}-selected sample 
of 155 galaxies spanning all morphological types with distances less
than 25 Mpc. 
We have used new, large-area CO J=3-2 maps from the NGLS 
to examine the molecular gas
mass fraction and the correlation between far-infrared and CO
luminosity using 47 galaxies drawn from the SINGS sample \citep{k03}. We
find a good correlation of the CO J=3-2 luminosity with the CO J=2-1
luminosity \citep{l09} and the CO J=1-0 luminosity \citep{k07}, with
average global line ratios of $0.18 \pm 0.02$ for the CO J=3-2/J=1-0
line ratio and $0.36\pm 0.04$ for the CO J=3-2/2-1 line ratio. The
galaxies in our sample span a wide range of ISM mass and molecular gas
mass fraction, with 21\% of the galaxies in the sample having more molecular
than atomic gas and 60\% having molecular to atomic mass ratios less
than 0.5.

We explore the correlation of the far-infrared
luminosity, which traces star formation, with the CO luminosity, which
traces the molecular gas mass. We find that the more luminous galaxies
in our sample (with $\log L_{FIR} > 9.5$ or $\log L_{CO(3-2)} > 7.5$)
have a uniform $L_{FIR}/L_{CO(3-2)}$ ratio of $62\pm 5$. The lower
luminosity galaxies show a wider range of
$L_{FIR}/L_{CO(3-2)}$ ratio and many of these galaxies are not
detected in the CO J=3-2 line. We can convert this luminosity ratio to
a molecular gas depletion time and find a value of $3.0\pm 0.3$ Gyr,
in good agreement with recent estimates from \citet{b11}.
We also find a correlation of the $L_{FIR}/L_{CO(3-2)}$ ratio
with the mass ratio of atomic to molecular gas, which suggests
that some of the far-infrared emission originates from dust associated
with the atomic gas. This effect is particularly important in galaxies
where the gas is predominantly atomic.

By comparing the NGLS data with merging
galaxies at low and high redshift from the sample of \citet{i09},
which have also been observed in the 
CO J=3-2 line, we show that the $L_{FIR}/L_{CO(3-2)}$ ratio 
shows a significant trend with luminosity and thus that the
correlation of $L_{FIR}$ with L$_{CO(3-2)}$ is not as linear a trend as
found by \citet{i09}. Taking into account differences in the CO
J=3-2/J=1-0 line ratio and CO-to-H$_2$ conversion factor between the
mergers and the normal disk galaxies, this trend in the
$L_{FIR}/L_{CO(3-2)}$ ratio is
consistent with a molecular gas depletion time of only 50 Myr in the
merger sample, roughly 60 times shorter than in the nearby normal galaxies. 

\section*{Acknowledgments}

The James Clerk Maxwell Telescope is operated by The Joint Astronomy
Centre on behalf of the Science and Technology Facilities Council of
the United Kingdom, the Netherlands Organisation for Scientific
Research, and the National Research Council of Canada. The research of
P.B., P. C., S. C., M.F., M.H., J.I. E.R., E.S., K.S., T.W., and
C.D.W. is supported by grants from NSERC 
(Canada). A.U. has been supported through a Post Doctoral Research
Assistantship from 
the UK Science \& Technology Facilities Council.
Travel support for
B.E.W. and T.W. was supplied by the National Research Council (Canada).
We acknowledge the usage of the HyperLeda database
(http://leda.univ-lyon1.fr). 
C.V. received support from the ALMA-CONICYT Fund for the
Development of Chilean Astronomy (Project 31090013) and from the
Center of Excellence in Astrophysics and Associated Technologies (PBF
06).
This research has made use of the 
NASA/IPAC Extragalactic Database (NED) which is operated by the Jet
Propulsion Laboratory, California Institute of Technology, under
contract with the National Aeronautics and Space Administration. 


\appendix

\section{Field and Virgo galaxy membership in the NGLS}\label{members}

The galaxies chosen to constitute the Field and Virgo samples are
listed in Tables~\ref{tbl-field}-\ref{tbl-virgo}. Note that not every
Virgo galaxy with an H\,{\sc i} detection will be included in our sample
because of the H\,{\sc i} flux limit and the constraints on the size of the
sample. See \S~\ref{subsec-sample} for more details.

\renewcommand{\thefootnote}{\alph{footnote}}
\begin{table*}
 \centering
 \begin{minipage}{140mm}
  \caption{Field galaxies in the NGLS\label{tbl-field}}
  \begin{tabular}{lccc}
  \hline 
Name & $\alpha$(J2000.0)\footnotemark[1] & $\delta$(J2000.0)\footnotemark[1] & $V_{hel}$\footnotemark[2] \\
& (h:m:s) & ($^\circ$:$'$:$''$) & (km s$^{-1}$) \\
\\
 \hline
ESO 538$-$024 & 00:10:17.9 & $-$18:15:54 & 1551 \\
NGC 0210 & 00:40:35.1 & $-$13:52:26 & 1663 \\
NGC 0216 & 00:41:26.8 & $-$21:02:45 & 1541 \\
IC 0051 & 00:46:24.3 & $-$13:26:32 & 1753 \\
NGC 0274 & 00:51:01.7 & $-$07:03:22 & 1773 \\
NGC 0404 & 01:09:27.0 & +35:43:04 & 18 \\
NGC 0450 & 01:15:30.9 & $-$00:51:38 & 1829 \\
NGC 0473 & 01:19:55.1 & +16:32:40 & 2222 \\
NGC 0615 & 01:35:05.7 & $-$07:20:25 & 1948 \\
ESO 477$-$016 & 01:56:16.0 & $-$22:54:04 & 1646 \\
NGC 1036 & 02:40:29.0 & +19:17:49 & 802 \\
NGC 1140 & 02:54:33.6 & $-$10:01:42 & 1492 \\
NGC 1156 & 02:59:42.8 & +25:14:27 & 383 \\
ESO 481$-$019 & 03:18:43.3 & $-$23:46:55 & 1480 \\
NGC 1325 & 03:24:25.6 & $-$21:32:38 & 1646 \\
NGC 2146A & 06:23:54.3 & +78:31:50 & 1386 \\
NGC 2742 & 09:07:33.5 & +60:28:45 & 1285 \\
NGC 2787 & 09:19:19.1 & +69:12:12 & 695 \\
UGC 05272 & 09:50:22.4 & +31:29:16 & 524 \\
NGC 3077 & 10:03:20.1 & +68:44:01 & $-$14 \\
NGC 3162 & 10:13:31.6 & +22:44:15 & 1384 \\
NGC 3227 & 10:23:30.7 & +19:51:54 & 1131 \\
NGC 3254 & 10:29:20.1 & +29:29:32 & 1262 \\
NGC 3353 & 10:45:22.4 & +55:57:37 & 941 \\
NGC 3413 & 10:51:20.7 & +32:45:59 & 667 \\
NGC 3447B & 10:53:29.8 & +16:47:02 & 1023 \\
UGC 06029 & 10:55:02.3 & +49:43:33 & 1403 \\
NGC 3507 & 11:03:25.4 & +18:08:07 & 967 \\
UGC 06161 & 11:06:49.9 & +43:43:25 & 804 \\
ESO 570$-$019 & 11:20:12.2 & $-$21:28:14 & 1249 \\
UGC 06378 & 11:22:08.3 & +69:37:54 & 1309 \\
UGC 06566 & 11:35:43.6 & +58:11:33 & 1249 \\
NGC 3741 & 11:36:05.8 & +45:17:03 & 223 \\
UGC 06578 & 11:36:36.8 & +00:49:00 & 1178 \\
NGC 3782 & 11:39:20.6 & +46:30:51 & 748 \\
UGC 06792 & 11:49:23.3 & +39:46:15 & 832 \\
NGC 3931 & 11:51:13.4 & +52:00:03 & 929 \\
NGC 3928 & 11:51:47.7 & +48:40:59 & 951 \\
NGC 3998 & 11:57:56.2 & +55:27:13 & 1099 \\
NGC 4013 & 11:58:31.5 & +43:56:49 & 744 \\
IC 0750 & 11:58:52.0 & +42:43:19 & 694 \\
UGC 07009 & 12:01:44.0 & +62:19:40 & 1097 \\
NGC 4041 & 12:02:12.2 & +62:08:14 & 1221 \\
NGC 4117 & 12:07:46.2 & +43:07:35 & 886 \\
NGC 4138 & 12:09:29.8 & +43:41:07 & 908 \\
NGC 4190 & 12:13:44.2 & +36:37:53 & 192 \\
IC 3105 & 12:17:33.7 & +12:23:14 & $-$159 \\
NGC 4288 & 12:20:38.3 & +46:17:31 & 556 \\
UGC 07428 & 12:22:02.7 & +32:05:41 & 1170 \\
UGC 07512 & 12:25:41.1 & +02:09:35 & 1505 \\
NGC 4504 & 12:32:17.5 & $-$07:33:48 & 1000 \\
NGC 4550 & 12:35:30.7 & +12:13:15 & 409 \\
UGC 07827 & 12:39:39.0 & +44:49:12 & 552 \\
PGC 043211 & 12:47:59.9 & +10:58:32 & 1140 \\
NGC 4772 & 12:53:29.1 & +02:10:05 & 998 \\
 \hline
\end{tabular}
\begin{tabular}{l}
\footnotemark[1] From RC3 \citep{deV91} \\
\footnotemark[2] Systemic velocity from the H\,{\sc i} line (heliocentric)\\
\end{tabular}
\end{minipage}
\end{table*}

\renewcommand{\thefootnote}{\alph{footnote}}
\begin{table*}
 \centering
 \begin{minipage}{140mm}
  \caption{Field galaxies in the NGLS (continued)\label{tbl-field2}}
  \begin{tabular}{lccc}
  \hline
Name & $\alpha$(J2000.0)\footnotemark[1] & $\delta$(J2000.0)\footnotemark[1] & $V_{hel}$\footnotemark[2] \\
& (h:m:s) & ($^\circ$:$'$:$''$) & (km s$^{-1}$) \\
\\
 \hline
NGC 4941 & 13:04:13.0 & $-$05:33:06 & 1108 \\
PGC 045195 & 13:04:31.2 & $-$03:34:20 & 1362 \\
UGC 08303 & 13:13:17.9 & +36:12:56 & 929 \\
ESO 508$-$030 & 13:14:55.2 & $-$23:08:44 & 1510 \\
NGC 5477 & 14:05:33.2 & +54:27:40 & 328 \\
NGC 5486 & 14:07:25.1 & +55:06:10 & 1368 \\
PGC 140287 & 14:16:57.3 & +03:50:03 & 1497 \\
IC 1024 & 14:31:27.1 & +03:00:29 & 1411 \\
NGC 5701 & 14:39:11.1 & +05:21:49 & 1524 \\
IC 1066 & 14:53:02.9 & +03:17:45 & 1581 \\
PGC 057723 & 16:17:15.8 & $-$11:43:54 & 934 \\
NGC 6140 & 16:20:58.0 & +65:23:23 & 866 \\
NGC 6118 & 16:21:48.6 & $-$02:17:02 & 1611 \\
PGC 058661 & 16:38:08.9 & $-$04:49:23 & 1581 \\
IC 1254 & 17:11:33.5 & +72:24:07 & 1283 \\
NGC 7465 & 23:02:01.0 & +15:57:53 & 1972 \\
NGC 7742 & 23:44:15.8 & +10:46:01 & 1677 \\
 \hline
\end{tabular}
\begin{tabular}{l}
\footnotemark[1] From RC3 \citep{deV91} \\
\footnotemark[2] Systemic velocity from the H\,{\sc i} line (heliocentric)\\
\end{tabular}
\end{minipage}
\end{table*}

\renewcommand{\thefootnote}{\alph{footnote}}
\begin{table*}
 \centering
 \begin{minipage}{140mm}
  \caption{Small Virgo galaxies in the NGLS\label{tbl-virgo}}
  \begin{tabular}{lccc}
  \hline
Name & $\alpha$(J2000.0)\footnotemark[1] & $\delta$(J2000.0)\footnotemark[1] & $V_{hel}$\footnotemark[2] \\
& (h:m:s) & ($^\circ$:':'') & (km s$^{-1}$) \\
\\
 \hline
PGC 039265 & 12:16:00.4 & 04:39:04 & 2198 \\
NGC 4241 & 12:18:00.1 & 06:39:07 & 733 \\
NGC 4262 & 12:19:30.6 & 14:52:39 & 1363 \\
IC 3155 & 12:19:45.3 & 06:00:21 & 2209 \\
NGC 4268 & 12:19:47.2 & 05:17:01 & 2169 \\
NGC 4270 & 12:19:49.5 & 05:27:48 & 2351 \\
NGC 4277 & 12:20:03.8 & 05:20:29 & 2398 \\
NGC 4298 & 12:21:32.8 & 14:36:22 & 1126 \\
NGC 4301 & 12:22:27.3 & 04:33:58 & 1275 \\
NGC 4318 & 12:22:43.4 & 08:11:54 & 1227 \\
NGC 4324 & 12:23:05.9 & 05:14:59 & 1685 \\
NGC 4376 & 12:25:18.1 & 05:44:28 & 1161 \\
NGC 4383 & 12:25:25.6 & 16:28:12 & 1641 \\
NGC 4390 & 12:25:50.7 & 10:27:32 & 1131 \\
NGC 4394 & 12:25:55.7 & 18:12:50 & 862 \\
NGC 4423 & 12:27:09.0 & 05:52:49 & 1104 \\
IC 3365 & 12:27:11.6 & 15:53:46 & 2375 \\
NGC 4430 & 12:27:26.2 & 06:15:46 & 1468 \\
UGC 07590 & 12:28:18.8 & 08:43:46 & 1118 \\
NGC 4468 & 12:29:30.9 & 14:02:57 & 908 \\
NGC 4470 & 12:29:37.8 & 07:49:23 & 2366 \\
NGC 4522 & 12:33:39.7 & 09:10:26 & 2342 \\
NGC 4532 & 12:34:19.5 & 06:28:02 & 2052 \\
UGC 07739 & 12:34:45.0 & 06:18:06 & 1997 \\
IC 3522 & 12:34:45.9 & 15:13:13 & 668 \\
NGC 4561 & 12:36:08.2 & 19:19:21 & 1430 \\
NGC 4567 & 12:36:32.8 & 11:15:28 & 2227 \\
NGC 4568 & 12:36:34.3 & 11:14:19 & 2256 \\
IC 3583 & 12:36:43.7 & 13:15:32 & 1039 \\
IC 3591 & 12:37:02.6 & 06:55:33 & 1635 \\
IC 3617 & 12:39:24.7 & 07:57:52 & 2079 \\
NGC 4595 & 12:39:51.9 & 15:17:52 & 604 \\
NGC 4639 & 12:42:52.5 & 13:15:23 & 977 \\
NGC 4640 & 12:42:57.8 & 12:17:12 & 2082 \\
NGC 4647 & 12:43:32.6 & 11:34:57 & 1334 \\
NGC 4651 & 12:43:42.7 & 16:23:36 & 804 \\
\hline
\end{tabular}
\begin{tabular}{l}
\footnotemark[1] From RC3 \citep{deV91} \\
\footnotemark[2] Systemic velocity from the H\,{\sc i} line (heliocentric)\\
\end{tabular}
\end{minipage}
\end{table*}

\section{Effect of shared offs on calculating uncertainty for
  $L_{CO(3-2)}$}\label{shared-offs}

Both the jiggle and raster mapping modes used in this survey make use
of a shared ``off'' position in order to increase the mapping
efficiency. For the jiggle observations, the off position is observed
after each complete cycle through the 16 different pointings that
produce a complete map. For a raster observation, the off
position is observe after the completion of a scan of a single
row. The result of using a shared off is that the noise is somewhat
correlated between different pixels in the map because the same off
position is used for many ``on'' observations. While the noise is
estimated properly for any individual pixel in the map, summing
many pixels to obtain the total luminosity of an extended object
requires a careful estimate of the resulting uncertainty. In effect,
if $\sigma_i$ is the measurement uncertainty in a single pixel, then
$\sqrt{\sum \sigma_i^2}$ will underestimate the uncertainty in the
summed luminosity because of the effect of the shared offs.

Consider first a very simple situation where we have $n$ ONs all
sharing the same single OFF measurement.
If we average $n$ ONs, the uncertainty in that average is given by
$$\sigma_{ON,avg} = \sqrt{\sum_{i=1}^n \sigma_{ON,i}^2}/n$$
But since the OFF is identical for all measurements, we have
$\sigma_{OFF,avg} = \sigma_{OFF}$ and the uncertainty in the average
difference ON-OFF is given by
$$\sigma_{ON-OFF,avg} = \sqrt{{\sum_{i=1}^n \sigma_{ON,i}^2}/n^2
+ \sigma_{OFF,j}^2}$$
We can rewrite this equation as
$$n\sigma_{ON-OFF,avg} = \sqrt{{\sum_{i=1}^n \sigma_{ON,i}^2}
+ n^2\sigma_{OFF,j}^2} \eqno {\rm (B1)}$$
to arrive at an expression for  $\sigma_L = n\sigma_{ON-OFF,avg}$, the
uncertainty in the sum of these $n$ pixels.

We now derive the appropriate formula for the uncertainty in the presence
of shared offs where not every pixel has the same off. Suppose we are
summing $N$ 
measurements, where each measurement can be represented as the
difference between an ON measurement and an OFF measurement (modulo
scaling factors such as system temperature and so on). So then
the luminosity can be represented by
$$ L = \sum_{i=1}^N ON_i - \sum_{j=1}^M N_j OFF_j$$
where there are $N$ ONs, $M$ OFFs, and $\sum_{j=1}^M N_j =N$. Let
the uncertainty in $ON_i$ be $\sigma_{ON,i}$ and
the uncertainty in $OFF_j$ be $\sigma_{OFF,j}$. 
The uncertainty in $L$, $\sigma_L$, is then given by
$$ \sigma_L^2 = \sum_{i=1}^N \sigma_{ON,i}^2 + \sum_{j=1}^M (N_j
\sigma_{OFF,j})^2 \eqno {\rm (B2)} $$ 
It is convenient to define 
$(N\overline{\sigma}_{OFF})^2 = \sum_{j=1}^M (N_j \sigma_{OFF,j})^2$ 
so that
$$ \sigma_L^2 = \sum_{i=1}^N \sigma_{ON,i}^2 + (N\overline{\sigma}_{OFF})^2$$

Now, we do not have access to the separate ON and OFF measurements,
merely the difference which gives us the signal in each individual
pixel. So we will have to be a little clever in getting at the
appropriate uncertainty for the sum of many pixels.
The measurement
uncertainty in any single pixel is given by
$ \sigma_{meas,i}^2 = \sigma_{ON,i}^2 +\sigma_{OFF,j}^2$ and so
$$ \sum_{i=1}^N\sigma_{meas,i}^2 = \sum_{i=1}^N \sigma_{ON,i}^2 +
\sum_{j=1}^M N_j \sigma_{OFF,j}^2$$
Now it is easy to show that
$$ \sigma_L^2 = \sum_{i=1}^N \sigma_{meas,i}^2 - 
\sum_{j=1}^M N_j \sigma_{OFF,j}^2 +
(N \overline{\sigma}_{OFF})^2 \eqno{\rm (B3)}$$

Now, for a given receptor in the detector array, 
$\sigma_{OFF,j}^2 = a \overline{\sigma_{meas}^2} = a
\sum_{j=1}^{N_j} \sigma_{meas,j}^2/N_j$
where $a = 1/(t_{off}/t_{on}+1)$ and $t_{off}$ and
$t_{on}$ are the integration times spent on the OFF and the ON
position, respectively. 
Similarly,
$\sigma_{OFF,j} = 
\sqrt{a} \sum_{j=1}^{N_j} \sigma_{meas,j}/N_j$  
and we can write equation (B3) as
$$ \sigma_L^2 = \sum_{i=1}^N \sigma_{meas,i}^2 - 
a \sum_{i=1}^N \sigma_{meas,i}^2  
+ a (\sum_{i=1}^N \sigma_{meas,i})^2  $$
and we arrive at our final equation for the uncertainty in the sum of
$N$ pixels in the presence of shared offs, 
$$ \sigma_L^2 =(1-a)\sum_{i=1}^N\sigma_{meas,i}^2 + a(\sum_{i=1}^N\sigma_{meas,i})^2 $$
For the jiggle map mode, $t_{off} = 4
t_{on}$ and so $a=0.2$. For the raster maps, the ratio of
$t_{off}/t_{on}$ varied with the size and dimensions of the map; we
have estimated an average value $a=1.07$ for the rasters and have used
this in the noise calculations for all the galaxies.

\section{Images of the SINGS galaxies from the NGLS}\label{images}

In this appendix, we present the images derived from the CO $J$=3-2
observations of the NGLS galaxies that are also members of the SINGS
sample \citep{k03}. Only galaxies with significant detections are
shown here; upper limits to the CO luminosity for the remaining galaxies
are given in Table~\ref{tbl-Ls}.

Two images are shown for each galaxy: the CO $J$=3-2
integrated intensity image, and the CO contours overlaid on an optical
image from the Digitized Sky Survey. For those galaxies with
sufficient signal-to-noise, we also show the first moment map derived
from the CO data cube, which traces the velocity field and
the second moment map, which traces the
velocity dispersion.
The second moment maps for nine galaxies 
with good signal to noise and inclinations less than 60$^o$ have been
presented and discussed in \citet{w11} using the same processing
techniques used in this paper and so are not reproduced here.

\begin{figure*}
\includegraphics[width=55mm]{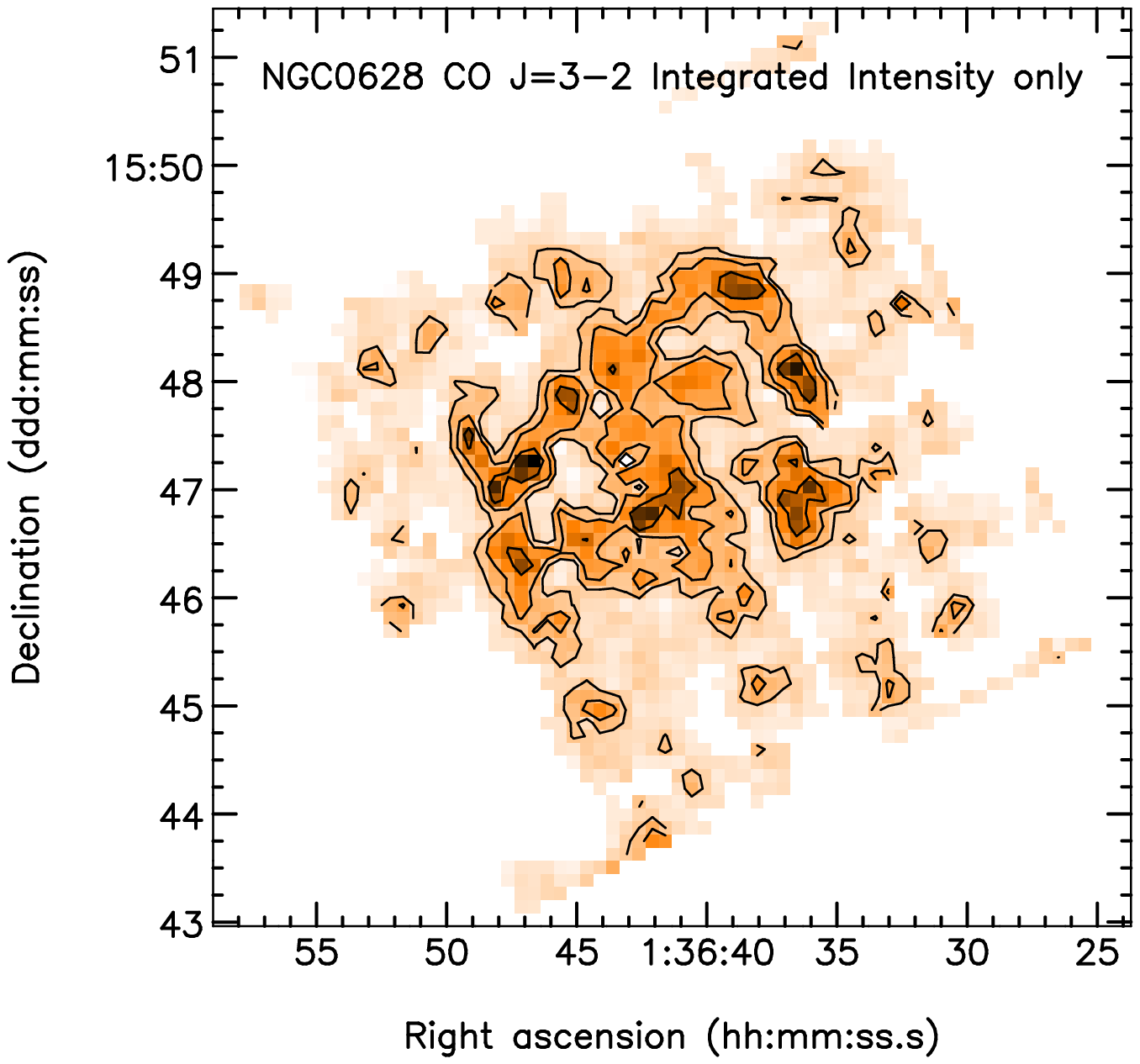}
\includegraphics[width=55mm]{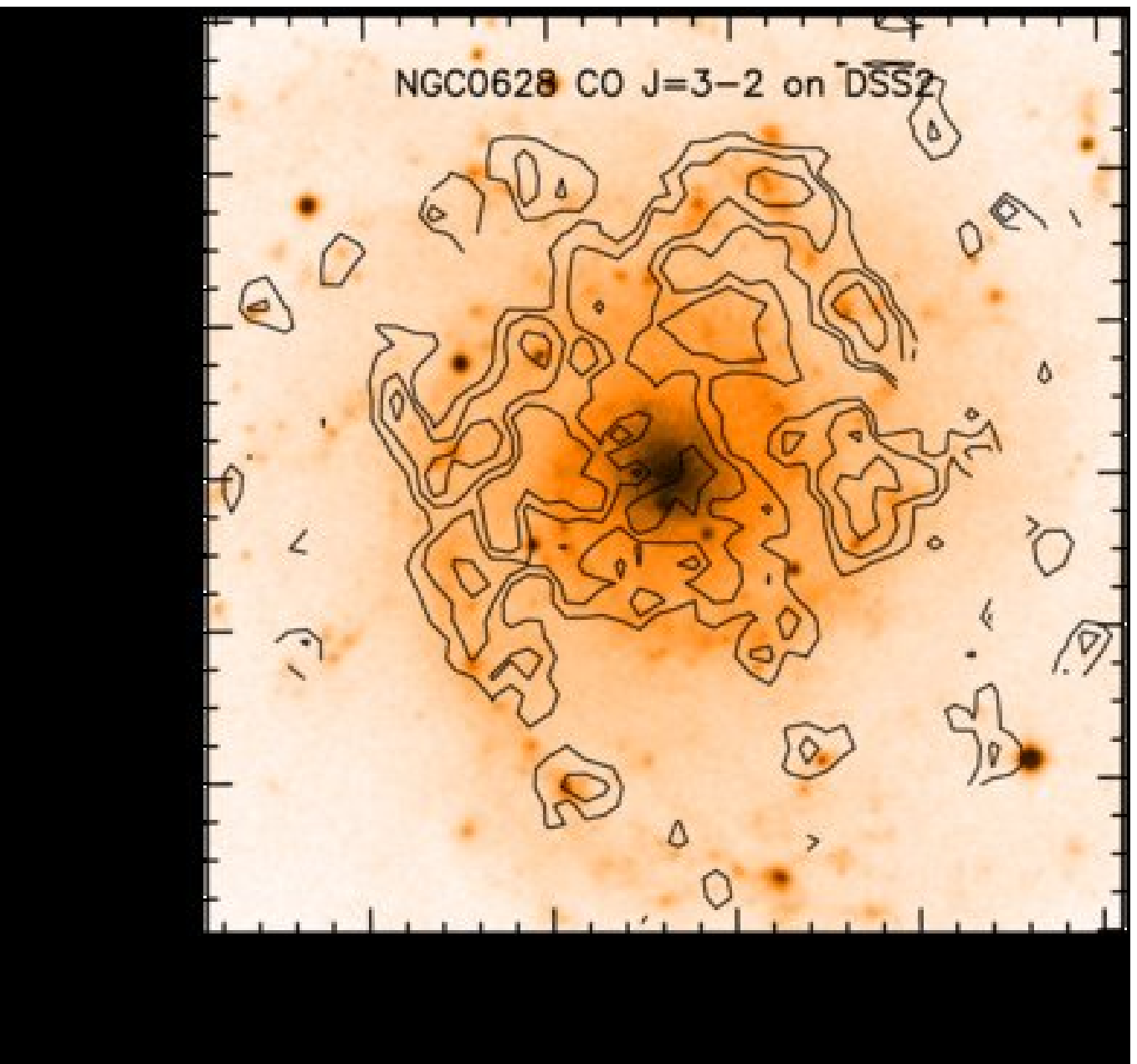}
\includegraphics[width=55mm]{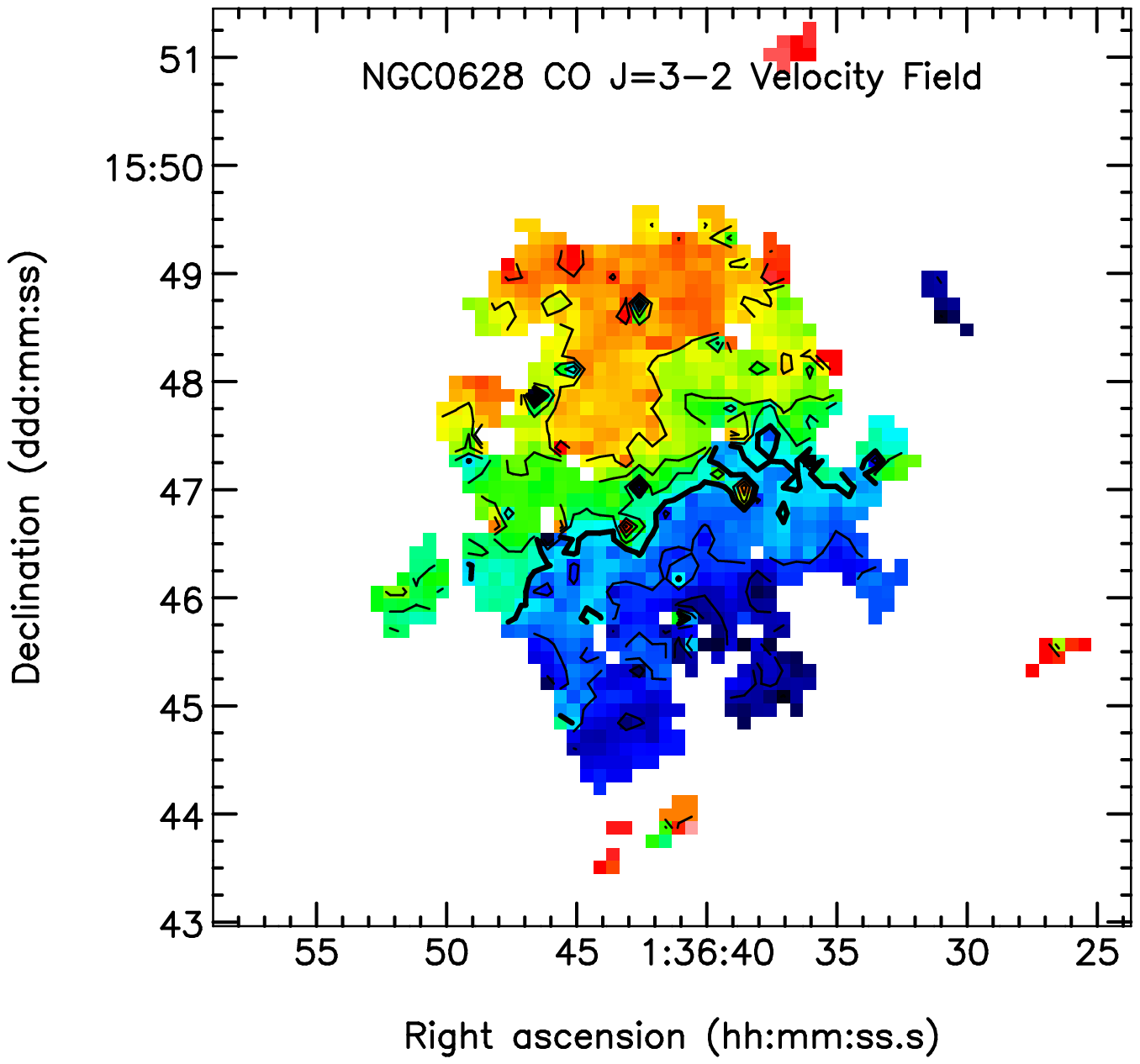}
\caption{CO $J$=3-2 images for NGC 0628. (a) CO $J$=3-2 integrated intensity
  image. Contours levels are (0.5,   1,   2)
K km s$^{-1}$ (T$_{MB}$).
(b) CO $J$=3-2 integrated intensity contours overlaid on an optical
image from the Digitized Sky Survey. (c) Velocity field as traced by the
CO $J$=3-2 first moment map. Contour levels are (624,   632,   640,
648,   656,   664,   672,   680) km s$^{-1}$. The thick line shows the
systemic velocity given in Table~\ref{tbl-large}.  
Similar images derived from the same data have been
published in \citet{warren10}. 
The velocity dispersion map has been published in \citet{w11}.
\label{fig-ngc628}}
\end{figure*}

\begin{figure*}
\includegraphics[width=60mm]{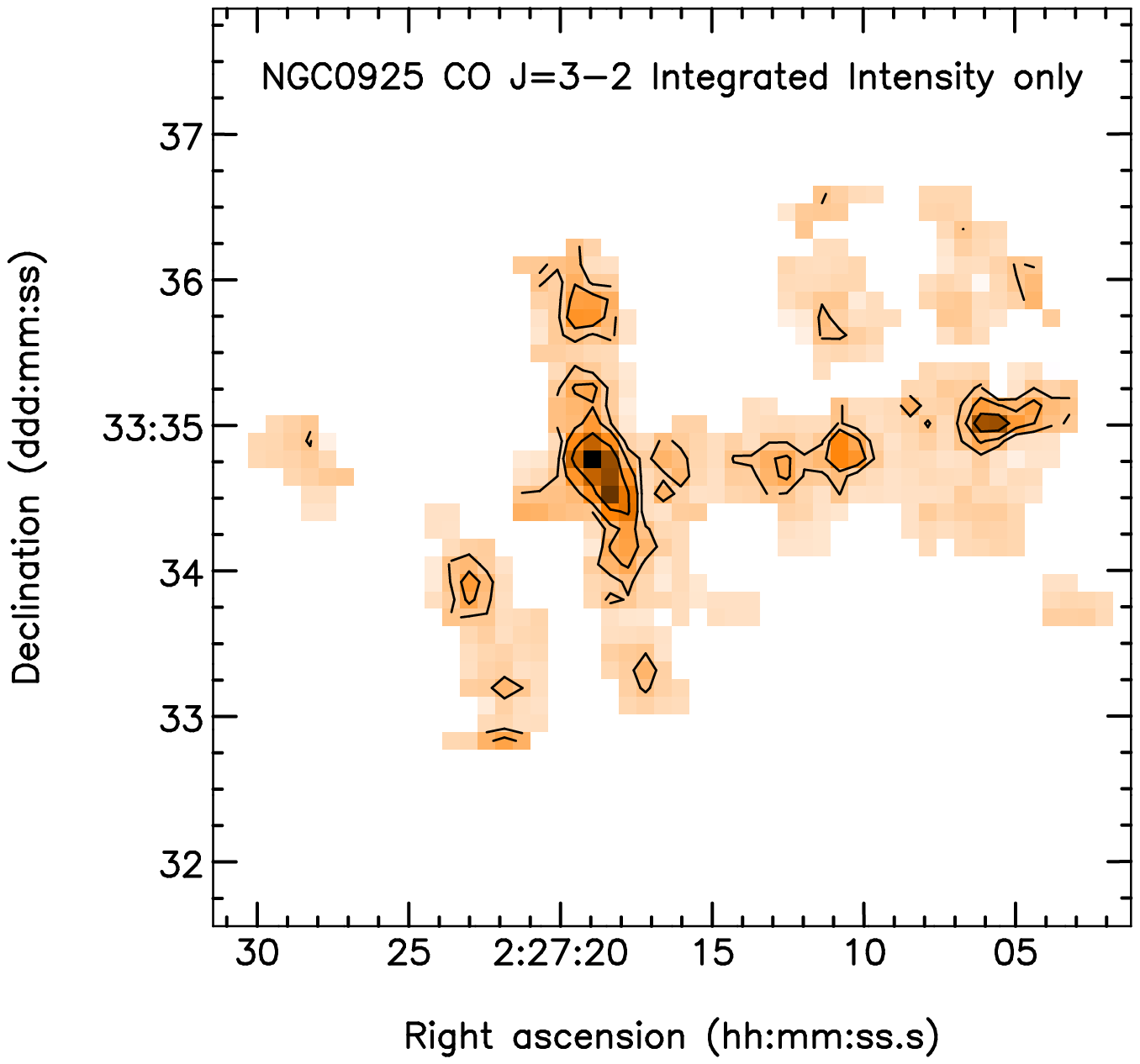}
\includegraphics[width=60mm]{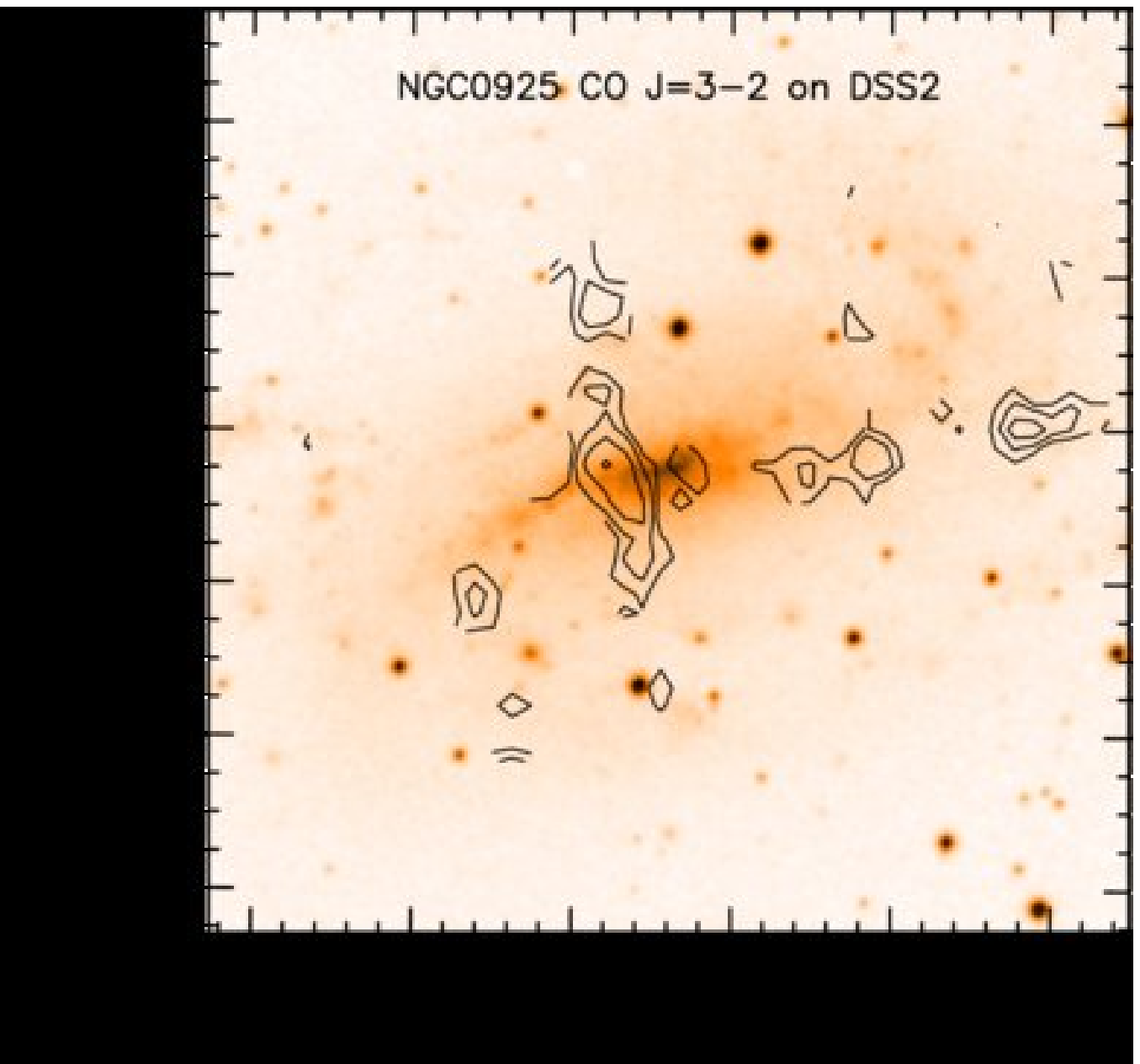}
\includegraphics[width=60mm]{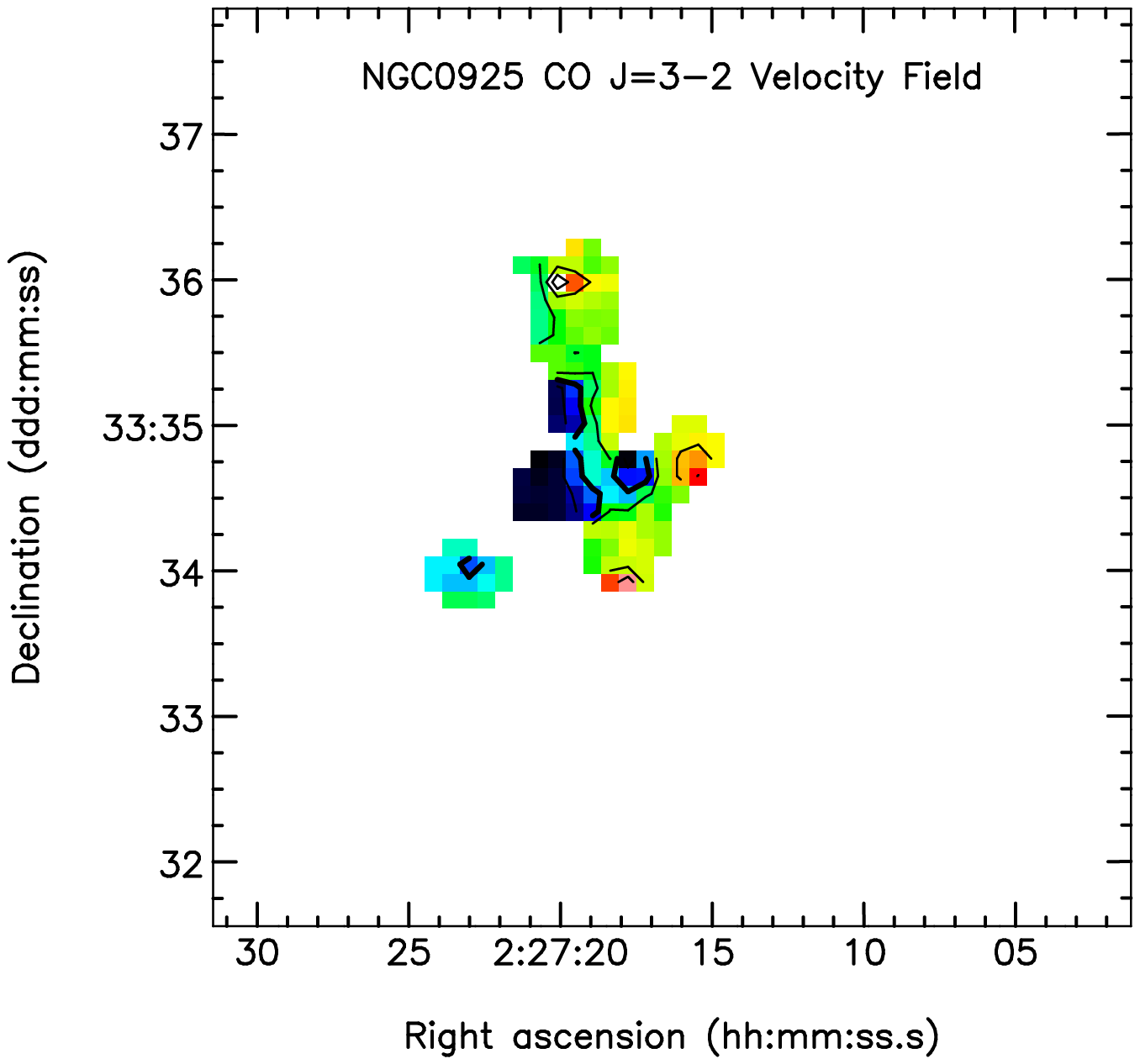}
\includegraphics[width=60mm]{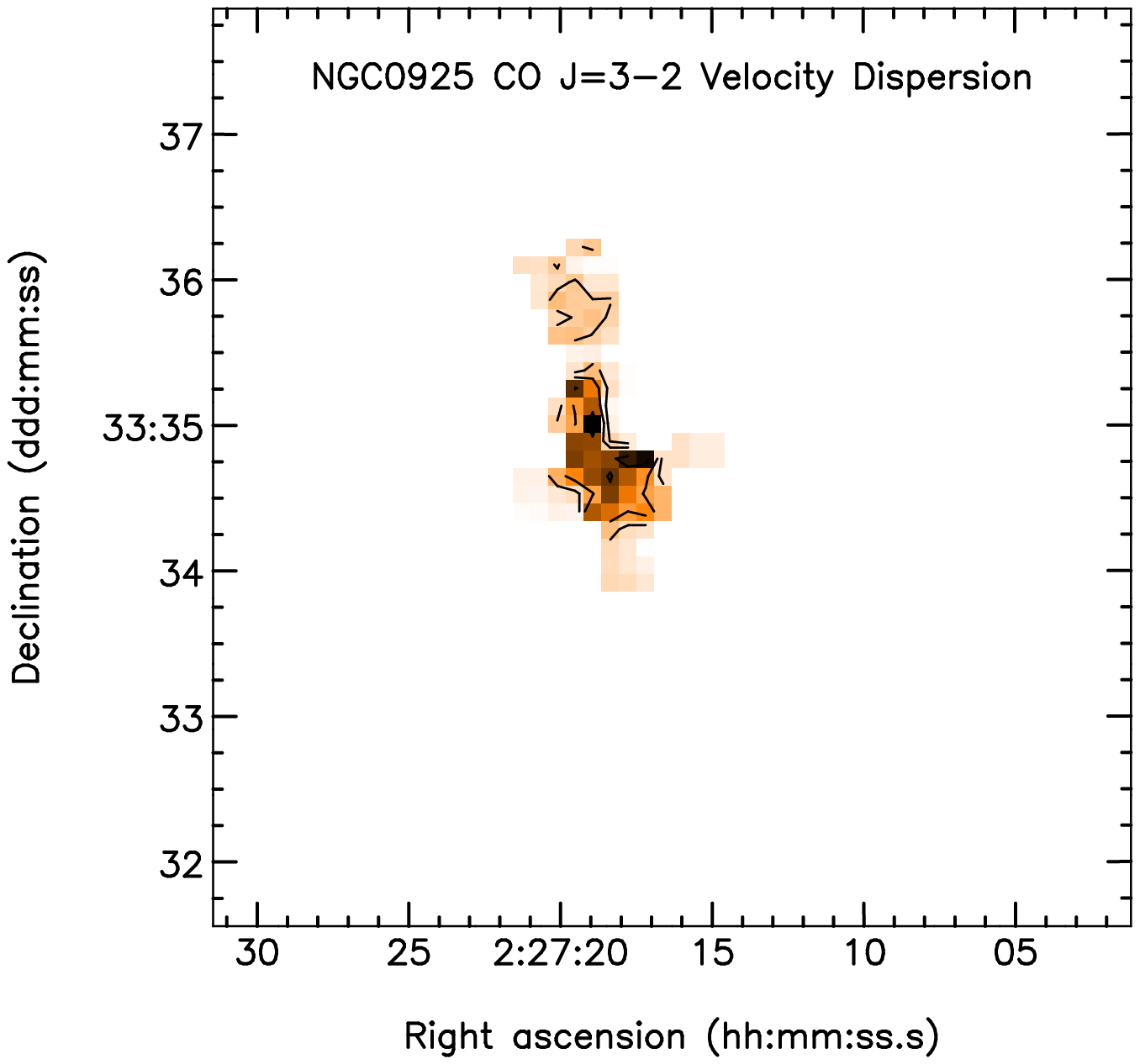}
\caption{CO $J$=3-2 images for NGC 0925.  (a) CO $J$=3-2 integrated intensity
  image. Contours levels are (0.5,   1,   2,   4) K km s$^{-1}$ (T$_{MB}$).
(b) CO $J$=3-2 overlaid on a
Digitized Sky Survey image. (c) Velocity field. Contour levels are
(525,   540,   555,   570,   585) 
km s$^{-1}$. 
(d) The velocity dispersion $\sigma_v$  as traced by the 
CO $J$=3-2 second moment map.  Contour levels are
(4,   8,   16)
km s$^{-1}$. 
\label{fig-ngc0925}}
\end{figure*}

\begin{figure*}
\includegraphics[width=55mm]{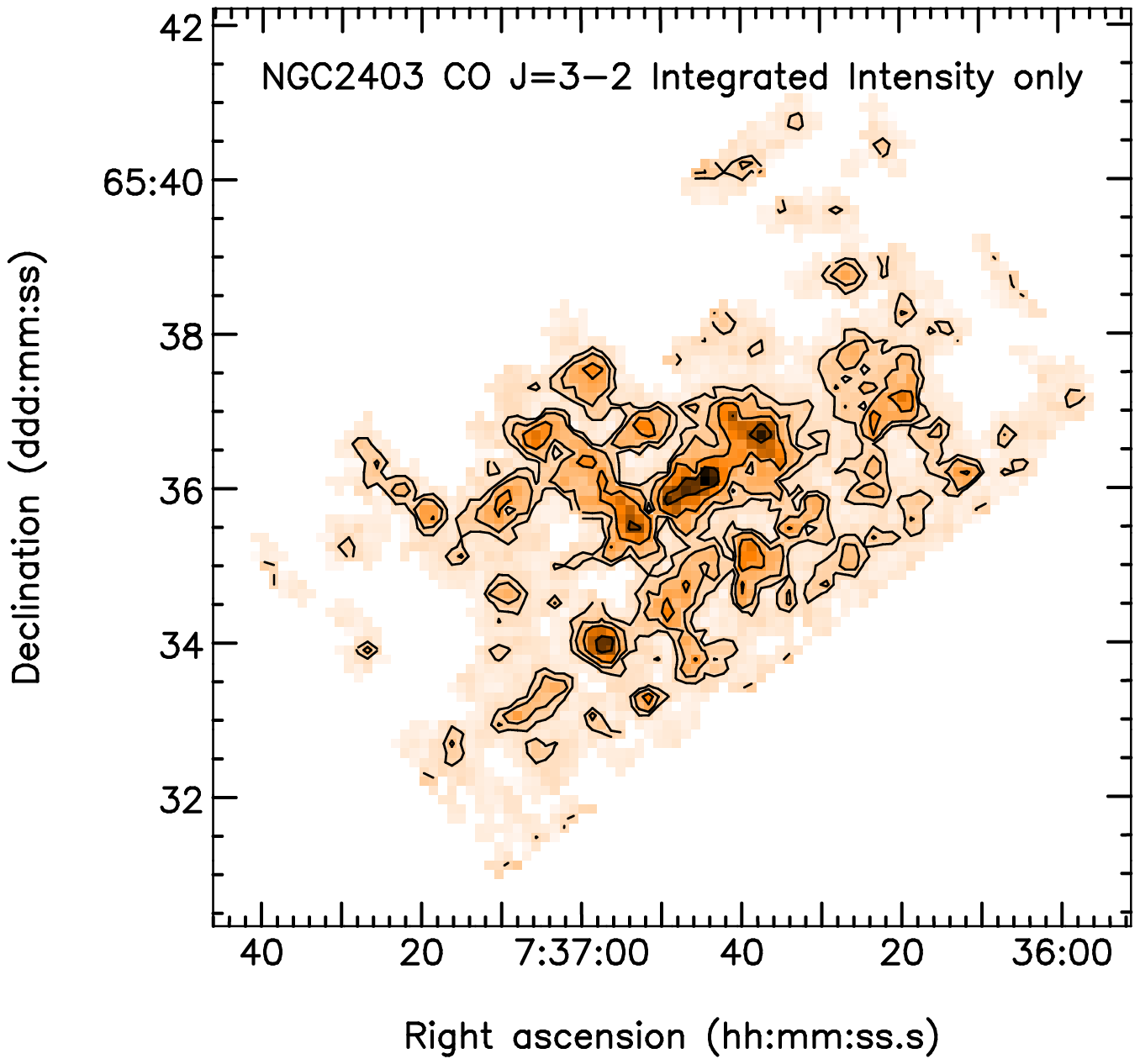}
\includegraphics[width=55mm]{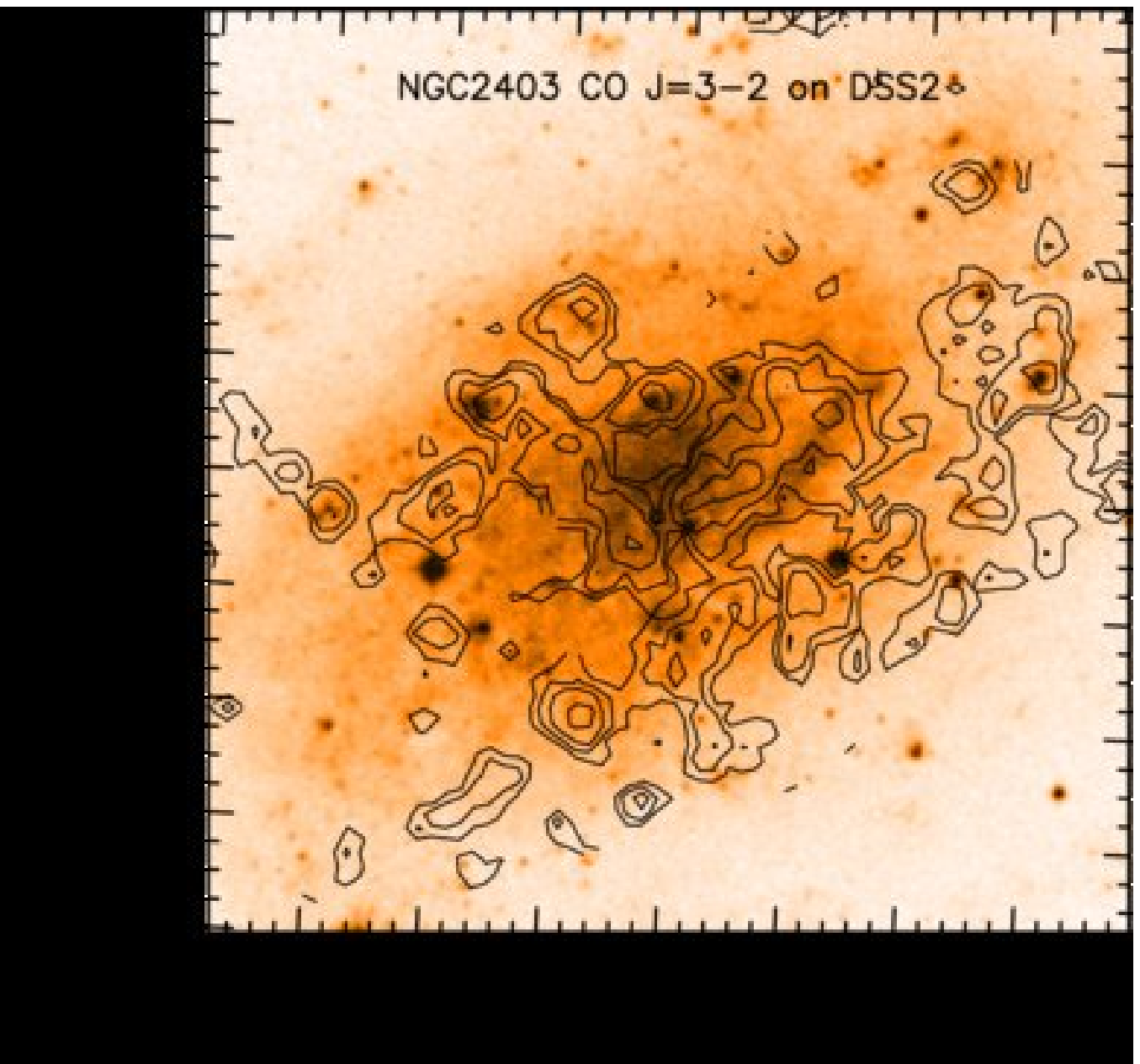}
\includegraphics[width=55mm]{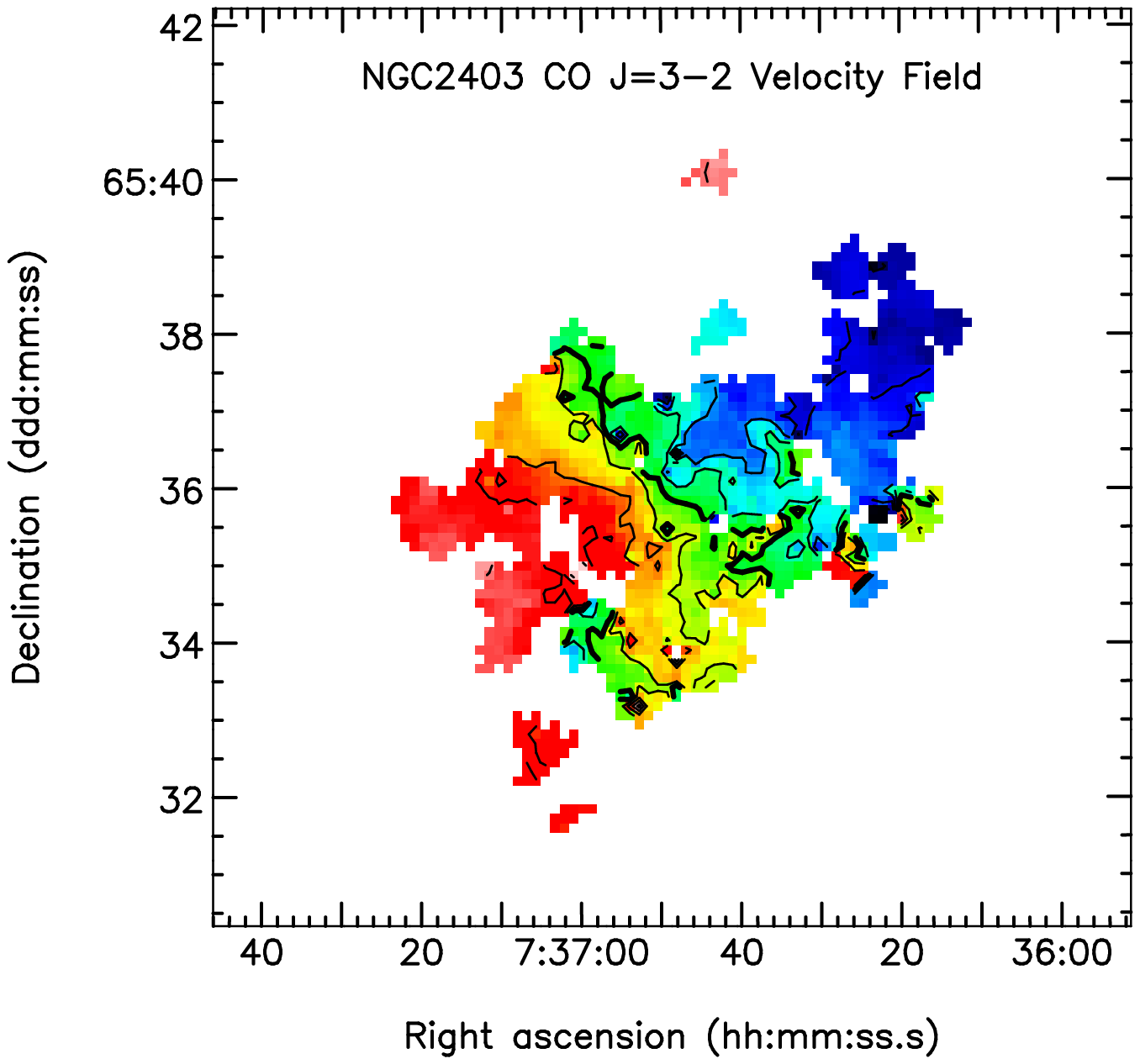}
\caption{CO $J$=3-2 images for NGC 2403. (a) CO $J$=3-2 integrated intensity
  image. Contours levels are (0.5,   1,   2,   4)
K km s$^{-1}$ (T$_{MB}$).
(b) CO $J$=3-2 integrated intensity contours overlaid on an optical
image from the Digitized Sky Survey. (c) Velocity field as traced by the
CO $J$=3-2 first moment map. Contour levels are (21,   46,   71,   96,   121,   146,   171,   196,   221) km s$^{-1}$.
The velocity dispersion map has been published in \citet{w11}.
\label{fig-ngc2403}}
\end{figure*}

\begin{figure*}
\includegraphics[width=60mm]{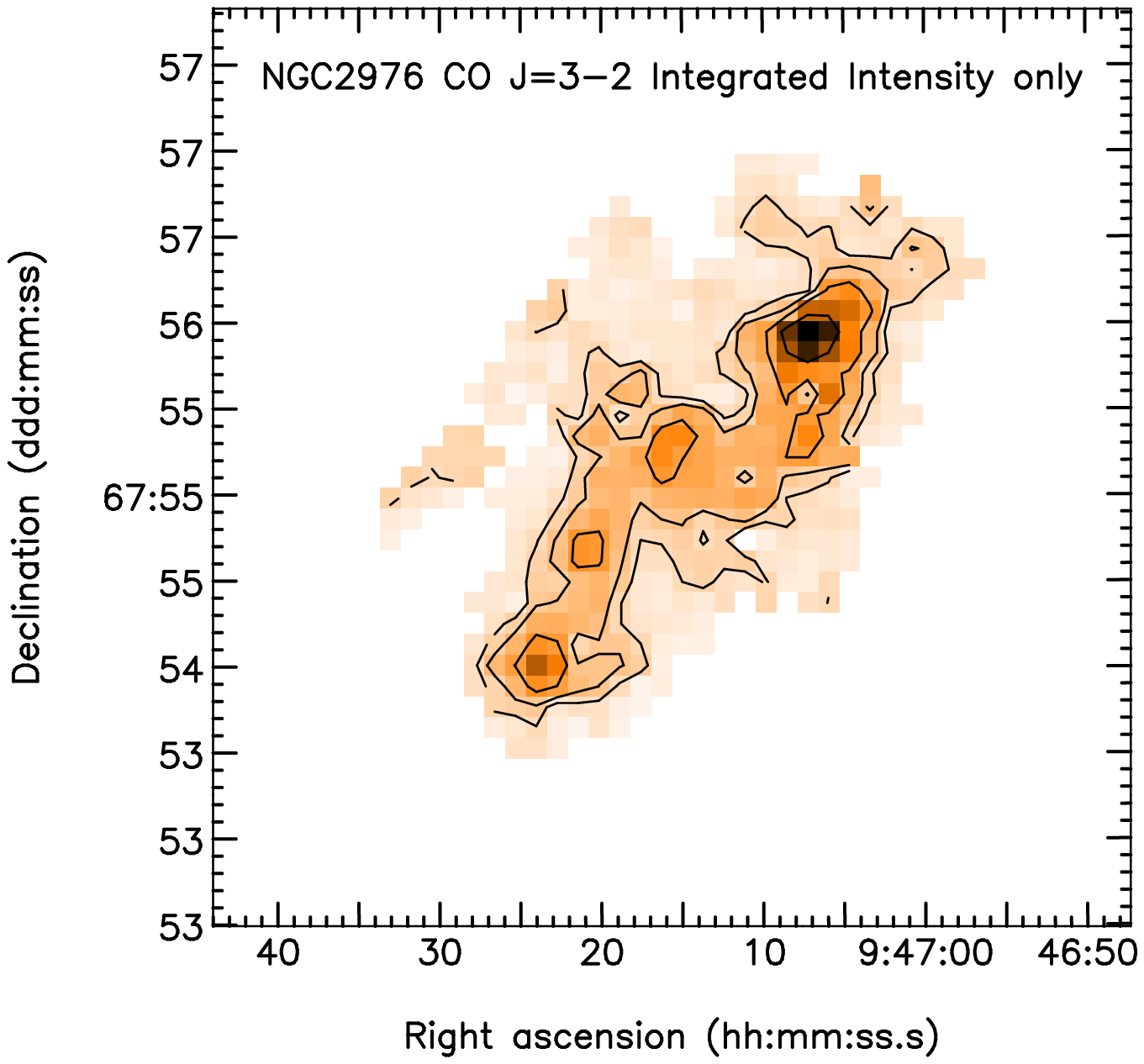}
\includegraphics[width=60mm]{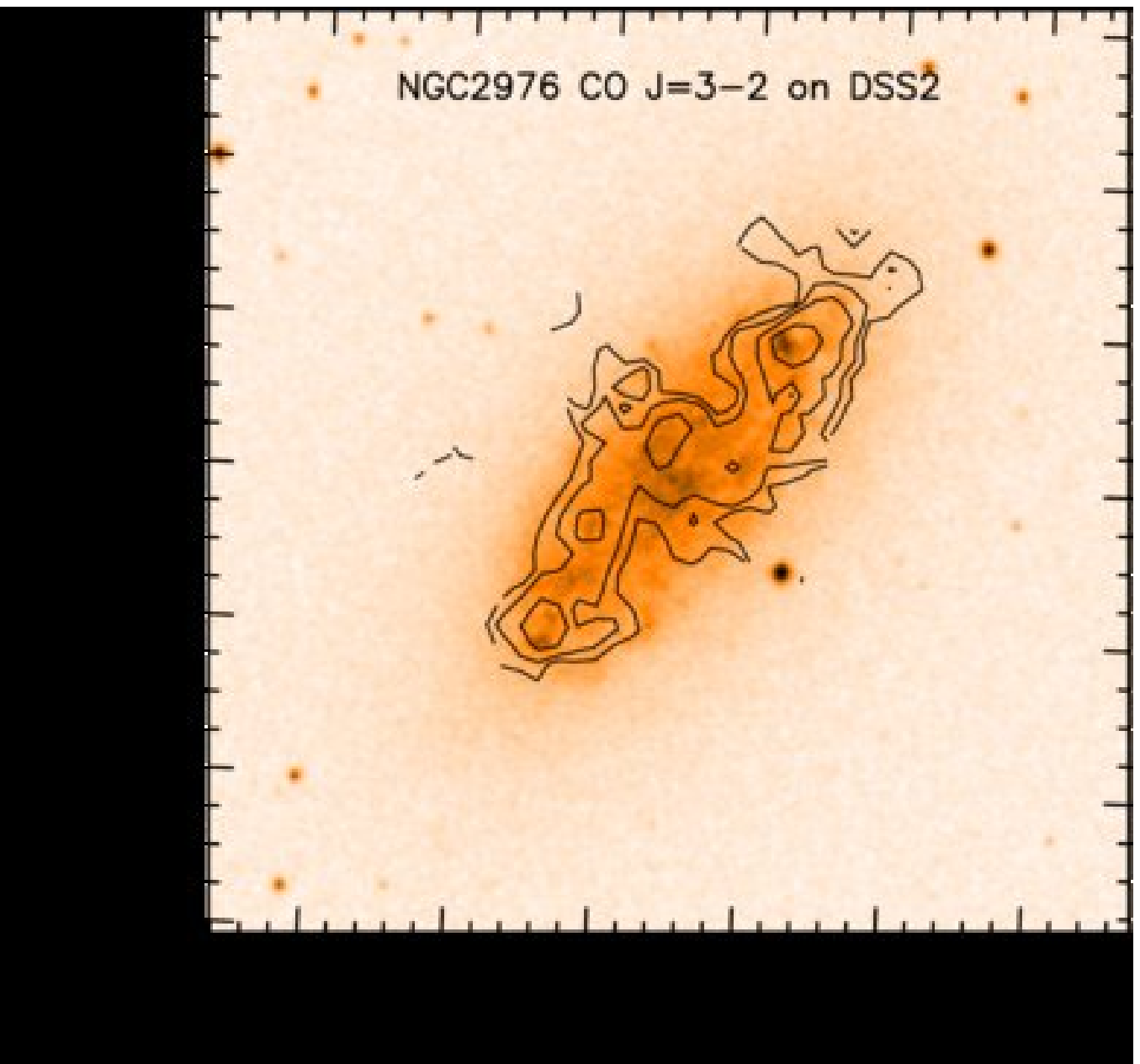}
\includegraphics[width=60mm]{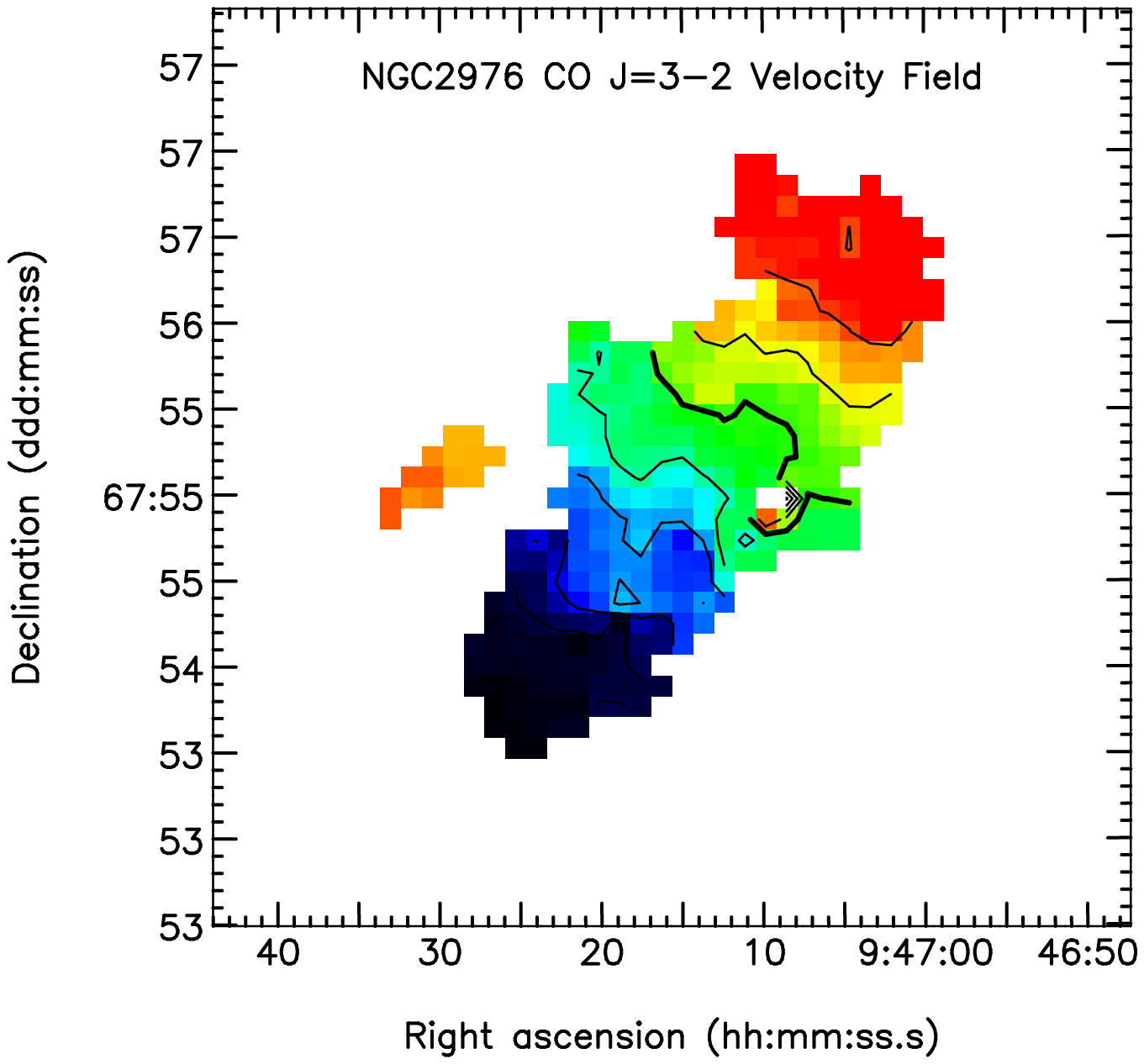}
\includegraphics[width=60mm]{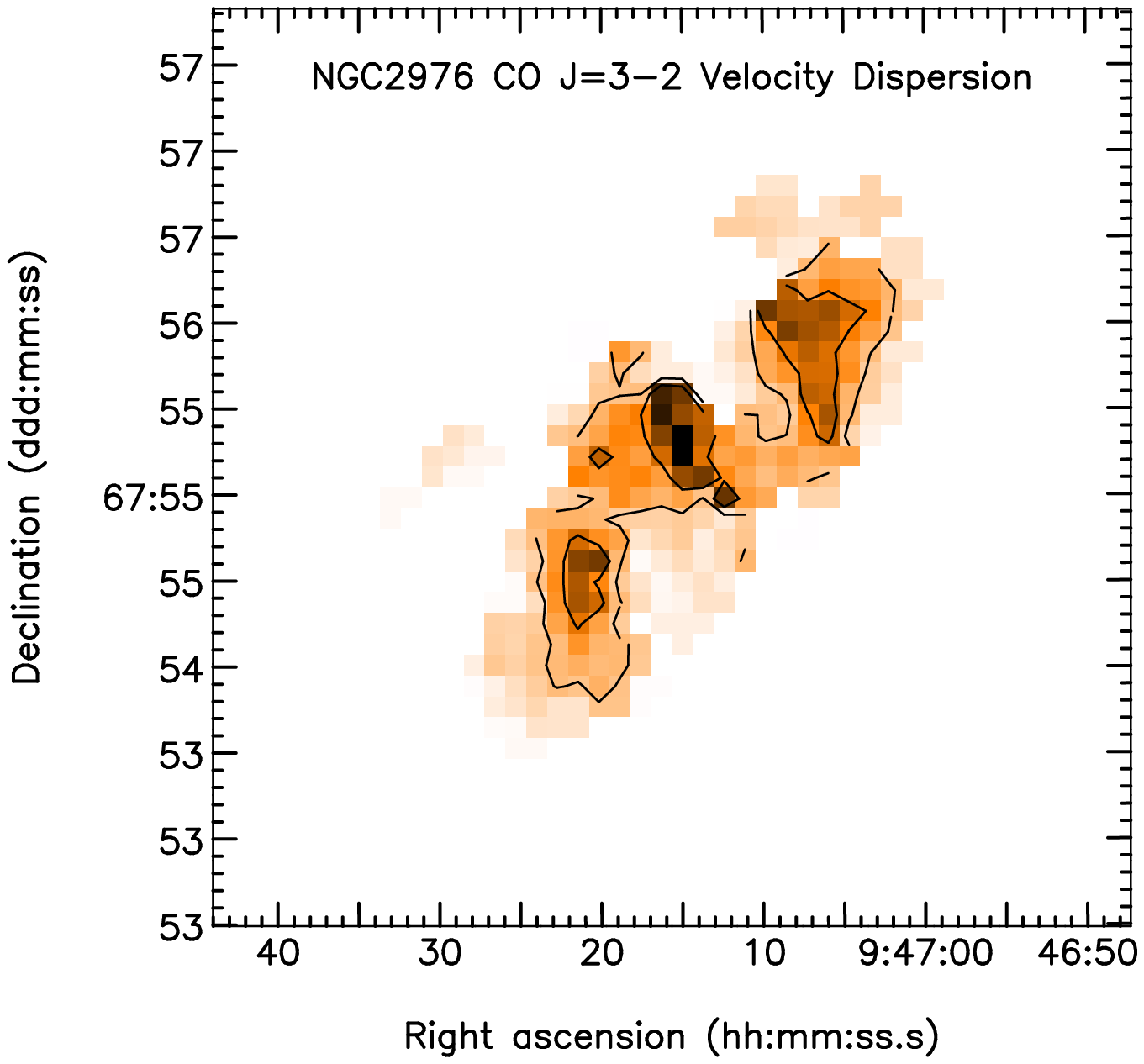}
\caption{CO $J$=3-2 images for NGC 2976.  (a) CO $J$=3-2 integrated intensity
  image. Contours levels are (0.5,   1,   2,   4) K km s$^{-1}$ (T$_{MB}$).
(b) CO $J$=3-2 overlaid on a
Digitized Sky Survey image. (c) Velocity field. Contour levels are
(-50,   -32,   -14,   4,   22,   40,   58,   76,   94) 
km s$^{-1}$. 
(d) The velocity dispersion $\sigma_v$  as traced by the 
CO $J$=3-2 second moment map.  Contour levels are
(4,   8)
km s$^{-1}$.
\label{fig-ngc2976}}
\end{figure*}

\begin{figure*}
\includegraphics[width=60mm]{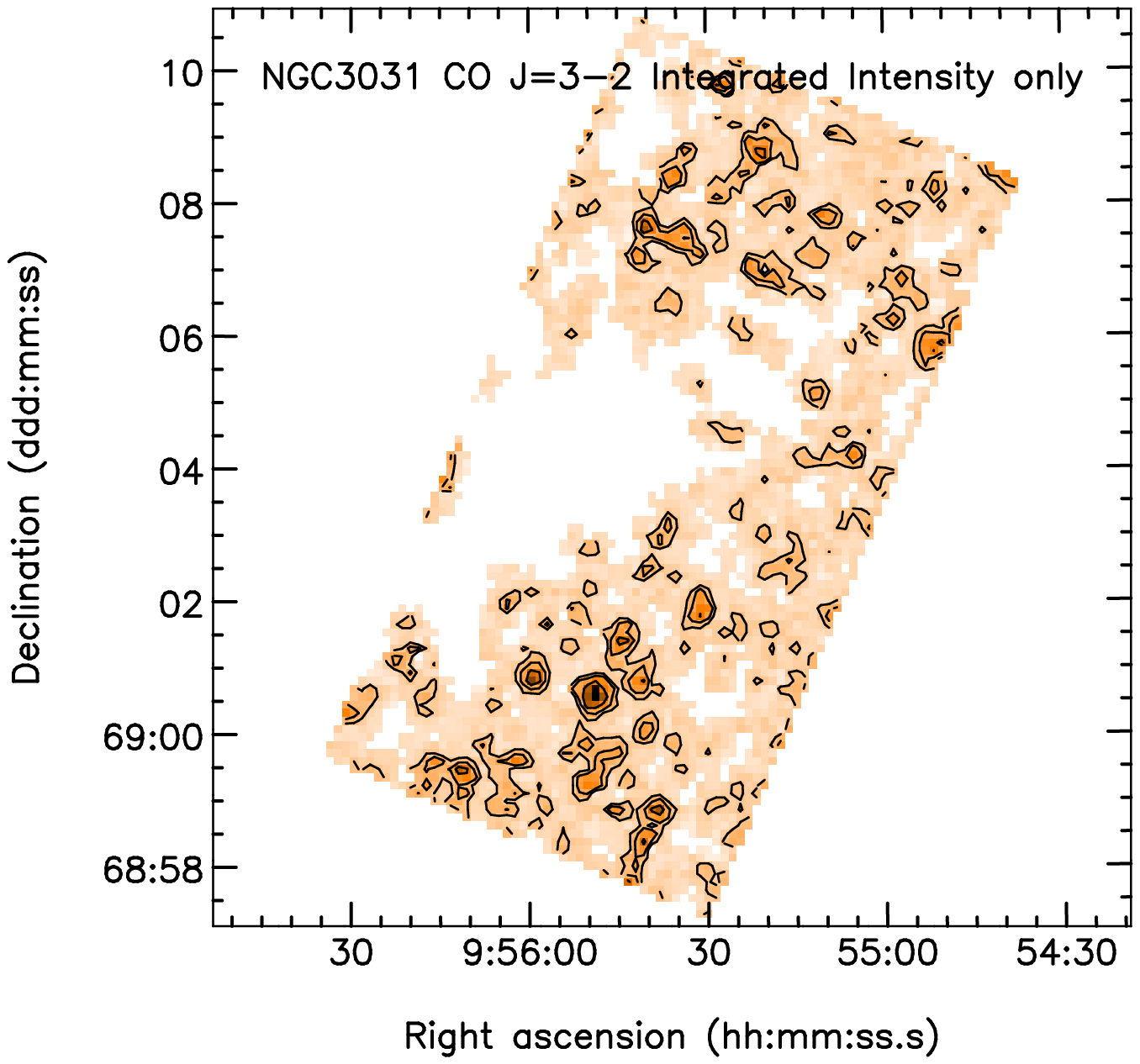}
\includegraphics[width=60mm]{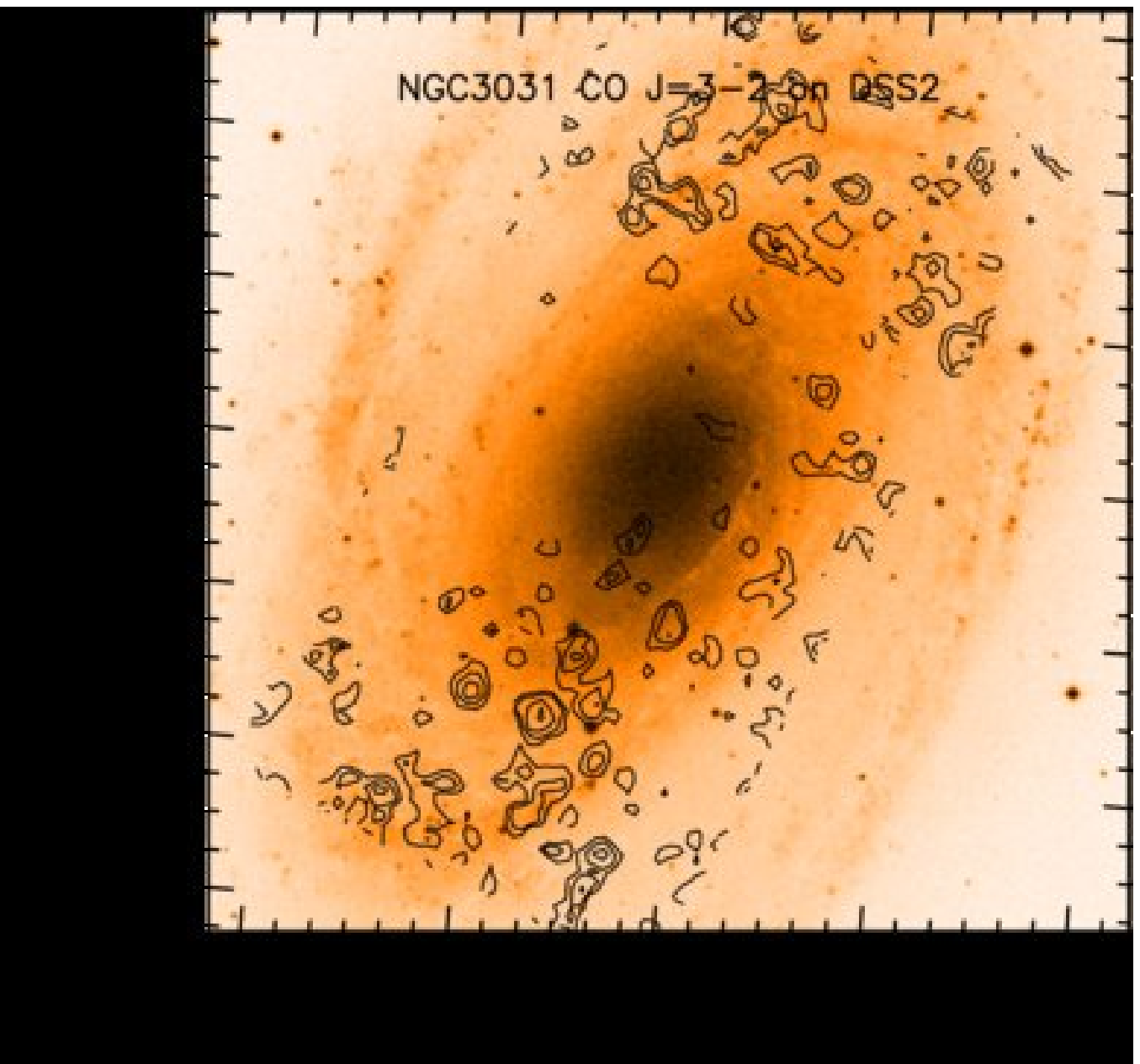}
\includegraphics[width=60mm]{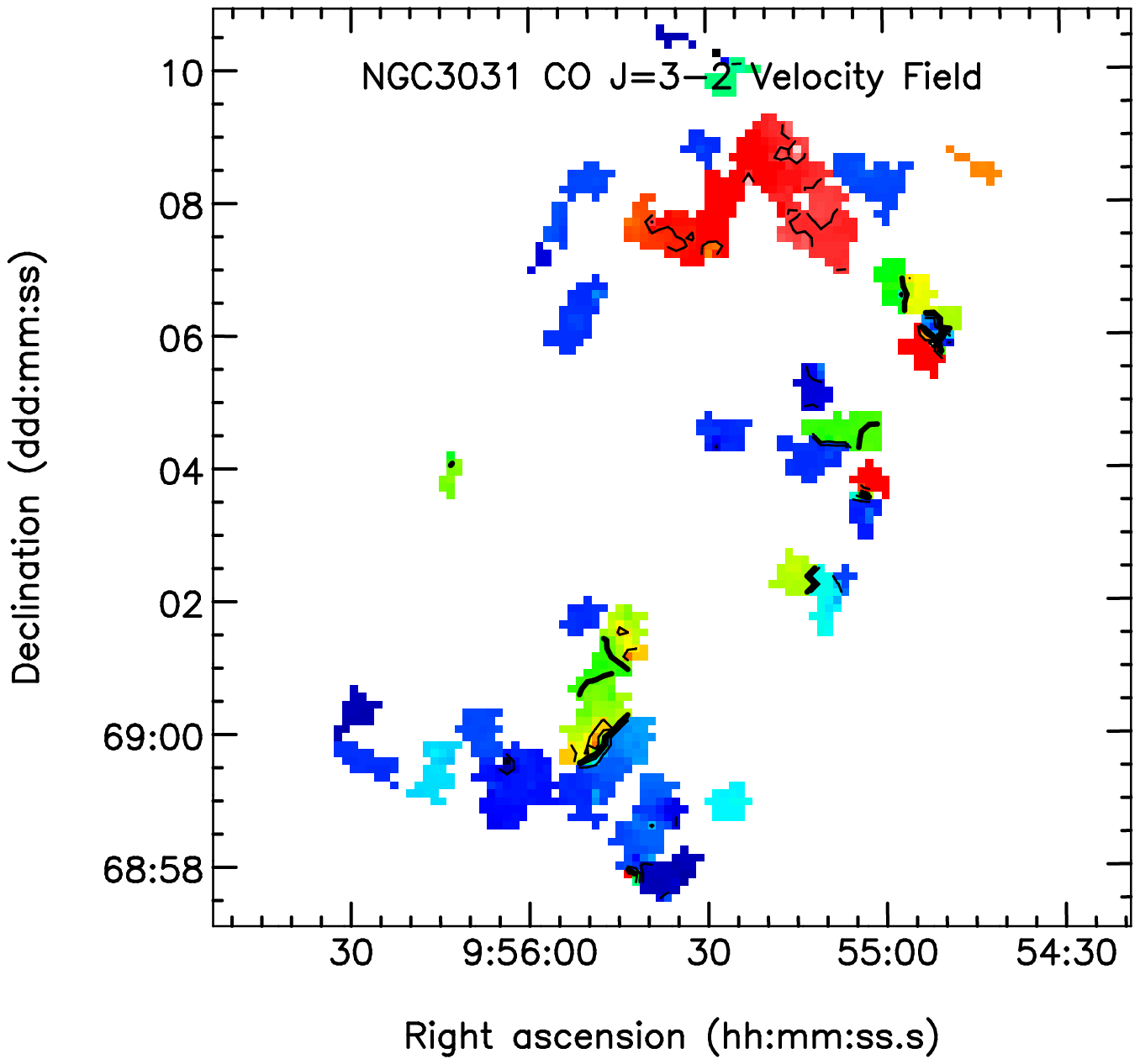}
\includegraphics[width=60mm]{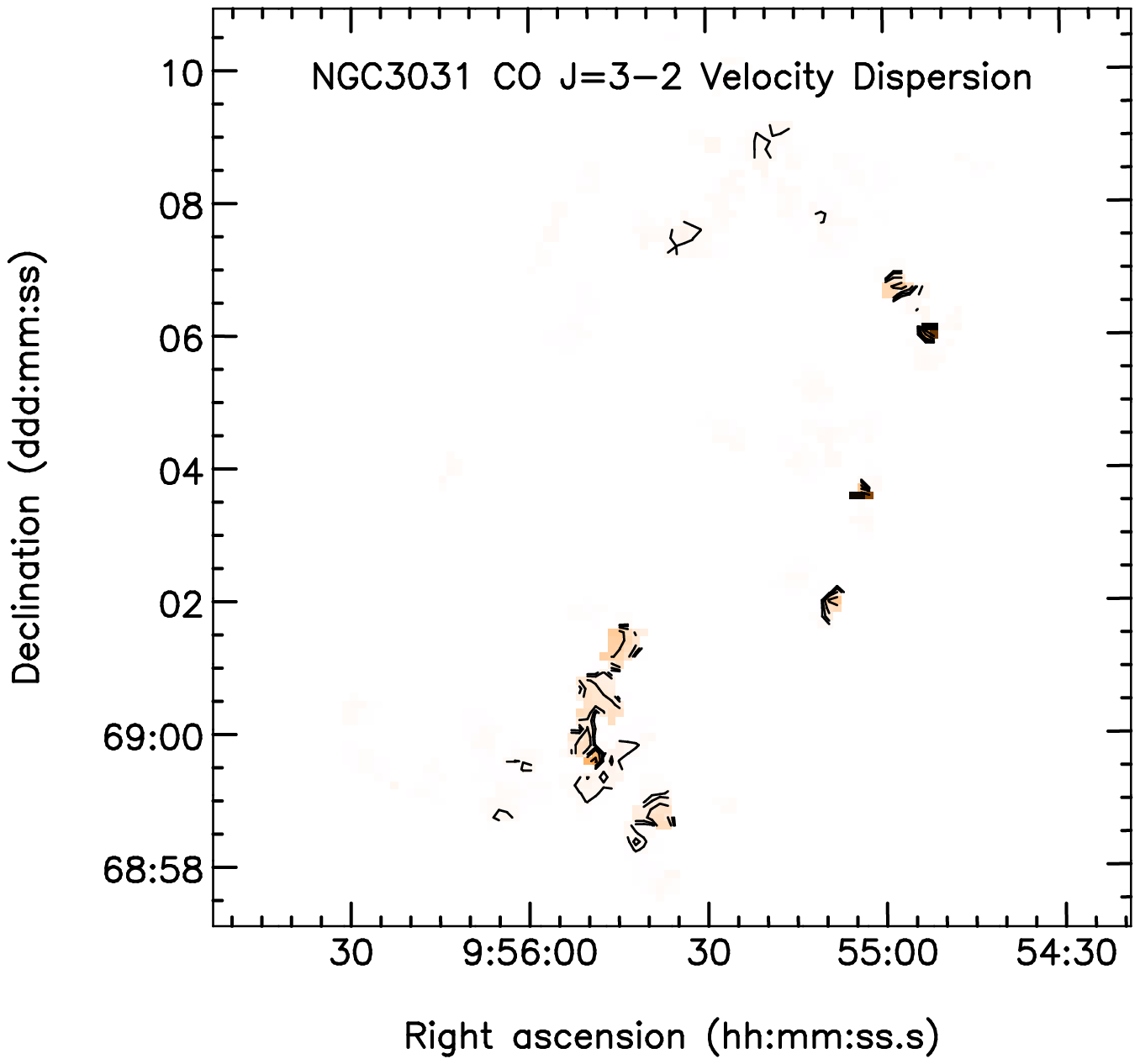}
\caption{CO $J$=3-2 images for NGC 3031.  (a) CO $J$=3-2 integrated intensity
  image. Contours levels are (0.5,   1,   2,   4) K km s$^{-1}$ (T$_{MB}$).
(b) CO $J$=3-2 overlaid on a
Digitized Sky Survey image. (c) Velocity field. Contour levels are
(-328,   -258,   -188,   -118,   -48,   22,   92,   162) 
km s$^{-1}$. 
(d) The velocity dispersion $\sigma_v$  as traced by the 
CO $J$=3-2 second moment map.  Contour levels are
(4,   8,   16,   32,   64,   128)
km s$^{-1}$.
Images derived from the same data using careful flagging and analysis
to detect weak but real features have been
published in \citet{s-g11}.
\label{fig-ngc3031}}
\end{figure*}

\begin{figure*}
\includegraphics[width=60mm]{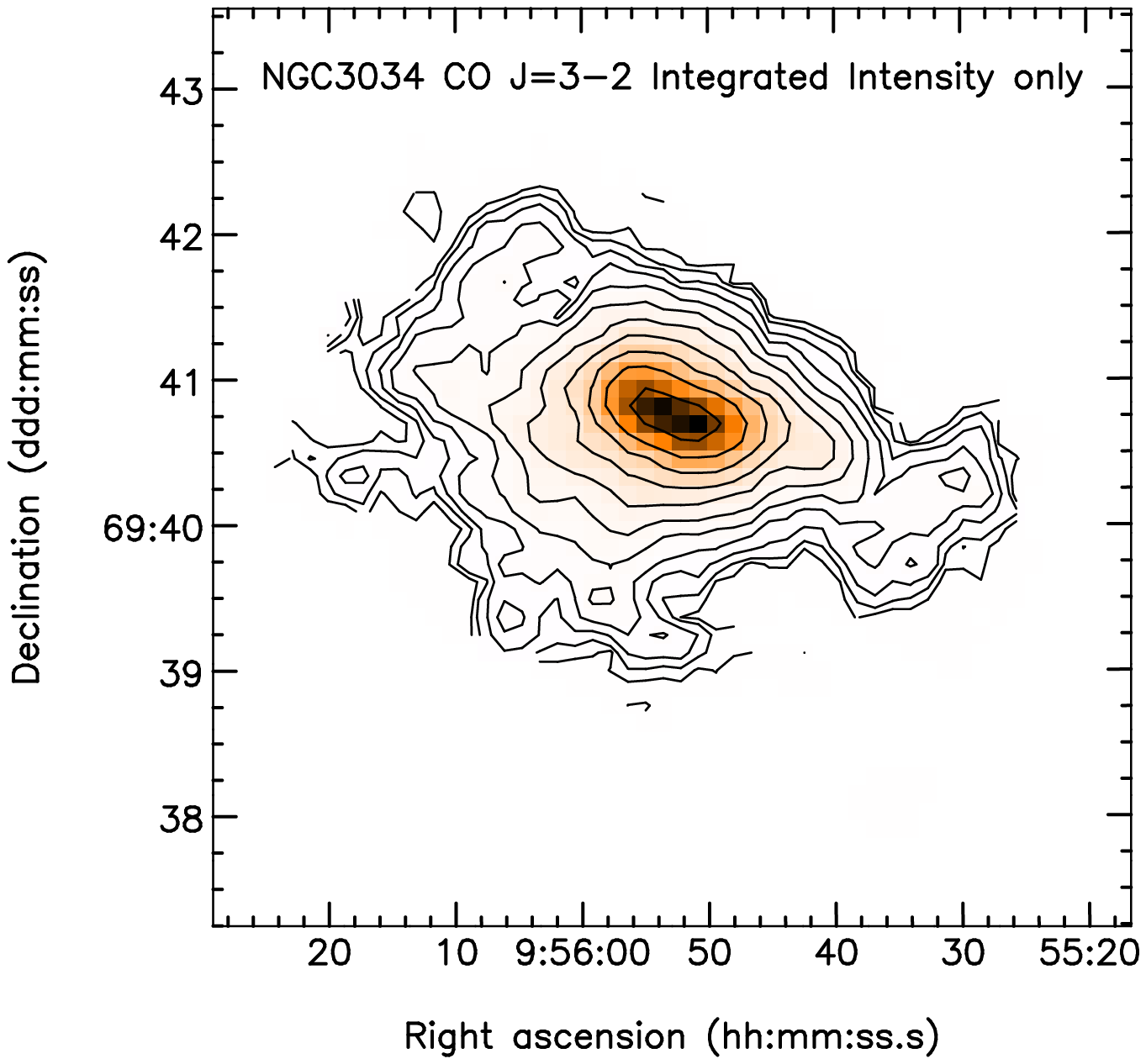}
\includegraphics[width=60mm]{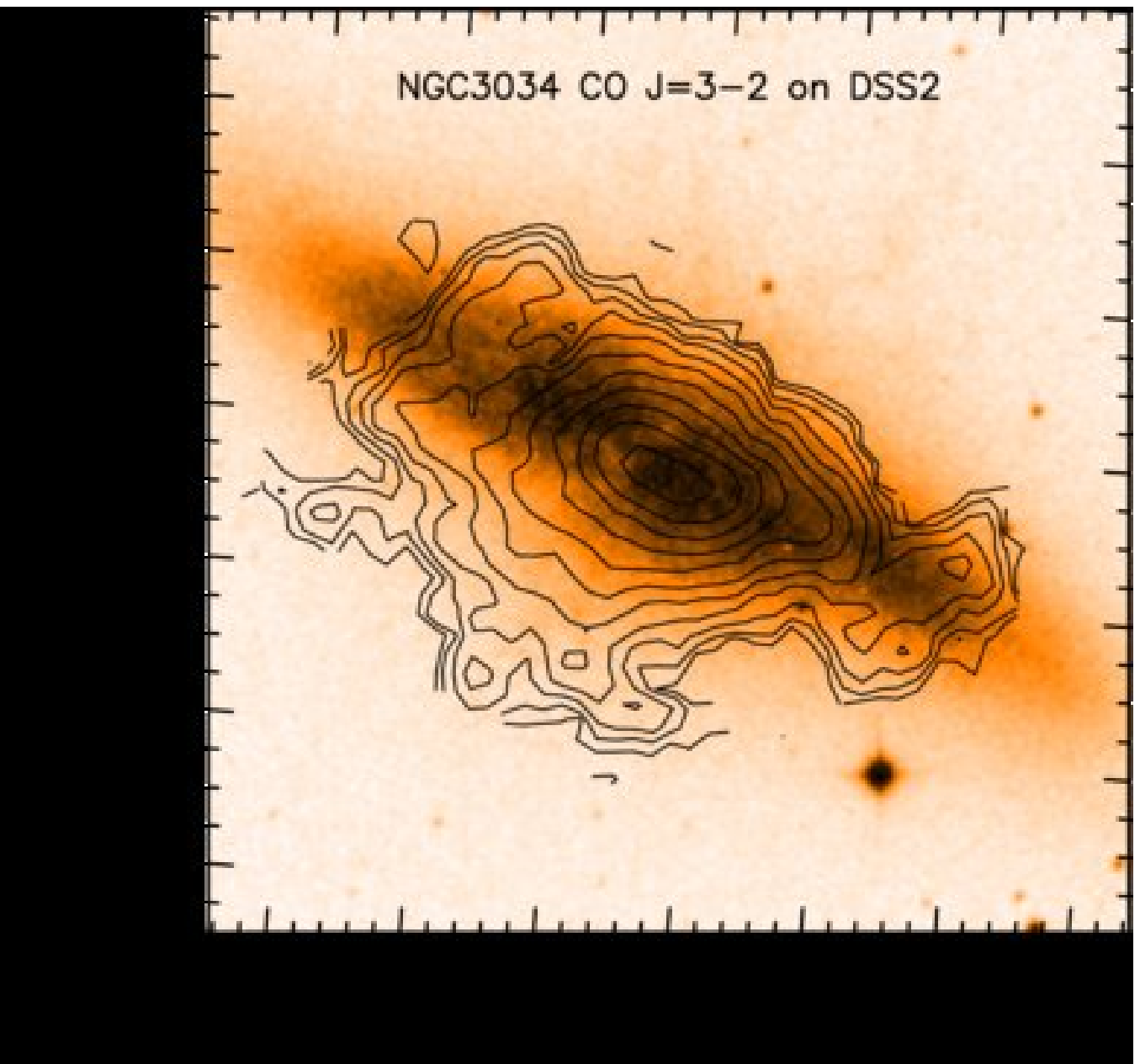}
\includegraphics[width=60mm]{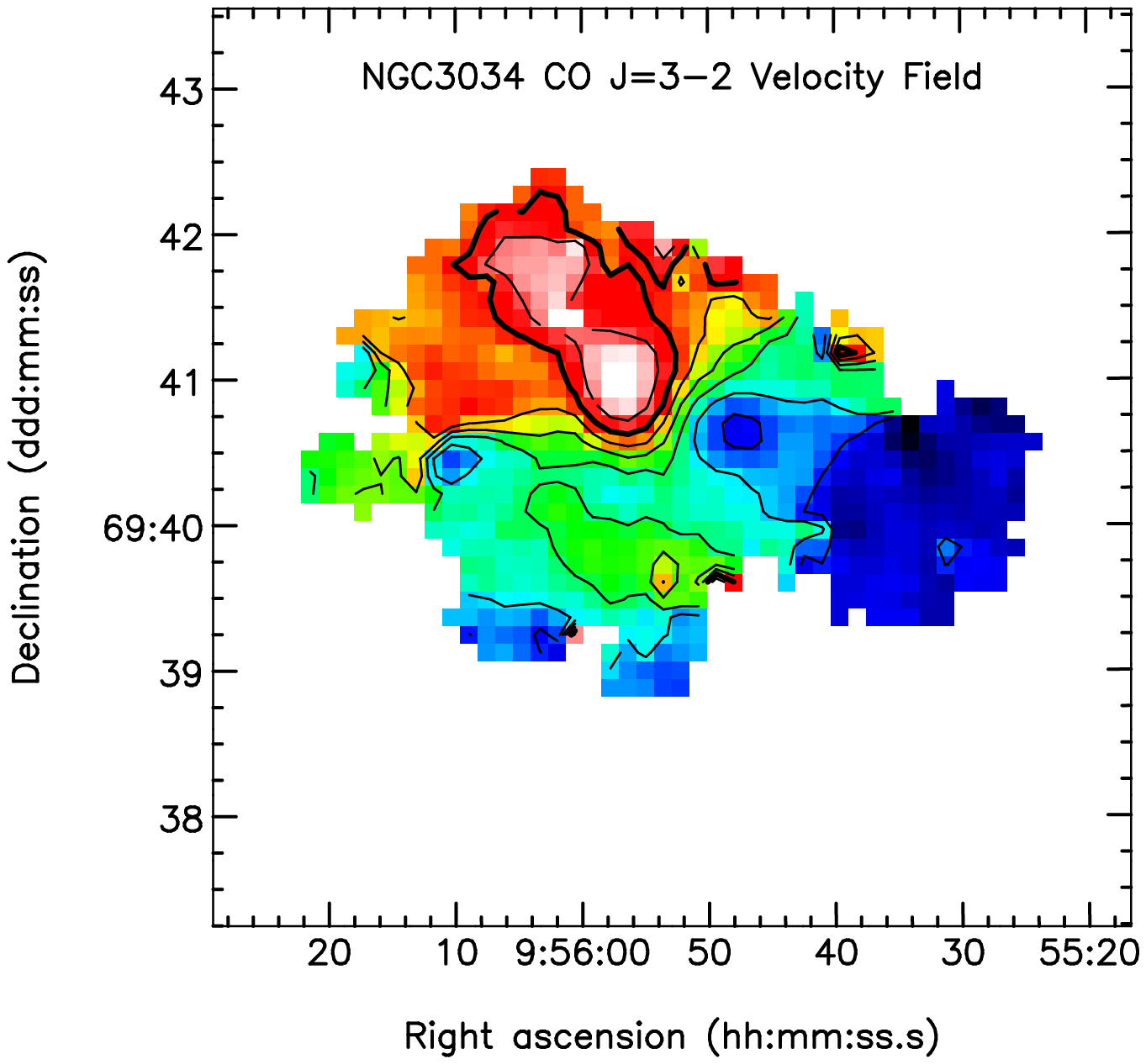}
\includegraphics[width=60mm]{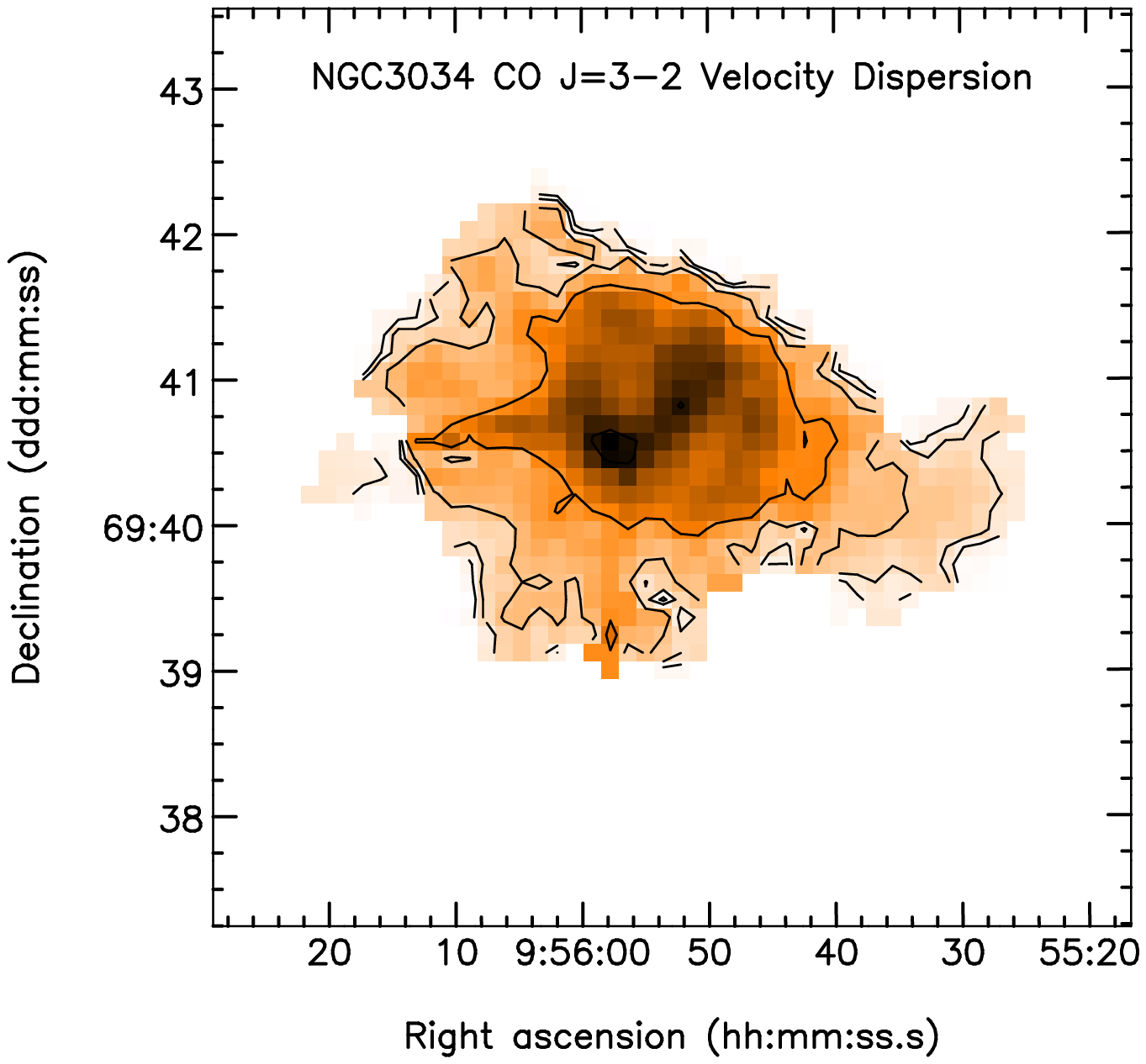}
\caption{CO $J$=3-2 images for NGC 3034.  (a) CO $J$=3-2 integrated intensity
  image. Contours levels are (0.5,   1,   2,   4,   8,   16,   32,
  64,   128,   256,   512) K km s$^{-1}$ (T$_{MB}$).
(b) CO $J$=3-2 overlaid on a
Digitized Sky Survey image. (c) Velocity field. Contour levels are
(129,   154,   179,   204,   229,   254,   279) 
km s$^{-1}$. 
(d) The velocity dispersion $\sigma_v$  as traced by the 
CO $J$=3-2 second moment map.  Contour levels are
(4,   8,   16,   32,   64)
km s$^{-1}$.
\label{fig-ngc3034}}
\end{figure*}

\begin{figure*}
\includegraphics[width=55mm]{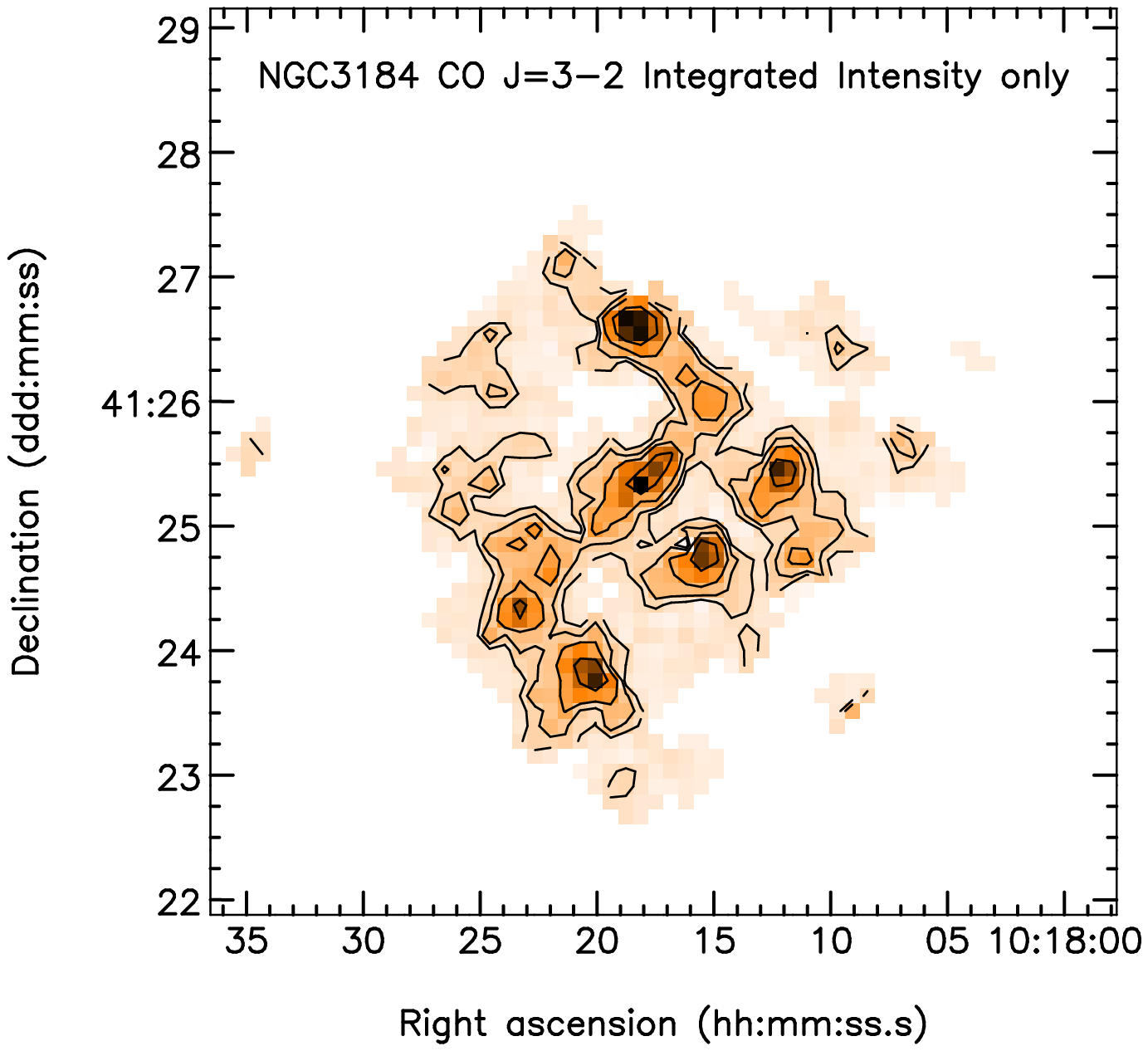}
\includegraphics[width=55mm]{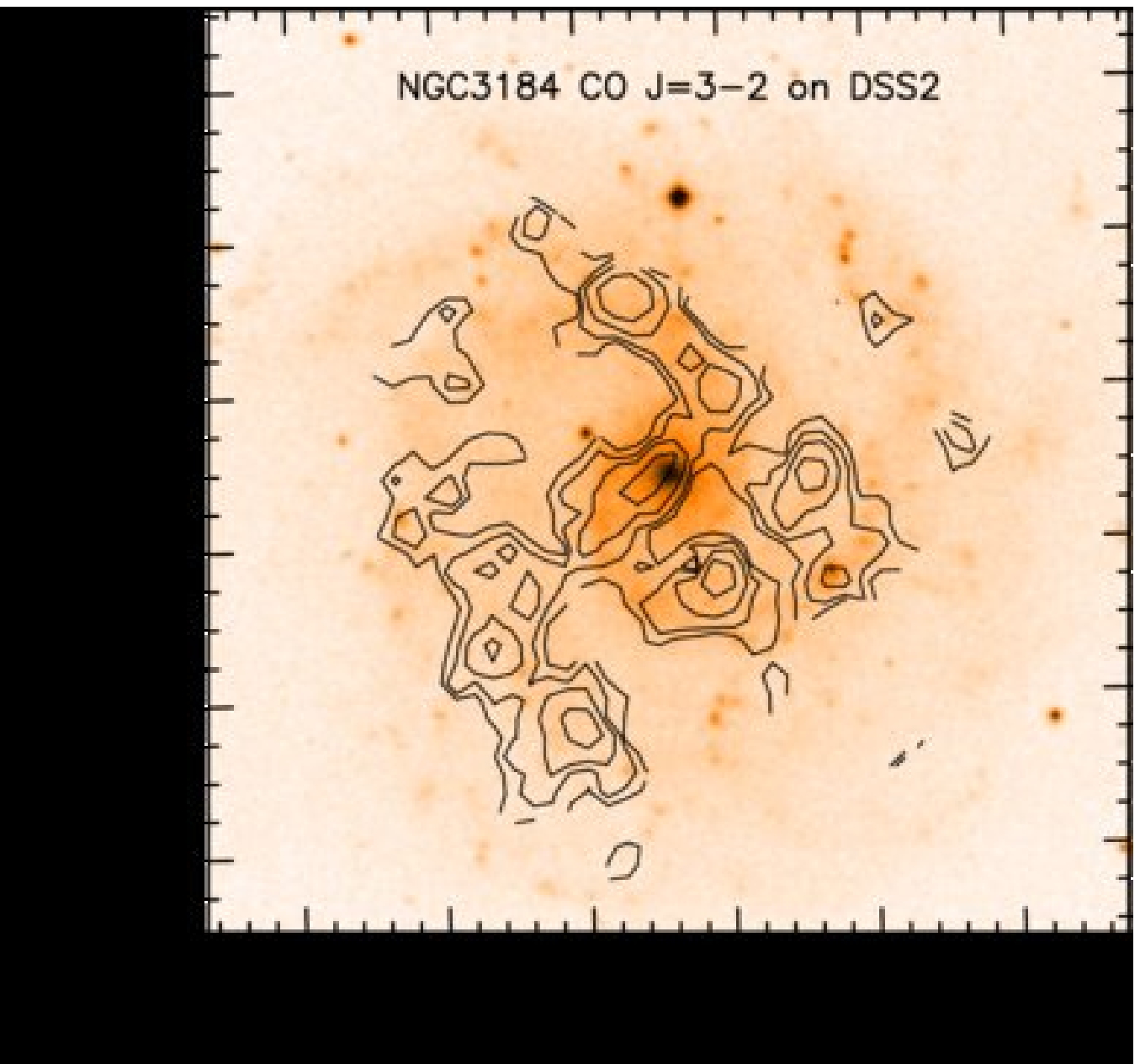}
\includegraphics[width=55mm]{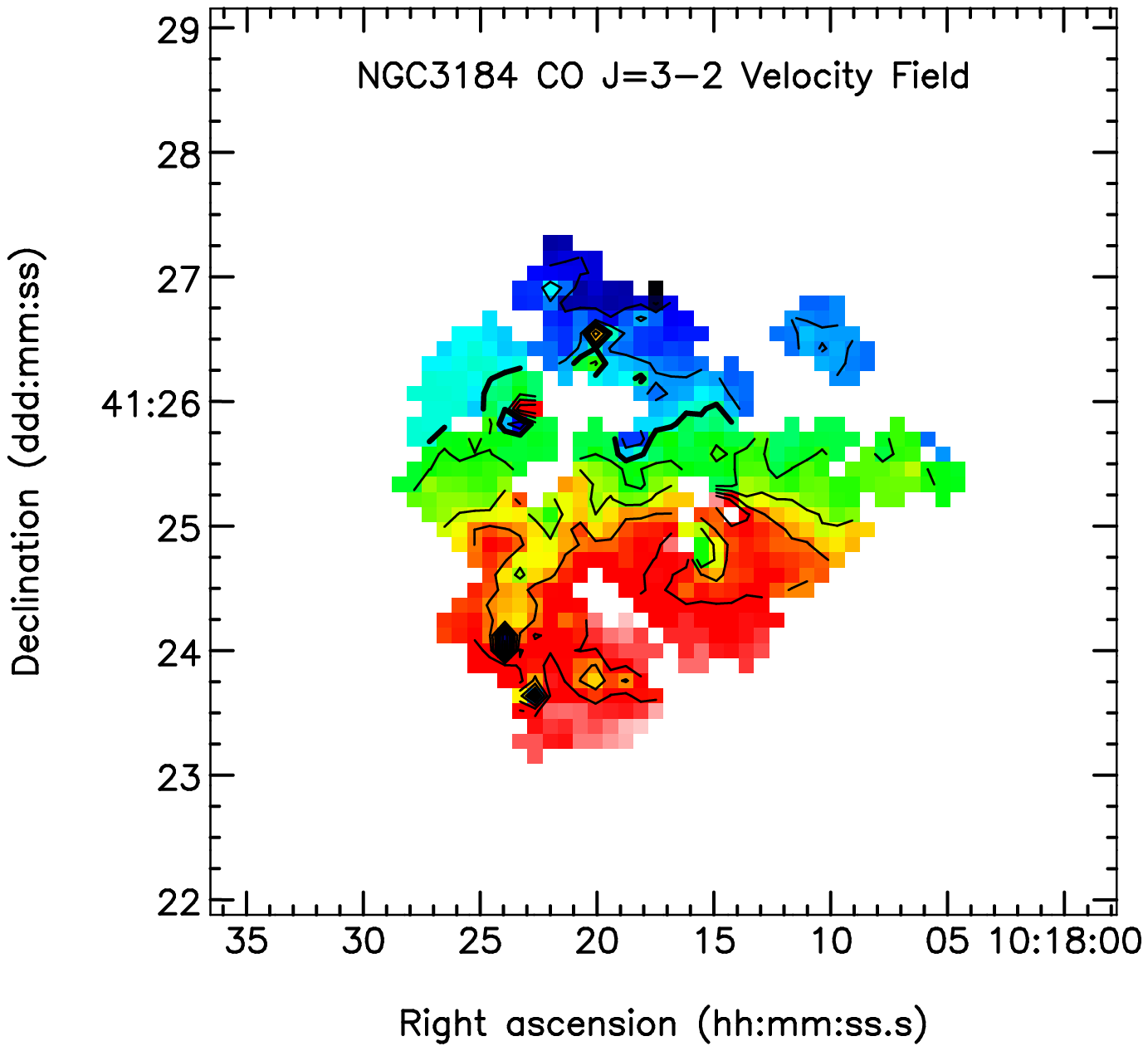}
\caption{CO $J$=3-2 images for NGC 3184. (a) CO $J$=3-2 integrated intensity
  image. Contours levels are (0.5,   1,   2,   4) 
K km s$^{-1}$ (T$_{MB}$).
(b) CO $J$=3-2 integrated intensity contours overlaid on an optical
image from the Digitized Sky Survey. (c) Velocity field as traced by the
CO $J$=3-2 first moment map. Contour levels are (532,   546,   560,   574,   588,   602,   616,   630) km s$^{-1}$.
The velocity dispersion map has been published in \citet{w11}.
\label{fig-ngc3184}}
\end{figure*}

\begin{figure*}
\includegraphics[width=60mm]{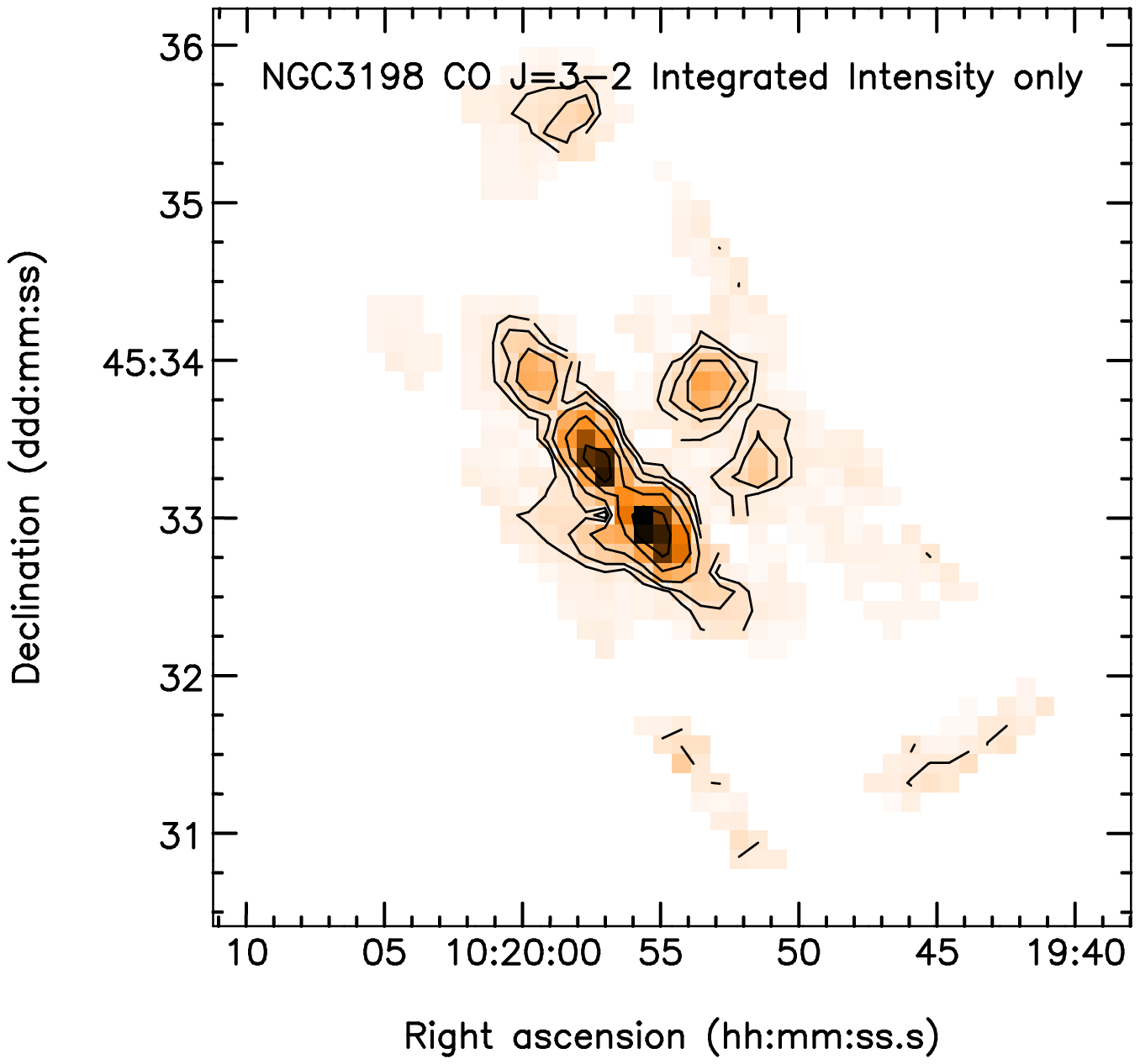}
\includegraphics[width=60mm]{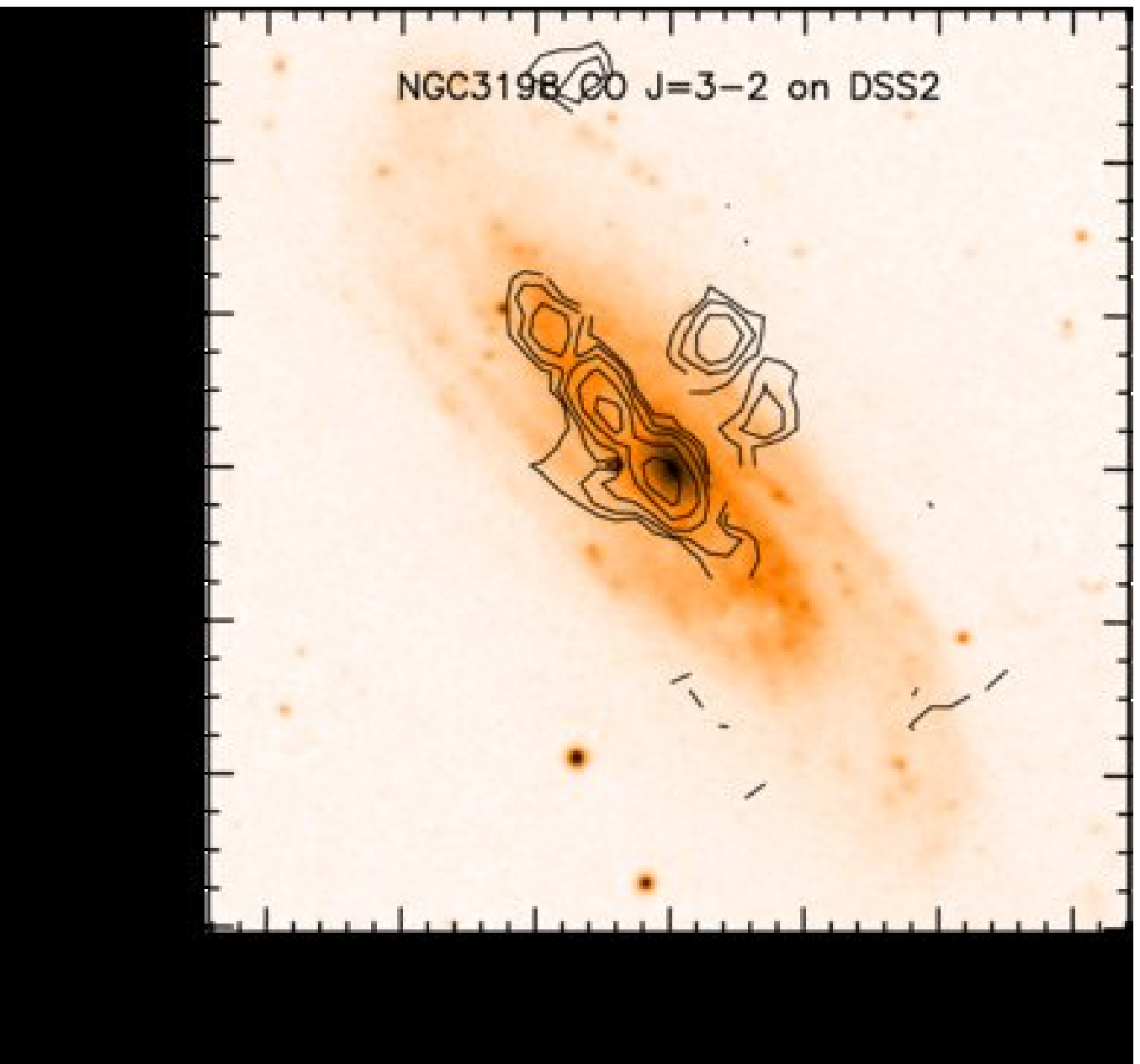}
\includegraphics[width=60mm]{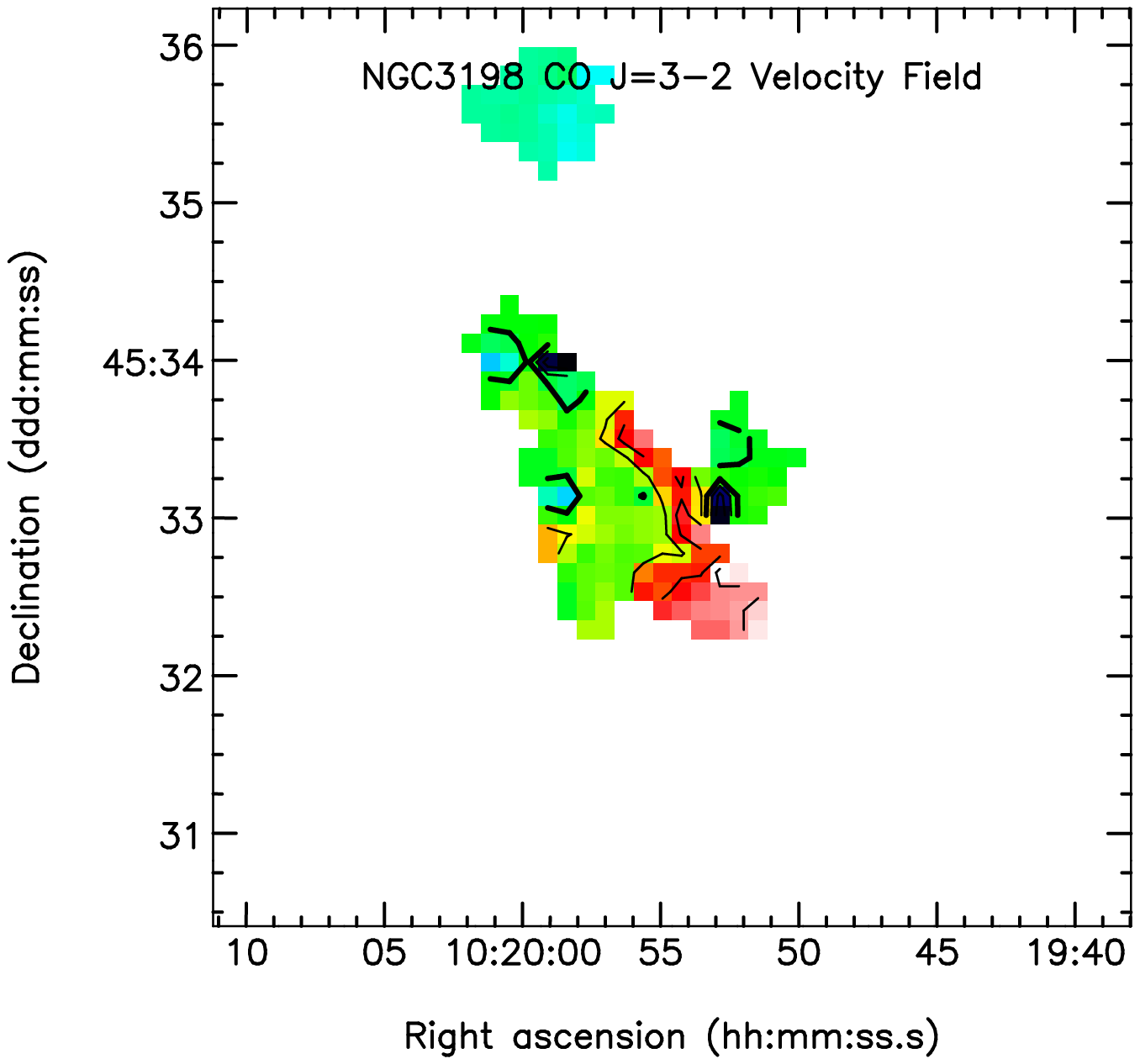}
\includegraphics[width=60mm]{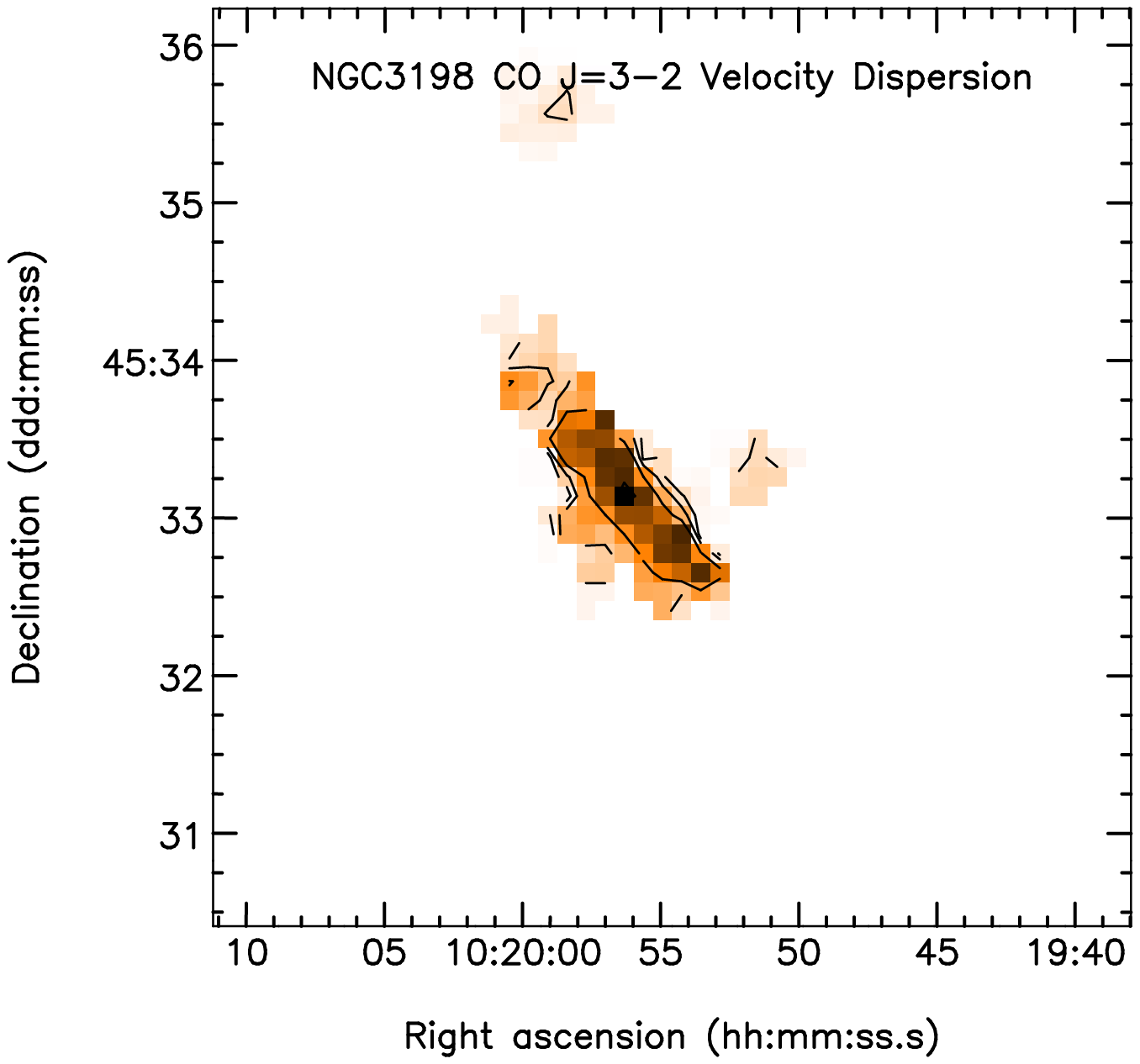}
\caption{CO $J$=3-2 images for NGC 3198.  (a) CO $J$=3-2 integrated intensity
  image. Contours levels are (0.5,   1,   2,   4,   8) K km s$^{-1}$ (T$_{MB}$).
(b) CO $J$=3-2 overlaid on a
Digitized Sky Survey image. (c) Velocity field. Contour levels are
(606,   636,   666,   696,   726,   756) 
km s$^{-1}$. 
(d) The velocity dispersion $\sigma_v$  as traced by the 
CO $J$=3-2 second moment map.  Contour levels are
(4,   8,   16,   32)
km s$^{-1}$.
\label{fig-ngc3198}}
\end{figure*}

\begin{figure*}
\includegraphics[width=60mm]{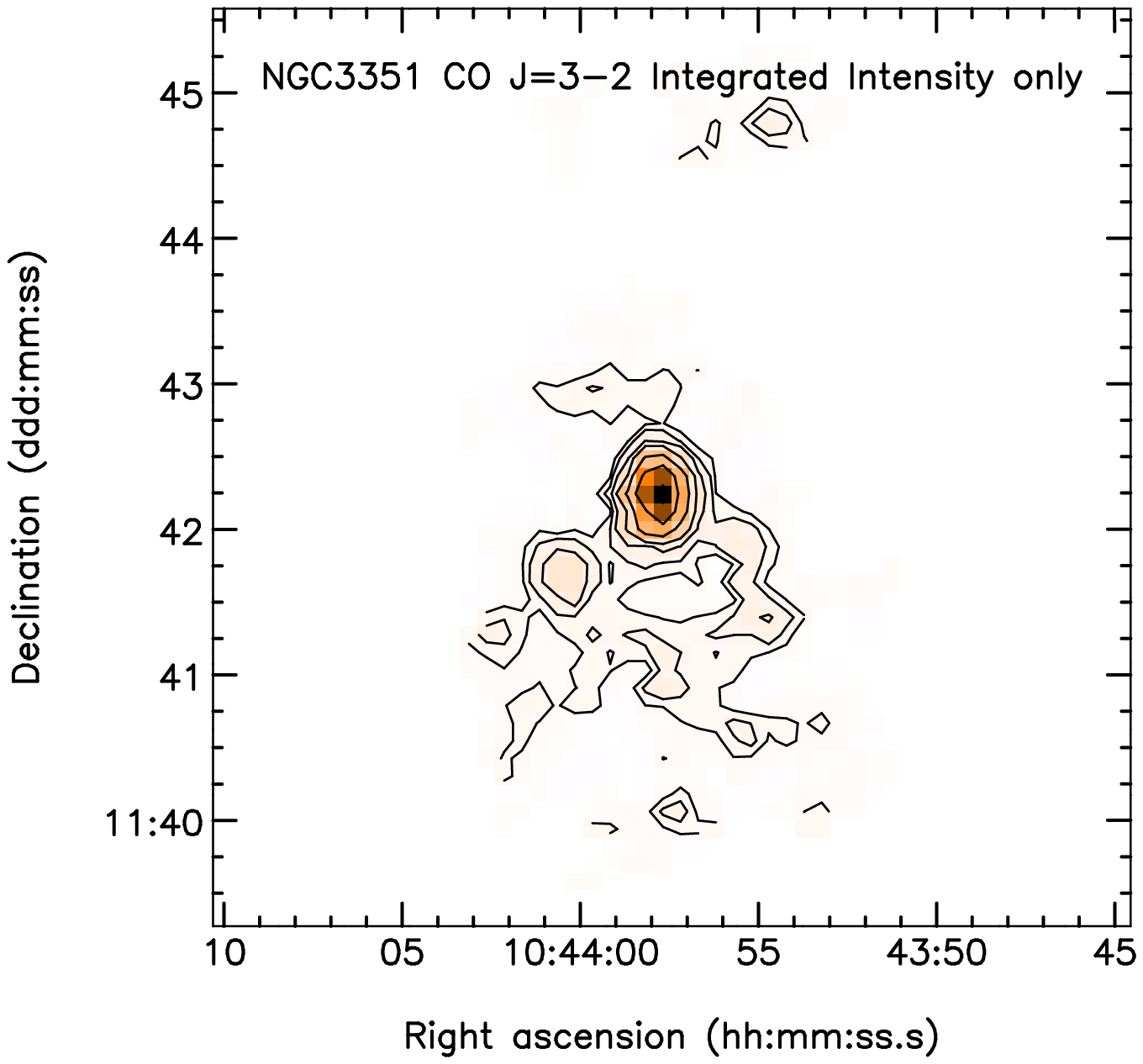}
\includegraphics[width=60mm]{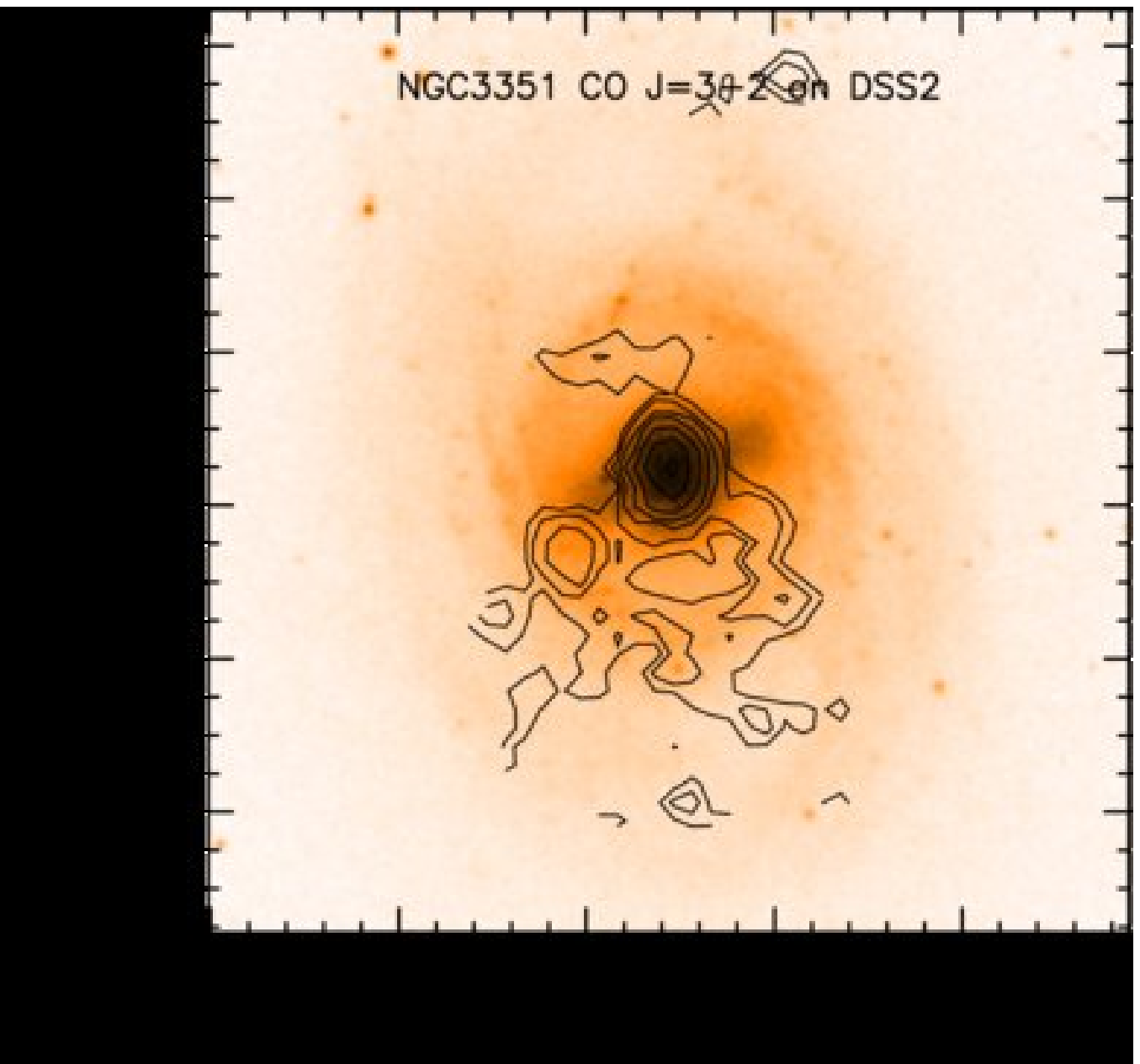}
\includegraphics[width=60mm]{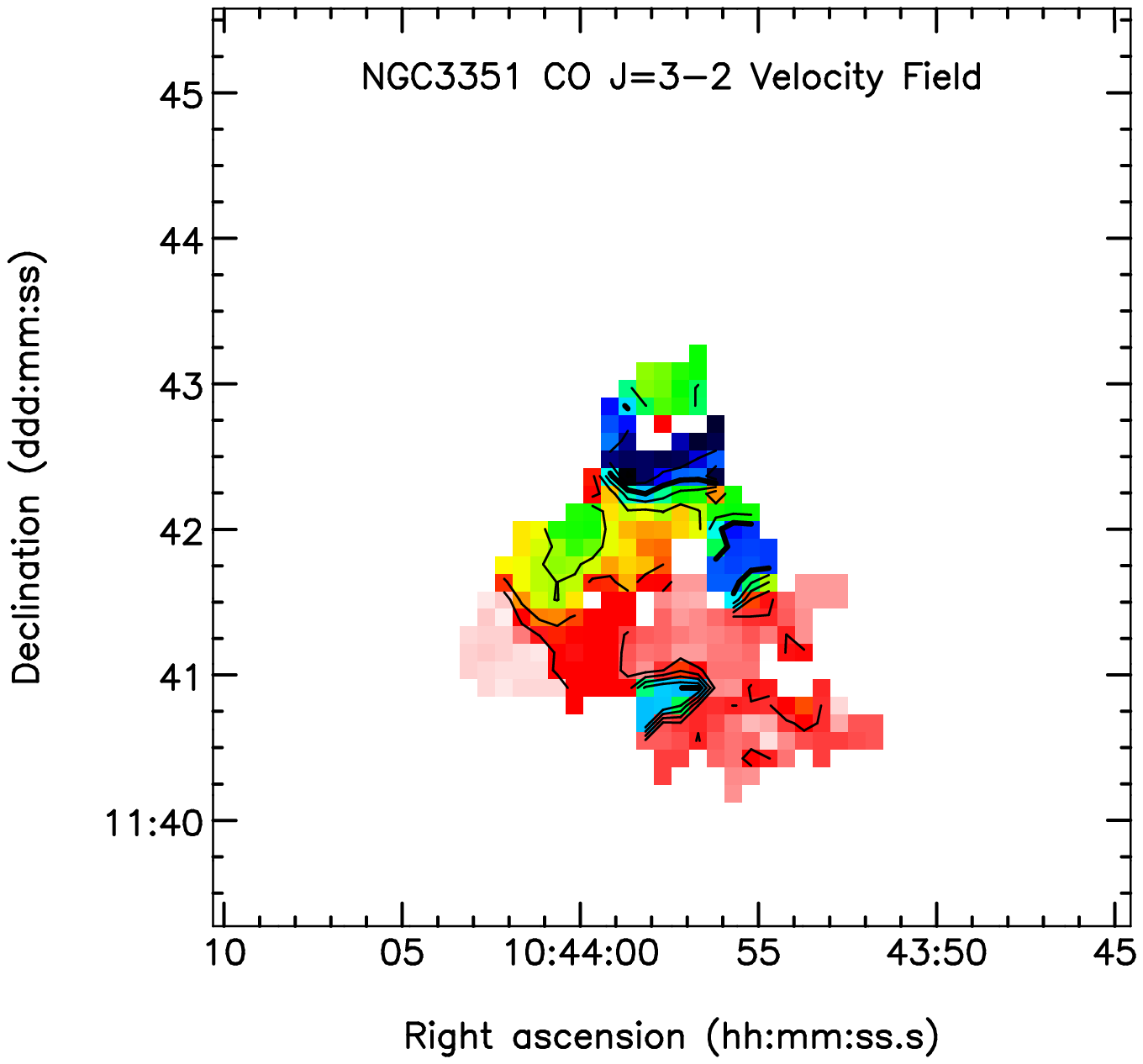}
\includegraphics[width=60mm]{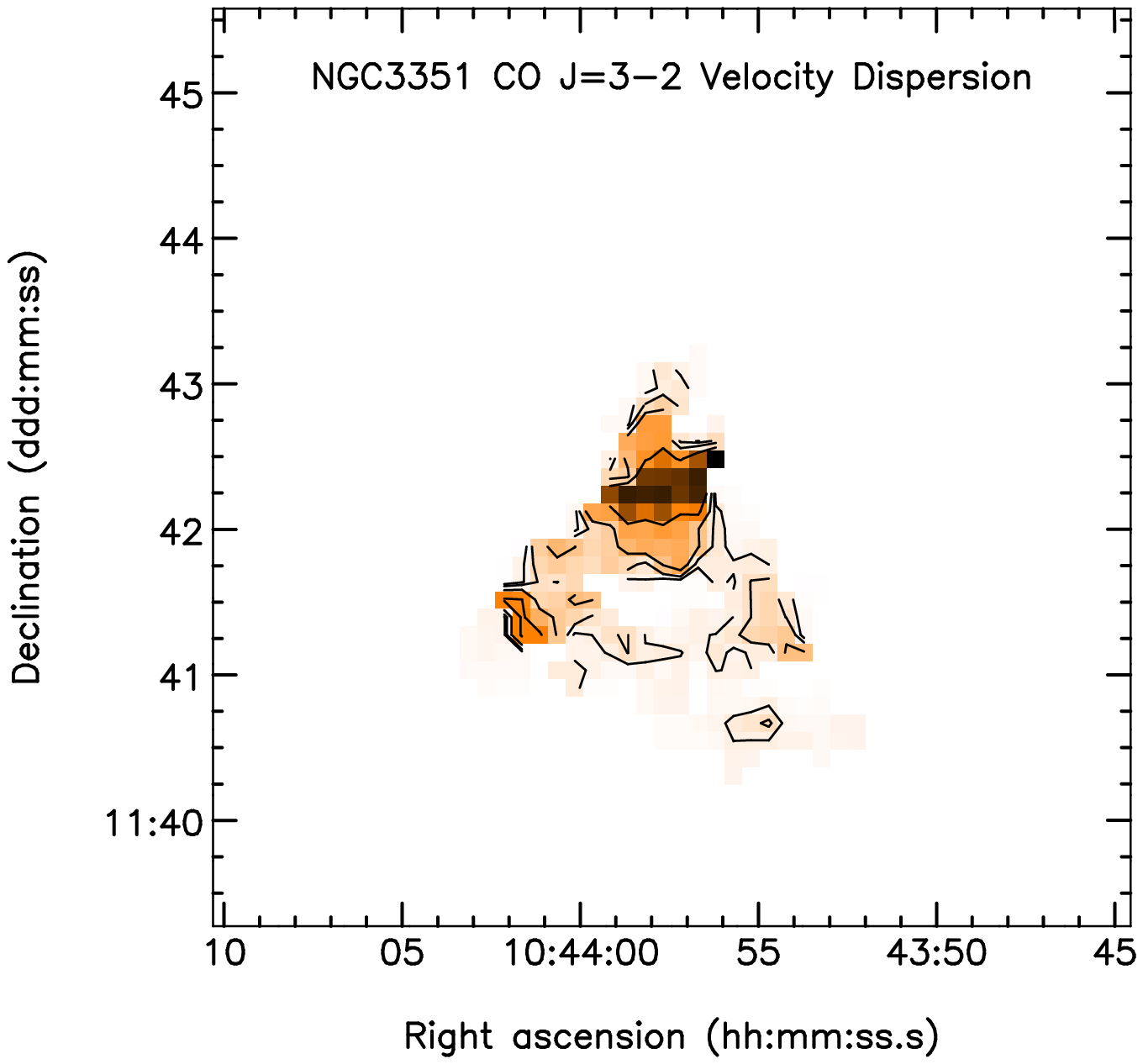}
\caption{CO $J$=3-2 images for NGC 3351.  (a) CO $J$=3-2 integrated intensity
  image. Contours levels are (0.5,   1,   2,   4,   8, 16, 32) K km s$^{-1}$ (T$_{MB}$).
(b) CO $J$=3-2 overlaid on a
Digitized Sky Survey image. (c) Velocity field. Contour levels are
(702,   735,   768,   801,   834,   867,   900) 
km s$^{-1}$. 
(d) The velocity dispersion $\sigma_v$  as traced by the 
CO $J$=3-2 second moment map.  Contour levels are
(4,   8,   16,   32, 64)
km s$^{-1}$.
\label{fig-ngc3351}}
\end{figure*}

\begin{figure*}
\includegraphics[width=60mm]{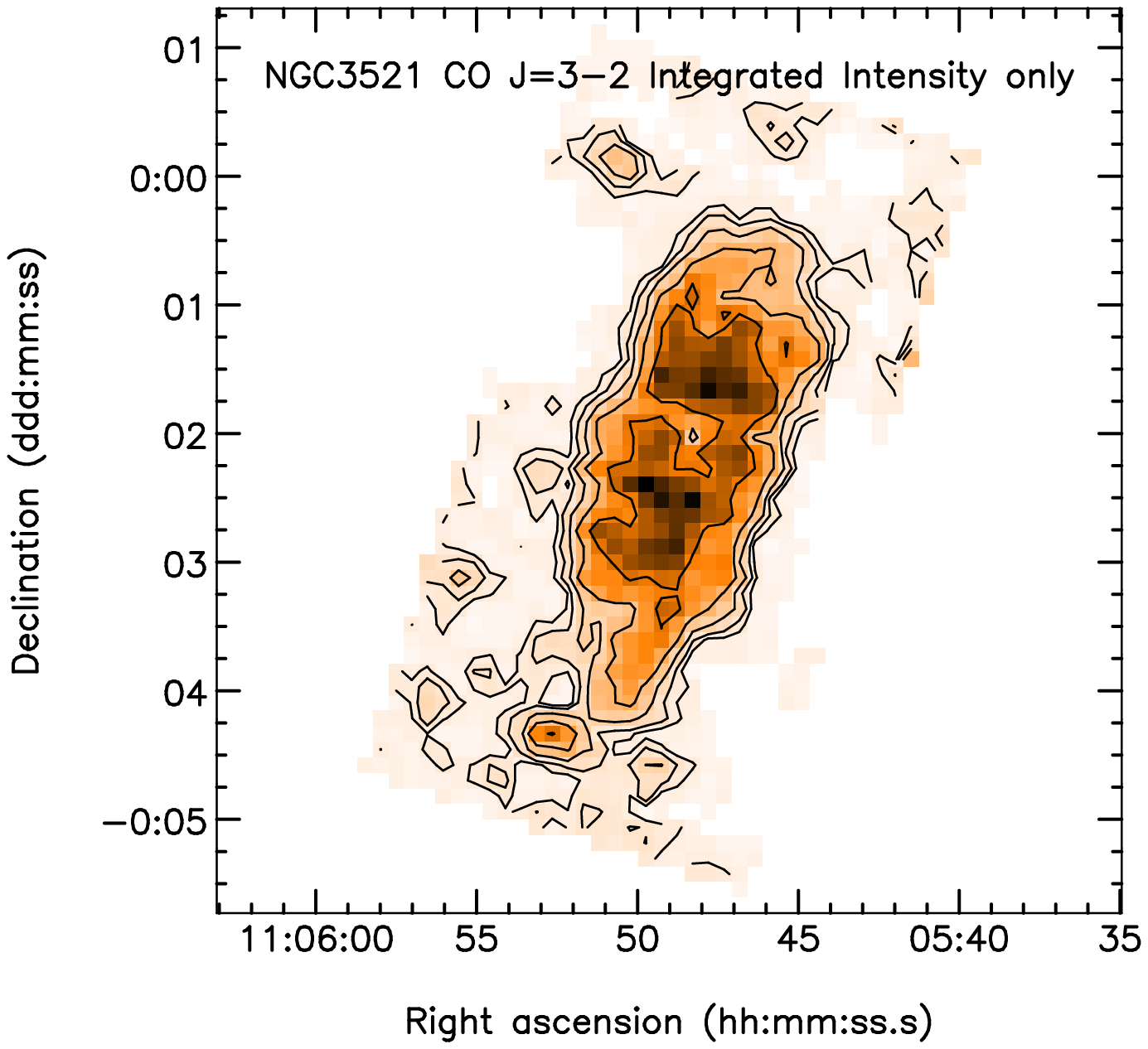}
\includegraphics[width=60mm]{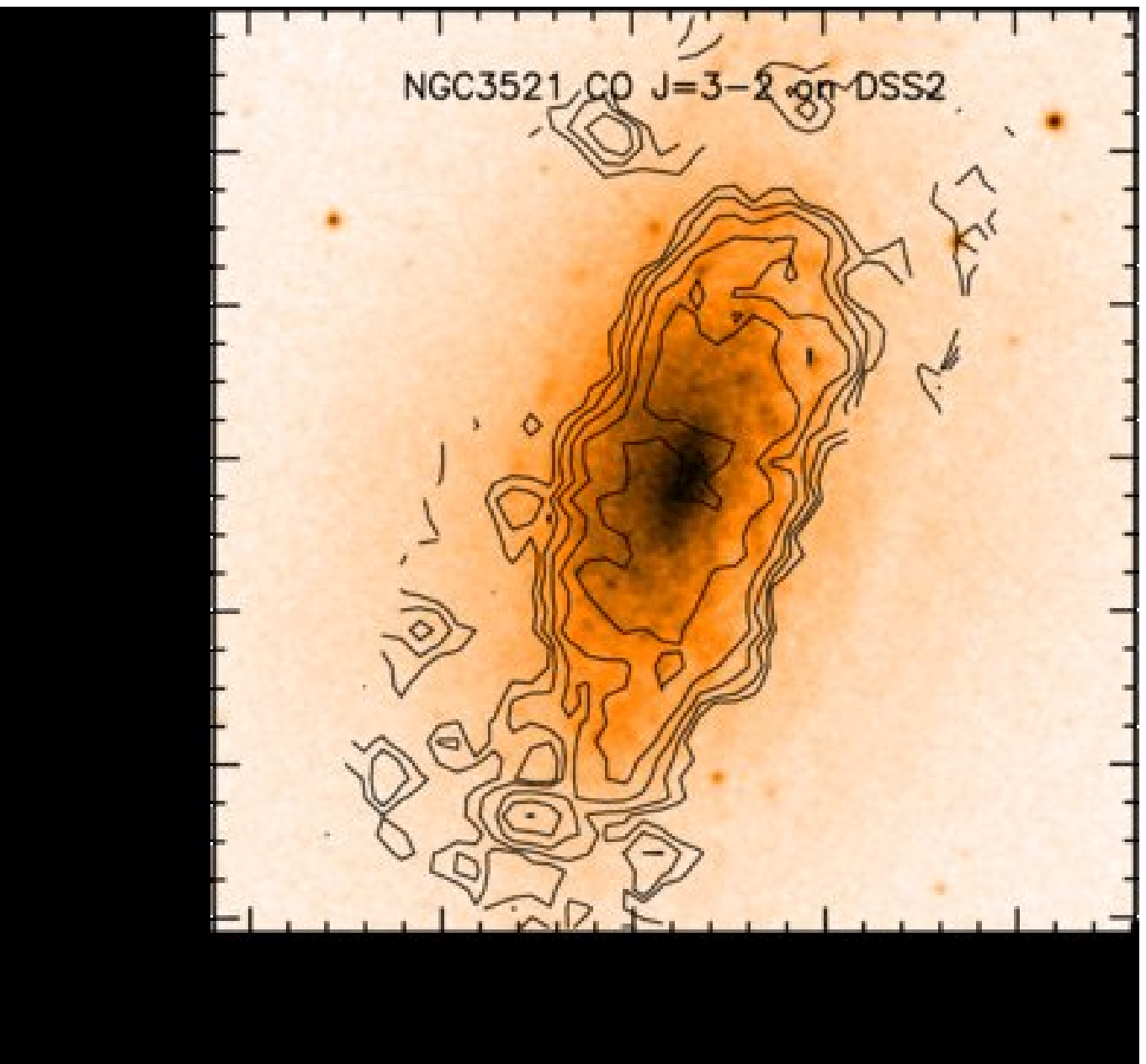}
\includegraphics[width=60mm]{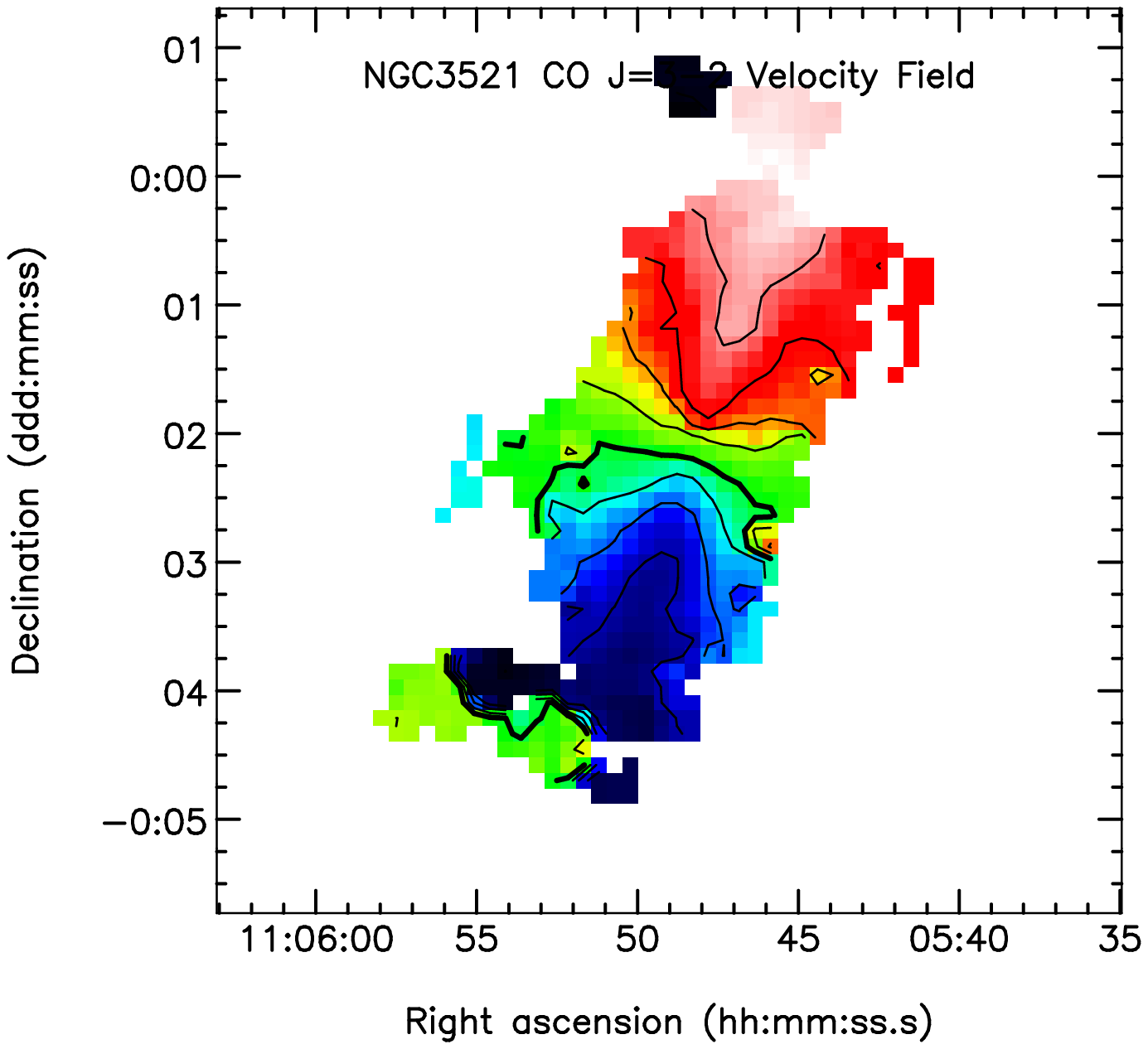}
\includegraphics[width=60mm]{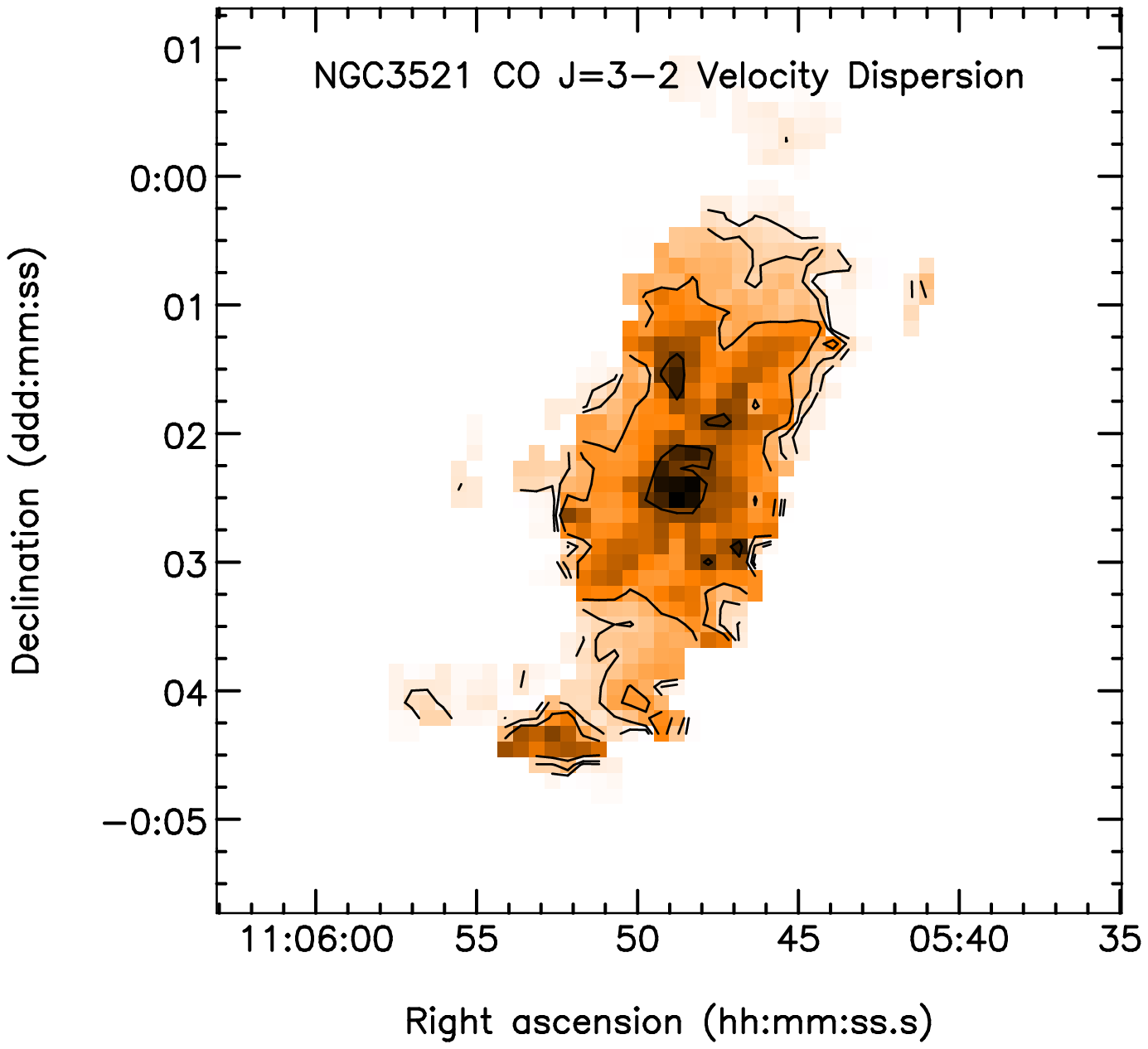}
\caption{CO $J$=3-2 images for NGC 3521.  (a) CO $J$=3-2 integrated intensity
  image. Contours levels are (0.5,   1,   2,   4,   8) K km s$^{-1}$ (T$_{MB}$).
(b) CO $J$=3-2 overlaid on a
Digitized Sky Survey image. (c) Velocity field. Contour levels are
(558,   613,   668,   723,   778,   833,   888,   943,   998) 
km s$^{-1}$. 
(d) The velocity dispersion $\sigma_v$  as traced by the 
CO $J$=3-2 second moment map.  Contour levels are
(4,   8,   16,   32)
km s$^{-1}$. Similar images derived from the same data have been
published in \citet{warren10}.
\label{fig-ngc3521}}
\end{figure*}

\begin{figure*}
\includegraphics[width=60mm]{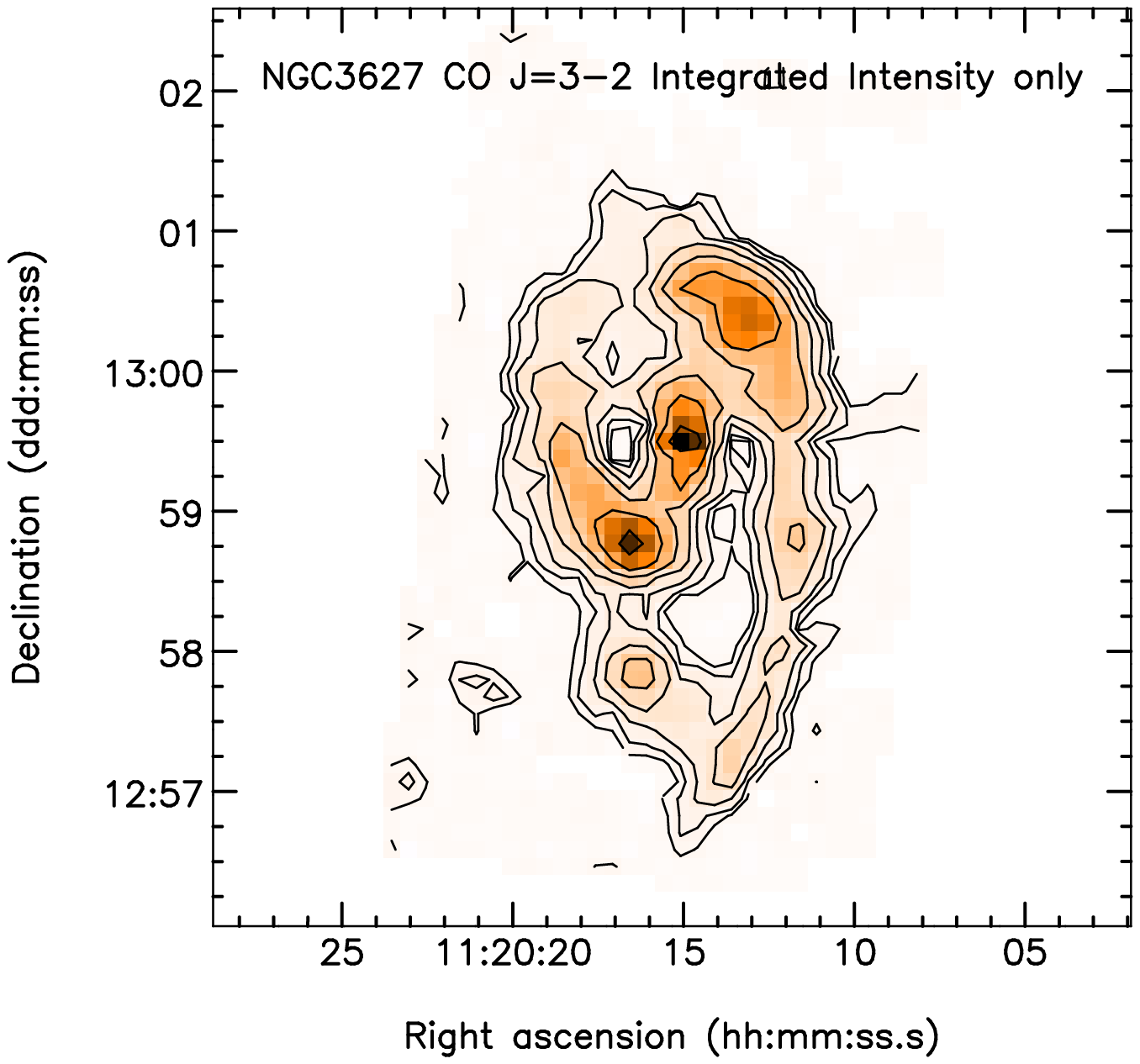}
\includegraphics[width=60mm]{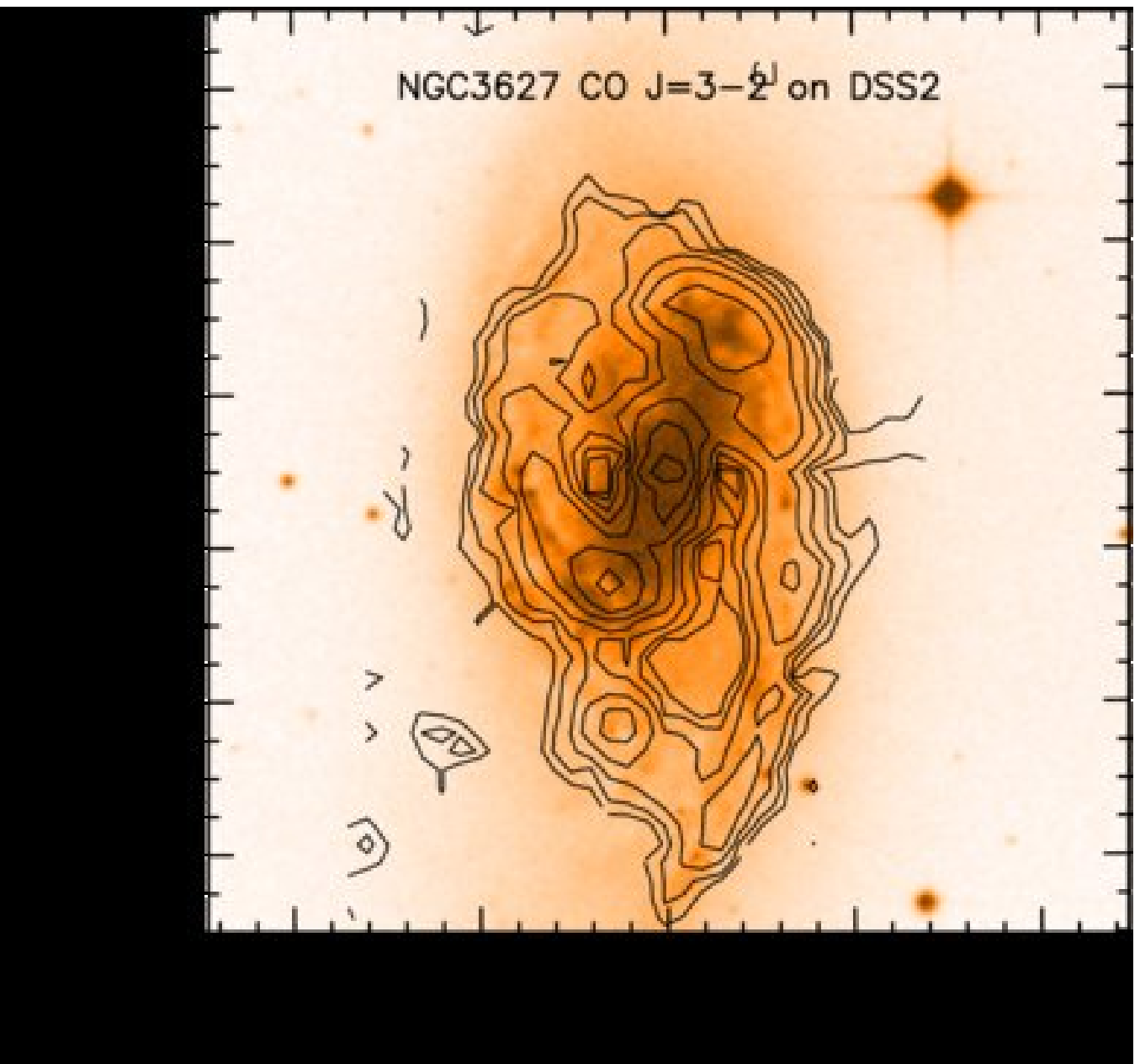}
\includegraphics[width=60mm]{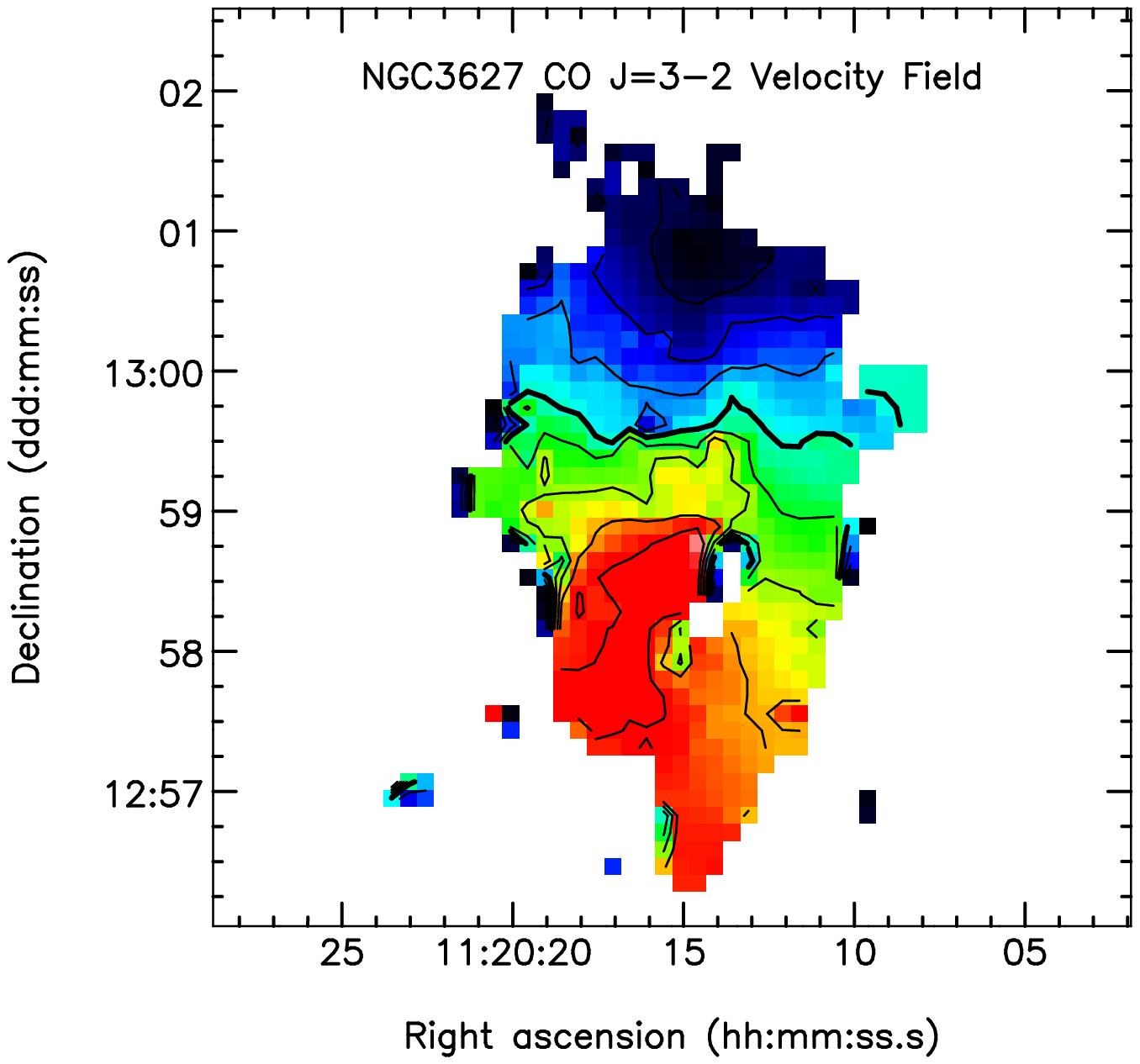}
\includegraphics[width=60mm]{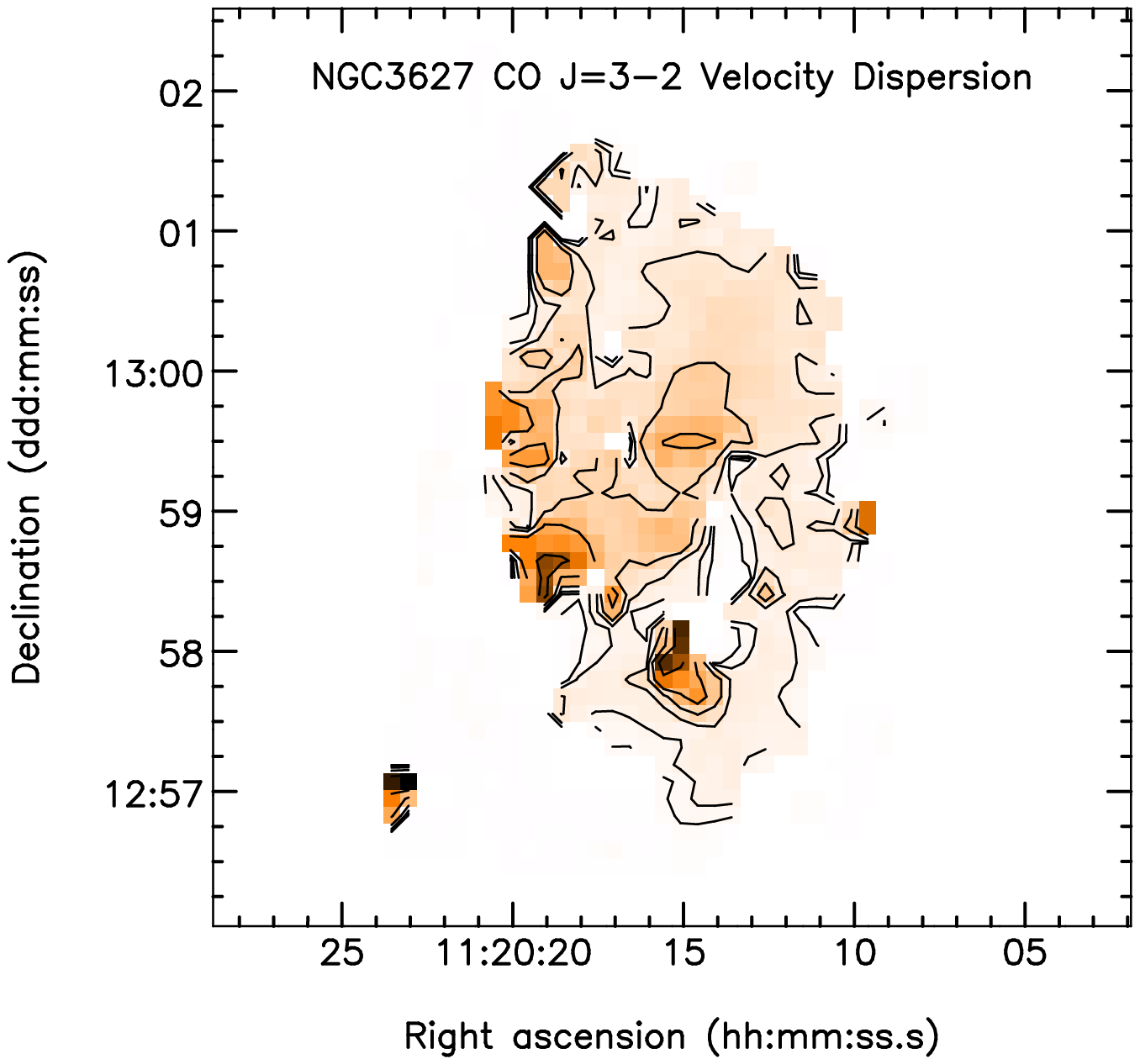}
\caption{CO $J$=3-2 images for NGC 3627.  (a) CO $J$=3-2 integrated intensity
  image. Contours levels are (0.5,   1,   2,   4,   8, 16, 32) K km s$^{-1}$ (T$_{MB}$).
(b) CO $J$=3-2 overlaid on a
Digitized Sky Survey image. (c) Velocity field. Contour levels are
(559,   605,   651,   697,   743,   789,   835,   881) 
km s$^{-1}$. 
(d) The velocity dispersion $\sigma_v$  as traced by the 
CO $J$=3-2 second moment map.  Contour levels are
(4,   8,   16,   32, 64, 128)
km s$^{-1}$. Similar images derived from the same data have been
published in \citet{warren10}.
\label{fig-ngc3627}}
\end{figure*}

\clearpage

\begin{figure*}
\includegraphics[width=55mm]{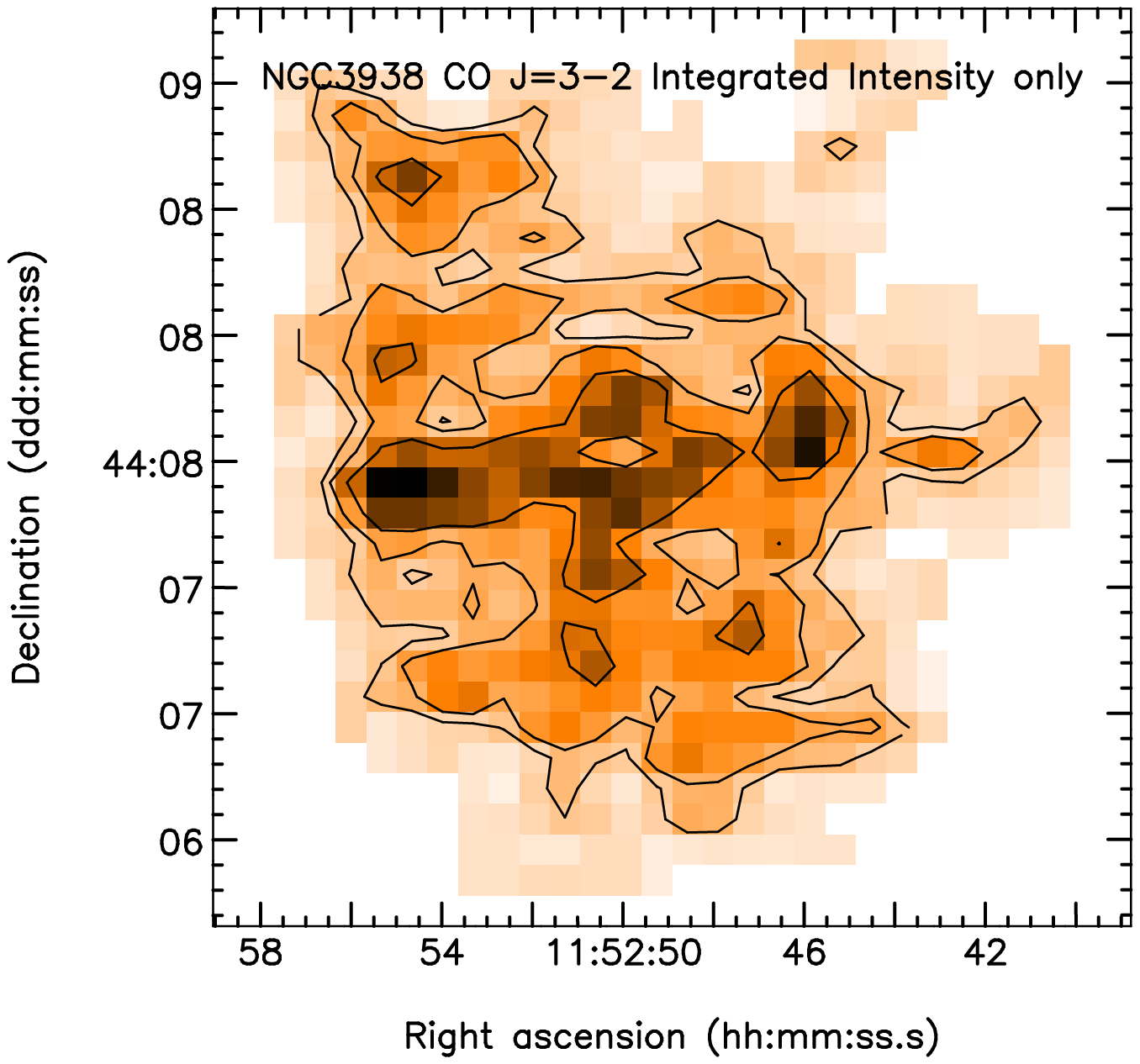}
\includegraphics[width=55mm]{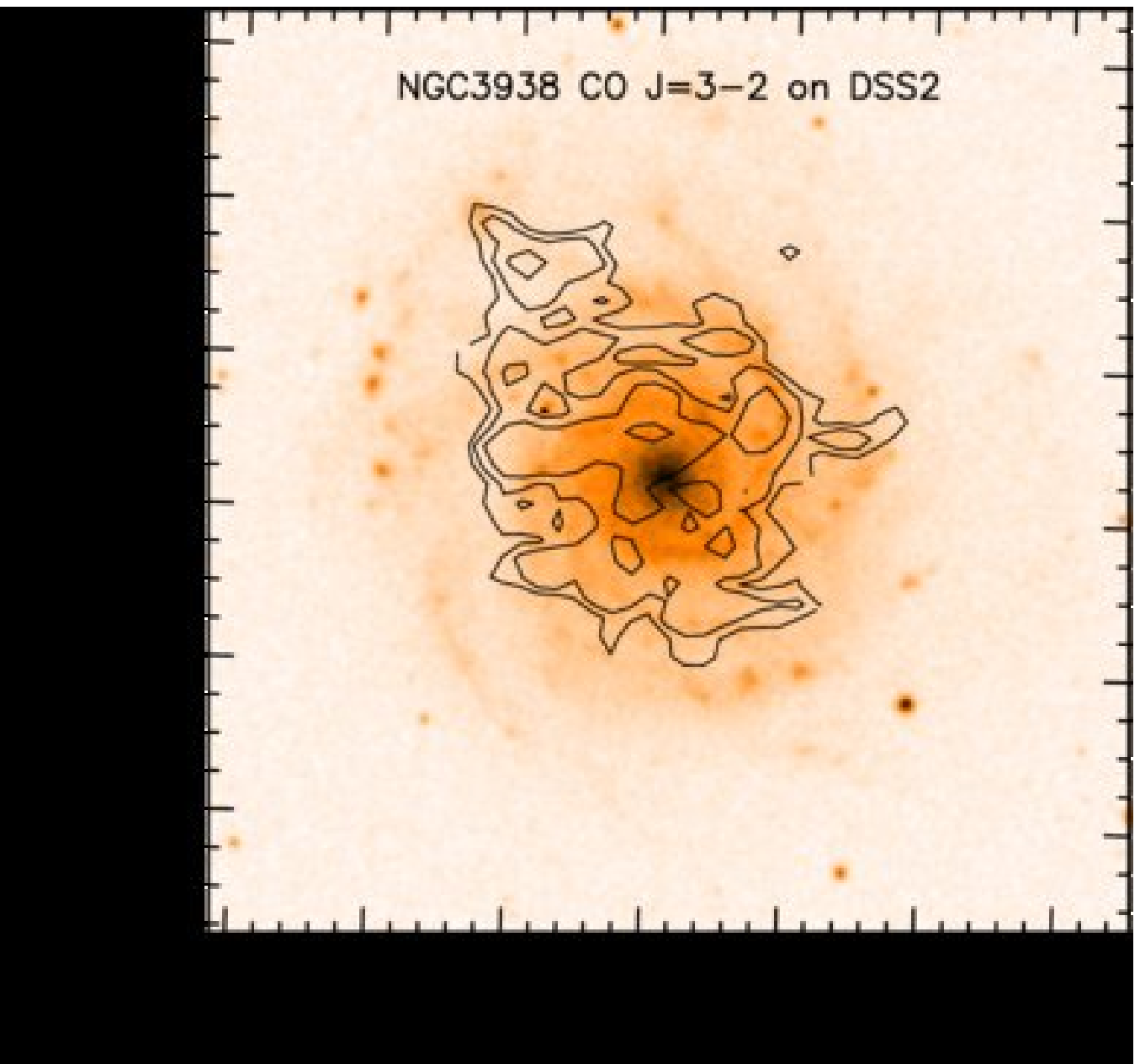}
\includegraphics[width=55mm]{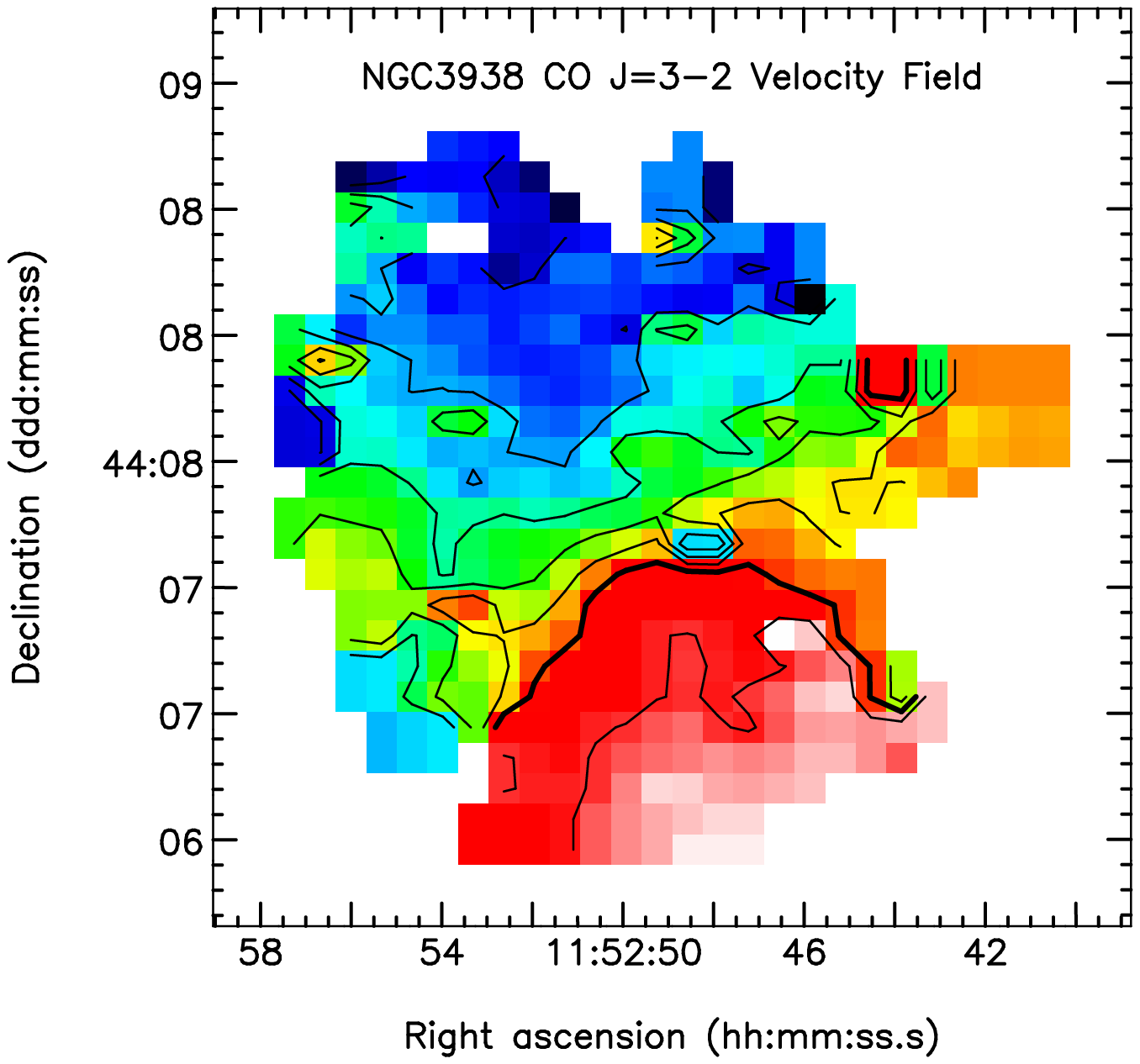}
\caption{CO $J$=3-2 images for NGC 3938. (a) CO $J$=3-2 integrated intensity
  image. Contours levels are (0.5,   1,   2)
K km s$^{-1}$ (T$_{MB}$).
(b) CO $J$=3-2 integrated intensity contours overlaid on an optical
image from the Digitized Sky Survey. (c) Velocity field as traced by the
CO $J$=3-2 first moment map. Contour levels are (769,   781,   793,
805,   817,   829,   841) km s$^{-1}$. 
The velocity dispersion map has been published in \citet{w11}.
\label{fig-ngc3938}}
\end{figure*}

\begin{figure*}
\includegraphics[width=60mm]{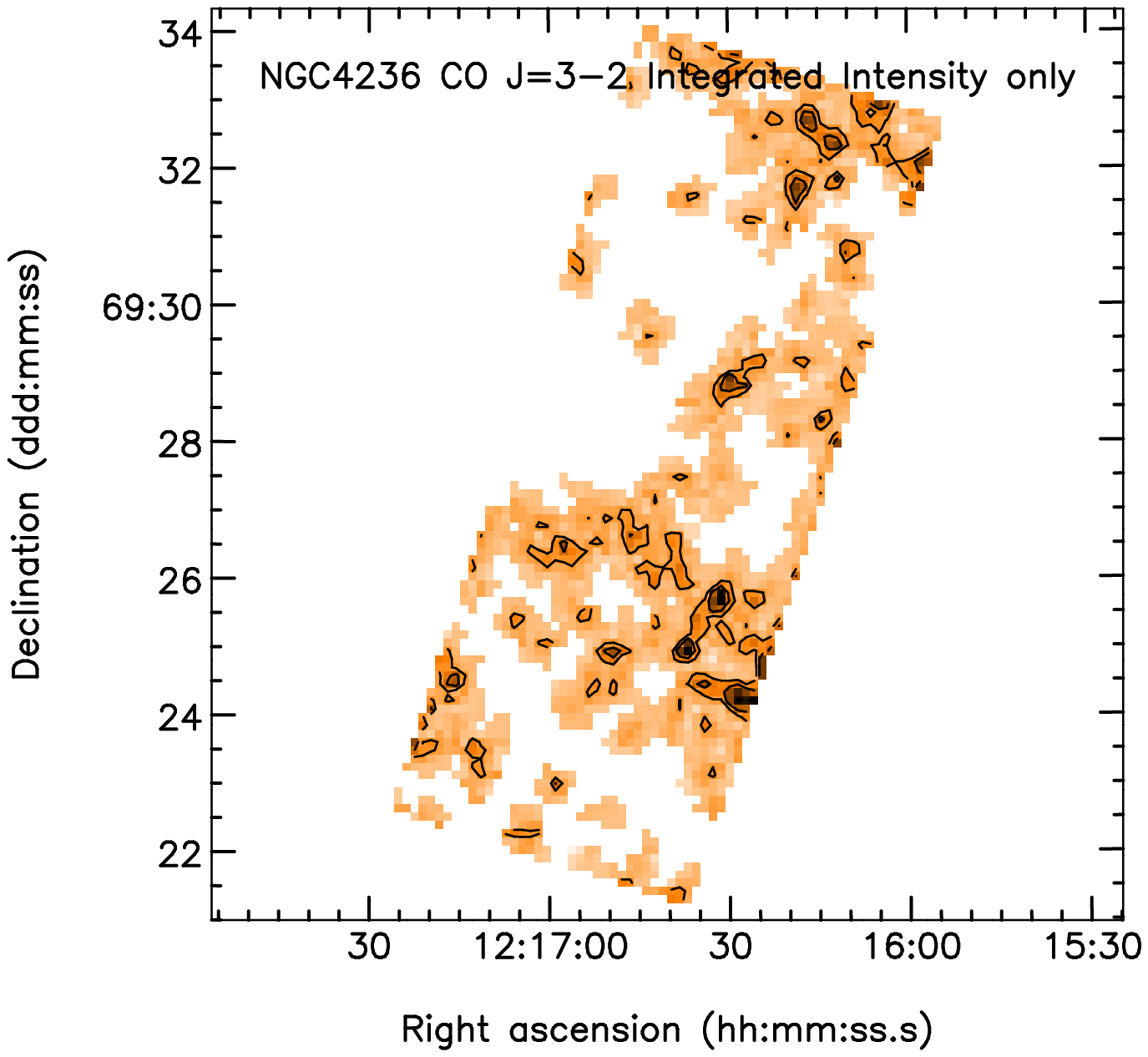}
\includegraphics[width=60mm]{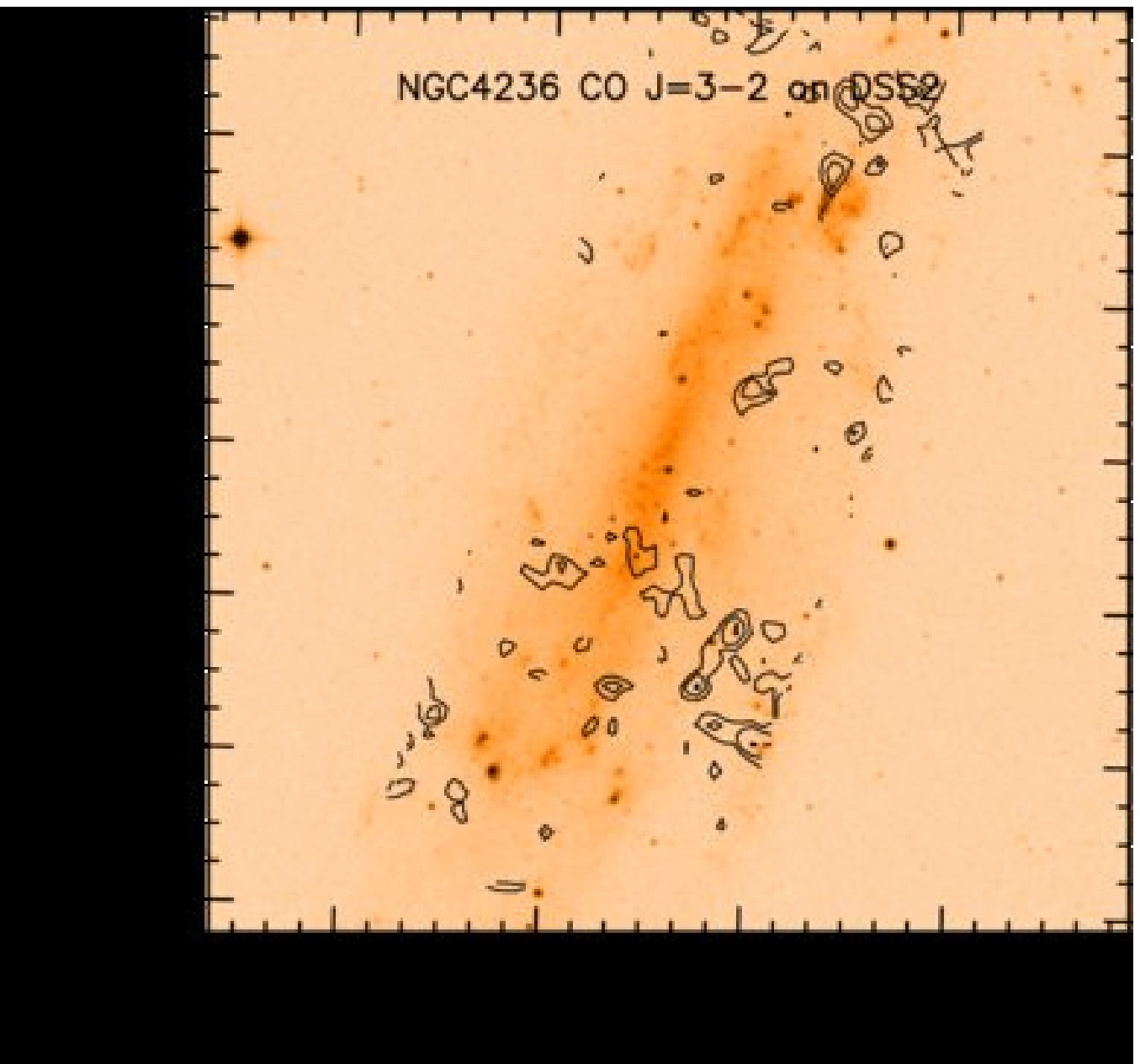}
\includegraphics[width=60mm]{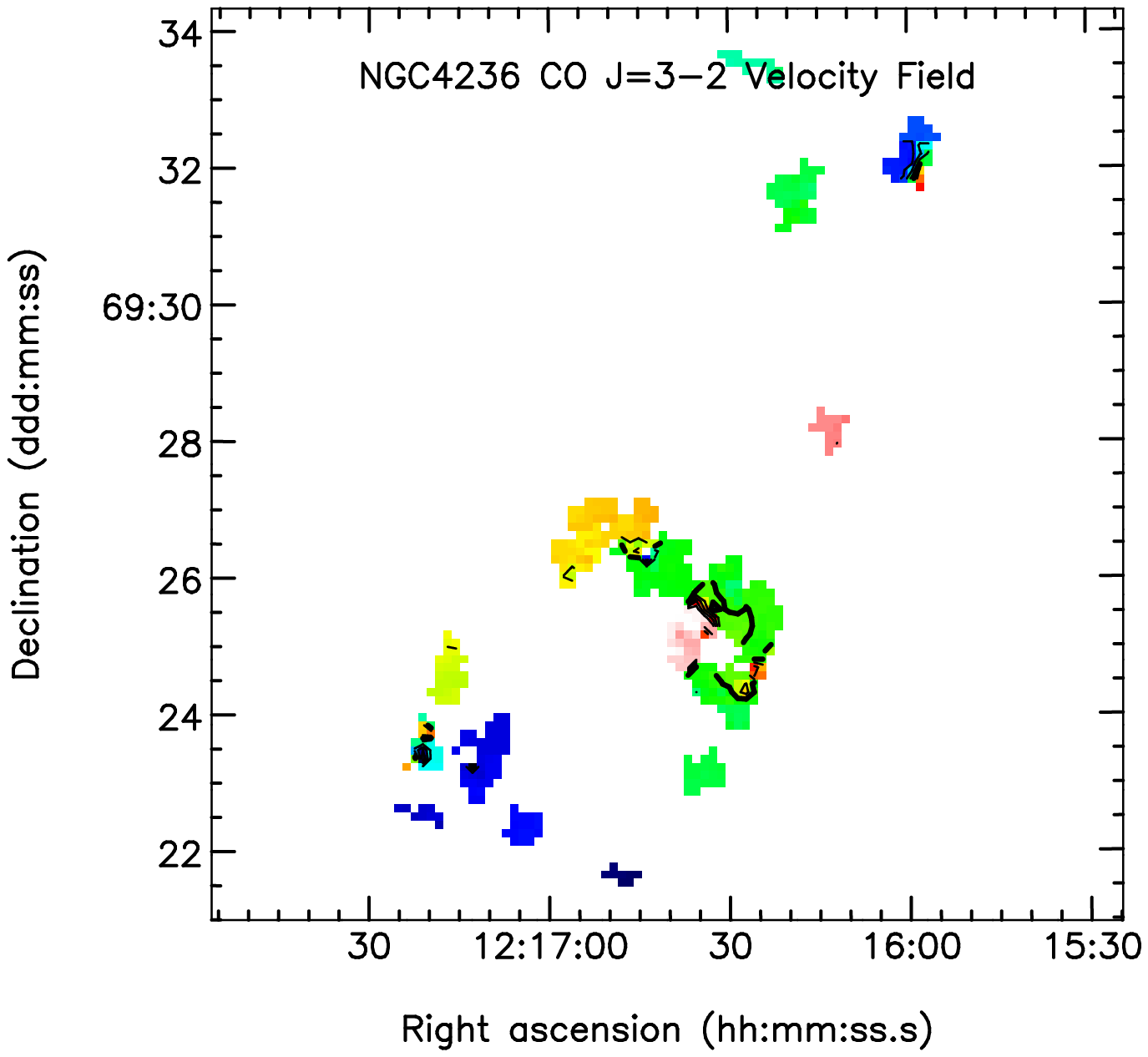}
\includegraphics[width=60mm]{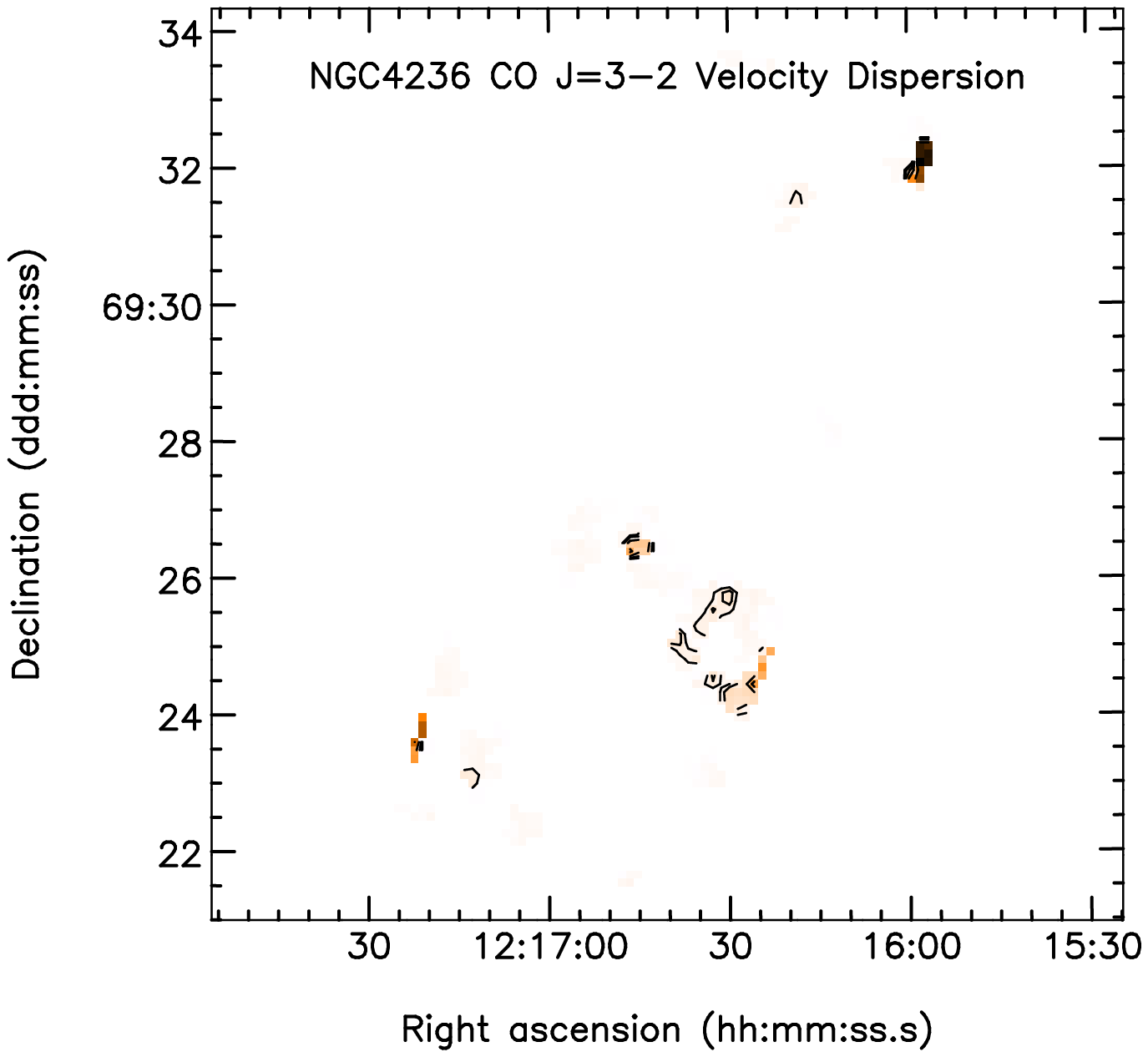}
\caption{CO $J$=3-2 images for NGC 4236.  (a) CO $J$=3-2 integrated intensity
  image. Contours levels are (0.5,   1,   2) K km s$^{-1}$ (T$_{MB}$).
(b) CO $J$=3-2 overlaid on a
Digitized Sky Survey image. (c) Velocity field. Contour levels are
(-190,   -150,   -110,   -70,   -30,   10,   50,   90,   130,
170) 
km s$^{-1}$. 
(d) The velocity dispersion $\sigma_v$ as traced by the 
CO $J$=3-2 second moment map.  Contour levels are
(4,   8,   16,   32, 64)
km s$^{-1}$.
\label{fig-ngc4236}}
\end{figure*}

\begin{figure*}
\includegraphics[width=55mm]{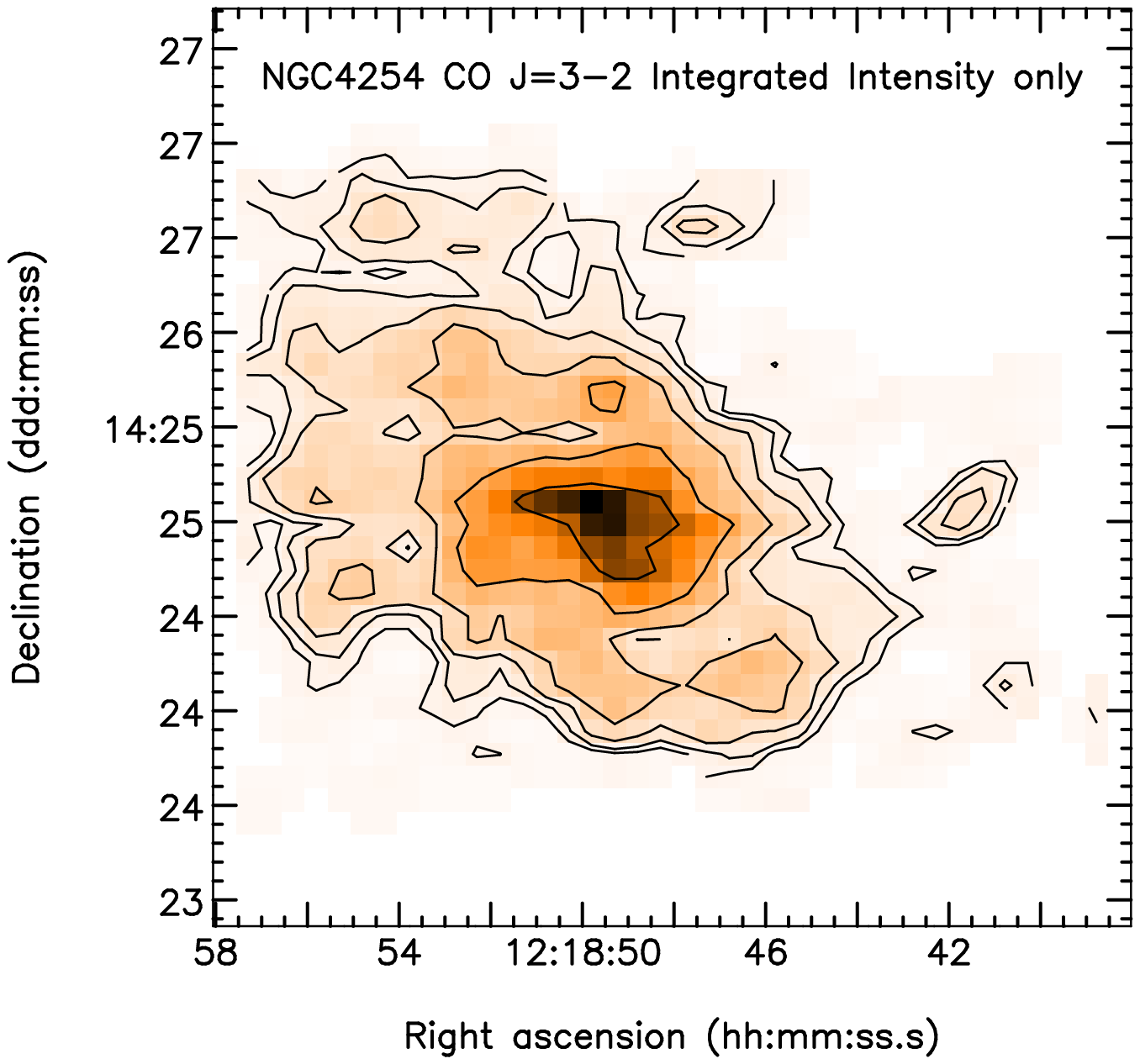}
\includegraphics[width=55mm]{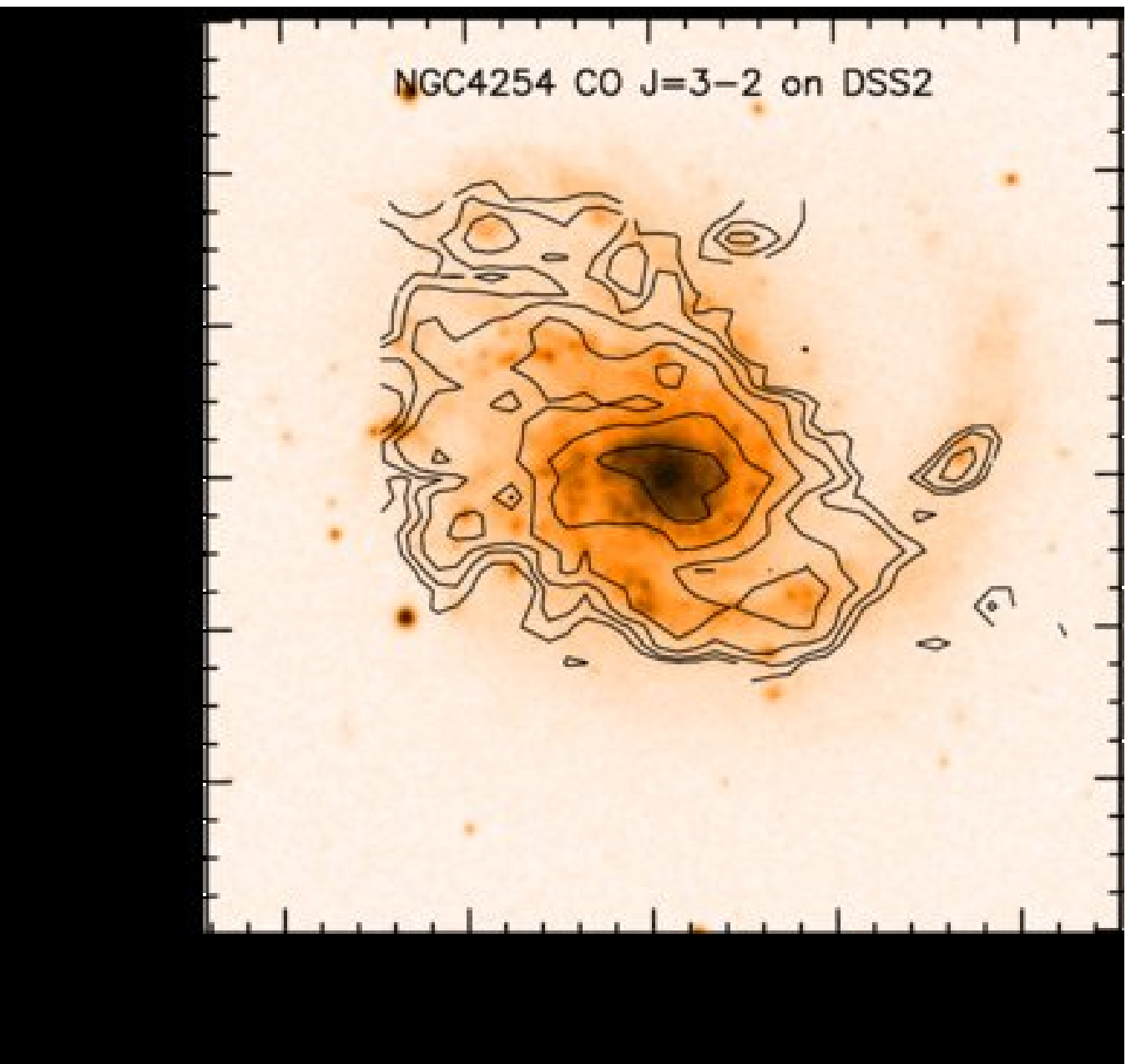}
\includegraphics[width=55mm]{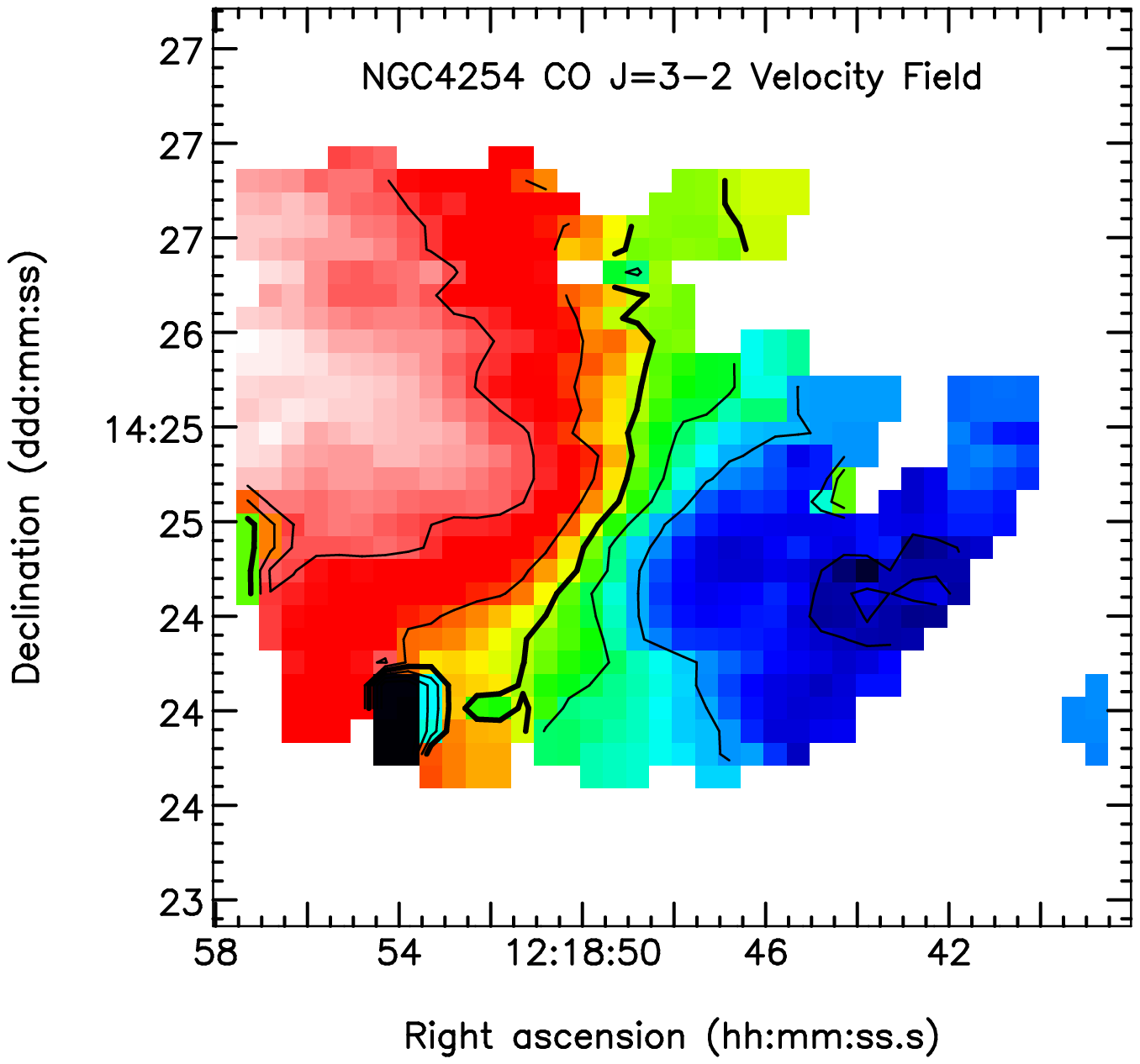}
\caption{CO $J$=3-2 images for NGC 4254. (a) CO $J$=3-2 integrated intensity
  image. Contours levels are (0.5,   1,   2,   4,   8,   16)
K km s$^{-1}$ (T$_{MB}$).
(b) CO $J$=3-2 integrated intensity contours overlaid on an optical
image from the Digitized Sky Survey. (c) Velocity field as traced by the
CO $J$=3-2 first moment map. Contour levels are (2304,   2340,   2376,
2412,   2448,   2484) km s$^{-1}$. 
Similar images derived from the same data have been
published in \citet{w09}.
The velocity dispersion map has been published in \citet{w11}.
\label{fig-ngc4254}}
\end{figure*}
 
\begin{figure*}
\includegraphics[width=55mm]{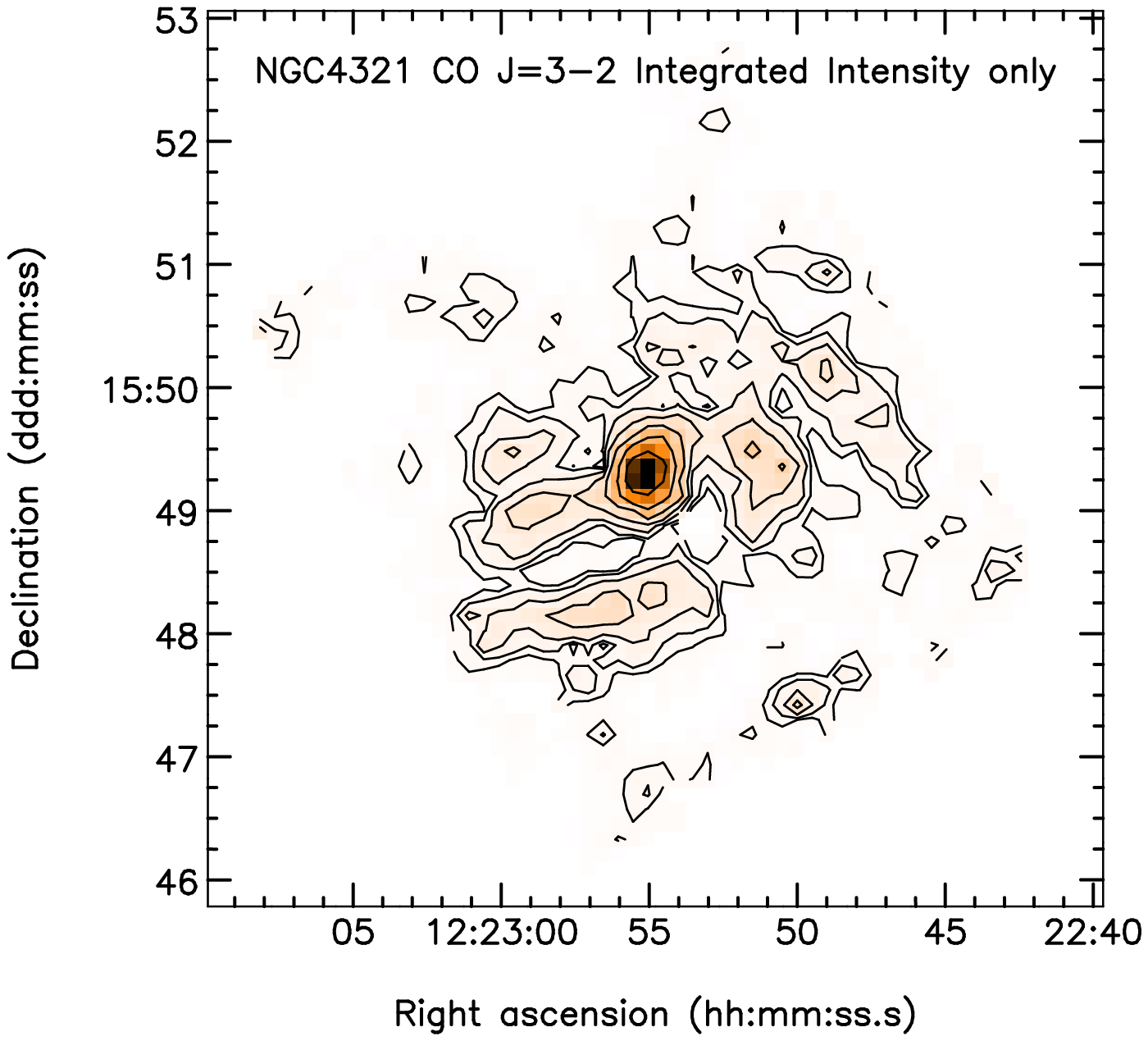}
\includegraphics[width=55mm]{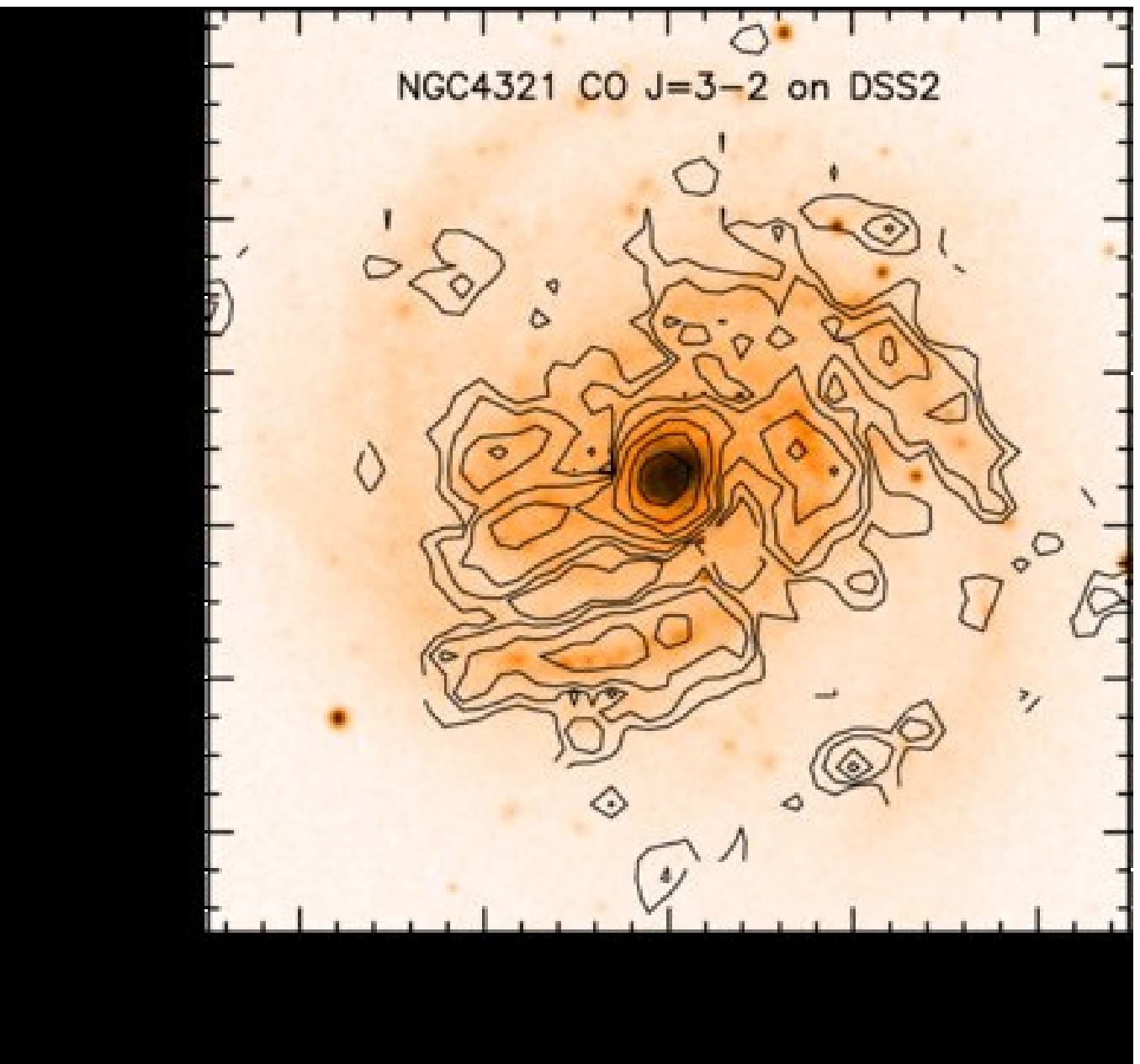}
\includegraphics[width=55mm]{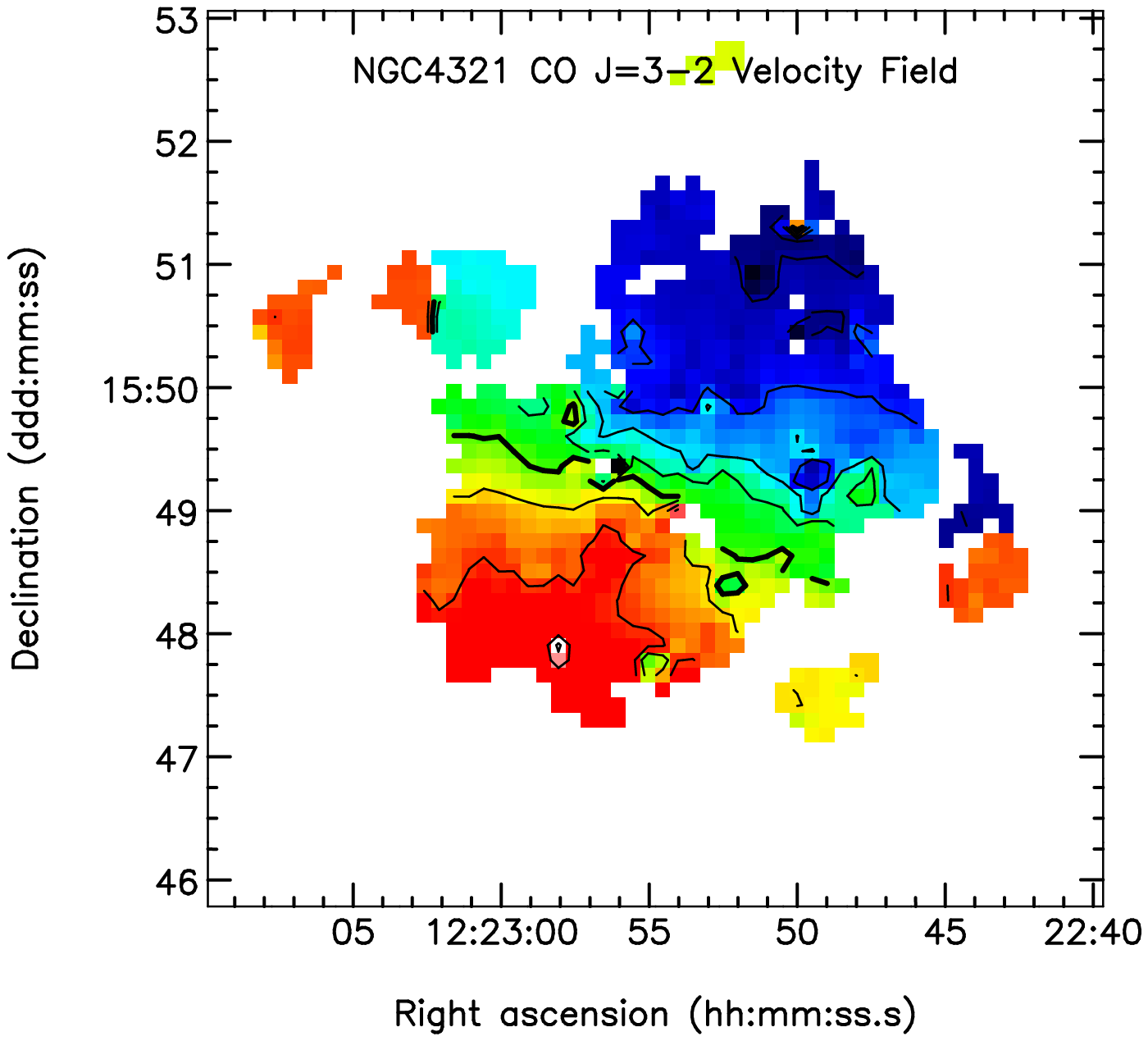}
\caption{CO $J$=3-2 images for NGC 4321. (a) CO $J$=3-2 integrated intensity
  image. Contours levels are (0.5,   1,   2,   4,   8,   16, 32)
K km s$^{-1}$ (T$_{MB}$).
(b) CO $J$=3-2 integrated intensity contours overlaid on an optical
image from the Digitized Sky Survey. (c) Velocity field as traced by the
CO $J$=3-2 first moment map. Contour levels are (1424,   1459,   1494,
1529,   1564,   1599,   1634,   1669, 
1704,   1739) km s$^{-1}$.
Similar images derived from the same data have been
published in \citet{w09}.
The velocity dispersion map has been published in \citet{w11}.
\label{fig-ngc4321}}
\end{figure*}

\begin{figure*}
\includegraphics[width=60mm]{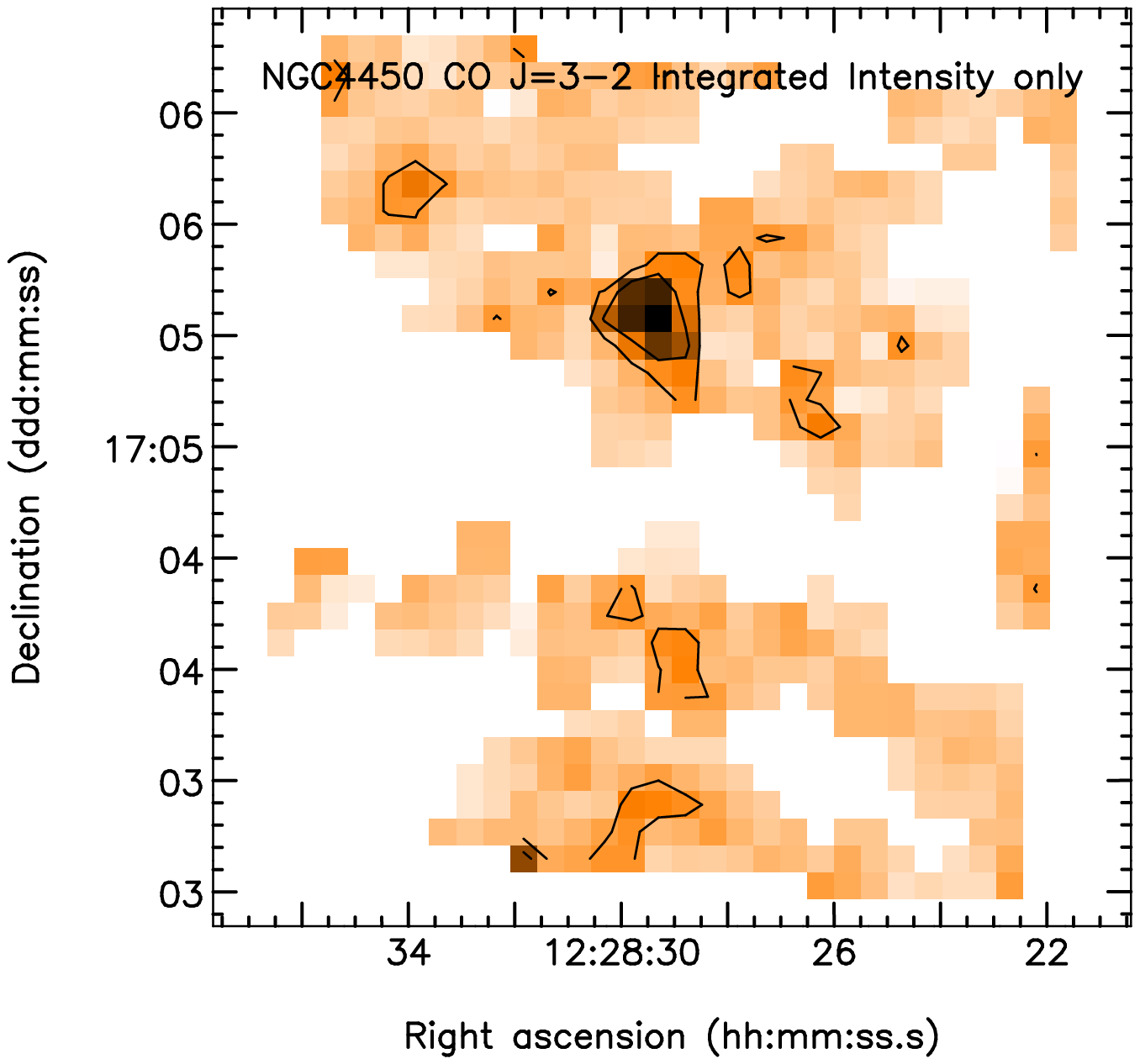}
\includegraphics[width=60mm]{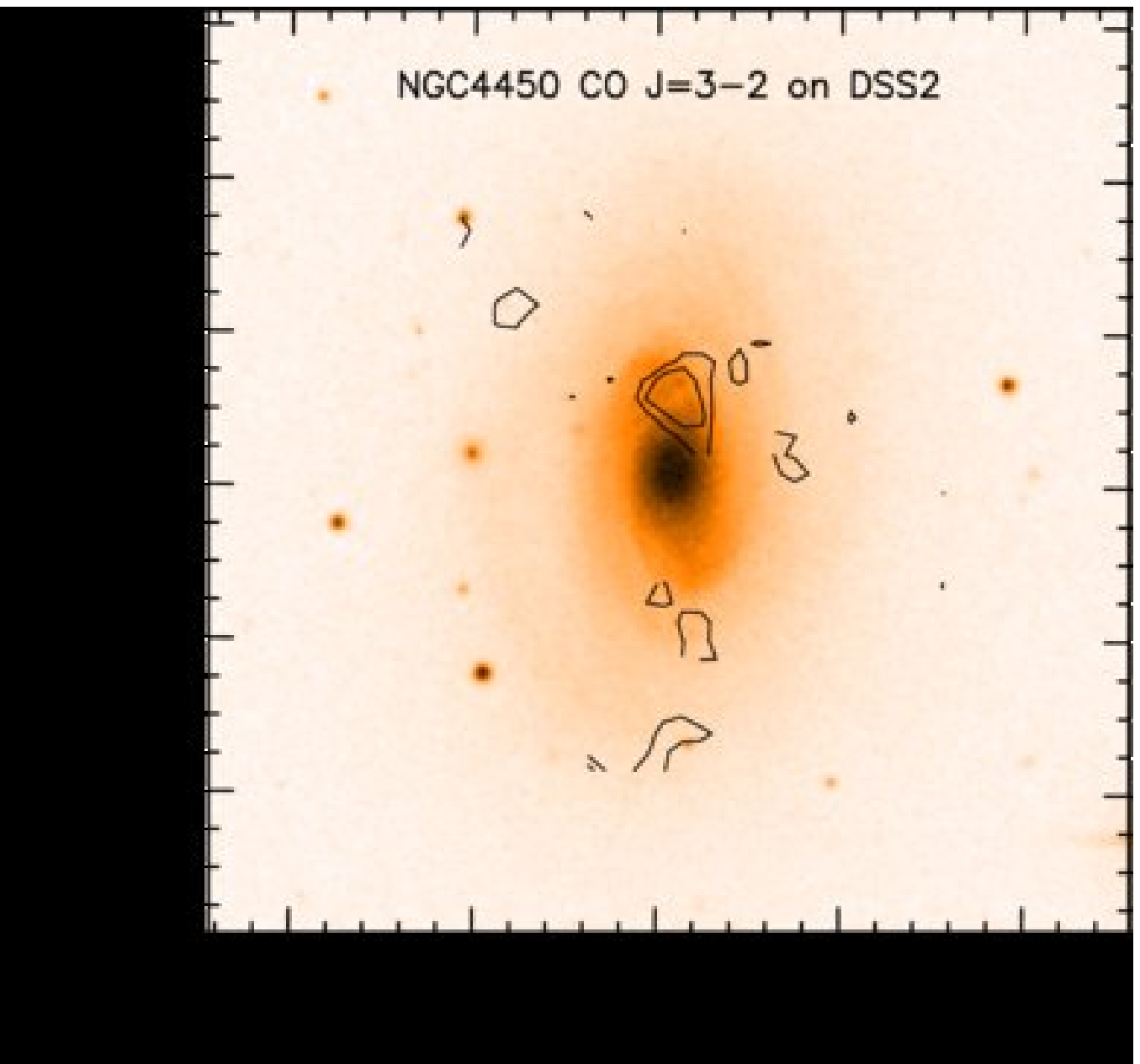}
\caption{CO $J$=3-2 images for NGC 4450.  (a) CO $J$=3-2 integrated intensity
  image. Contours levels are (0.5,   1) K km s$^{-1}$ (T$_{MB}$).
(b) CO $J$=3-2 overlaid on a
Digitized Sky Survey image. 
\label{fig-ngc4450}}
\end{figure*}

\begin{figure*}
\includegraphics[width=60mm]{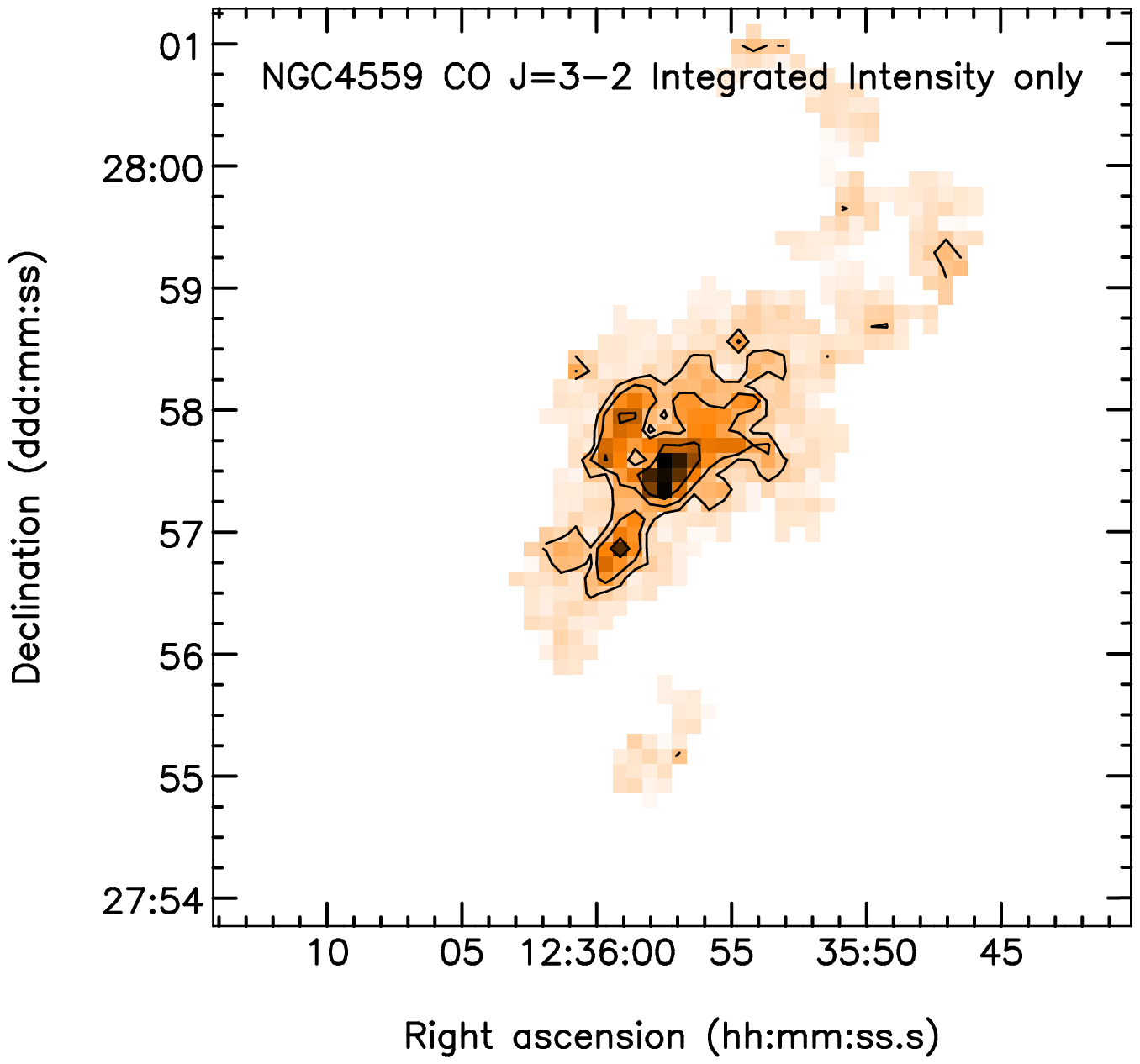}
\includegraphics[width=60mm]{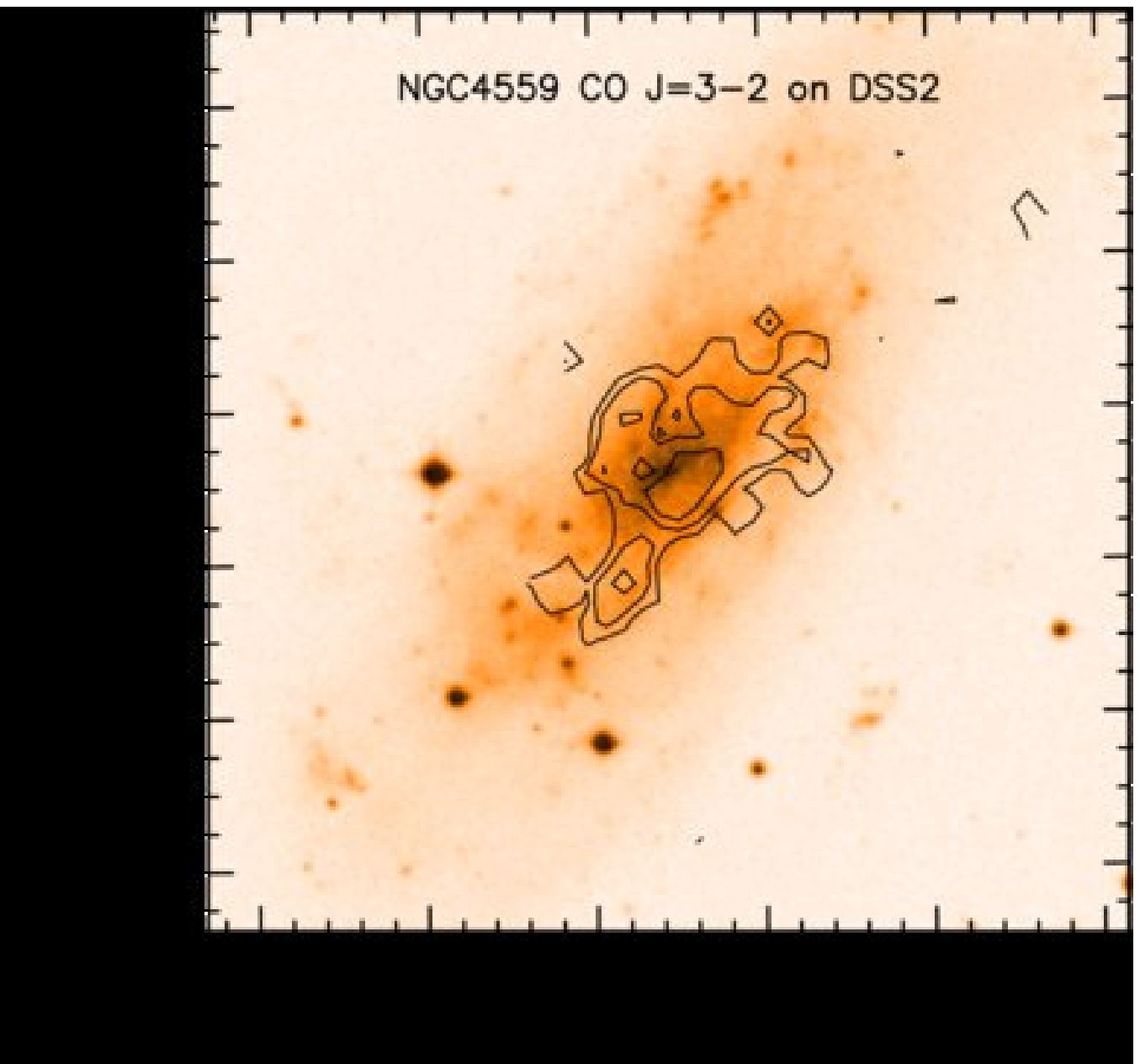}
\includegraphics[width=60mm]{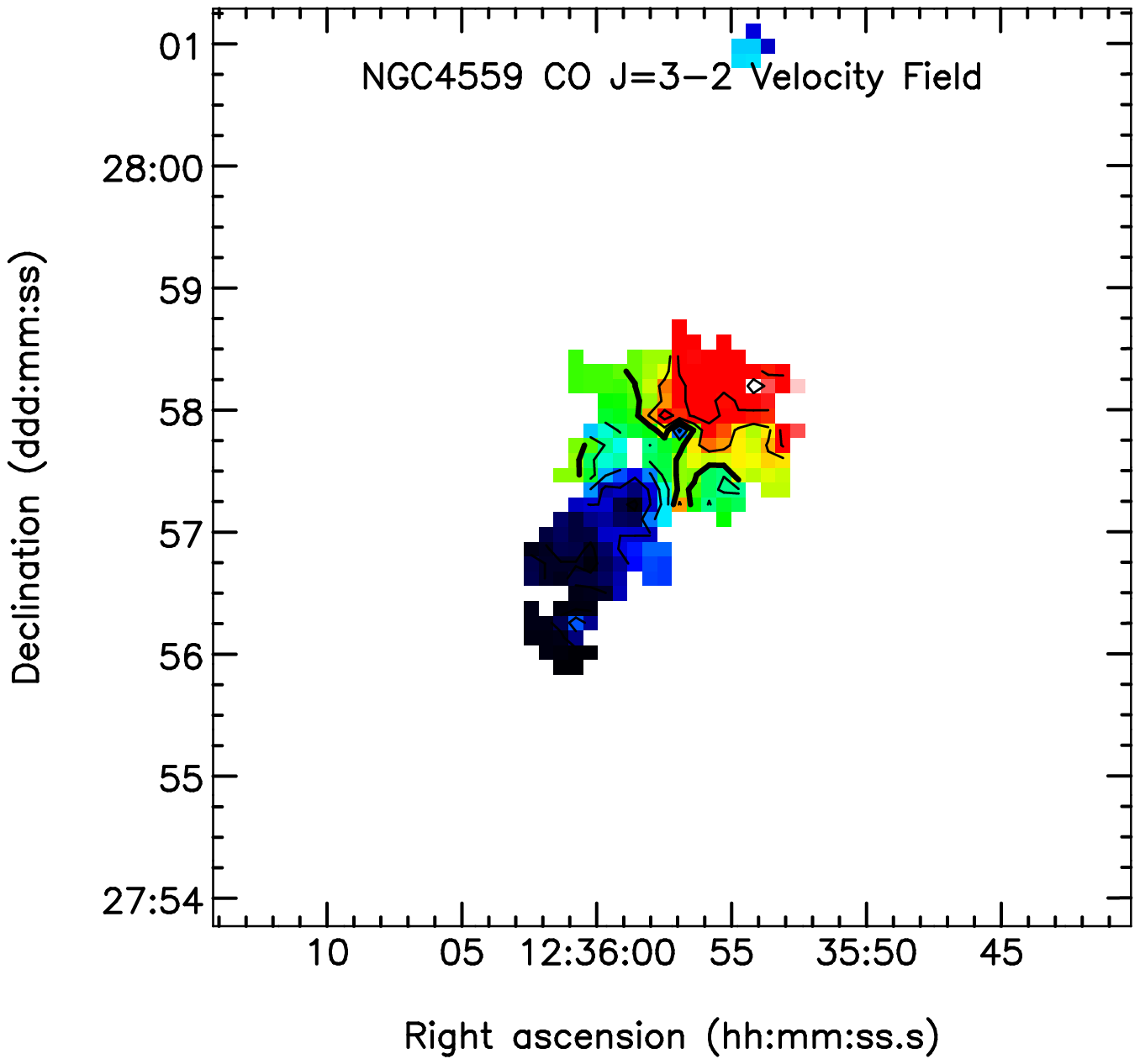}
\includegraphics[width=60mm]{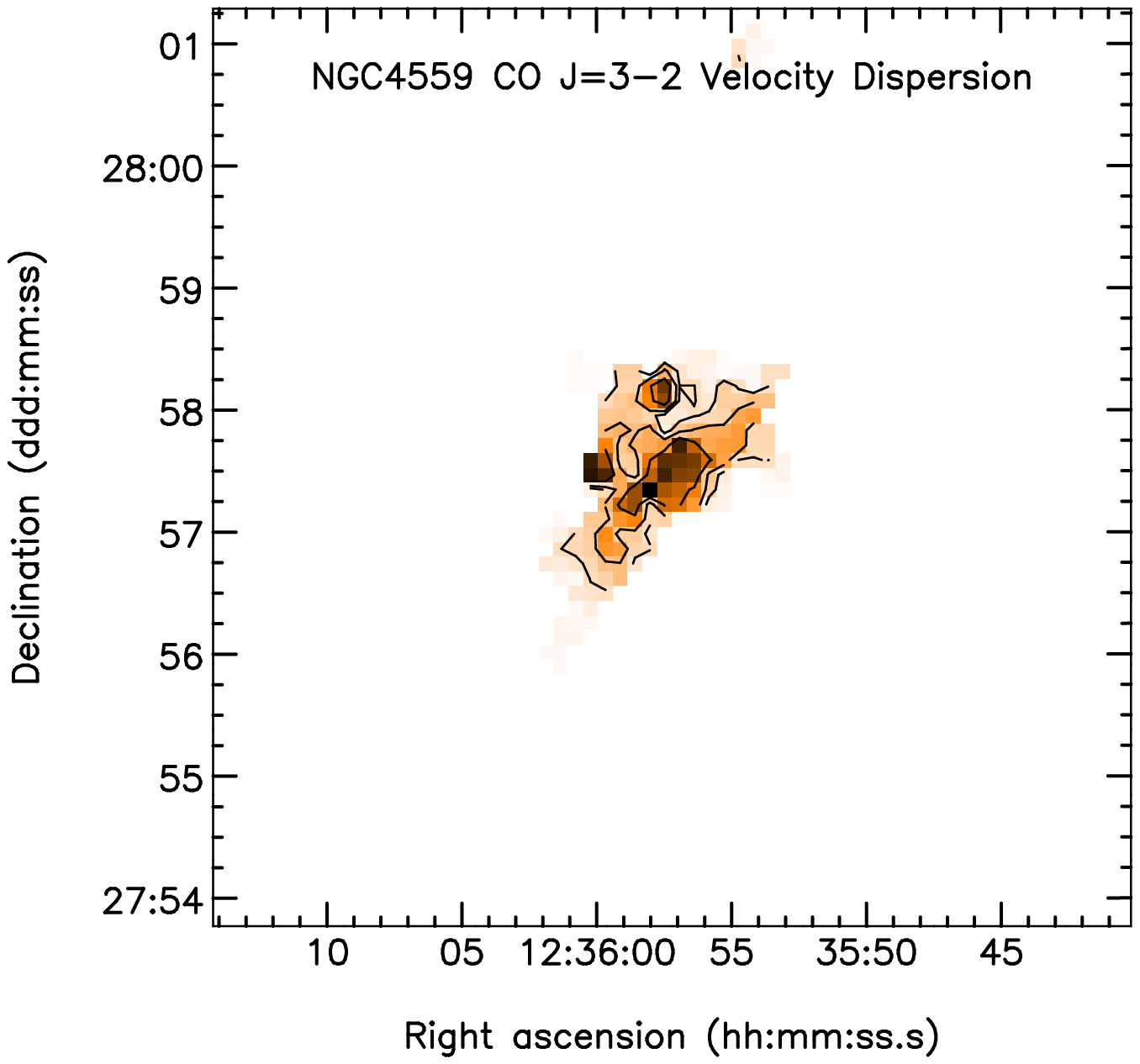}
\caption{CO $J$=3-2 images for NGC 4559.  (a) CO $J$=3-2 integrated intensity
  image. Contours levels are (0.5,   1,   2) K km s$^{-1}$ (T$_{MB}$).
(b) CO $J$=3-2 overlaid on a
Digitized Sky Survey image. (c) Velocity field. Contour levels are
(735,   758,   781,   804,   827,   850,   873,   896)
km s$^{-1}$. 
(d) The velocity dispersion $\sigma_v$ as traced by the 
CO $J$=3-2 second moment map.  Contour levels are
(4,   8,   16)
km s$^{-1}$.
\label{fig-ngc4559}}
\end{figure*}

\begin{figure*}
\includegraphics[width=60mm]{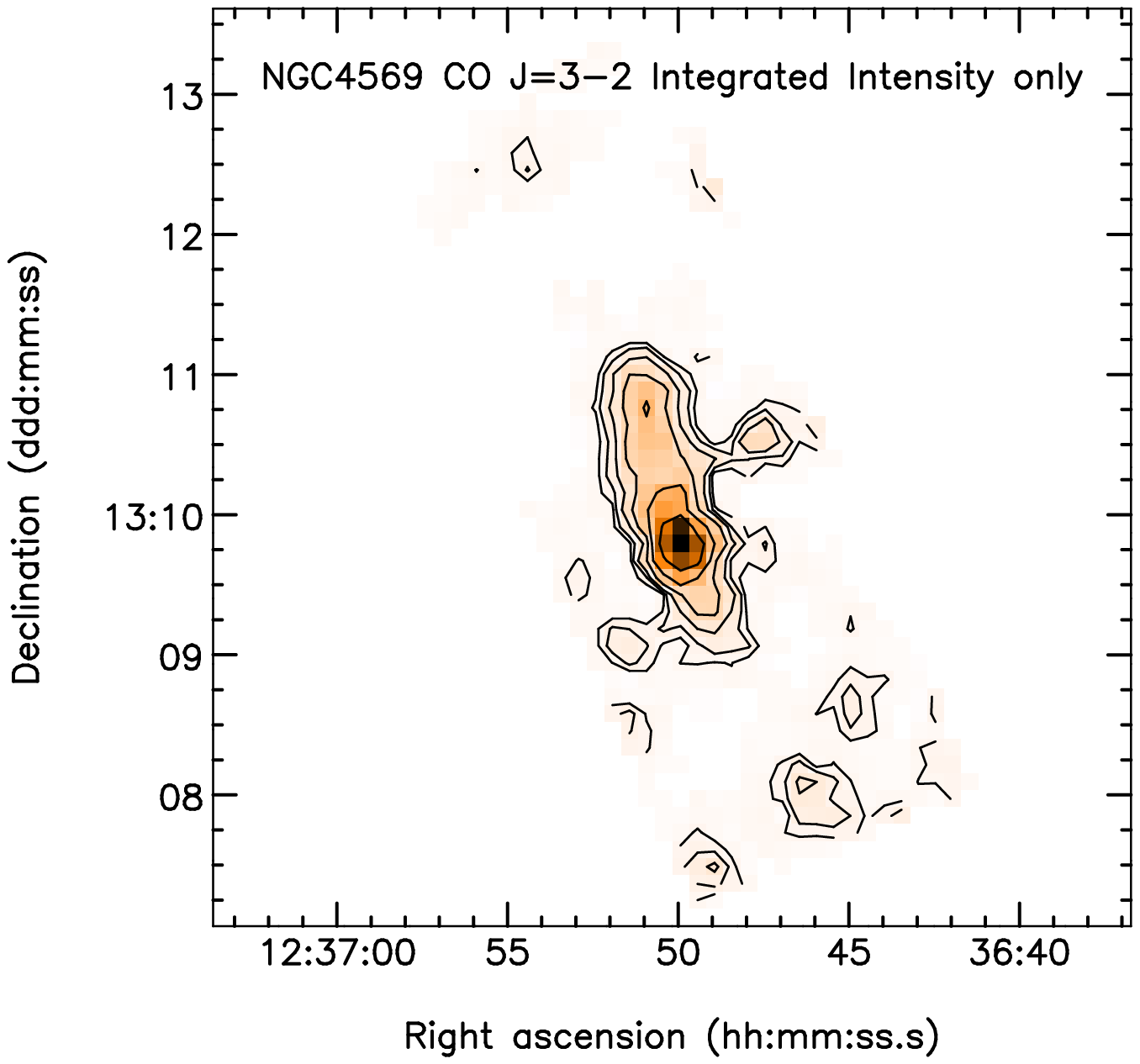}
\includegraphics[width=60mm]{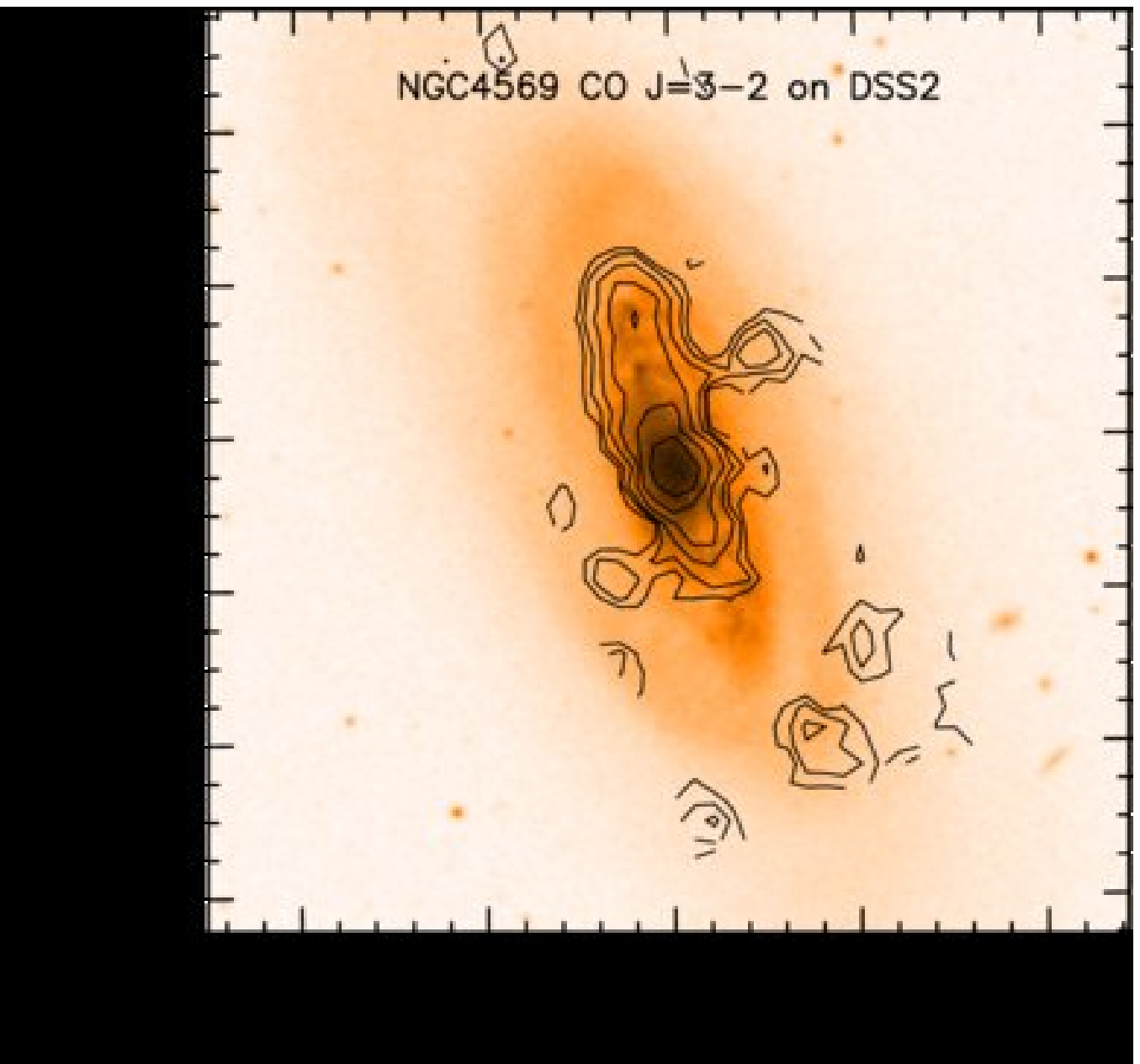}
\includegraphics[width=60mm]{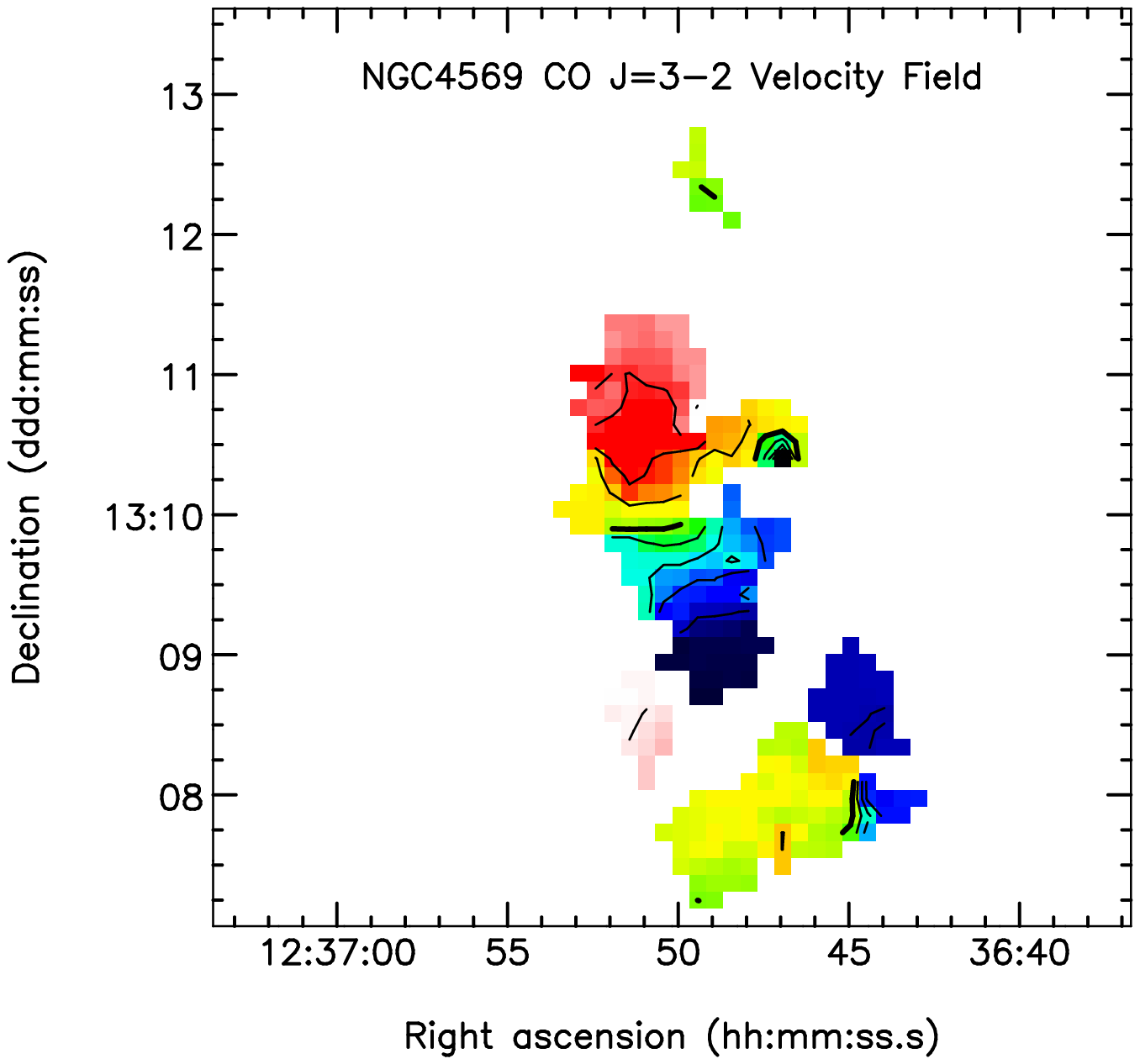}
\includegraphics[width=60mm]{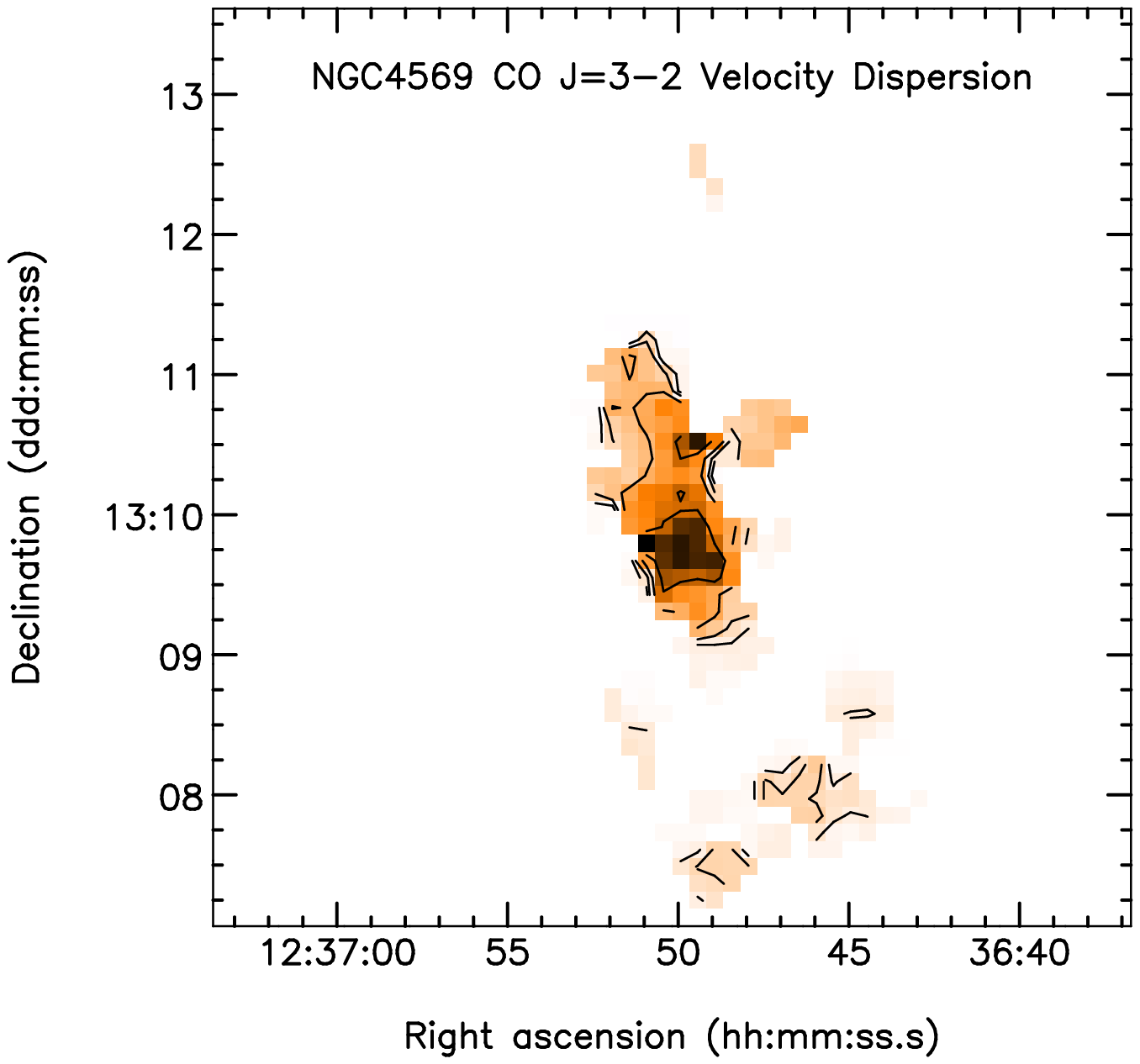}
\caption{CO $J$=3-2 images for NGC 4569.  (a) CO $J$=3-2 integrated intensity
  image. Contours levels are (0.5,   1,   2, 4,   8,   16) K km
  s$^{-1}$ (T$_{MB}$). 
(b) CO $J$=3-2 overlaid on a
Digitized Sky Survey image. (c) Velocity field. Contour levels are
(-379,   -339,   -299,   -259,   -219,   -179,   -139,   -99,
-59,   -19) 
km s$^{-1}$. 
(d) The velocity dispersion $\sigma_v$ as traced by the 
CO $J$=3-2 second moment map.  Contour levels are
(4,   8,   16,   32)
km s$^{-1}$.
Similar images derived from the same data have been
published in \citet{w09}.
\label{fig-ngc4569}}
\end{figure*}

\begin{figure*}
\includegraphics[width=60mm]{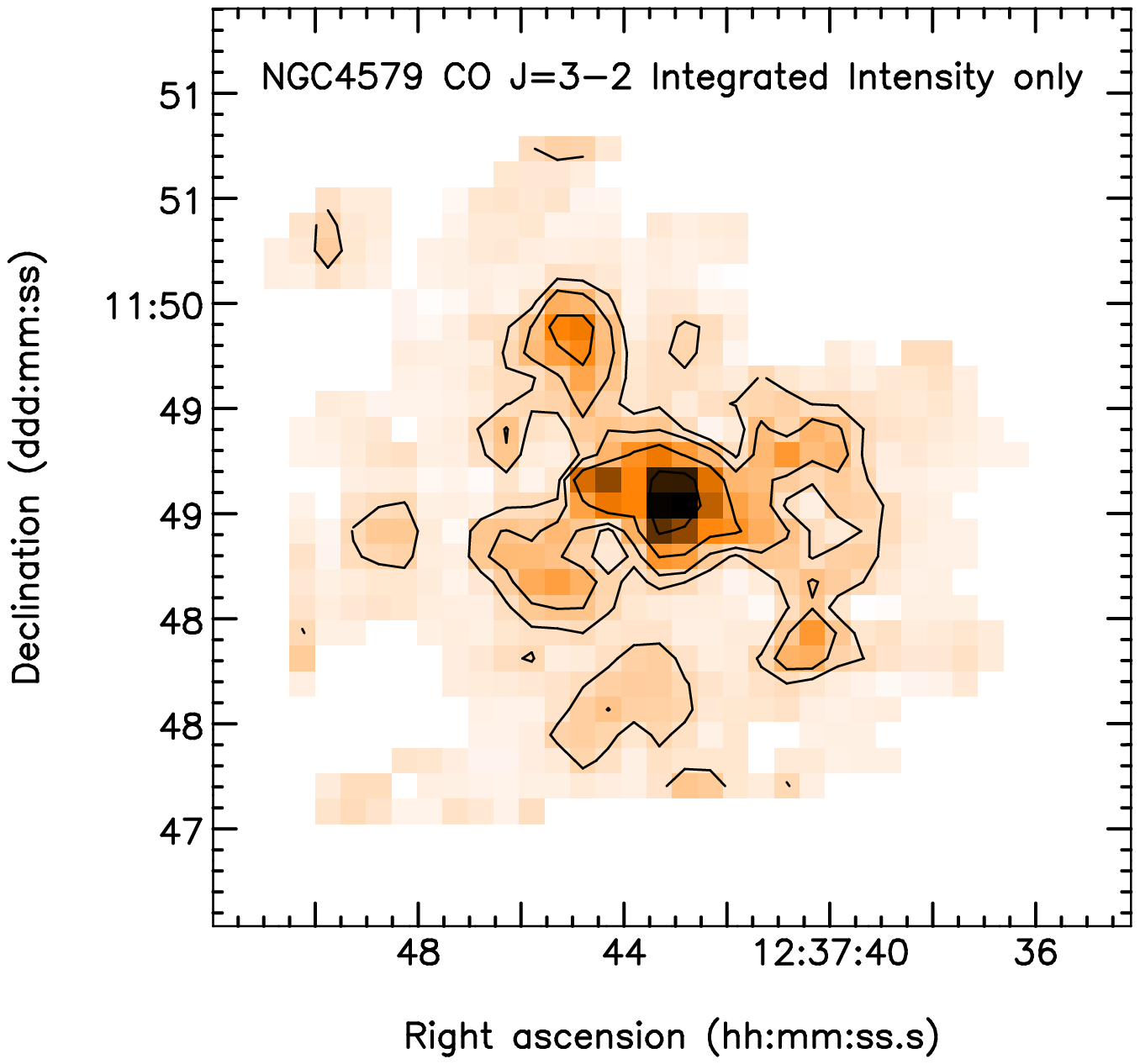}
\includegraphics[width=60mm]{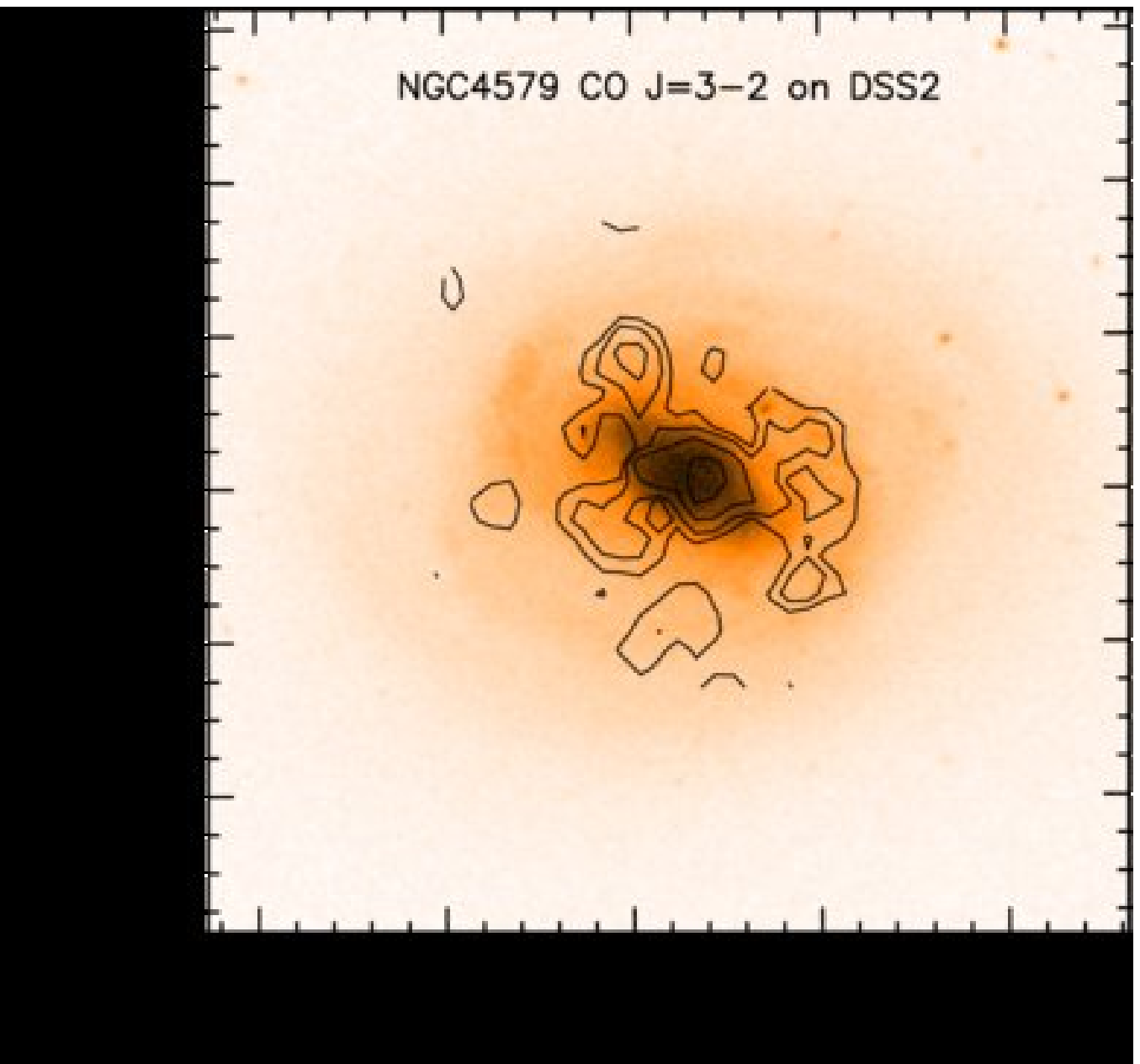}
\includegraphics[width=60mm]{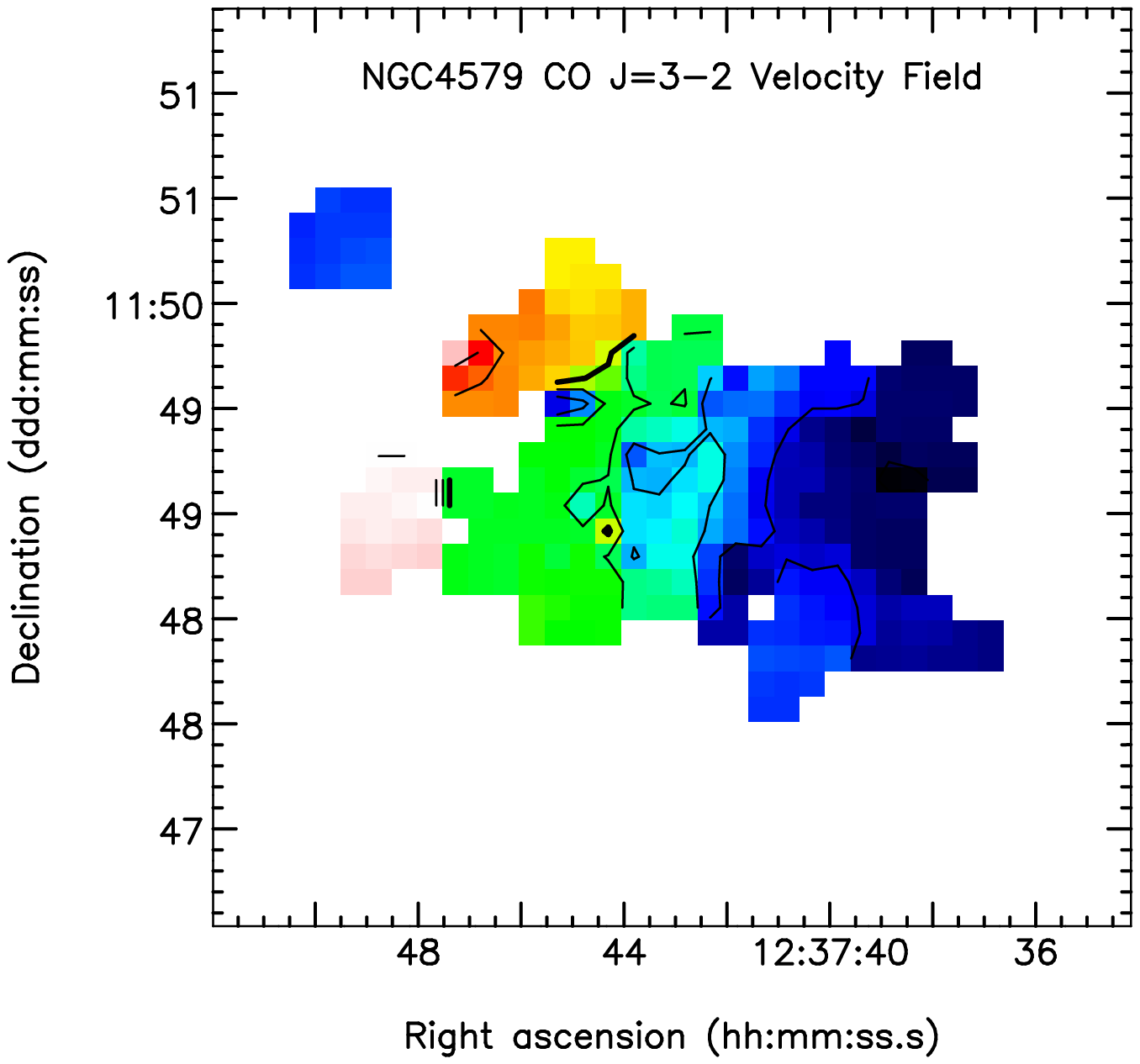}
\includegraphics[width=60mm]{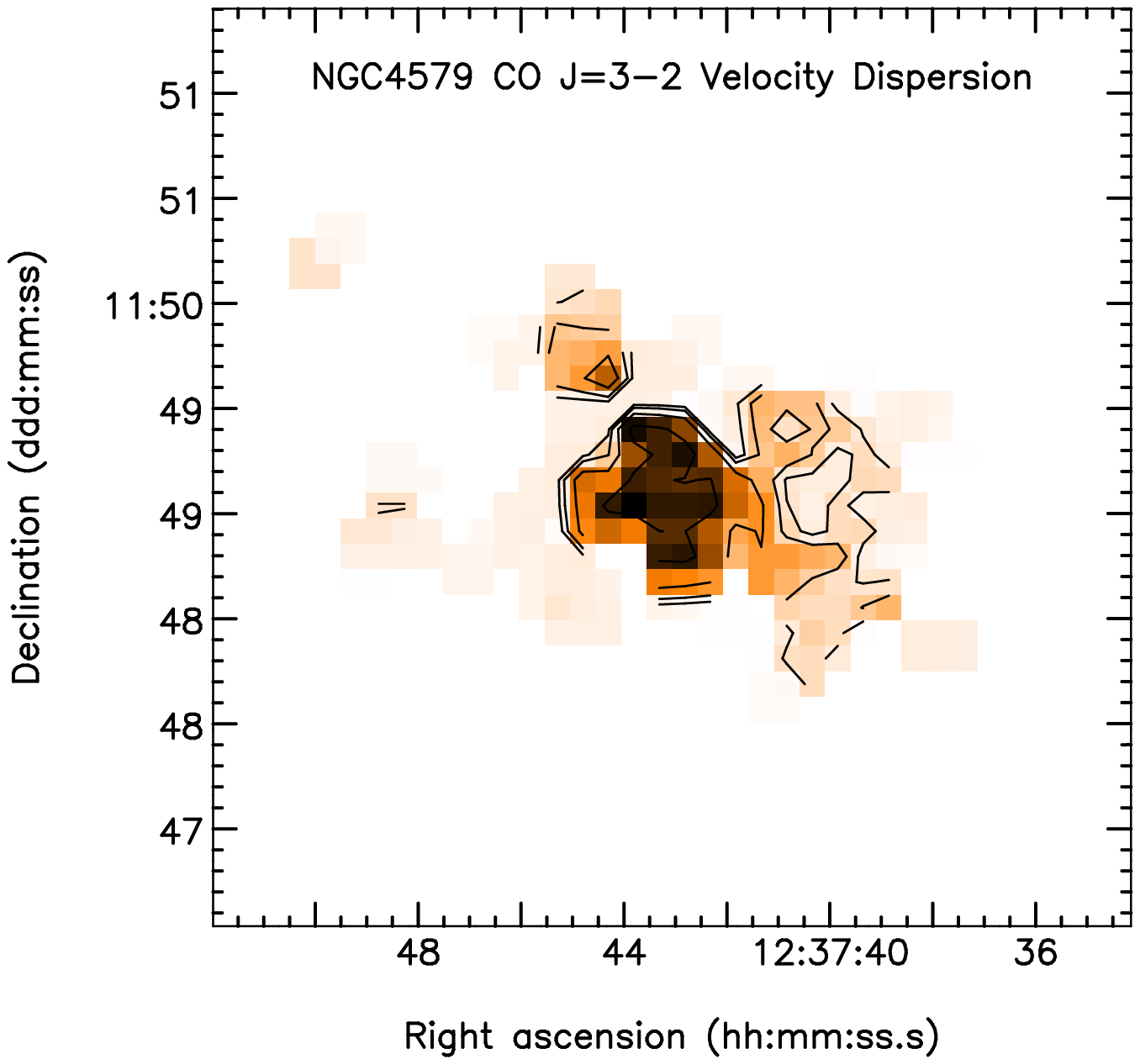}
\caption{CO $J$=3-2 images for NGC 4579.  (a) CO $J$=3-2 integrated intensity
  image. Contours levels are (0.5,   1,   2, 4) K km
  s$^{-1}$ (T$_{MB}$). 
(b) CO $J$=3-2 overlaid on a
Digitized Sky Survey image. (c) Velocity field. Contour levels are
(1340,   1390,   1440,   1490,   1540,   1590,   1640,   1690) 
km s$^{-1}$. 
(d) The velocity dispersion $\sigma_v$ as traced by the 
CO $J$=3-2 second moment map.  Contour levels are
(4,   8,   16,   32)
km s$^{-1}$.
Images of this galaxy derived from early observations in the survey
have been published in \citet{w09}. These images are made using
later data which had a higher quality.
\label{fig-ngc4579}}
\end{figure*}

\begin{figure*}
\includegraphics[width=60mm]{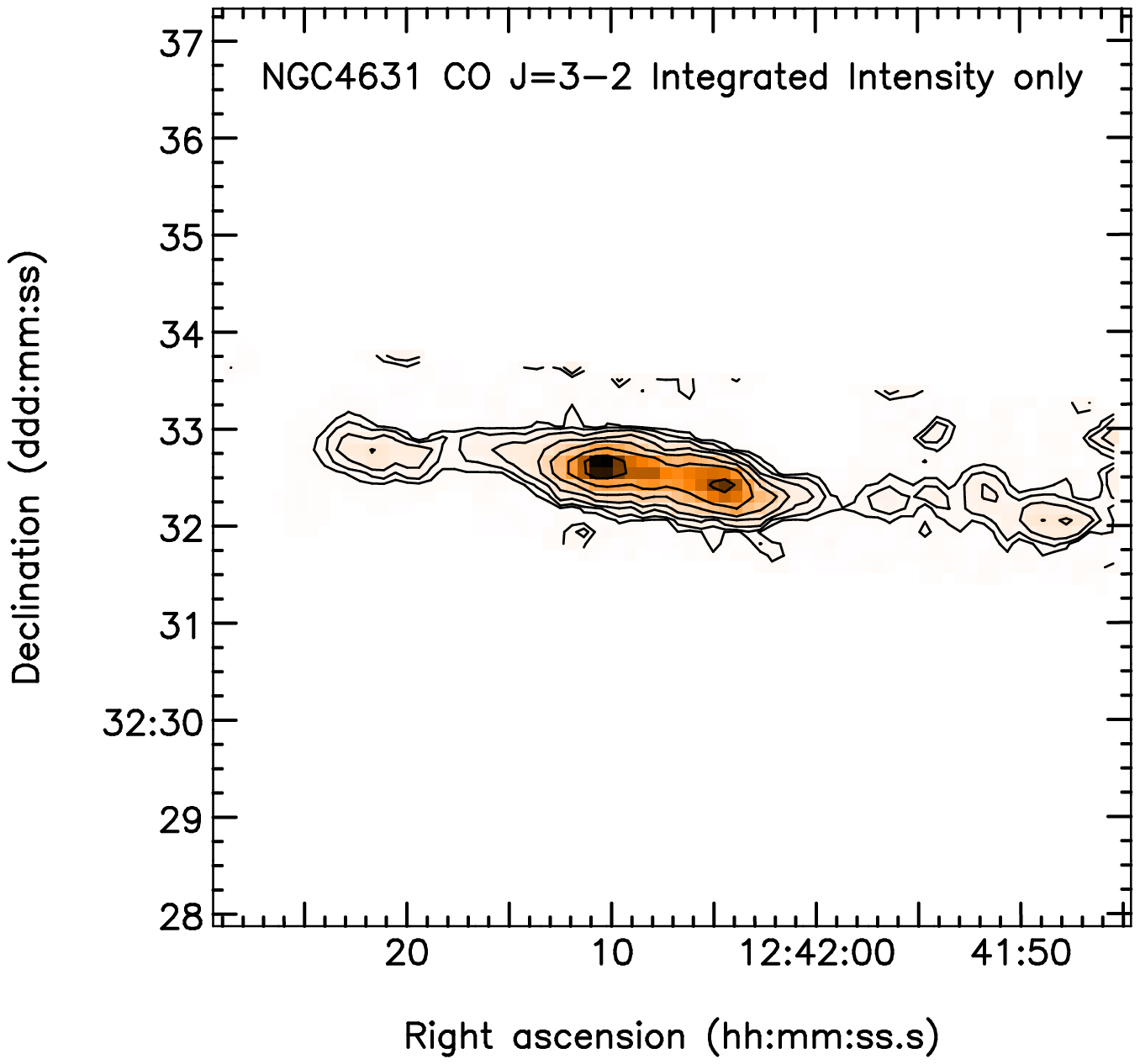}
\includegraphics[width=60mm]{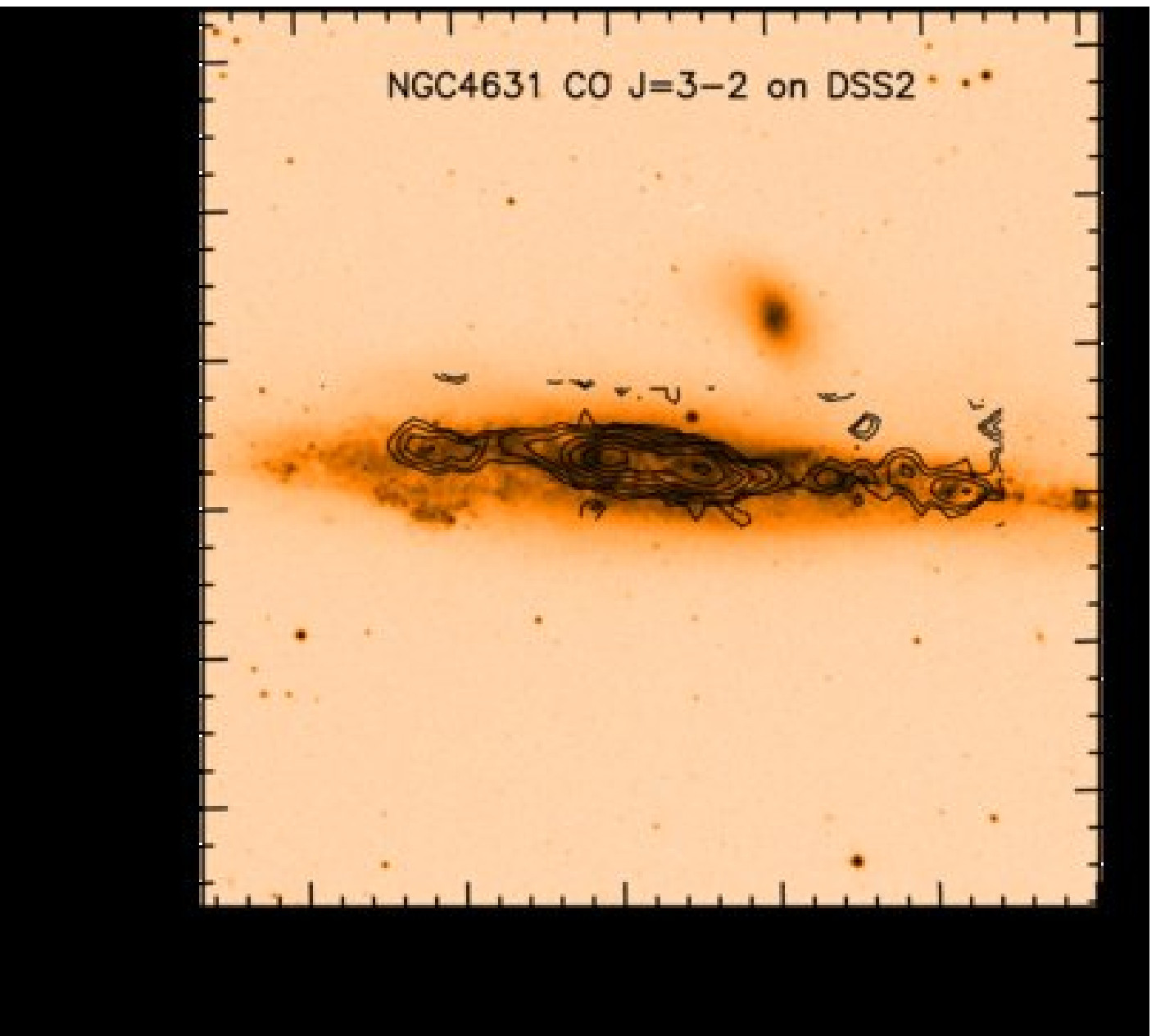}
\includegraphics[width=60mm]{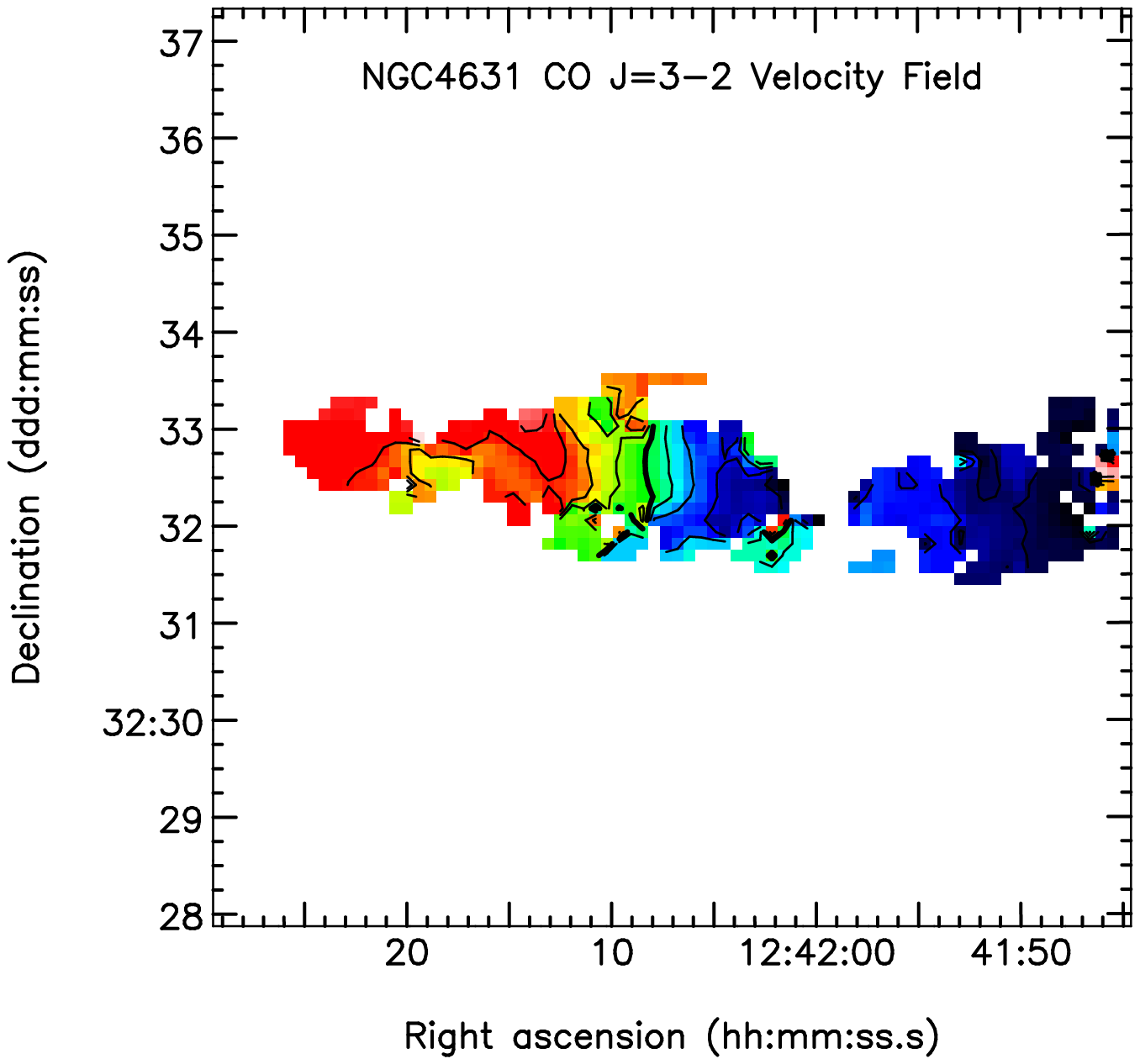}
\includegraphics[width=60mm]{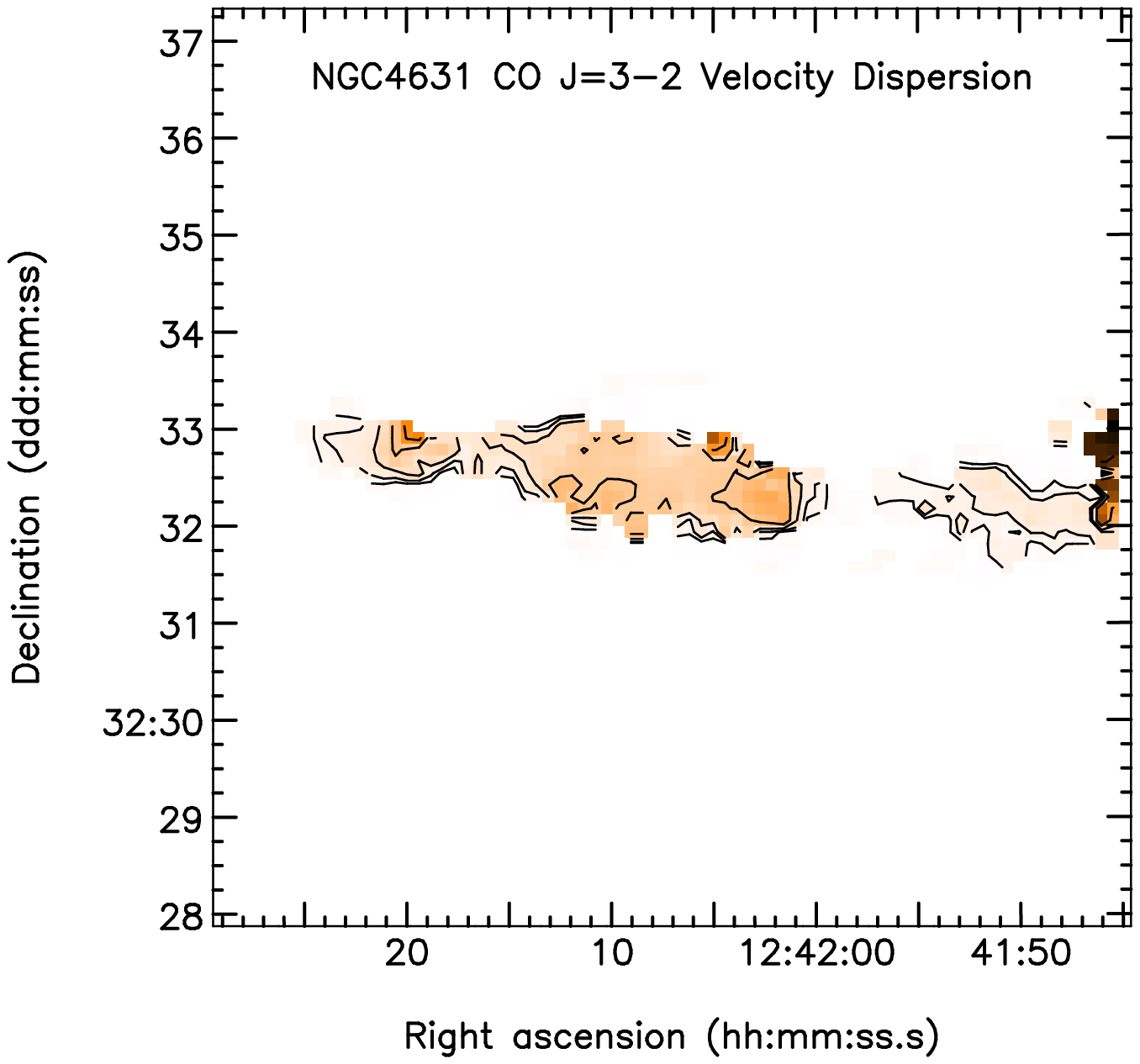}
\caption{CO $J$=3-2 images for NGC 4631.  (a) CO $J$=3-2 integrated intensity
  image. Contours levels are (0.5,   1,   2, 4,   8,   16,   32) K km
  s$^{-1}$ (T$_{MB}$). 
(b) CO $J$=3-2 overlaid on a
Digitized Sky Survey image. (c) Velocity field. Contour levels are
(520,   550,   580,   610,   640,   670,   700,   730,   760) 
km s$^{-1}$. 
(d) The velocity dispersion $\sigma_v$ as traced by the 
CO $J$=3-2 second moment map.  Contour levels are
(4,   8,   16,   32,   64,   128)
km s$^{-1}$.
Similar images derived from the same data have been
published in \citet{i11}.
\label{fig-ngc4631}}
\end{figure*}

\begin{figure*}
\includegraphics[width=55mm]{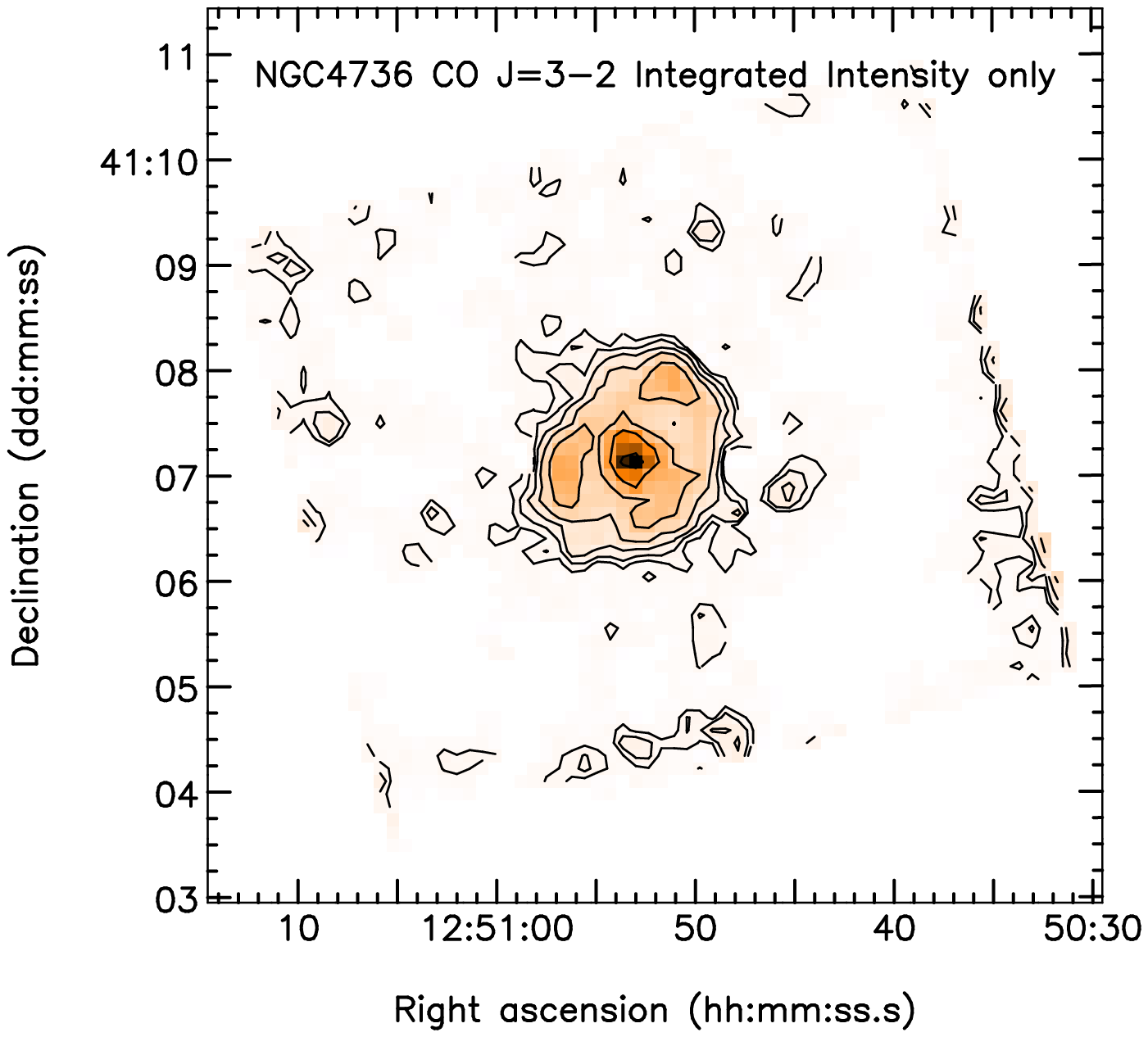}
\includegraphics[width=55mm]{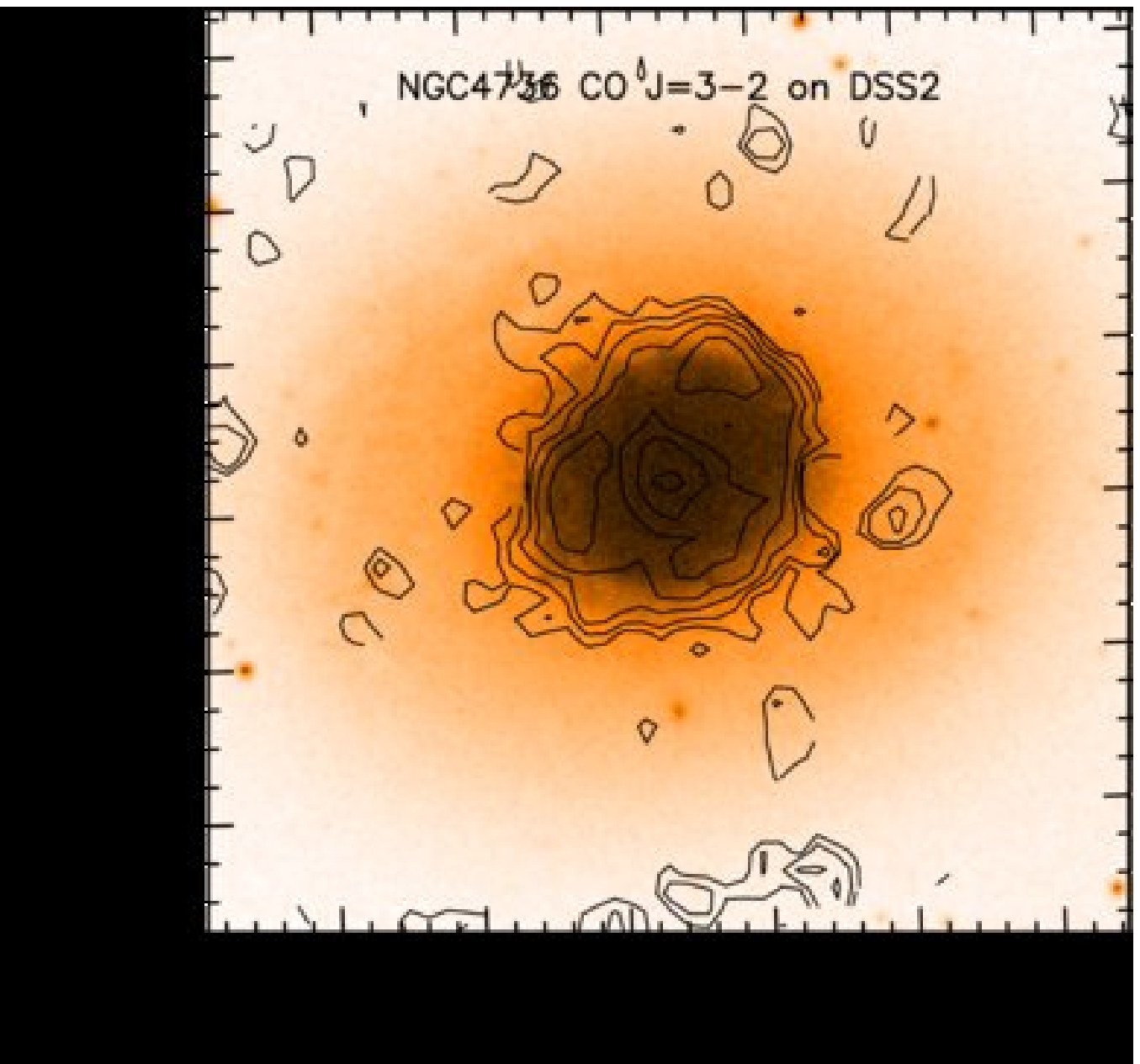}
\includegraphics[width=55mm]{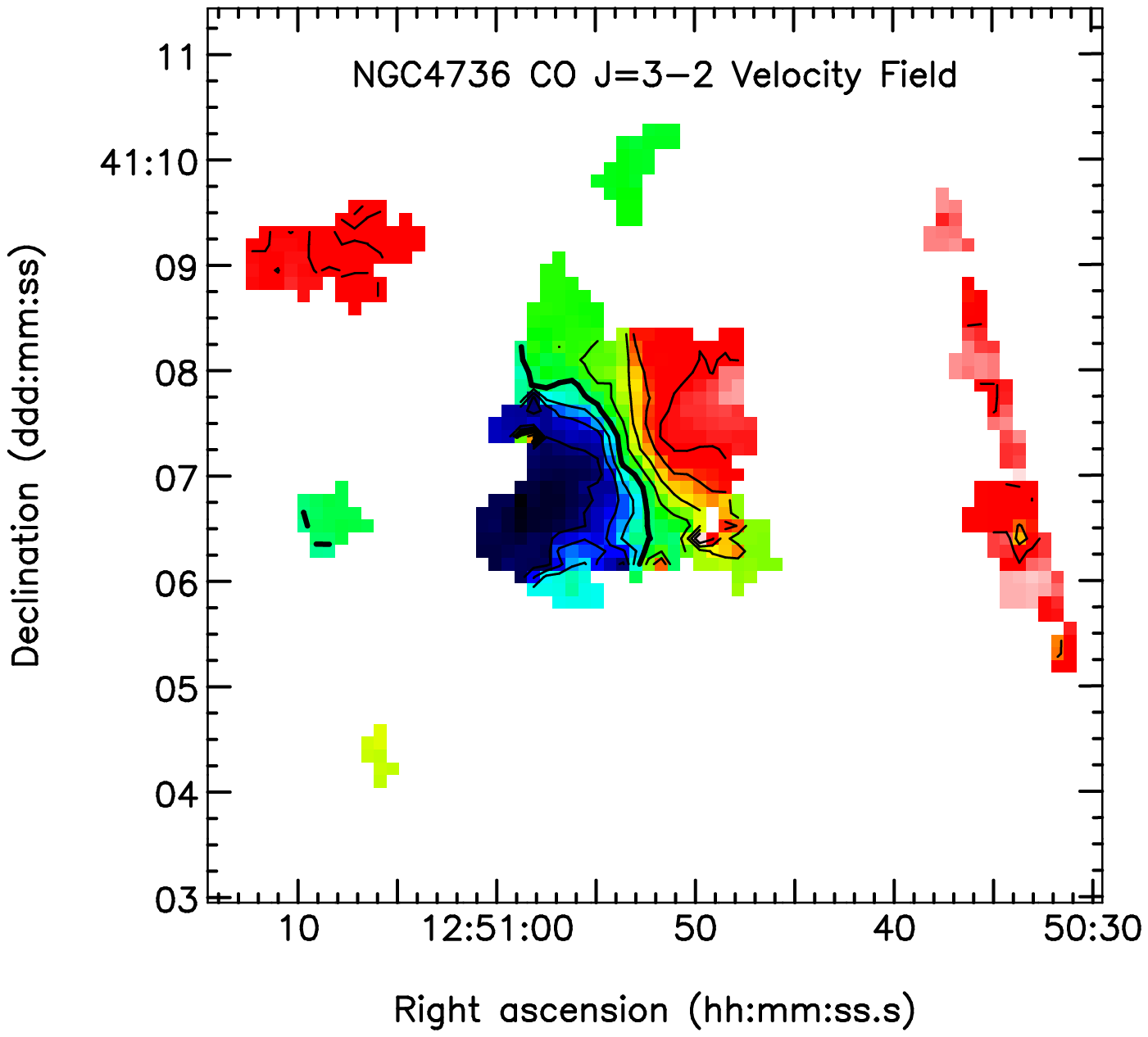}
\caption{CO $J$=3-2 images for NGC 4736. (a) CO $J$=3-2 integrated intensity
  image. Contours levels are (0.5,   1,   2,   4,   8,   16, 32)
K km s$^{-1}$ (T$_{MB}$).
(b) CO $J$=3-2 integrated intensity contours overlaid on an optical
image from the Digitized Sky Survey. (c) Velocity field as traced by the
CO $J$=3-2 first moment map. Contour levels are (214,   241,   268,
295,   322,   349,   376,   403) km s$^{-1}$. 
The velocity dispersion map has been published in \citet{w11}.
\label{fig-ngc4736}}
\end{figure*}

\begin{figure*}
\includegraphics[width=55mm]{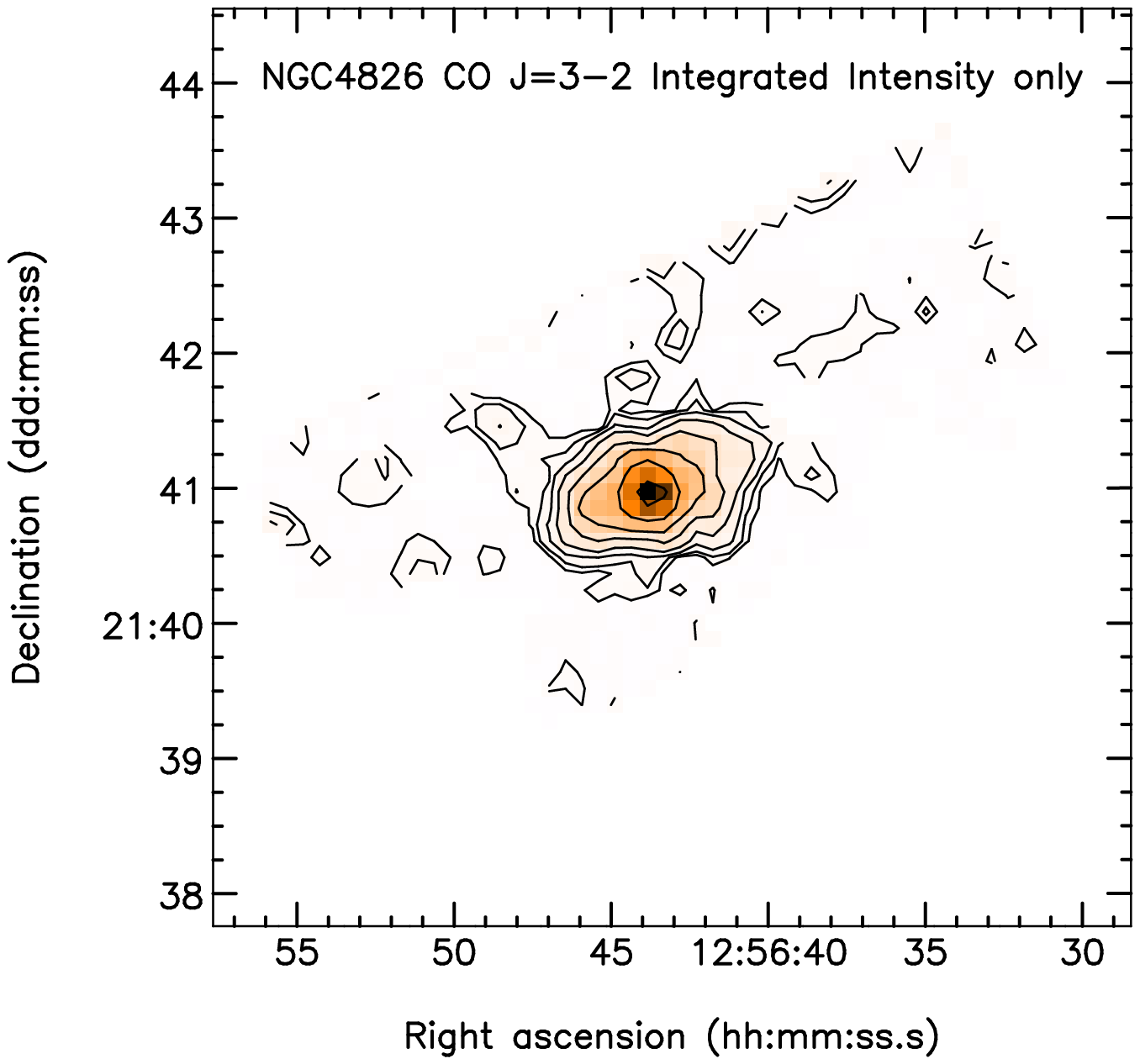}
\includegraphics[width=55mm]{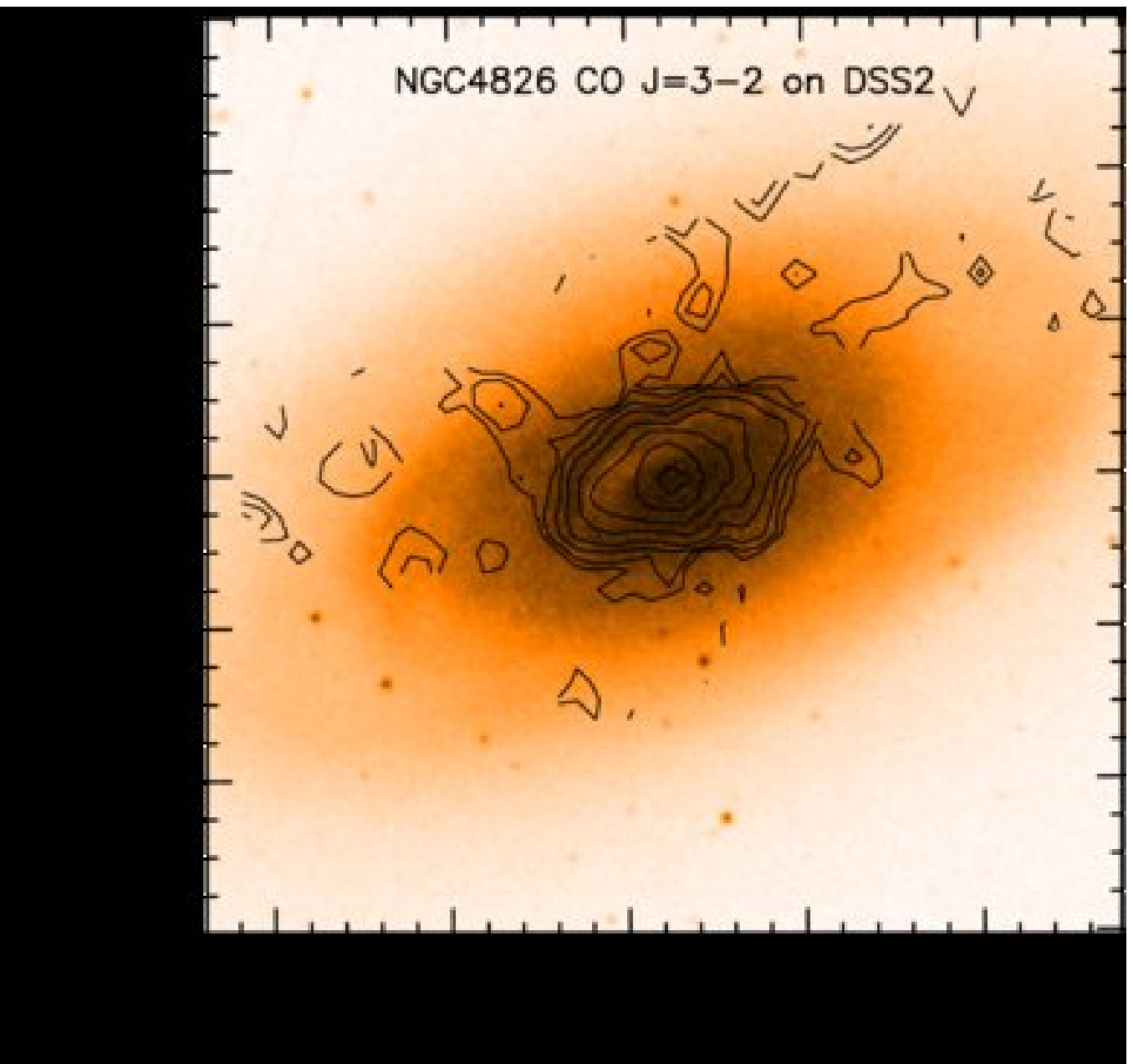}
\includegraphics[width=55mm]{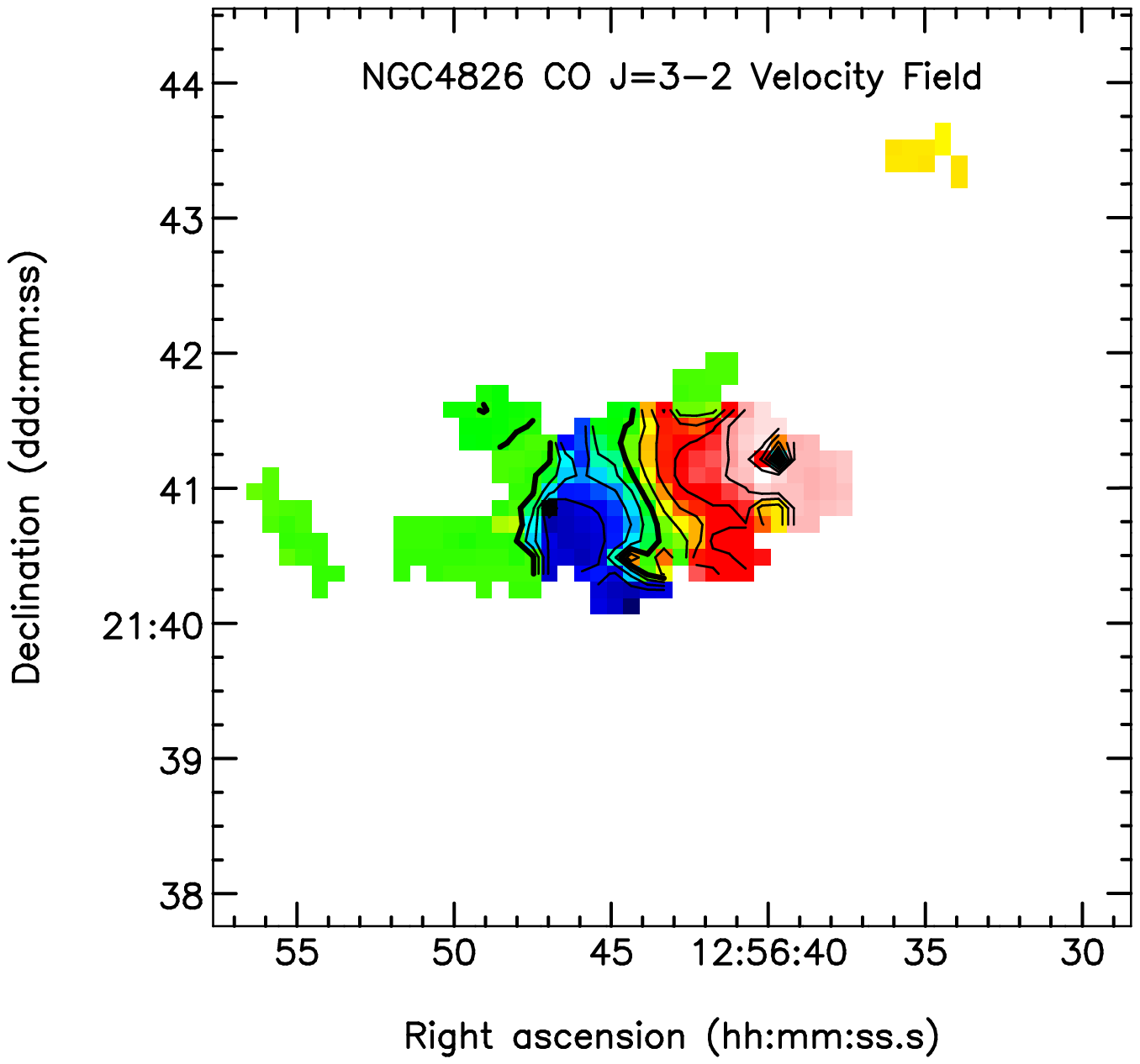}
\caption{CO $J$=3-2 images for NGC 4826. (a) CO $J$=3-2 integrated intensity
  image. Contours levels are (0.5,   1,   2,   4,   8,   16, 32, 64)
K km s$^{-1}$ (T$_{MB}$).
(b) CO $J$=3-2 integrated intensity contours overlaid on an optical
image from the Digitized Sky Survey. (c) Velocity field as traced by the
CO $J$=3-2 first moment map. Contour levels are (230,   270,   310,
350,   390,   430,   470,   510,   550) 
 km s$^{-1}$.
The velocity dispersion map has been published in \citet{w11}.
\label{fig-ngc4826}}
\end{figure*}

\begin{figure*}
\includegraphics[width=60mm]{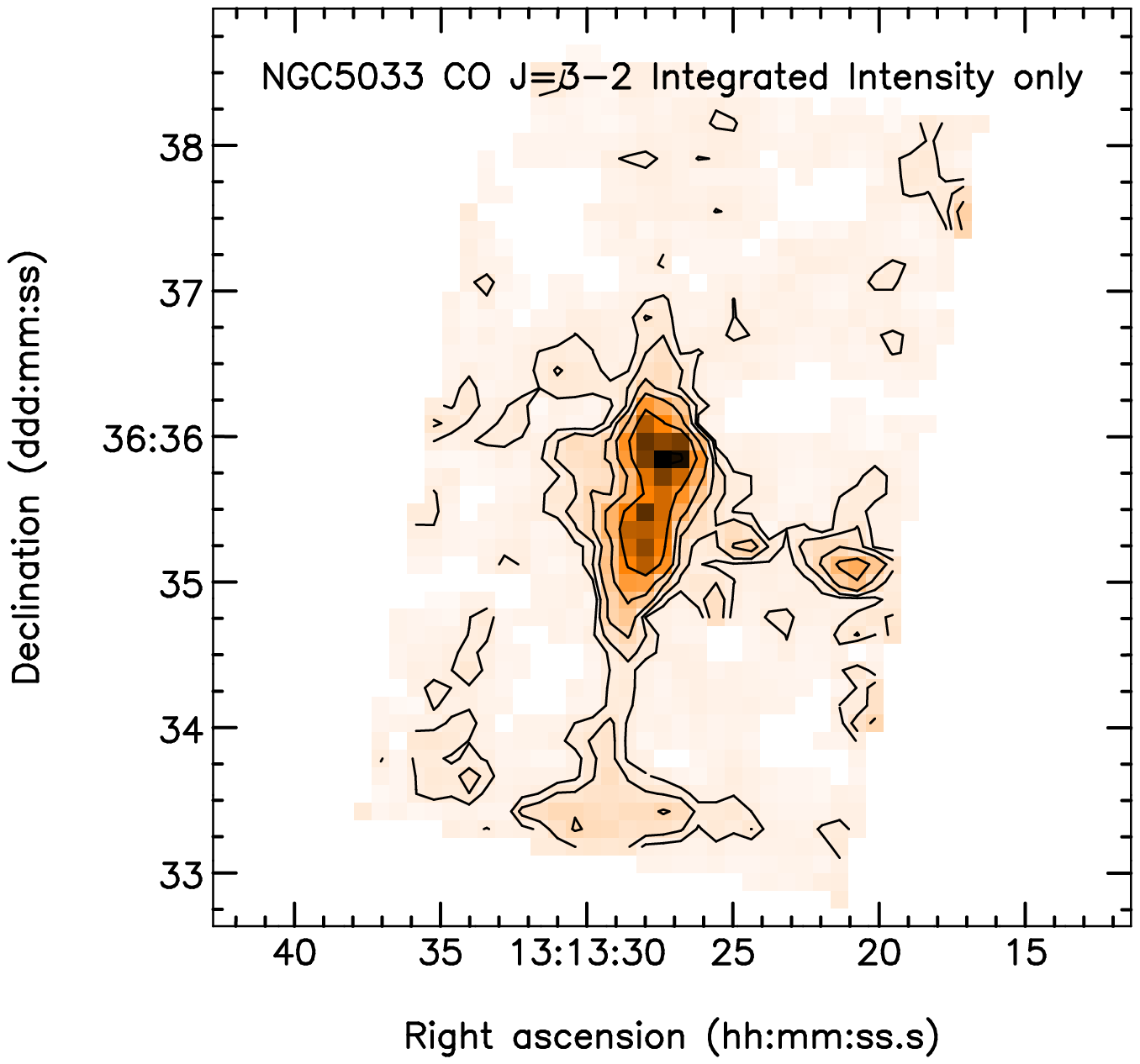}
\includegraphics[width=60mm]{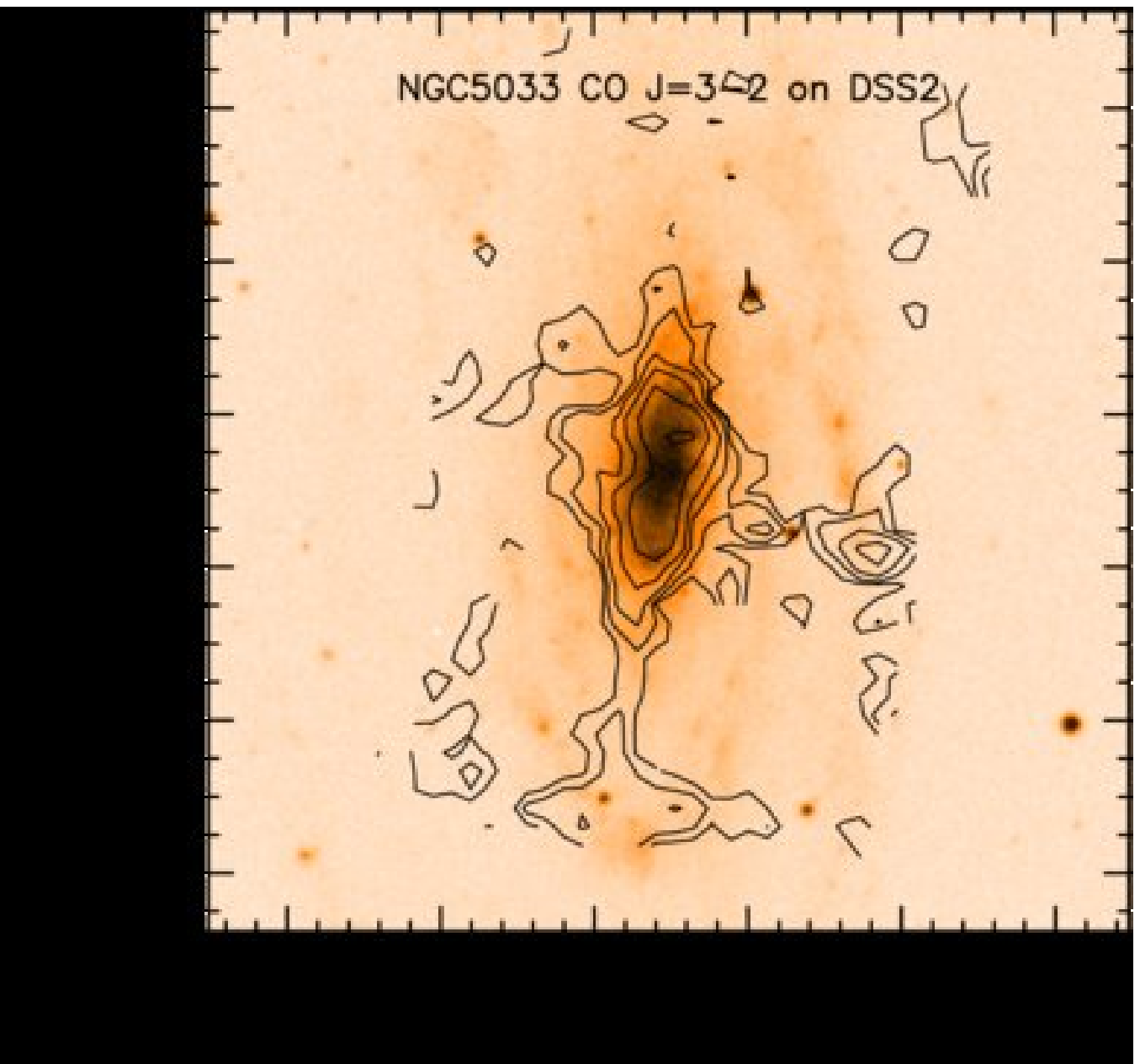}
\includegraphics[width=60mm]{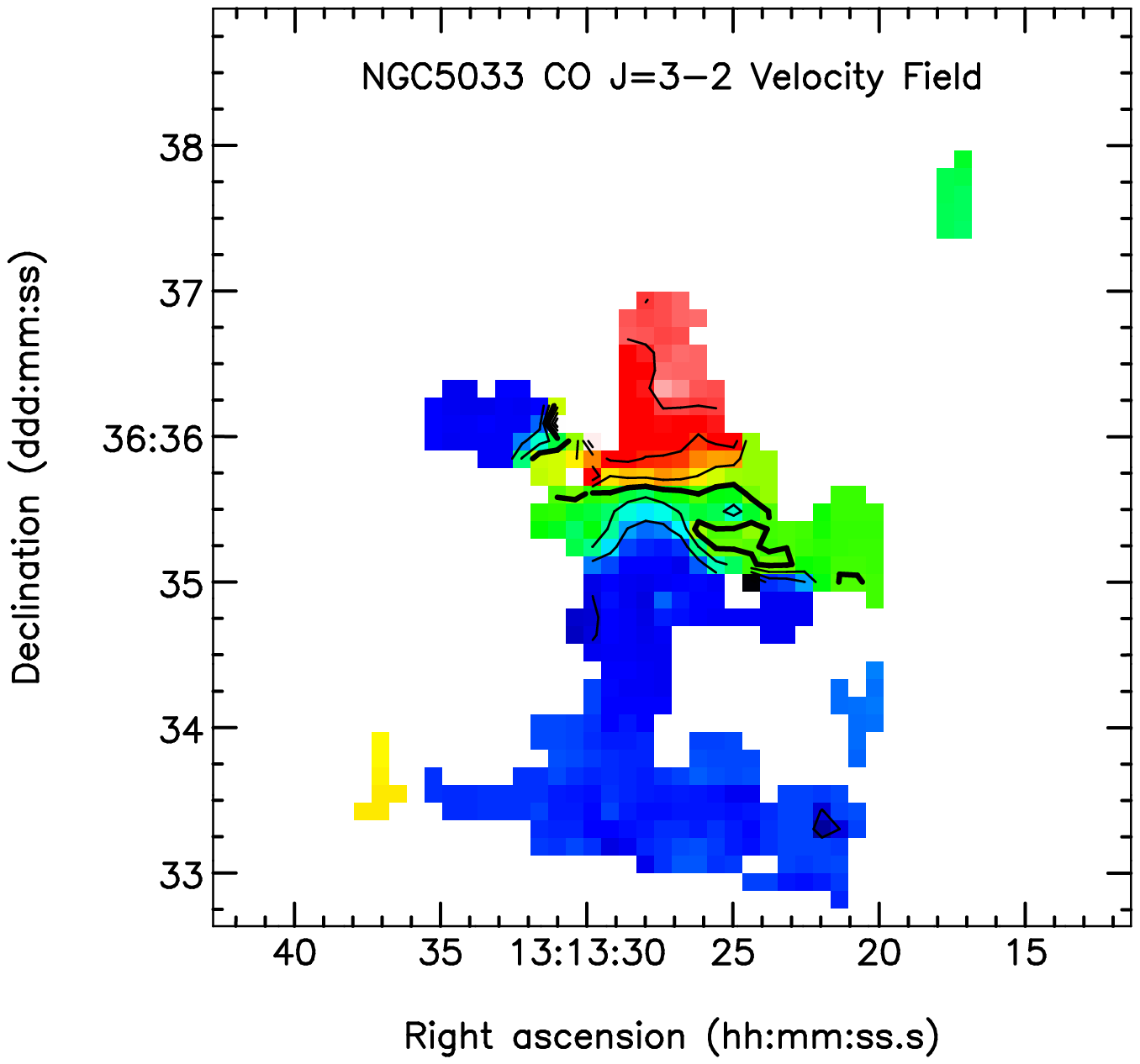}
\includegraphics[width=60mm]{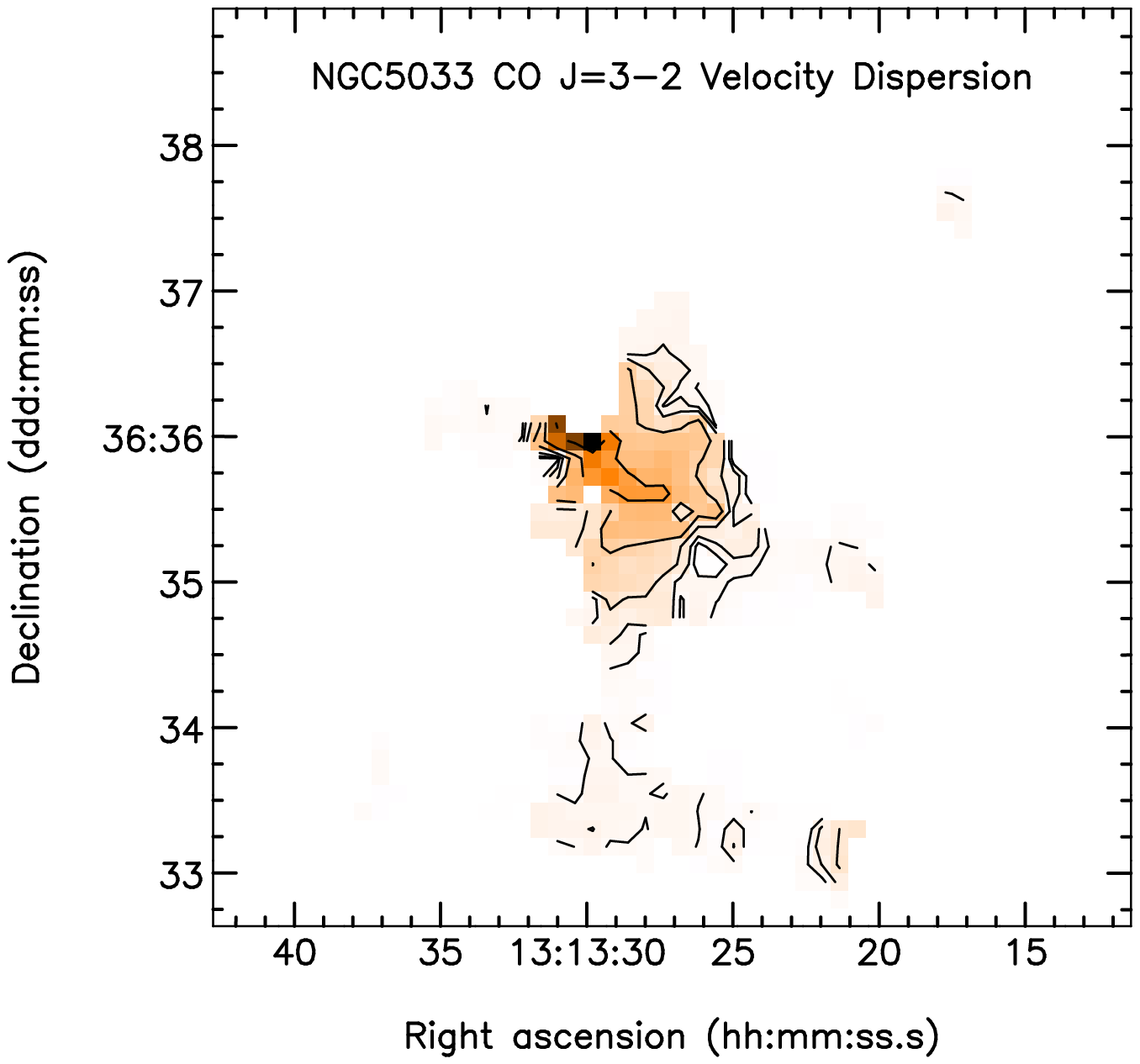}
\caption{CO $J$=3-2 images for NGC 5033.  (a) CO $J$=3-2 integrated intensity
  image. Contours levels are (0.5,   1,   2, 4,   8,   16) K km
  s$^{-1}$ (T$_{MB}$). 
(b) CO $J$=3-2 overlaid on a
Digitized Sky Survey image. (c) Velocity field. Contour levels are
(593,   663,   733,   803,   873,   943,   1013,   1083) 
km s$^{-1}$. 
(d) The velocity dispersion $\sigma_v$ as traced by the 
CO $J$=3-2 second moment map.  Contour levels are
(4,   8,   16,   32,   64,   128)
km s$^{-1}$.
\label{fig-ngc5033}}
\end{figure*}

\begin{figure*}
\includegraphics[width=55mm]{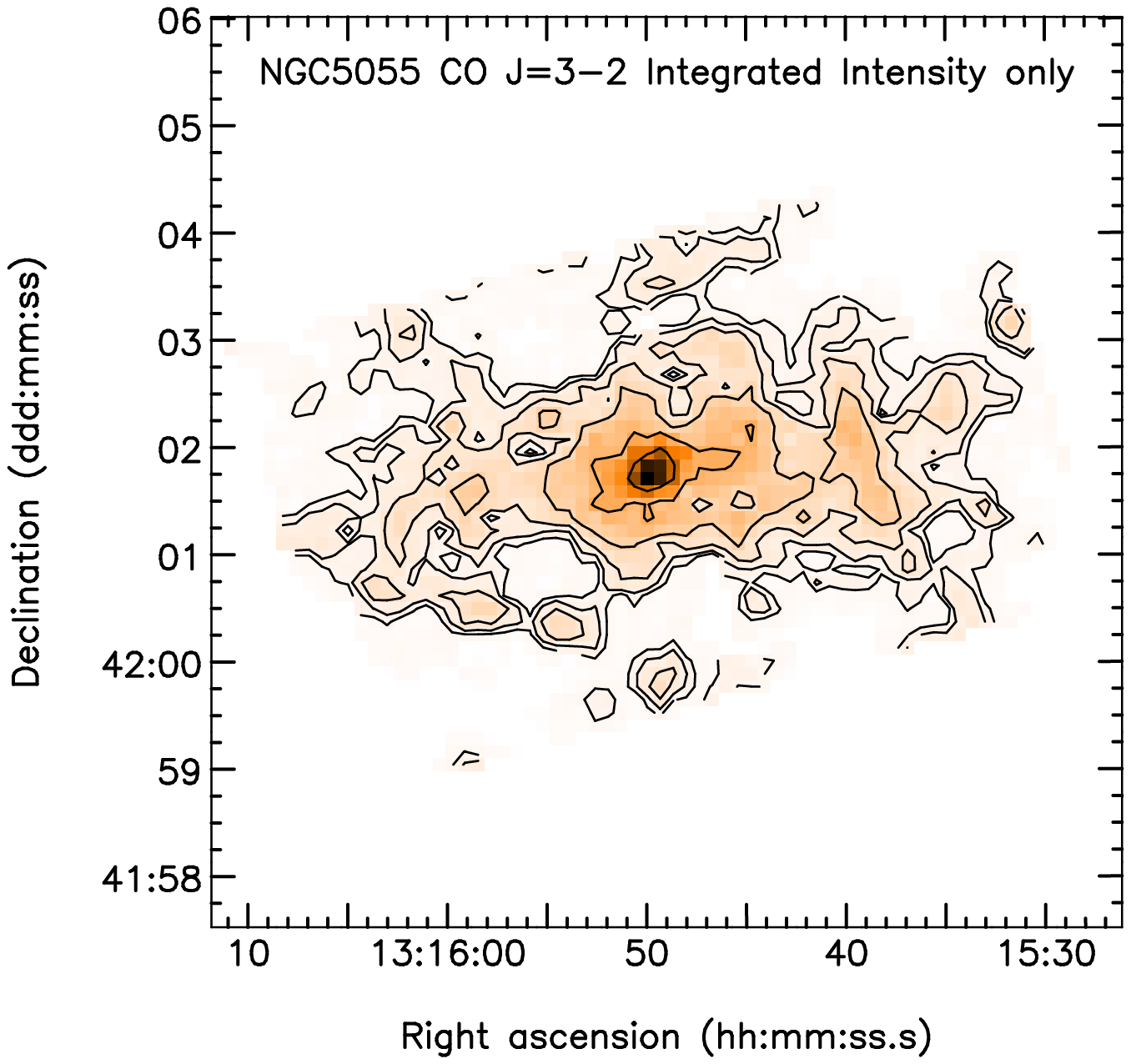}
\includegraphics[width=55mm]{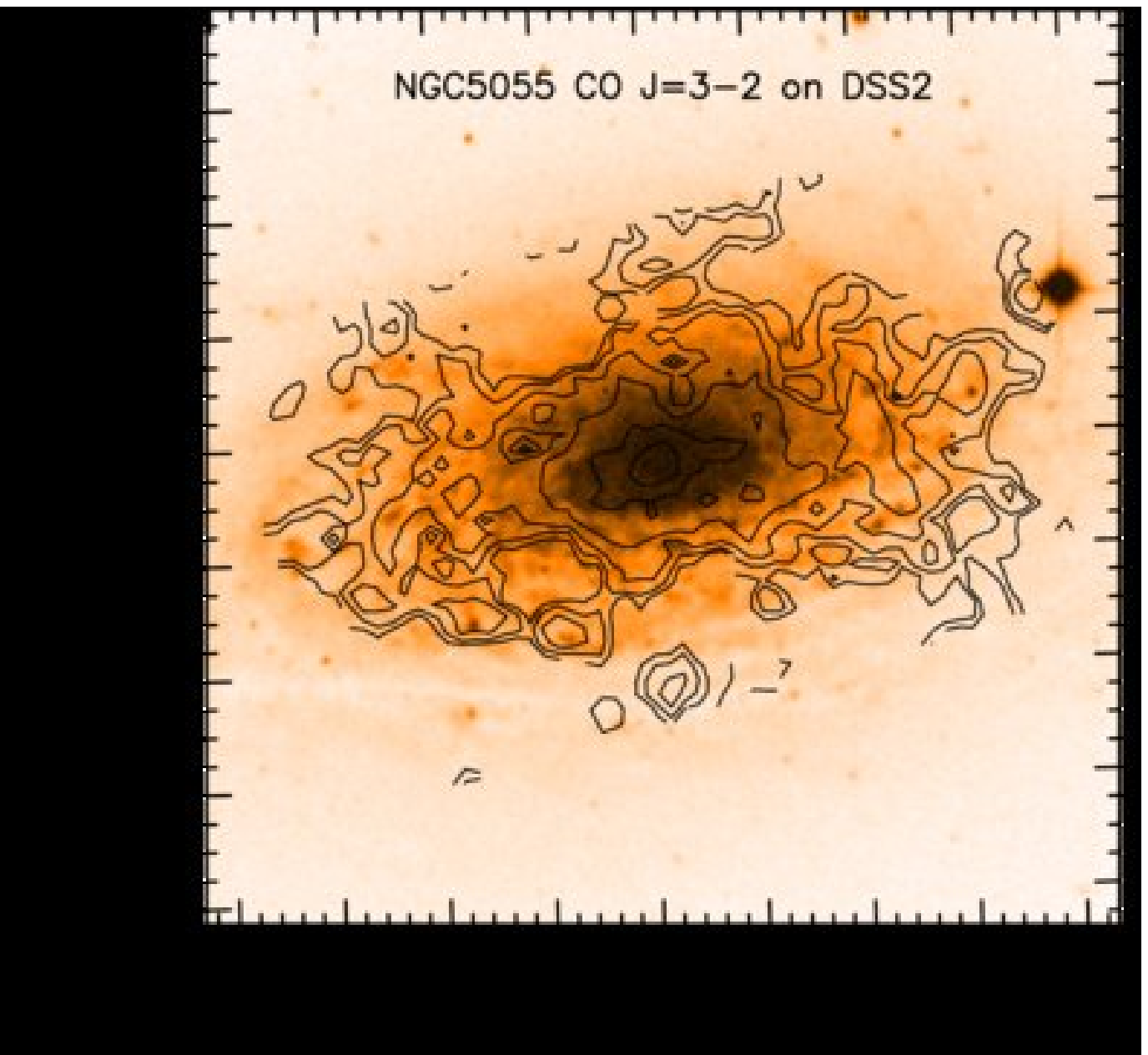}
\includegraphics[width=55mm]{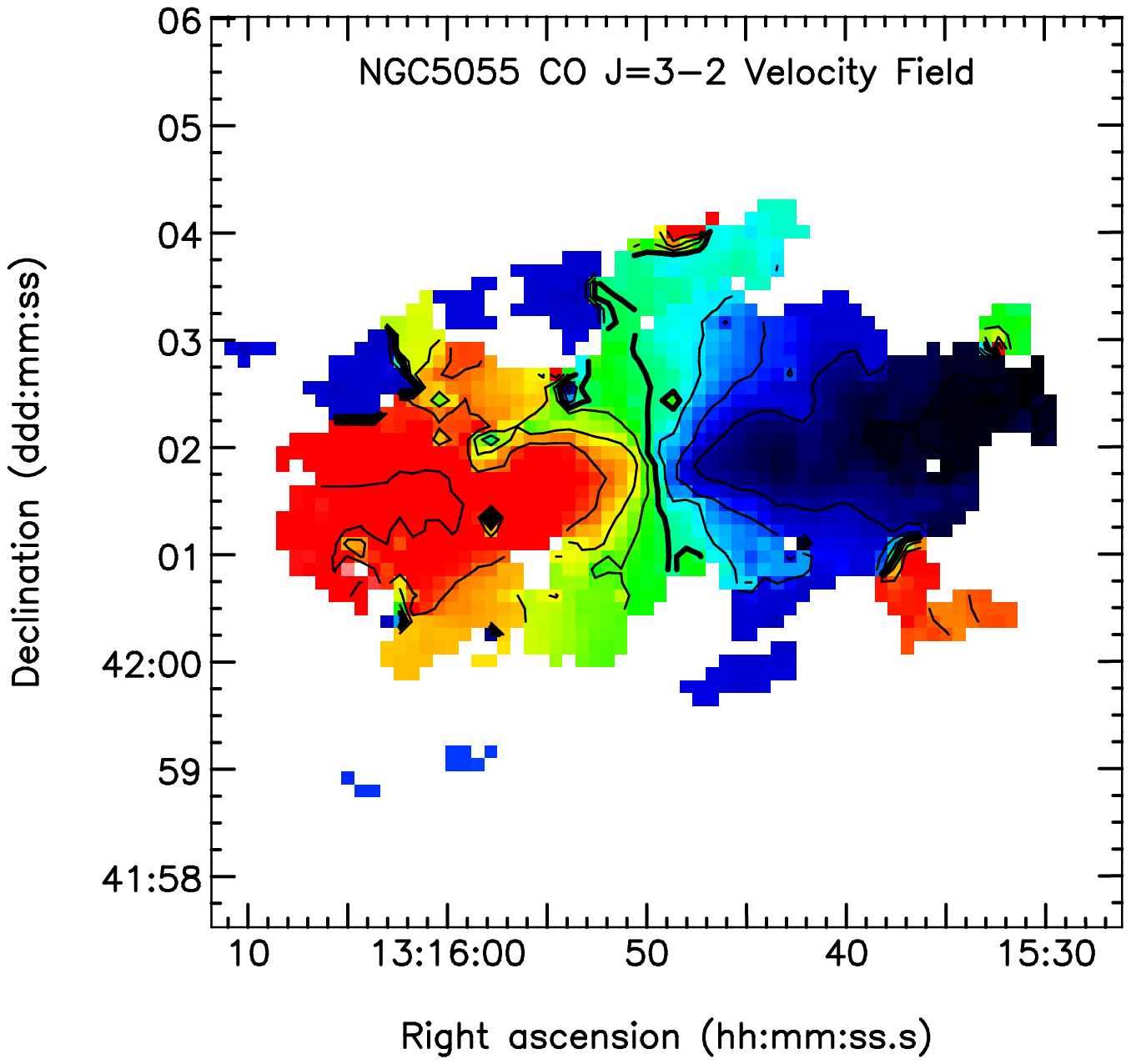}
\caption{CO $J$=3-2 images for NGC 5055. (a) CO $J$=3-2 integrated intensity
  image. Contours levels are (0.5,   1,   2,   4,   8,   16)
K km s$^{-1}$ (T$_{MB}$).
(b) CO $J$=3-2 integrated intensity contours overlaid on an optical
image from the Digitized Sky Survey. (c) Velocity field as traced by the
CO $J$=3-2 first moment map. Contour levels are (315,   360,   405,
450,   495,   540,   585,   630,   675) km s$^{-1}$.
The velocity dispersion map has been published in \citet{w11}.
\label{fig-ngc5055}}
\end{figure*}

\begin{figure*}
\includegraphics[width=60mm]{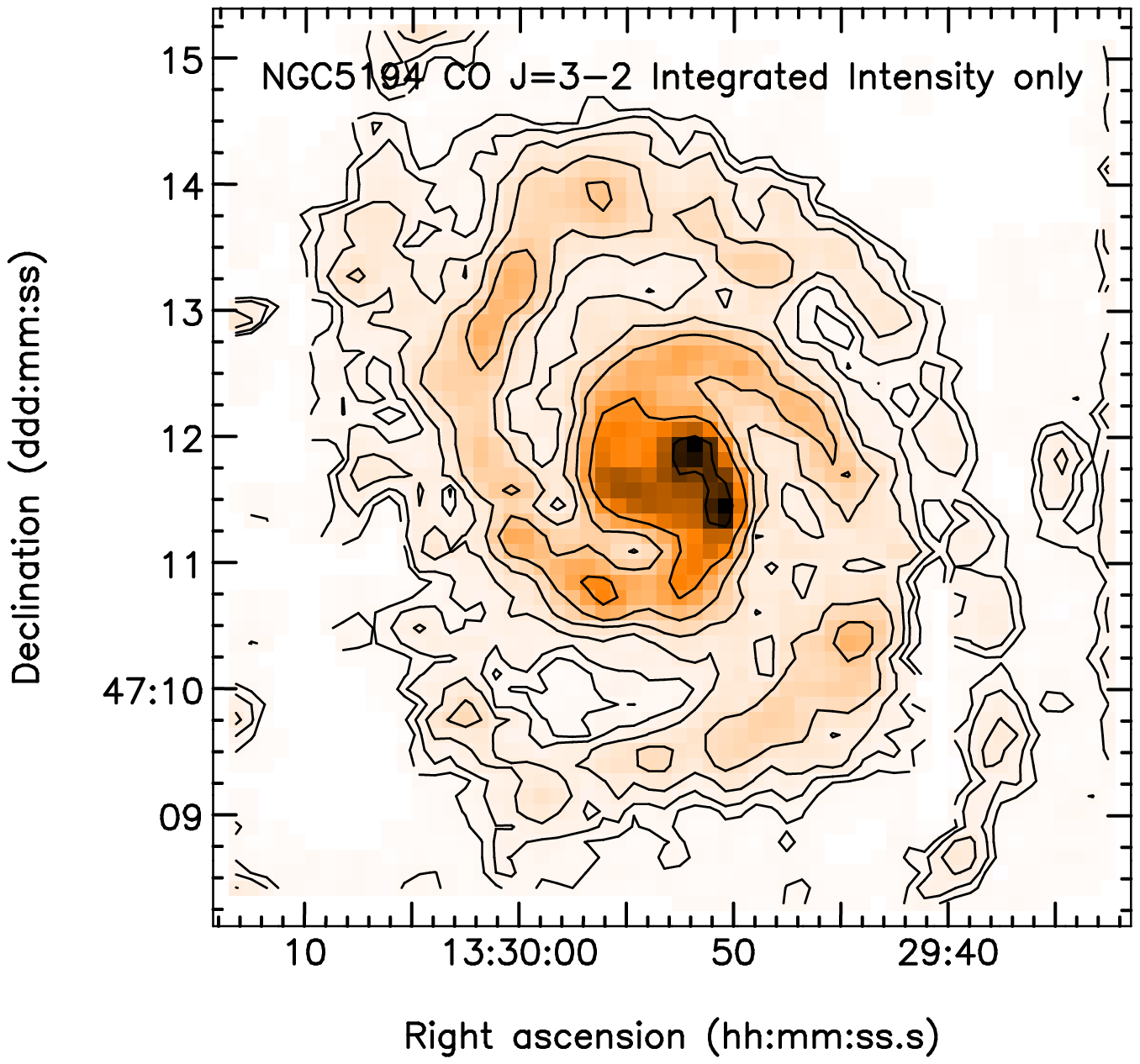}
\includegraphics[width=60mm]{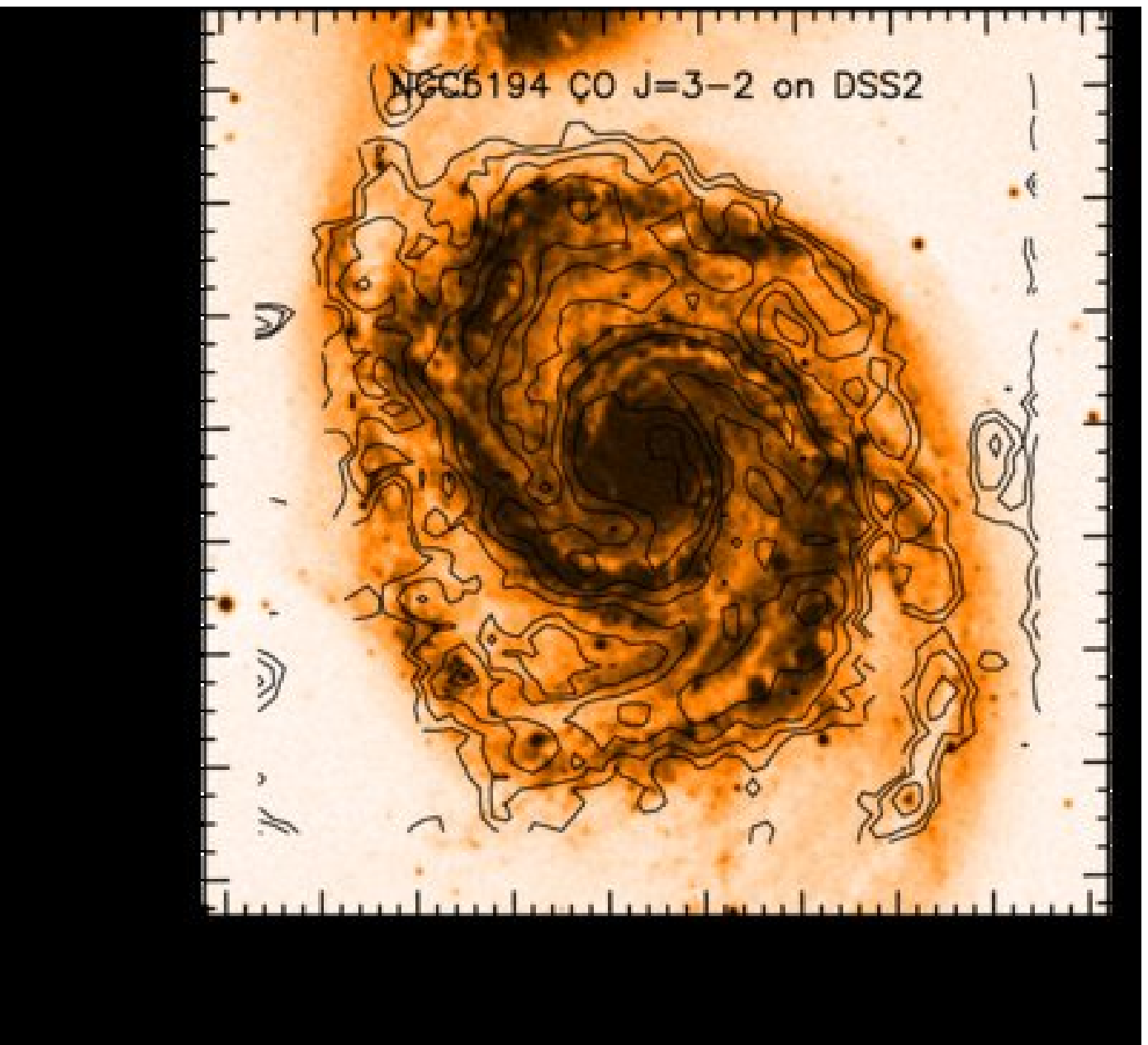}
\includegraphics[width=60mm]{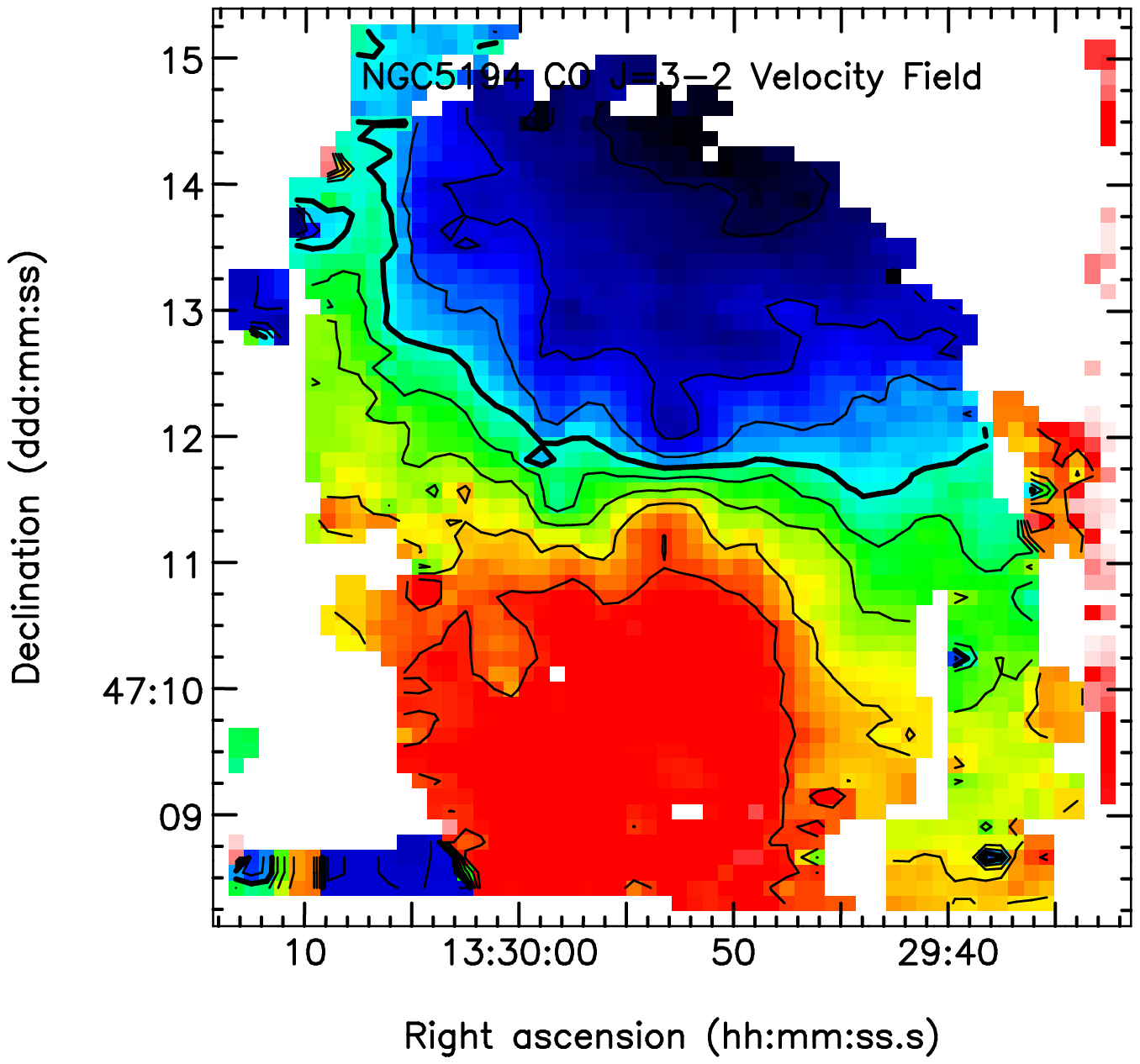}
\includegraphics[width=60mm]{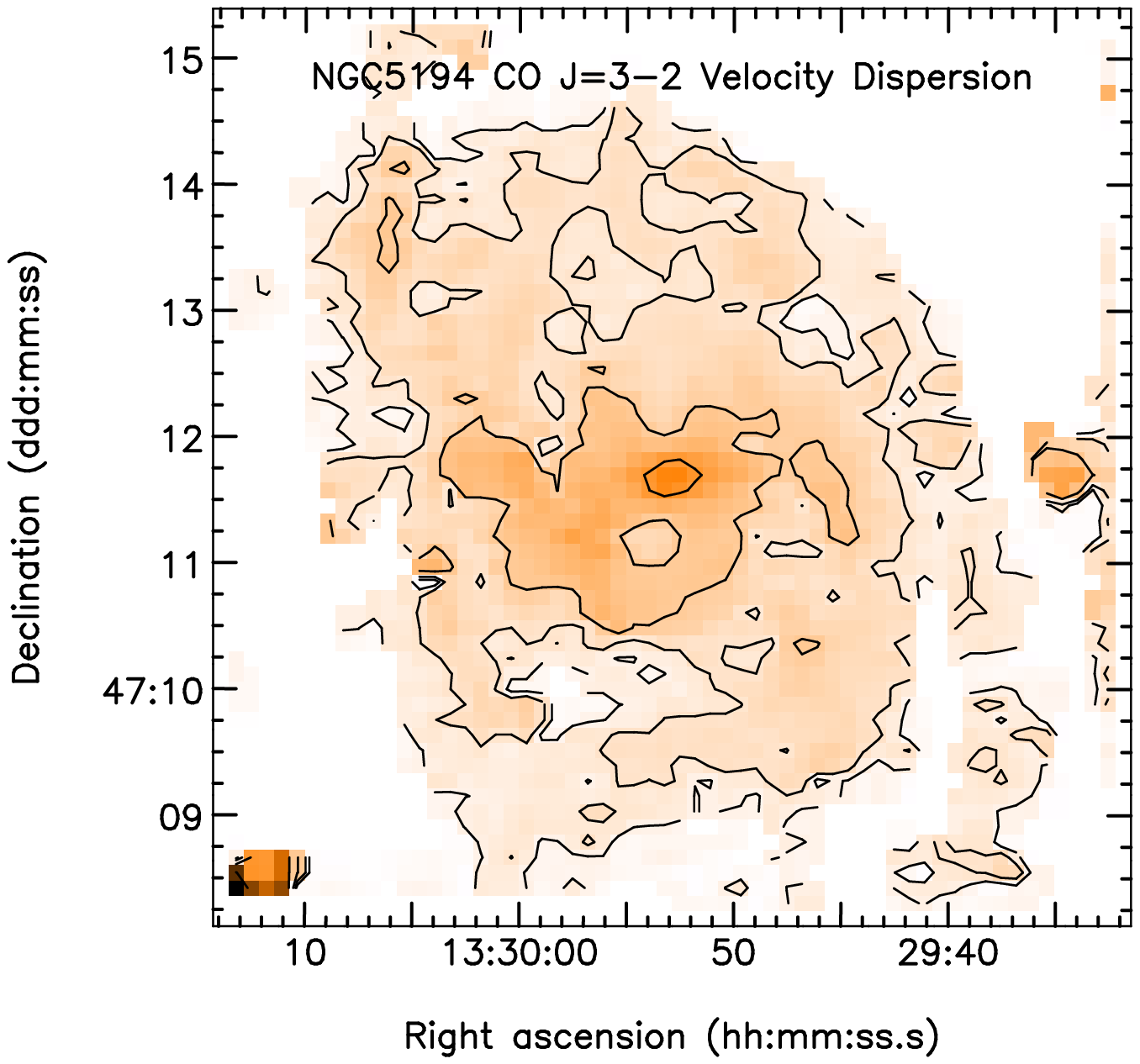}
\caption{CO $J$=3-2 images for NGC 5194.  (a) CO $J$=3-2 integrated intensity
  image. Contours levels are (0.5,   1,   2, 4,   8,   16, 32) K km
  s$^{-1}$ (T$_{MB}$). 
(b) CO $J$=3-2 overlaid on a
Digitized Sky Survey image. (c) Velocity field. Contour levels are
(392,   412,   432,   452,   472,   492,   512,   532) 
km s$^{-1}$. 
(d) The velocity dispersion $\sigma_v$ as traced by the 
CO $J$=3-2 second moment map.  Contour levels are
(4,   8,   16,   32,   64)
km s$^{-1}$. A more complete analysis fo the data for NGC 5194 will be
published in Vlahakis et al., in preparation.
\label{fig-ngc5194}}
\end{figure*}

\begin{figure*}
\includegraphics[width=40mm]{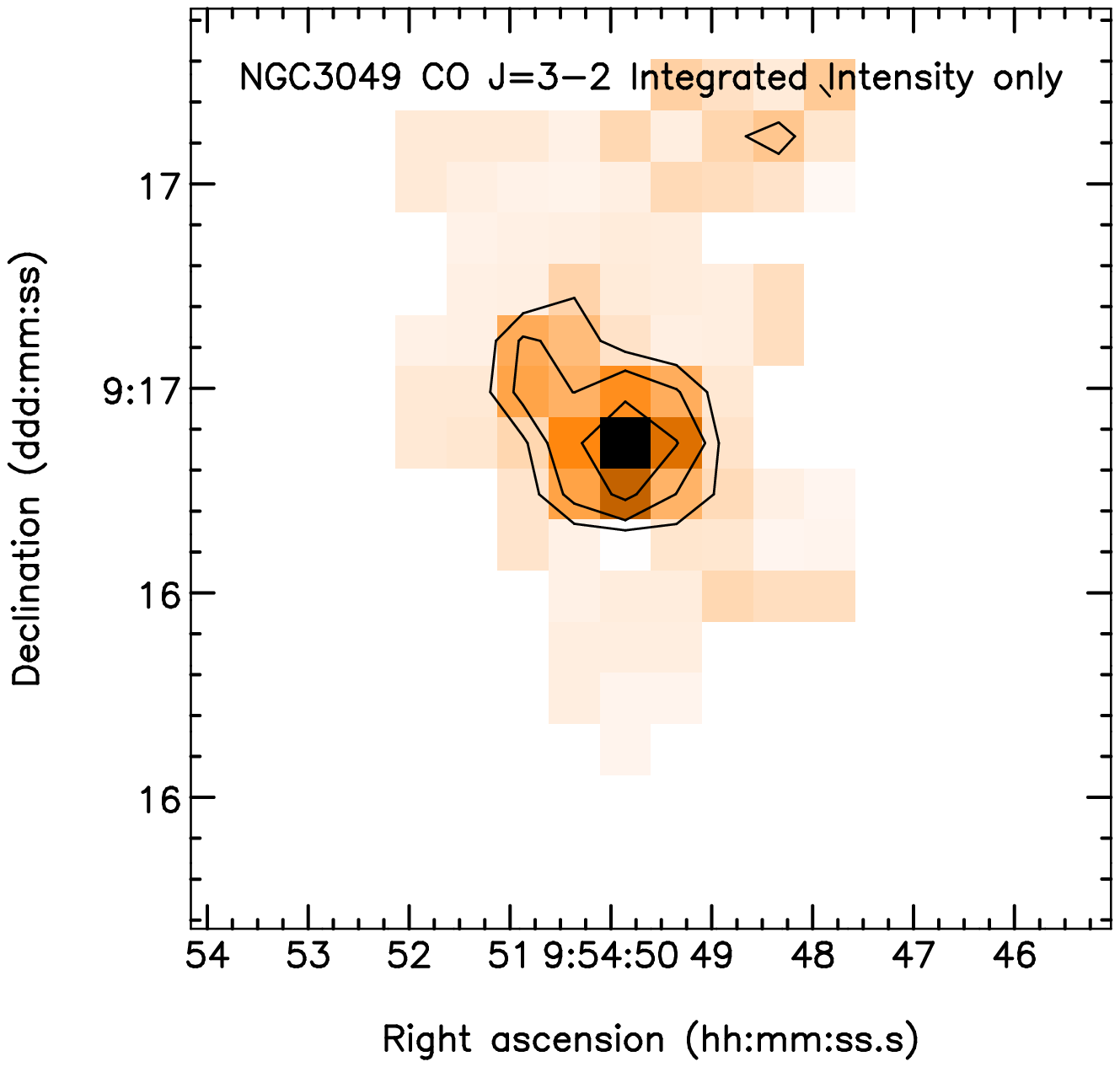}
\includegraphics[width=40mm]{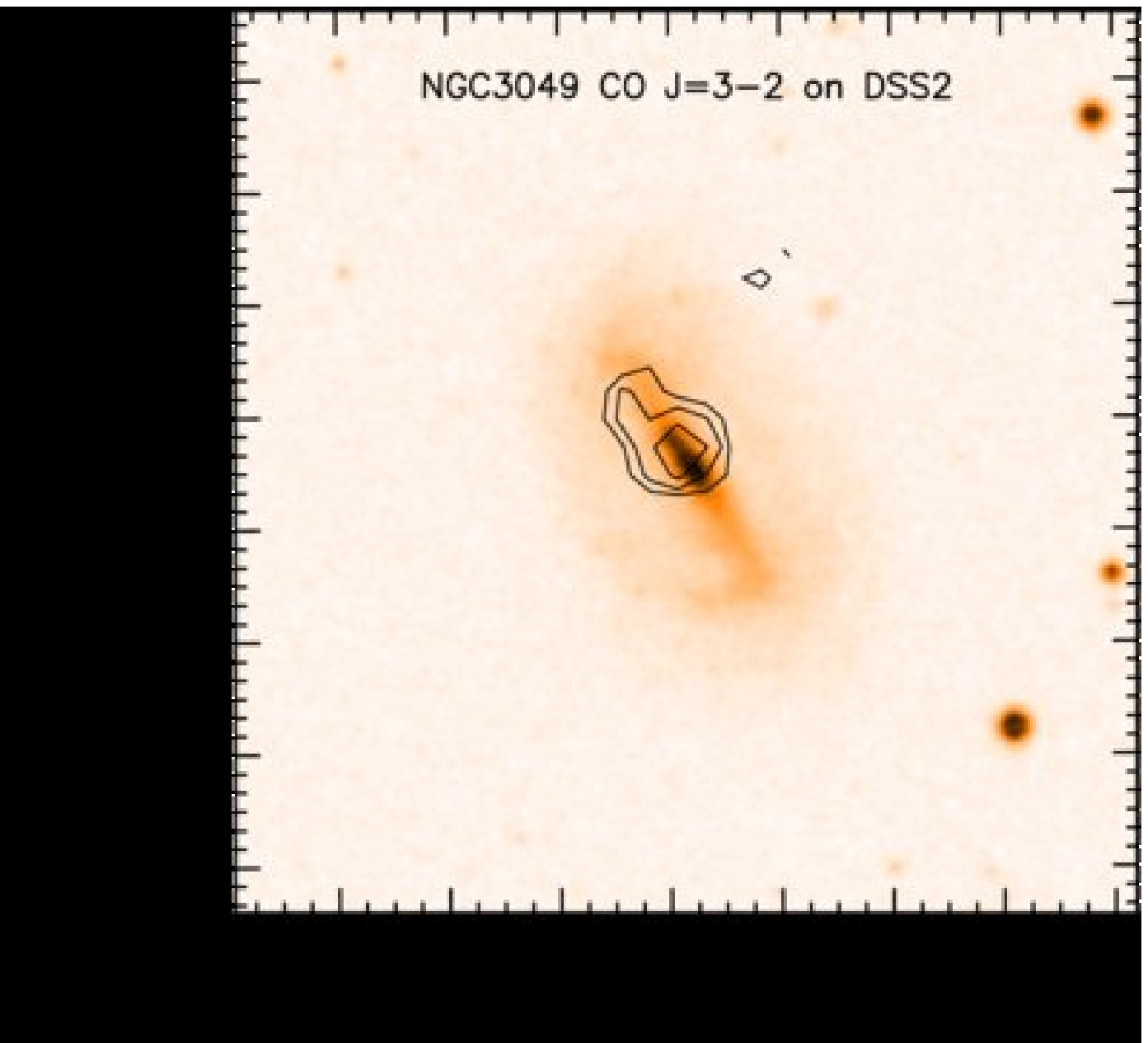}
\includegraphics[width=40mm]{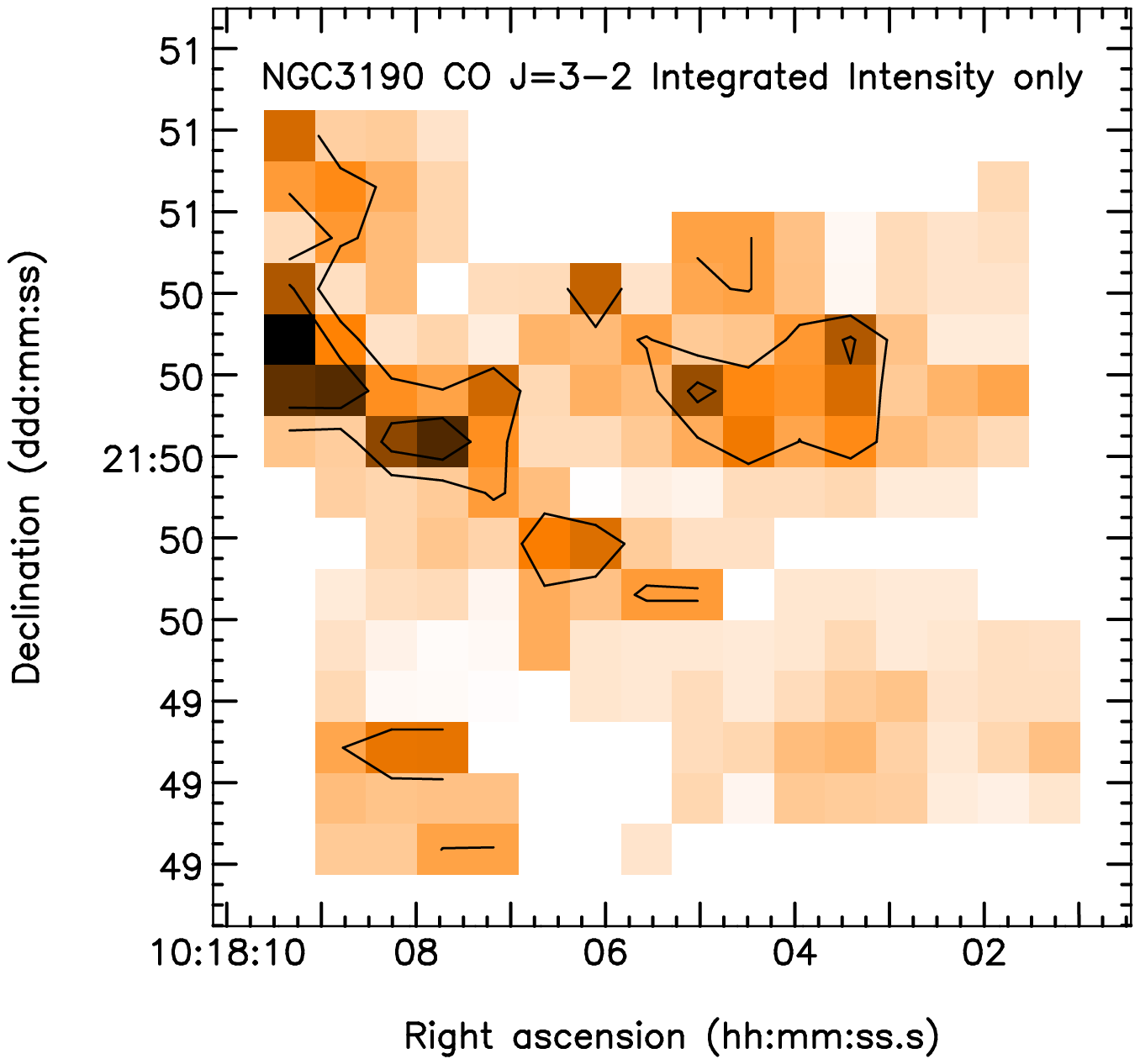}
\includegraphics[width=40mm]{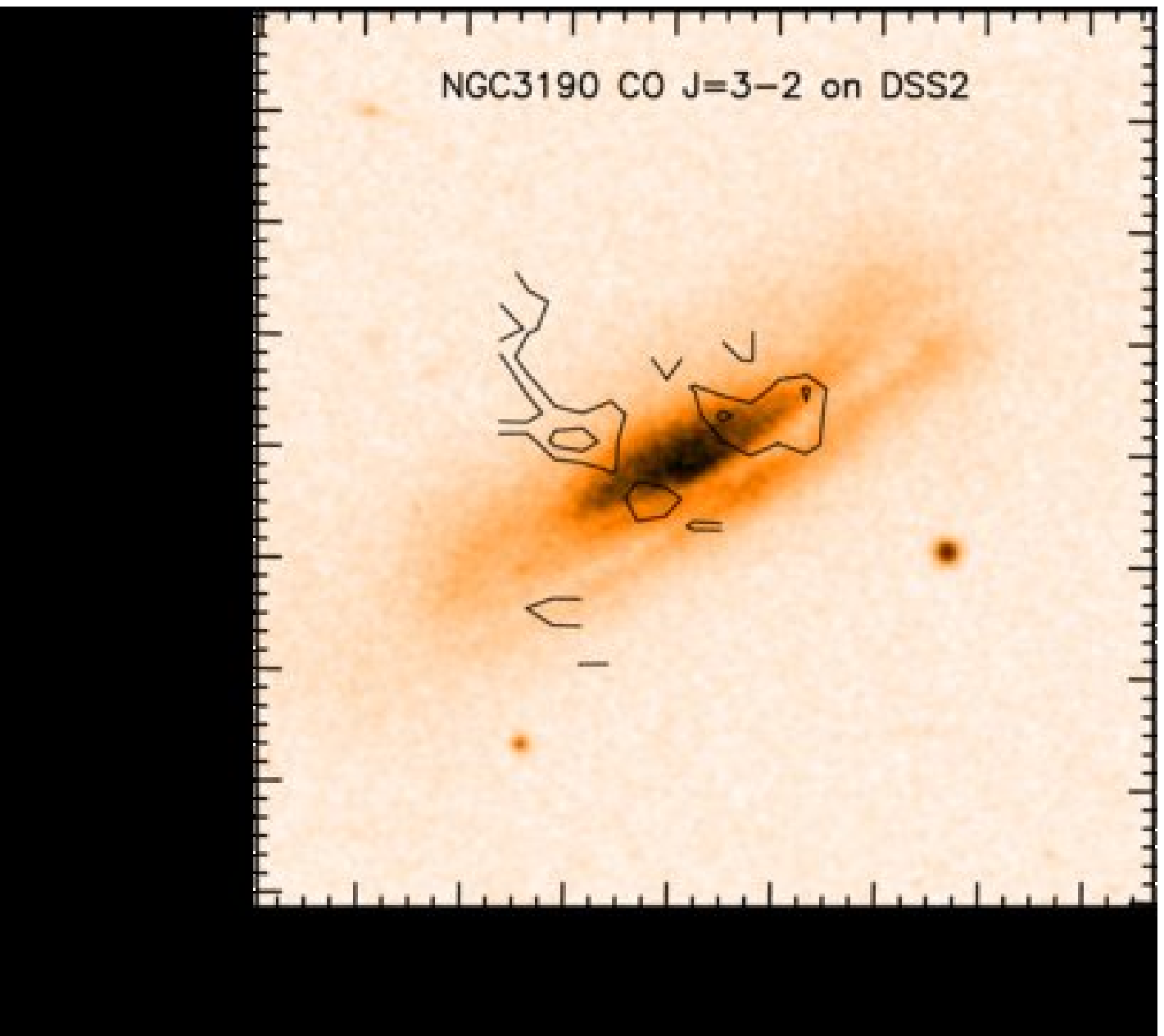}
\includegraphics[width=40mm]{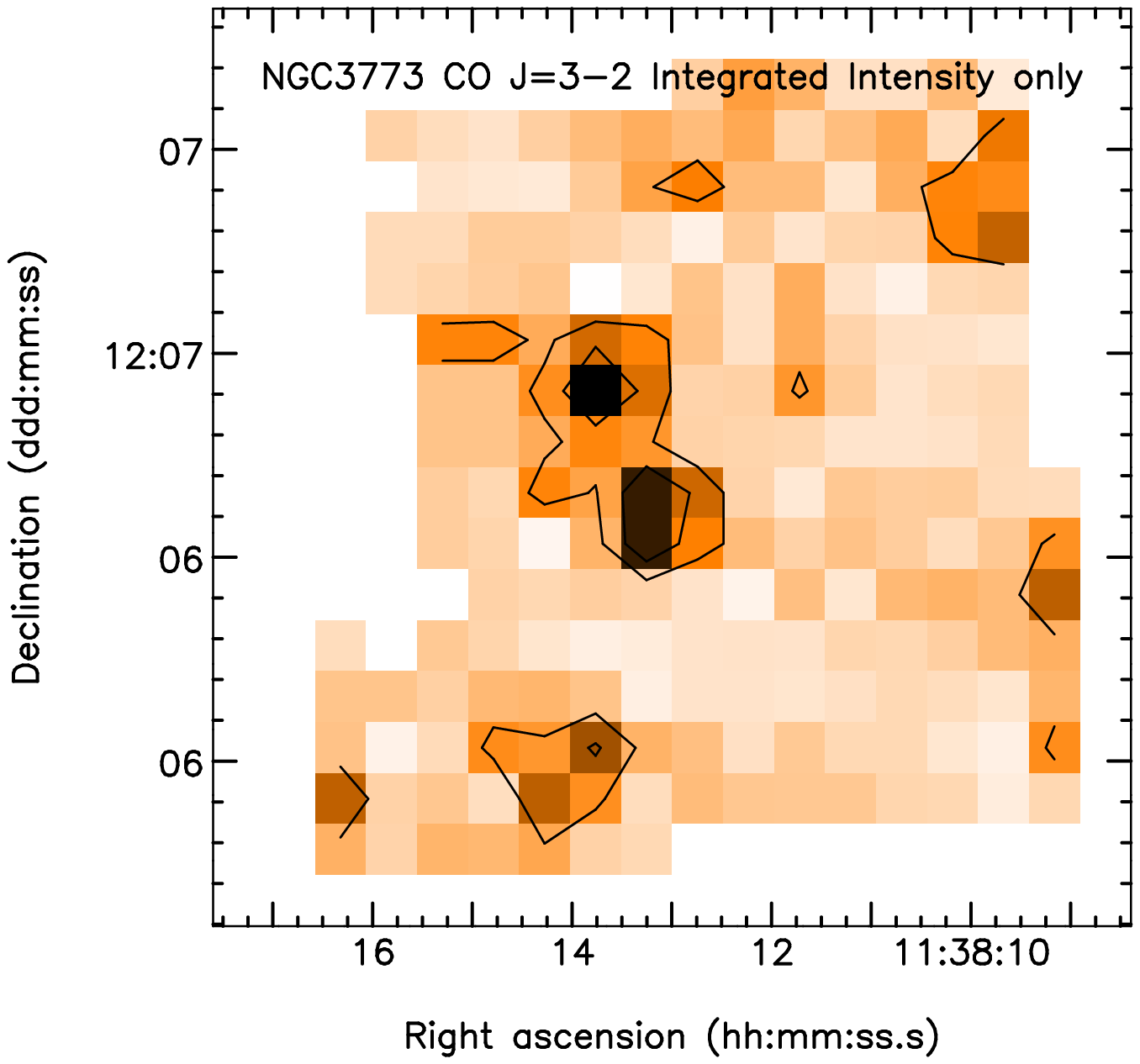}
\includegraphics[width=40mm]{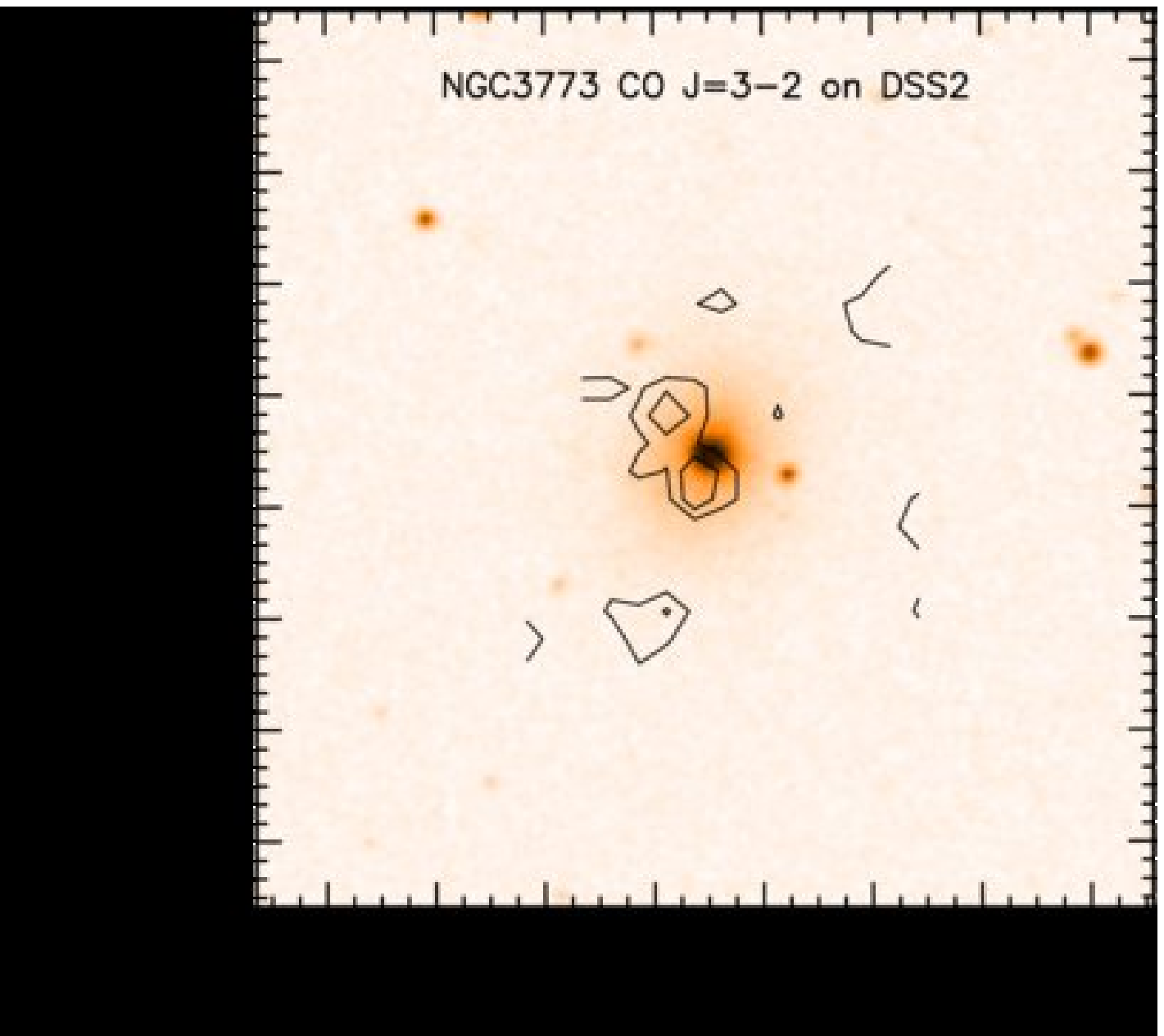}
\includegraphics[width=40mm]{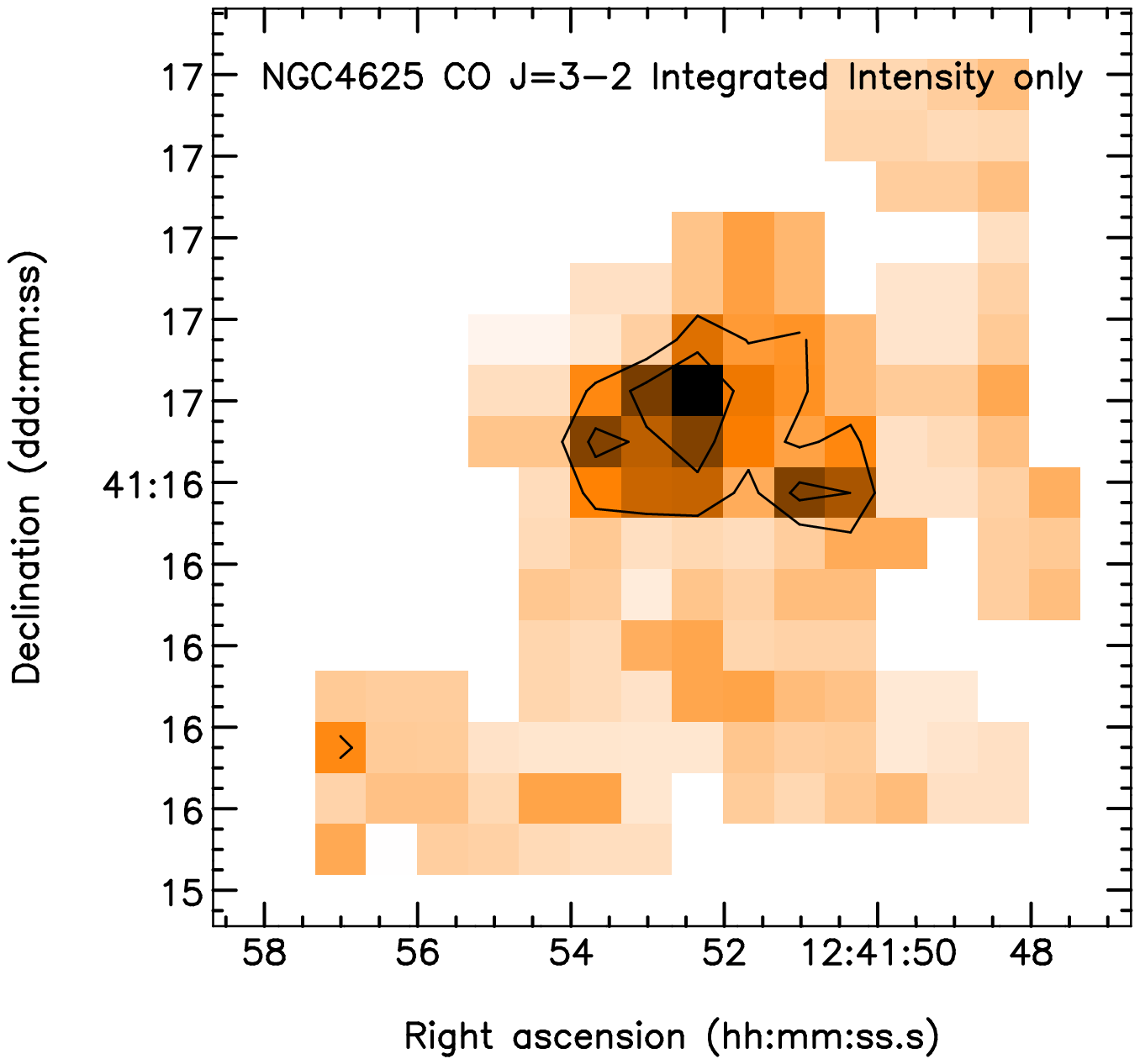}
\includegraphics[width=40mm]{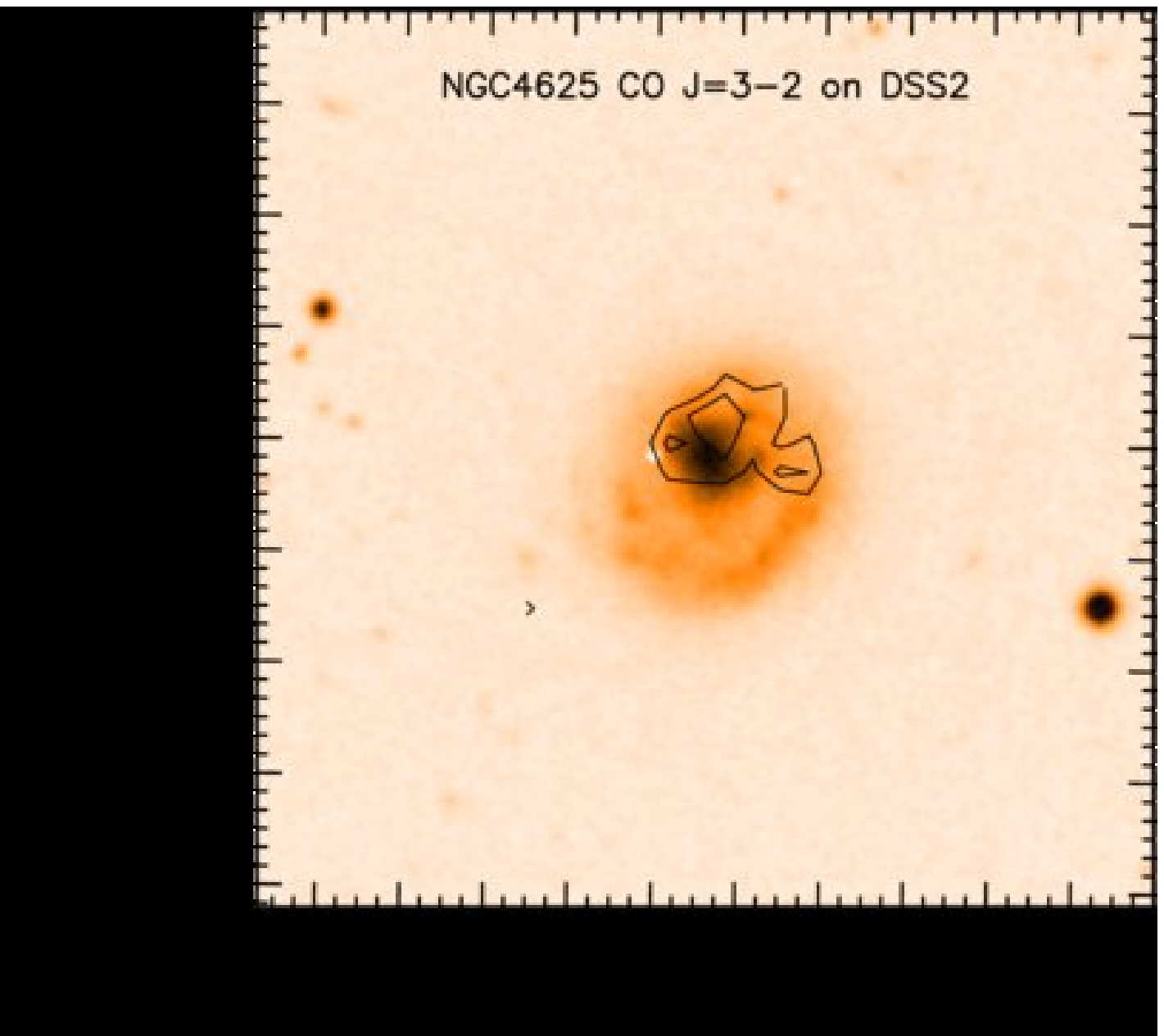}
\includegraphics[width=40mm]{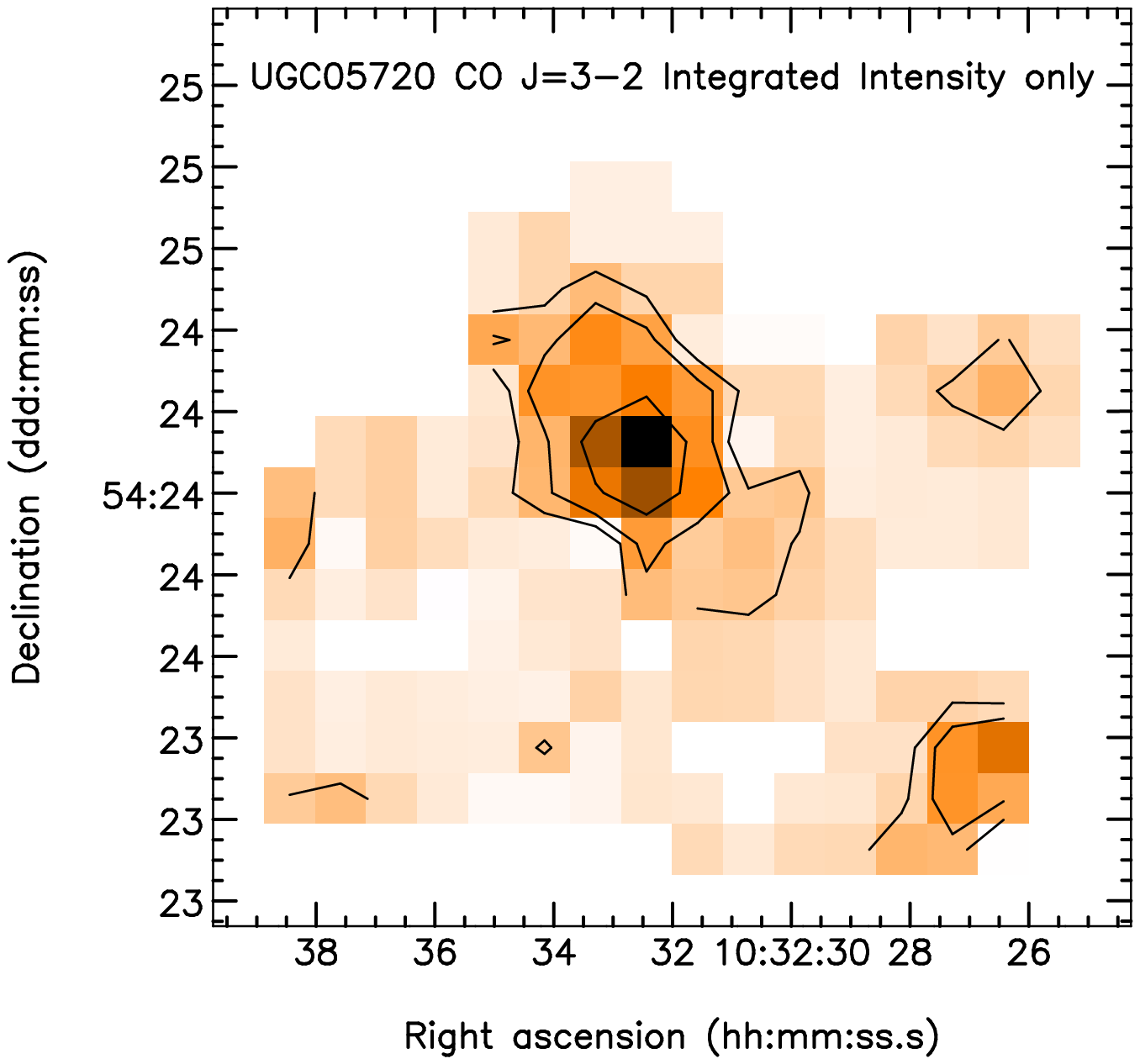}
\includegraphics[width=40mm]{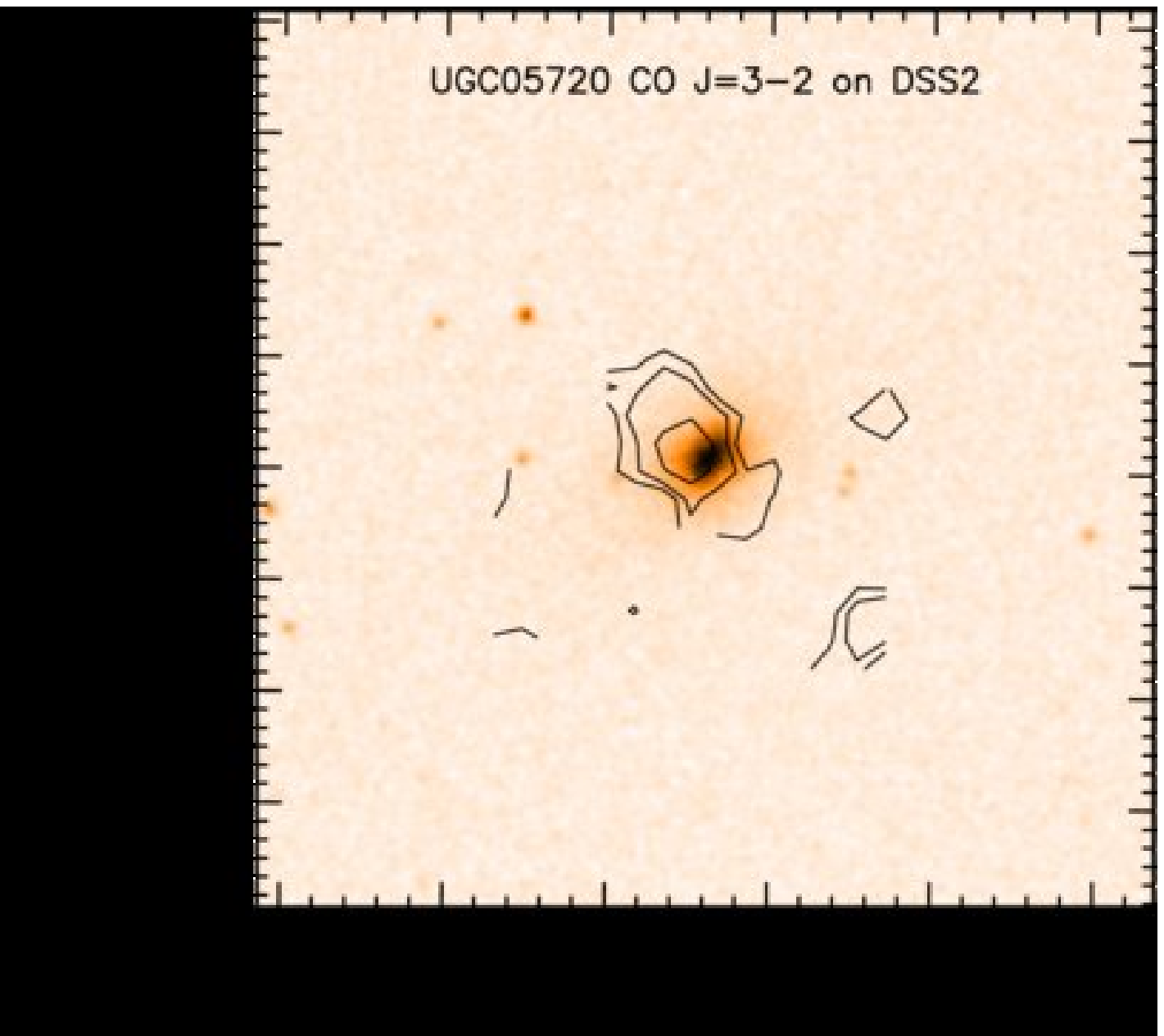}
\caption{ CO $J$=3-2 images for the small SINGS galaxies detected in
  jiggle map mode. See
  Figure~\ref{fig-ngc628} for more details. (left of pair) CO $J$=3-2 integrated intensity
  image. Contours levels are (0.5, 1, 2, 4) K km s$^{-1}$ (T$_{MB}$).
(right of pair) CO $J$=3-2 overlaid on a Digitized Sky Survey image. 
\label{fig-jiggles}}
\end{figure*}


\end{document}